%% file: main.tex
\documentclass[twocolumn,twocolappendix]{aastex631}

\usepackage{xspace}
\usepackage{amssymb}
\usepackage{amsmath}
\usepackage{natbib}

\usepackage{hyperref}

\newcommand{\mathsc}[1]{{\normalfont\textsc{#1}}}
\usepackage{url}
\usepackage{graphicx}
\usepackage{color}
\usepackage{txfonts}

\usepackage{multirow}

\DeclareRobustCommand{\ION}[2]{%
\relax\ifmmode
\ifx\testbx\f@series
{\mathbf{#1\,\mathsc{#2}}}\else
{\mathrm{#1\,\mathsc{#2}}}\fi
\else\textup{#1\,{\mdseries\textsc{#2}}}%
\fi}

\newcommand{\SII}{[\ION{S}{ii}]/H$\alpha$}

\newcommand{\EWa}{EW(H$\alpha$)}

\newcommand{\HII}{\ION{H}{ii}}

\newcommand{\ns}{$n_{\rm s}$}

\newcommand{\ageLW}{\ifmmode \mathcal{A}_{\star,L} \else $\mathcal{A}_{\star,L}$\fi\xspace}
\newcommand{\age}{\ifmmode \mathcal{A}_\star \else $\mathcal{A}_\star$\fi\xspace}
\newcommand{\met}{\ifmmode \mathcal{Z}_\star \else $\mathcal{Z}_\star$\fi\xspace}
\newcommand{\metLW}{\ifmmode \mathcal{Z}_{\star,L} \else $\mathcal{Z}_{\star,L}$\fi\xspace}
\newcommand{\ageMW}{\ifmmode \mathcal{A}_{\star,M} \else $\mathcal{A}_{\star,M}$\fi\xspace}
\newcommand{\metMW}{\ifmmode \mathcal{Z}_{\star,M} \else $\mathcal{Z}_{\star,M}$\fi\xspace}

\newcommand{\flux}{erg/s/cm$^2$}

\newcommand{\funitsI}{10$^{-16}$\,erg/s/\AA/cm$^2$}
\newcommand{\funits}{10$^{-16}$\,erg/s/\AA/cm$^2$}

\def\pyf{\texttt{pyFIT3D}\xspace}
\def\pyp{\texttt{pyPipe3D}\xspace}

\newcommand{\cb}{C\&B\xspace}
\newcommand{\chmk}{\checkmark}
\newcommand{\inft}{$\infty$}
\newcommand{\starlight}{{\sc starlight}}                % STARLIGHT

\shorttitle{pyPipe3D analysis of MaNGA data}
\shortauthors{S\'anchez et al.}
\graphicspath{{./}{figures/}}
\begin{document}

\title[]{SDSS-IV MaNGA: pyPipe3D analysis release for 10,000 galaxies}

\author[0000-0001-6444-9307]{S.F.~S\'anchez}
\author[0000-0003-2405-7258]{J.K.~Barrera-Ballesteros}
\author[0000-0001-7231-7953]{E.~Lacerda}
\author[0000-0002-8931-2398]{A.~Mej\'\i a-Narvaez}
\affiliation{Instituto de Astronom\'ia, Universidad Nacional Aut\'onoma de  M\'exico, A.~P. 70-264, C.P. 04510, M\'exico, D.F., Mexico}
\author[0000-0002-2555-1074]{A.~Camps-Fari\~na}
\affiliation{Instituto de Astronom\'ia, Universidad Nacional Aut\'onoma de  M\'exico, A.~P. 70-264, C.P. 04510, M\'exico, D.F., Mexico}
\affiliation{ Departamento de Astrof{\'i}sica y CC. de la Atm{\'o}sfera, Universidad Complutense de Madrid, E-28040, Madrid, Spain}
\author[0000-0002-6971-5755]{Gustavo Bruzual}
\affiliation{Instituto de Radioastronom\'ia y Astrof\'isica, Universidad Nacional Aut\'onoma de M\'exico, Campus Morelia, C.P. 58089, Michoac\'an, M\'exico}
\author[0000-0002-9658-8886]{C.~Espinosa-Ponce}
\author[0000-0002-0170-5358]{A.~Rodr\'\i guez-Puebla}
\author[0000-0001-8798-2542]{A.~R.~Calette}
\affiliation{Instituto de Astronom\'ia, Universidad Nacional Aut\'onoma de  M\'exico, A.~P. 70-264, C.P. 04510, M\'exico, D.F., Mexico}
\author[0000-0002-9790-6313]{H.~Ibarra-Medel}
\affiliation{University of Illinois Urbana-Champaign, Department of Astronomy, 1002 W Green St, Urbana, Illinois, 61801, United States}
\author[0000-0002-3461-2342]{V.~Avila-Reese}
\author{H.~Hernandez-Toledo}
\affiliation{Instituto de Astronom\'ia, Universidad Nacional Aut\'onoma de  M\'exico, A.~P. 70-264, C.P. 04510, M\'exico, D.F., Mexico}
\author[0000-0002-3131-4374]{M.~A.~Bershady}
\affiliation{Department of Astronomy, University of Wisconsin-Madison, 475N. Charter St., Madison WI 53703, USA}
\affiliation{South African Astronomical Observatory, PO Box 9, Observa- tory 7935, Cape Town, South Africa}
\affiliation{Department of Astronomy, University of Cape Town, Private Bag X3, Rondebosch 7701, South Africa}
\author[0000-0001-9553-8230]{M.~Cano-Diaz}
\author{A.M.~Munguia-Cordova}
\affiliation{Instituto de Astronom\'ia, Universidad Nacional Aut\'onoma de  M\'exico, A.~P. 70-264, C.P. 04510, M\'exico, D.F., Mexico}

\begin{abstract}

  We present here the analysis performed using the \pyp\ pipeline for the final MaNGA dataset included in the SDSS seventeenth data-release. This dataset comprises more than 10,000 individual datacubes, being the integral field spectroscopy galaxy survey with the largest number of galaxies. \pyp\ processes the IFS datacubes to extract spatially-resolved spectroscopic properties of both the stellar population and the ionized-gas emission lines. A brief summary of the properties of the sample and the characteristics of the analyzed data are included. The article provides details on (i) the performed analysis, (ii) a description of the pipeline, (iii) the adopted stellar population library, (iv) the morphological and photometric analysis, (v) the adopted datamodel for the derived spatially resolved properties and (vi) the individual integrated and characteristic galaxy properties included in a final catalog. Comparisons with results from a previous version of the pipeline for earlier data releases and from other tools using this dataset are included. 
  A practical example on how to use of the full dataset, and final catalog illustrates how to handle the delivered product. 
  Our full analysis can be accessed and downloaded { from the webpage \url{http://ifs.astroscu.unam.mx/MaNGA/Pipe3D_v3_1_1/}}.

\end{abstract}


\section{Introduction}
\label{sec:intro}

%$\mathcal{A_{\rm *}}$
%$\mathcal{A_{\rm *L}}$
%$\mathcal{A^*_{L}}$
%$\mathcal{T_*}$ 
%T$_*$
%$T_*$
%$\Tau_*$

Large imaging and spectroscopic galaxy surveys in the nearby universe, covering statistically well defined samples \citep[e.g., Sloan Digital Sky Survey, SDSS, Galaxy and Mass Assembly survey, GAMA,][respectively]{york2000,gamma} have changed our understanding of galaxy evolution in the last few decades \citep[e.g.][]{blanton+2017}. Their development and implementation have imposed substantial challenges, changing how we do science in astronomy. Among other things they have led to the development of automatic or semi-automatic software packages that handle the reduction and analysis of these large datasets, i.e., pipelines. These pipelines have been extremely important for the development of the recent Integral Field Spectroscopic galaxy surveys \citep[IFS-GS, for a recent review see][]{ARAA}. The large number of spectra involved and their unique characteristics (spatial continuity of the data) requires the implementation of new analysis procedures and methods to store and distribute the results of this explorations. We remind the reader that any of the recent IFS-GS data releases (DR) comprises millions of spectra \citep[e.g., SAMI,][]{sami} and, at the same time, they may contain between hundreds \citep[e.g., CALIFA, MaNGA,][]{califa, manga} and tens of thousands of individual spectra per galaxy \citep[e.g., AMUSSING++,][]{carlos20}. 
%This has lead to the developemnt of different reduction and analysis pipelines

%%%%%%%%%%%%%%%%%%%%%%%%%%%%%%%%%%%%%%%%%%%%%%%%%%%%%%%%%%%%%%%%%%%%%%%5
\begin{figure*}
 \minipage{0.99\textwidth}
 \includegraphics[width=8.5cm]{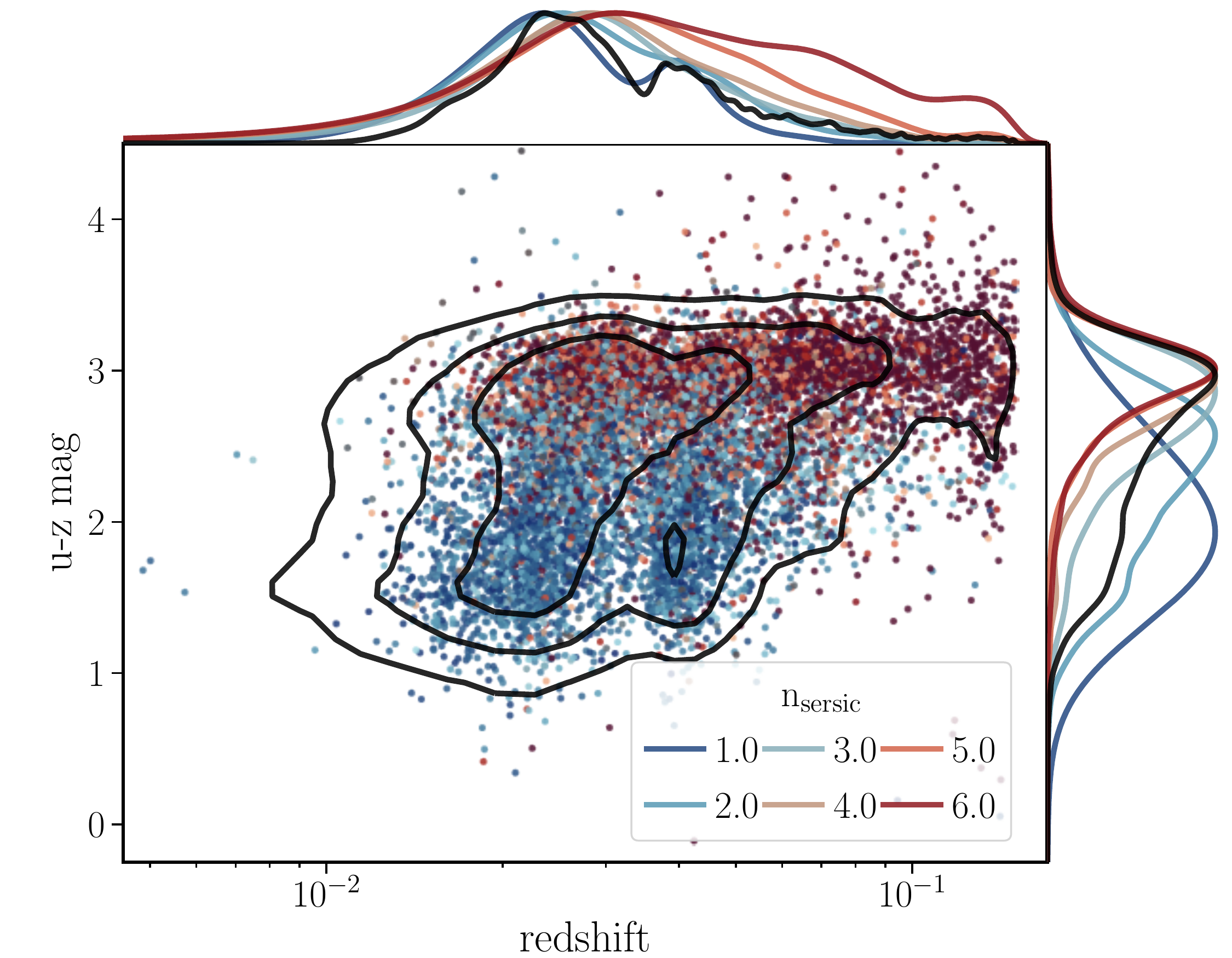}\includegraphics[width=8.5cm]{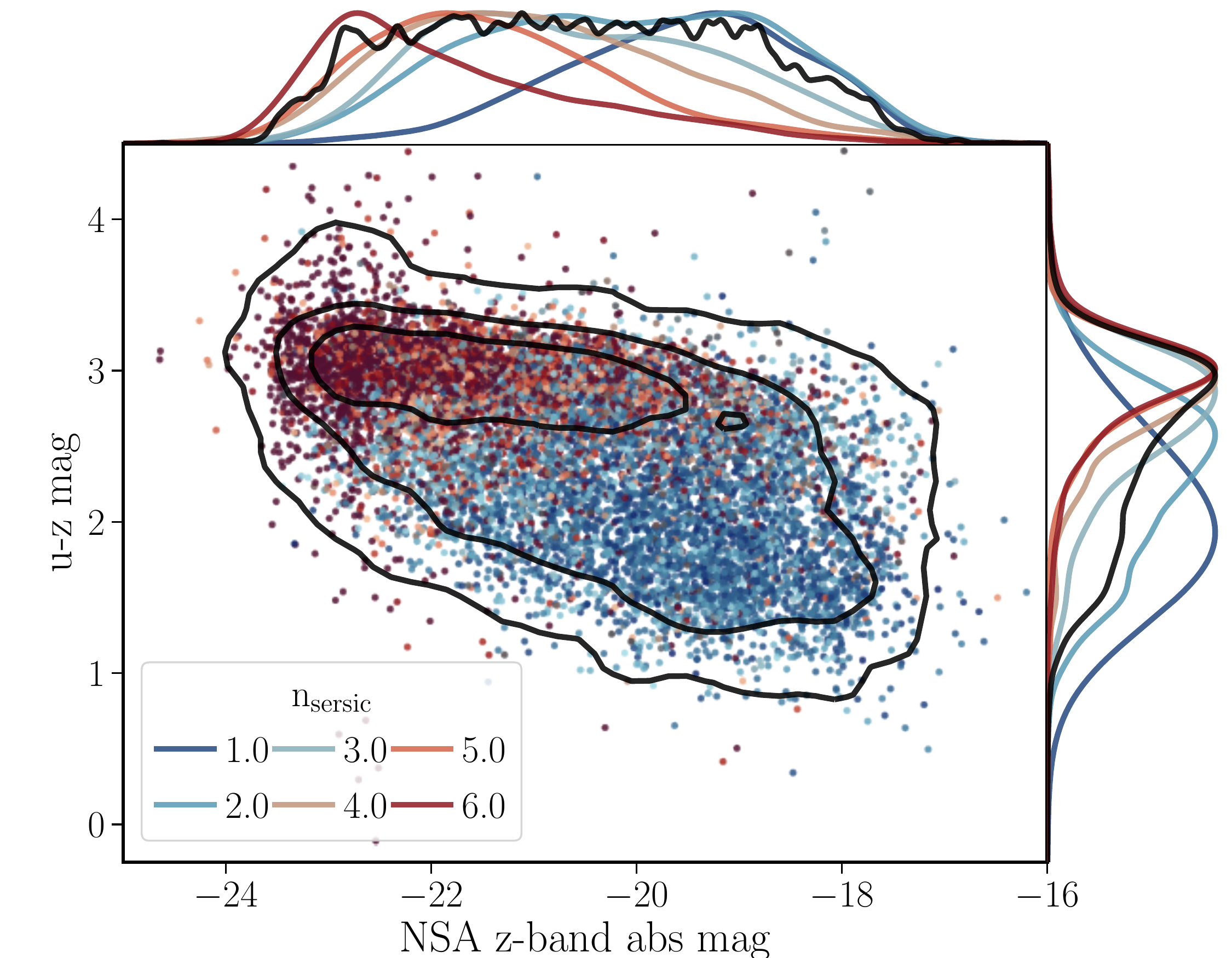}
 \endminipage 
 \caption{Distribution of galaxies in the MaNGA sample in the color-redshift diagram (left panel) and the color-magnitude diagram (right panel). Each solid dot corresponds to one of the analyzed objects, color coded by its 
 Sersic index. Contours represent the density distribution, encircling 90\%, 65\% and 40\% of the objects, respectively. Density distributions of the full sample (black line) and segregated in bins of Sersic index (color lines)
 as a function of redshift, absolute magnitude and color are represented in the upper and left edges of the figure. All parameters were extracted from the NSA catalog.}
 \label{fig:cmd}
\end{figure*}
%%%%%%%%%%%%%%%%%%%%%%%%%%%%%%%%%%%%%%%%%%%%%%%%%%%%%%%%%%%%%%%%%%%%%%%5

Different reduction and analysis pipelines were developed as IFS-GS appeared during the last decade. In some cases they adopted already available tools, like \starlight\ \citep{cid-fernandes05} or {\tt pPXF} \citep{cappellari04}, designed for the analysis of the spectra of stellar populations, in combination with ad-hoc tools tailored to explore the emission line component. This led to IFS analysis pipelines such as  {\tt PyCASSO} \citep{amorim17} and DAP, the MaNGA Data Analysis Pipeline \citep{dap,dap_elines}, developed to analyze specific datasets, CALIFA and MaNGA in this case. In other instances, new packages were developed from scratch, adopting existing algorithms to create multi-purpose pipelines that can handle data from different surveys, e.g., {\tt Pipe3D} \citep{pipe3d_ii}. The
{\tt Pipe3D} pipeline is based on the {\tt FIT3D} package routines and algorithms \citep{pipe3d}, aimed at extracting the properties of both the stellar populations and the ionized gas producing the emission lines from generic IFS data in the visible range. {\tt Pipe3D} has been used to analyze data from different IFS-GS, such as CALIFA \citep{laura16,espi20}, SAMI \citep{sanchez19} and AMUSSING++ \citep{laura18,carlos20}. We analyzed the different MaNGA DRs with {\tt Pipe3D}  \citep[e.g.][]{sanchez18}, resulting on sets of dataproducts for each galaxy (cube), delivered to the community as part of different Value Added Catalogs\footnote{\url{https://www.sdss.org/dr14/manga/manga-data/manga-pipe3d-value-added-catalog/} and \url{https://www.sdss.org/dr15/manga/manga-data/manga-pipe3d-value-added-catalog/}}.

In this paper we present the results of the analysis of the full MaNGA dataset, distributed as part of the SDSS DR17 \citep{DR17}, using an updated version of the {\tt Pipe3D} pipeline \citep[\pyp,][]{pypipe3d}.
The paper is structured as follows: (i) Sections~\ref{sec:sample} and \ref{sec:data} contain a brief description of the sample of galaxies and the main characteristics of the data analyzed in this work. (ii) A detailed description of our analysis and, in particular, the differences with previous implementations of this pipeline, are included in Sec.~\ref{sec:ana}. This section contains a brief description of the pipeline itself (Sec.~\ref{sec:pipe3d}), the adopted spectral library (Sec.~\ref{sec:ssp}), a newly introduced morphological classification (Sec.~\ref{sec:morp}), the photometric and structural galaxy properties extracted from the datacubes (Sec.~\ref{sec:phot}) and, finally, the quality control performed to validate our analysis (Sec.~\ref{sec:qc}). (iii) The results of our analysis are presented in Sec.~\ref{sec:res}, including a description of the {\tt Pipe3D} datamodel adopted to distribute the results (Sec.~\ref{sec:cubes}) and a practical example of its use (Sec.~\ref{sec:ex_cube}). Sec.~\ref{sec:int} contains the description of the integrated and characteristic properties of each galaxy extracted from this analysis included in the final delivered catalog. (iv) A comparison with previous results obtained with this and other tools is included in Sec.~\ref{sec:comp}. (v) Sec.~\ref{sec:agns} includes an example of the use of the final catalog, updating the selection criteria for candidate galaxies to host an AGN presented in \citet{sanchez18}. (vi) Sec.~\ref{sec:summary} summarizes our results. The full dataproducts and final catalog are freely available\footnote{\url{http://ifs.astroscu.unam.mx/MaNGA/Pipe3D_v3_1_1/}}.

%The need for (i) well-defined observing strategies and schedules that handle almost continuous observations for long periods of time, (ii) the development, continuous updates, and execution of reduction algorithms that handles large number of data, and (iii) the development of analysis algorithms that allows to extract scientifically useful parameters and physical properties from these data too (dataproducts), has lead to the development of new working structures and
%also changed the way we do science in astronomy, as they impose a challange
%TBW
%
% 10.02.22
%

\section{Sample}
\label{sec:sample}

We analyze the full dataset provided by the Mapping Nearby Galaxies at Apache Point Observatory (MaNGA) survey \citep{manga}, i.e.,  more than 10,000 galaxies observed in the period between April 2014 and August 2020. MaNGA is one of the three projects included in the 4th version of the Sloan Digital Sky Survey \citep[SDSS-IV,][]{blanton+2017}. The sample of galaxies was selected from the NASA-Sloan Atlas\footnote{\url{http://nsatlas.org/}} (NSA), a catalog of images and parameters of local galaxies ($z\lesssim$0.1), derived from the combination of ultraviolet (GALEX), optical (SDSS) and near-infrared (2MASS) images of galaxies with spectroscopic information provided by the SDSS survey itself. From the NSA, MaNGA selected a representative sample of the population of galaxies in the nearby universe, aiming at, (i) obtaining a flat distribution of stellar mass 
(adopting an absolute magnitude plus color proxy), (ii) sampling the optical extension of galaxies up to a certain desired level, and (iii) obtaining enough galaxies of any morphological type and in any environment to make possible statistically significant comparisons between different sub-samples \citep{manga}. 
The final sample contains three main subsamples: 
(i) a primary sample, comprising $\sim$60\% of the objects, selected so that the field-of-view (FoV) of the adopted integral field units (IFUs) covers at least 1.5 galaxy effective radius (r$_e$); 
(ii) a secondary sample, comprising $\sim$30\% of the objects, such that the FoV
of the IFUs covers at least 2.5 r$_e$; and
(iii) a color-enhanced sample, comprising $\sim$10\% of the galaxies, aiming to over-populate the green-valley between star-forming and retired galaxies, in order to have a statistically large-enough number of galaxies in this regime to allow comparisons with the other two groups.
Finally, a small number of galaxies was cherry-picked to use un-allocated IFUs due to limitations of the coverage and availability of galaxies in the three previous sub-samples at a certain observing period. The latter galaxies correspond to a small percentage of the total sample. Details on the final sample selection are given in \citet{wake17}. The complex sample selection and the final implementation of the observing program imply that the observed sample does not reproduce the expected distribution of galaxies in the nearby universe unless a proper volume correction is applied \citep[e.g.,][]{sanchez18b}.

Figure~\ref{fig:cmd} shows the distribution of $u-z$ color of the galaxies in the final sample as a function of redshift and $z-$band absolute magnitude, color coded by their Sersic index (\ns). All parameters were extracted from the NSA catalog, without any additional processing. As expected from the implicit diameter selection of both primary and secondary sub-samples, there is a trend to select intrinsically brighter galaxies at higher redshifts than at lower ones \citep[e.g.,][]{walcher14}. This is reflected in the prevalence of earlier type, redder (higher \ns) and, in general, more massive galaxies at larger cosmological distances. Two trends with redshift, created by the diameter selection of the sub-samples, are clearly seen in the left panel of the figure. 
The wide redshift range covered by the MaNGA survey in comparison with other IFS galaxy surveys \citep[e.g., PHANGS or CALIFA,][]{roso19,califa} is worth noticing, although it is slightly narrower than the range covered by other surveys like SAMI \citep{sami}. Then, (earlier type/more massive) higher redshift galaxies are not observed at exactly the same cosmic epoch as (later type/less massive) lower redshift ones. 
As stated in previous studies \citep[e.g.,][]{ibarra16,sanchez18b}, this is relevant when exploring galaxy properties that present a clear cosmological evolution (e.g., age, metallicity, star-formation rate, etc). Although this feature is not unique to the MaNGA sample selection, it is particularly relevant in our study because the MaNGA redshift range corresponds to $\sim$1.2 Gyr of cosmic time. For the same reason, galaxies at different redshifts are not observed at the same physical resolution, as will be discussed later on. For some particular studies, it is recommended to perform a thoughtful redshift selection to avoid spurious results { (Barrera-Ballesteros et al., submitted)}.
Another interesting feature is that the galaxy distribution in the color-magnitude diagram of Figure~\ref{fig:cmd} is smoother than usually: the green-valley and the bimodallity between red-sequence and blue-cloud galaxies is less marked. This is a consequence of the addition of the color-enhanced sub-sample. As discussed below, this behavior disappears once we apply the proper volume correction.

\section{Data} 
\label{sec:data}

All galaxies were observed using the MaNGA IFU \citep{drory15}, attached to two double-arm twin spectrographs \citep{smee13} that allows to cover the wavelength range between $\sim$3600-10000\AA, with a resolution of R$\sim$2000. The IFU system comprises 17 fiber bundles of different size (12-32$\arcsec$/diameter) and number of fibers, each of them following an hexagonal pattern. Each set of fiber bundles is connected to a plate chart located at the focal plane of the telescope, which covers the entire FoV of the 2.5m SDSS telescope \citep[$\sim$1$^{\circ}$,][]{gunn06}, following a configuration that optimizes the number of observed galaxies at each visited location.
In this way each plate defines a set of observed objects and, therefore, each observed cube is unambiguously defined by the combination of the {\tt plate} number, the number of fibers in the bundle (19,37,61,91 or 127) and an index defining the bundle number (01,02...). The two latter numbers are combined into a single index named {\tt ifudsgn} (e.g., 12701). In summary, each cube is identified by the combination {\tt manga-plate-ifudsgn} (e.g., manga-7443-12701), that will be used as the primary index in this article. 

In order to cover the gaps between adjacent fibers in each bundle, observations were performed following a minimum of three dithering exposures with a fixed exposure time, a procedure adopted in previous IFS surveys \citep[][]{califa}. Since the survey strategy required to reach a certain minimum depth for all targets \citep{renbin16b}, the same field/plate could be re-visited several times along the survey to achieve this goal. Individual visits were then combined and the final reduced dataset comprises just a single frame for each plate and IFU, irrespective of the number of times that it was (re)-observed. In addition to the science IFUs, each plate includes a set of 12 micro-IFUs, comprising 7-fibers that point towards field stars, and a total of 15 sky fibers. These additional observations are used in the flux calibration and sky subtraction procedures, described in detail in \citet{renbin16} and \citet{law15}, respectively.

Data reduction was performed using version 3.1.1 of the MaNGA Data Reduction Package \citep[DRP,][]{law16}. This package comprises the usual steps in the reduction of fiber-feed IFS data \citep{sanchez06a}, including: (1) tracing the location of the spectrum corresponding to each individual fiber in the CCD; (2) extraction of these spectra; (3) wavelength calibration; (4) homogenization of the spectrophotometric transmission of each individual fiber (fiberflat); (5) sky subtraction; (6) combination of several dithering exposures, re-observations of the same plate, and re-sampling of the data into a regular grid datacube (a step that requires correcting for differential atmospheric refraction when needed); and (7) flux calibration. The result of the data reduction for each combination of {\tt plate} and {\tt ifudsgn} is a single datacube, in which the X and Y axis correspond to the spatial coordinates (RA and DEC) and the Z axis to the spectral information. The flux intensity at each spatial location and wavelength is stored in each (X,Y,Z) entry in this 3D array or cube. In this way, each channel in the Z axis corresponds to a monocromatic image, and each pixel at location (X,Y) comprises a single spectrum (for this reason pixels in IFU datacubes are known as spaxels, i.e., spectral-pixels). The current version of the DRP provides three different versions of the reduced data, a row-stacked spectra (plus a position table), comprising the individual spectra before resampling to a final datacube, and two versions of the datacubes, one with a logarithmic sampling of the spectral range (i.e., $\Delta z={\Delta \lambda}/{\lambda}$ and another with a linear sampling ($\Delta z=\Delta \lambda$). Along this paper we will use the linear sampling version of the datacube. Additional information provided by the DRP includes (i) an error cube, comprising the propagated error associated with each spectrum at each location in the (X,Y) plane; (ii) a mask of the bad pixels (cosmic rays, CCD problems, etc); (iii) a cube comprising the spectral resolution at each location and wavelength; and (iv) additional information, such as reconstructed broad-band images at different bands, or the Galactic dust extinction at the location of each target. The delivered products of the DRP are stored in a single fits file, with a different extension containing each product, and a header that includes relevant information regarding the observing procedure (atmospheric conditions, number of visits, etc).

Prior to any analysis, we perform a pre-processing of the linear wavelength sampling datacubes provided by the MaNGA DRP. %(i.e., those which wavelength solution consist of a linear fixed step per spectral pixel in the $z$ axis of the cubes). 
This procedure transforms the original data to the format needed by our analysis pipeline, a fits file with at least three extensions: (1) one datacube comprising the spectral information (flux intensity), corrected by Galactic extinction, in units of \funits, the wavelength calibrated in a linear step with a normalized spectral resolution in FWHM; (2) one datacube comprising the error in the flux intensity in the same format and units as the first extension; and (3) one datacube comprising the mask of bad pixels, in the same format as the two previous extensions. The Galactic extinction correction was performed using the dust extinction in the $V$-band (A$_{\rm V}$) estimated from the E(B-V) parameter included in the header of the original MaNGA datacube  \citep[{\tt EBVGAL} keyword, extracted from ][]{schlegel98}, adopting a canonical value of R$_{\rm V}$=3.1 for the Milky Way, and the \citet{cardelli89} extinction law. The wavelength resolution was normalized to a value of FWHM$=$3.7\AA\ by convolving each spectrum at each wavelength with a Gaussian function comprising the differential resolution at the wavelengths where the original resolution is better than this value. This imposes a small degradation of the spectral resolution according to Fig. 18 of \citet{law16}, at the advantage of non requiring to normalize the resolution in each step of the spectral fitting procedure (as we discuss later on).

%\Com{MaNGA description of MPL-11}
%\section{Data}
%\label{sec:data}

\section{Analysis}
\label{sec:ana}

The final MaNGA v3.1.1 data-set comprises 11,273 unique datacubes. Of them, 10,245 correspond to unique datacubes of objects with redshift in the NSA-catalog (i.e., galaxies). As indicated before, the same galaxy could be observed using a different combination of plate and IFU, which would correspond to a different datacube. As listed in Appendix \ref{app:rep}, a total of 44 galaxies were observed twice (corresponding to 88 datacubes). In this section we describe the analysis performed to this full dataset.

\subsection{Summary of Pipe3D}
\label{sec:pipe3d}

The analysis is performed using a new implementation of the Pipe3D pipeline \citep{pipe3d_ii}, fully transcribed to python \citep[pyPipe3D,][]{pypipe3d}. This version of the code uses similar algorithms and the same analysis sequence as the previous version, adapted to make use of the unique computational capabilities of the new adopted coding language, improving its performance, and correcting bugs when needed. Pipe3D has been extensively used to analyze IFS data of very different nature, in particular data from the CALIFA \citep[e.g.][]{mariana16,espi20}, MaNGA \citep[e.g.][]{ibarra16,jkbb18,sanchez18b,bluck19,laura19} and SAMI surveys \citep[e.g.][]{sanchez19}, and individual \citep{censushii} and large MUSE datasets \citep{laura18,carlos20}. It has been described in detail in different previous articles \citep[e.g.][]{pipe3d,pipe3d_ii,sanchez20}, and thoroughly tested using both ad-hoc simulations and mock datasets based on hydrodynamical simulations \citep{guidi18,ibarra19}. To avoid unnecessary repetition, we provide here a very brief summary, emphasizing the few novelties of the new code. 

In summary, the code separates the stellar component and the ionized-gas line-emission in each spectrum by fitting the former with a combination of simple stellar population (SSP) spectra. Prior to this decomposition, the SSP spectra are shifted to account for the stellar velocity ($vel_\star$), and convolved with a Gaussian function to account for the velocity dispersion ($\sigma_\star$). Finally, the SSPs are dust-attenuated adopting the \citet{cardelli89} extinction law. The parameters defining the kinematics ($vel_\star$ and $\sigma_\star$) and the intrinsic dust extinction (A$_{\rm V,\star}$) are estimated using a limited set of SSP templates that restricts the space of parameters to avoid or limit the intrinsic degeneracies between these parameters and the intrinsic properties of stellar populations \citep[e.g., metallicity/dispersion][]{patri11}. This two step procedure and its benefits is described in detail in \citep{pipe3d_ii} and \citet{pypipe3d}. The stellar decomposition provides the light-fraction or weight ($w_{\star,L}$) contributed by each SSP to the problem spectrum at a fixed spectral range (5450-5550\AA, similar to the $V-$band central wavelength), in addition to the $vel_\star$, $\sigma_\star$ and A$_{\rm V,\star}$ parameters obtained in the first step. Each observed spectrum ($S_{obs}$) is fitted to a model spectrum of the stellar component ($S_{mod}$) following the expression
\begin{equation}\label{eq:dec}
  \begin{aligned}
S_{obs}(\lambda) \approx\ & S_{mod}(\lambda) = \\ &\left[ \Sigma_{ssp} w_{ssp,\star,L} S_{ssp}(\lambda) \right]
10^{-0.4\ A_{\rm V,\star}\ E(\lambda)} \ast G(vel_\star,\sigma_\star),
  \end{aligned}
\end{equation}
where $S_{ssp}$ is the spectrum of each SSP in the template library, $E(\lambda)$ the adopted extinction law, and $G(vel_\star,\sigma_\star)$ the Gaussian function describing the line-of-sight stellar velocity distribution. %, and the remaining parameters have been described before.

After subtracting the best stellar population model from the observed spectrum, the fitting algorithm models the ionized gas emission with a set of individual Gaussian functions that are fitted to a set of pre-defined emission lines at known wavelengths, shifted by the observed velocity of the gas component. 
In this way we derive the flux intensity ($F_{el}$), velocity ($vel_{el}$) and velocity dispersion ($\sigma_{el}$) for each emission line ($el$). The procedure is repeated iteratively, with the emission line analysis performed after each step in which the stellar population spectrum is modeled and subtracted from the original spectrum. The figure of merit used to define the best fitting model (a reduced $\chi^2$) takes into account both the stellar and emission line models provided by the full analysis just described. %The fitting procedure is explained in detail in \citet{pypipe3d}. 
In summary, our procedure involves a combination of brute-force exploration of the non-linear parameters within a limited range ($vel_\star$, $\sigma_\star$ and A$_{\rm V,\star}$ for the stellar population, and $vel$ and $\sigma$ for each emission line), together with a pure linear inversion to derive the weights ($w_{ssp}$) of the decomposition of the stellar population, plus an iterative automatic selection of the SSP templates to be included in the decomposition based on the results of the previous fit.
%Monte Carlo (MC) iteration to derive the errors of the different parameters.

The procedure just outlined is then applied to each datacube. However, it is not applied spaxel by spaxel. Results from simulations indicate that there is a minimum signal-to-noise ratio (SN) required for the fit to the stellar population to provide reliable results \citep{pipe3d}. We thus perform a spatial binning, known as Continuum-Segmentation { \citep[$CS$, described in][]{pipe3d_ii}}, that groups adjacent spaxels whose SN is below a defined threshold to produce a higher SN average spectrum. In this procedure, contrary to other ones found in the literature \citep[e.g.][]{capp03}, we do not group the spectra of adjacent spaxels if their intensity differs by more than a pre-defined percentage. In this way, the spatial shape of the original galaxy is preserved (to some extent), while the SN is increased (although not always reaching our goal value). 
From this analysis we recover the parameters of the stellar population for each spatial bin (tessella or voxel), i.e., the weights of the decomposition of the SSPs (see Sec. \ref{sec:sfh_cube} below), the kinematic parameters and the dust extinction, together with the best spectral model. The model spectrum for each tesella is then scaled to the flux intensity of each spaxel by using the so-called deszonification parameter \citep[DZ,][]{cid-fernandes13}, which is the ratio between the flux intensity of each spaxel with respect to the average value in the tesella in which this spaxel was grouped. Restricting this procedure to model the stellar spectrum, we obtain a model cube of the stellar population. Subtracting this model from the original cube, we obtain the so-called pure-GAS cube, a cube comprising just the information of the ionized-gas emission-lines (plus noise and residuals from the imperfect subtraction of the stellar component).

The pure-GAS cube is then analyzed using two different procedures. In the first one, a limited set of strong emission lines is fitted with individual Gaussian functions for each individual spaxel, providing a set of maps with the flux velocity and velocity dispersion of each fitted emission line, and their corresponding errors (Sec. \ref{sec:elines_cube}). In a second step, a much larger set of emission lines is analyzed using weighted-moment analysis, deriving maps of the three parameters indicated before ($F_{el}$, $vel_{el}$ and $\sigma_{el}$), together with the spatial distribution of the equivalent width ($EW_{el}$) of the considered emission line, $el$ (Sec.~\ref{sec:flux_elines_cube}). 
Two main differences have been introduced in the analysis of the emission lines with respect to previous analysis of MaNGA data using Pipe3D \citep{sanchez18}: (1) The Gaussian fitting is now performed in two steps. The first step replicates the brute-force exploration of the range of parameters described in \citet{pipe3d}. This exploration avoids to fall into a local minimum. However, it is not very accurate. A second step is introduced by executing a Levenberg-Marquardt minimization algorithm that uses as a guess the results of the first exploration. This additional fitting increases significantly the accuracy and precision of the results \citep{pypipe3d}. (2) A new and larger set of emission lines, with an improved definition of wavelength, has been adopted for the weighted-momentum analysis, as will be explained below. The former version of the emission line treatment has been performed as well, for a simpler comparison with previous results.

%
% indices
%
Finally, the pipeline obtains the spatial distribution (map) of a set of stellar spectral indices (see \ref{sec:indices_cube}). To do so, we subtract the best model for the emission lines from the spectrum in each tessella, in order to generate a stellar spectrum without the contamination by the ionized gas contribution. Then, for each stellar spectral index, we derive its equivalent width by defining a wavelength range at which we measure the median flux intensity, and two adjacent wavelength ranges at which the continuum is estimated. We adopt the definition of the different stellar indices and the procedures described in \citet{Cardiel:2003p3435}. We note that we depart from the classical definition of D4000, derived using the flux intensity in units of frequency \citep[i.e., $F_\nu$][]{bruzual83}, and adopt the more convenient functional form proposed by \citet{gorgas99} (their Eq. 2). Like in the previous cases, the pipeline provides with maps of the spectral index and its estimated error.

%%%%%%%%%%%%%%%%%%%%%%%%%%%%%%%%%%%%%%%%%%%%%%%%%%%%%%%%%%%%%%%%%%%%%%%5
\begin{figure*}
%\captionsetup[subfigure]{labelformat=empty}
 \minipage{0.99\textwidth}
\includegraphics[clip,trim=10 0 40 10,width=17.5cm]{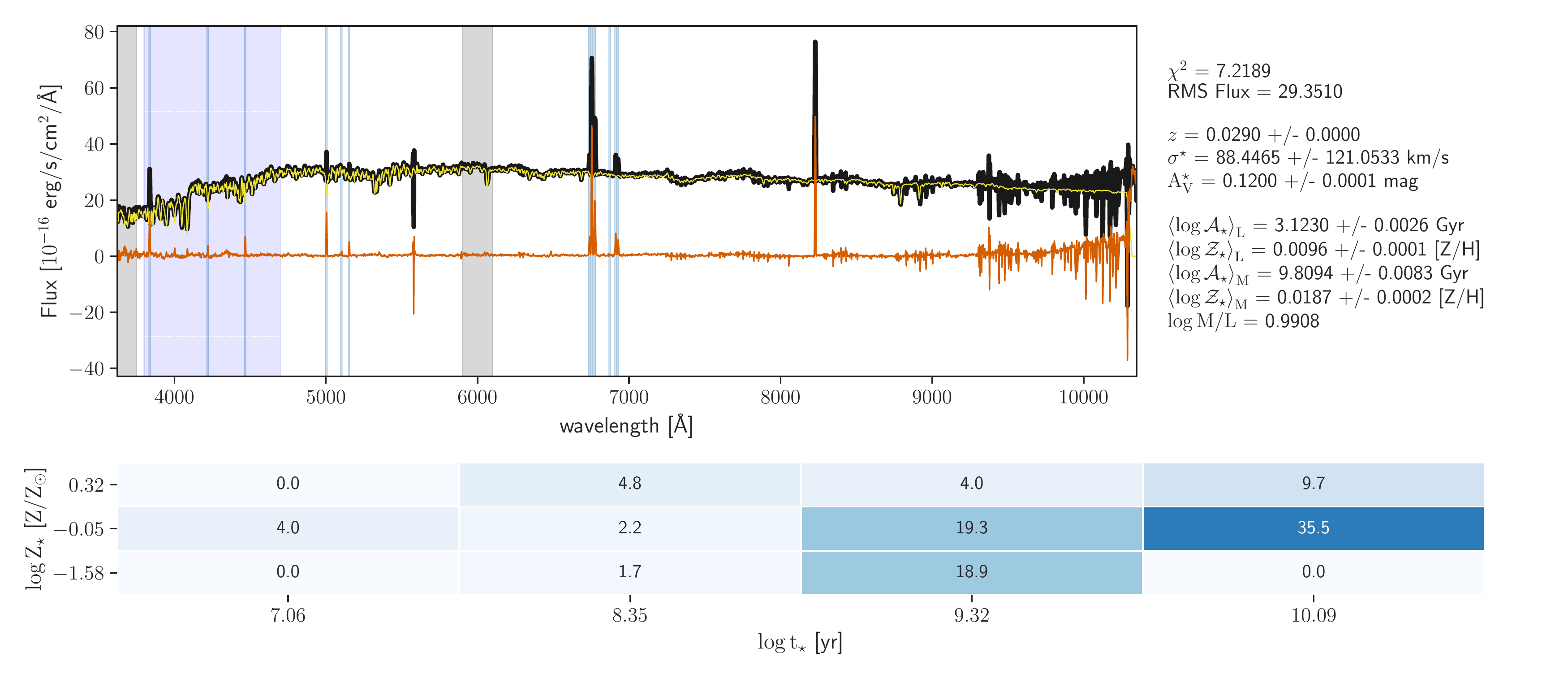}
 \endminipage
 \caption{Example of the information explored in the quality control process for the galaxy/cube manga-7495-12703. Top panel shows the central 5$\arcsec$/aperture spectrum of this galaxy (black-solid line), together with the best fitted model (yellow-solid line) using the simple SSP-template adopted in the derivation of the non-linear parameters of the stellar population \citep[Sec \ref{sec:pipe3d}][]{pypipe3d}, and the residual of the subtraction of this model to the original spectrum. Grey and blue shaded-areas correspond to masked regions along the full fitting process and in the first iteration of the analysis before subtracting the emission lines, respectively. The coefficients of the decomposition in this simple SSP library, i.e., the fraction of light at the normalization wavelength for each template in the library ($w_{ssp}$ in Eq. \ref{eq:dec}) are shown in the bottom panel as a heatmap running for the four ages and three metallicities of the considered library. The non-linear parameters (systemic velocity, velocity dispersion and dust extinction) and the luminosity- and mass-weighted age and metallicity are indicated in the top-right legend.}
 \label{fig:qc_spec}
\end{figure*}
%%%%%%%%%%%%%%%%%%%%%%%%%%%%%%%%%%%%%%%%%%%%%%%%%%%%%%%%%%%%%%%%%%%%%%%

The errors provided by Pipe3D are based on a set of Monte Carlo (MC) iterations in which the original spectrum is perturbed within the errors (a needed input in the pre-processed datacubes, see Sec.~\ref{sec:data}). 
Every time a stellar or an emission line model is subtracted from the original spectrum, the uncertainties in the model are also propagated into the errors. The final errors thus include both the noise level and the uncertainties in the modelling of the galaxy component. More details on this procedure are given in \citet{pypipe3d}. %Lacerda et al. in prep.

\subsection{Adopted Stellar library}
\label{sec:ssp}

One of the major changes with respect to previous versions of the delivered dataproducts, besides the use of an improved and transcribed version of the code, is the use of a new SSP spectral library in our analysis. 
For the SDSS DR14 and DR15 we delivered versions v2\_1\_2 and v2\_4\_3 of the MaNGA dataproducts, produced adopting the {\tt GSD156} SSP library \citep{cid-fernandes13}. This library results from the combination of two sets of SSP model spectra. 
{ For stellar populations older than 65 Myr, the {\tt GSD156} library uses the synthetic spectra from \citet{vazdekis10} and \citet{falcon-barroso:2011}, based on the MILES empirical stellar library \citep{sanchez-blazquez:2006}.
For stellar populations younger than 65 Myr (not included in the cited models), {\tt GSD156} uses the synthetic spectra from the GRANADA library \citep{martins:2005aa,gonzalezdelgado05}, based on theoretical stellar spectra.
We adopted the \cite{Salpeter:1955p3438} Initial Mass Function (IMF) for stellar masses between 0.1 and 100 M$_\odot$.
{\tt GSD156} comprises 156 SSP spectra, sampling 39 ages from 1 Myr to 14 Gyr (on a near logarithmic scale) and 4 metallicities (Z/Z$_\odot$~=~0.2, 0.4, 1, and 1.5).}

For the current implementation of the code we adopt a completely different library, that we call {\tt MaStar\_sLOG}. { This SSP library uses the recently delivered MaNGA stellar library \citep[MaStar,][]{yan19}, which includes 8646 spectra for 3321 unique stars.}
%, that in its last distribution comprises $\sim$24K good quality spectra \citep{mejia21}. 
This is a considerable increase in the number of stars and the range of sampled atmospheric properties over the previously adopted stellar library (MILES comprises a total of $\sim$1000 stars).
{ In Appendix~\ref{app:ssp} we present a summary of the updated set of the \citet[][hereafter BC03]{bruzual:2003} stellar population synthesis models that use several stellar libraries,
including MaStar. The full MaStar SSP library comprises a total of 3520 individual spectra, covering 220 ages and 16 metallicities} (assuming a solar [$\alpha$/Fe] ratio{, and a solar metallicity of [Fe/H]=0.017}). This library is by far too impractical to be applied as such in a stellar decomposition technique like the one performed by \pyf, outlined in Sec. \ref{sec:pipe3d}. Thus, we experiment with different combinations of SSPs extracted from this full library, adopting linear, logarithmic, and mixed samplings of both the age and the metallicity, comparing the results with previous ones and with theoretical expectations. In all cases, we repeat the full analysis described in this article for a subset of the full MaNGA dataset, comprising 9,500 datacubes (the so-called MPL-10 internal release). A detailed description of the experiments and their results will be presented elsewhere (Sánchez in prep.). In summary, we concluded that a sampling in age and metallicity in which the step between consecutive values increases multiplicative (i.e., a pseudo-logarithmic sampling) results in a good compromise between (i) an adequate sampling of the spectral properties, (ii) an efficient exploration (in terms of computing time), and (iii) the estimation of accurate and precise results. We must recall that based on previous experiments \citep{pipe3d}, an arbitrary increase of the sampling of the stellar parameters by an SSP-library is not feasible for data of limited signal-to-noise ratio. { The finally adopted {\tt MaStar\_sLOG} library was built following this scheme and comprises 273 SSP spectra, sampling 39 ages from 1 Myr to 13.5 Gyr and 7 metallicities ($Z$~=~0.0001, 0.0005, 0.002, 0.008, 0.017, 0.03 and 0.04) or
(Z/Z$_\odot$~=~0.006, 0.029, 0.118, 0.471, 1, 1.764, 2.353), as indicated in Tables~\ref{tab:sfh} and \ref{tab:parsec}.}

%\Com{SFS: What about a figure with the SSPs and a heatmap with the M/L to illustrate the coverage in age and metallicity?}

%
% Different explored
%

%\subsection{Photometric and structural properties}
%\label{sec:phot}
%In addition to the properties derived 

\subsection{Morphological Classification}
\label{sec:morp}

%{ C\&P, TBW}
Morphology is known to be a fundamental property that affects (and is affected by) galaxy evolution, being directly connected with the dynamical stage, the stellar content, the star-formation stage, the gas content, and even the presence of an active galactic nuclei. Therefore, it is essential to have a morphological classification of the galaxies as a basis to compare with the properties derived by our analysis.
%In order to study how morphology affects the properties of the galaxies we need a catalog for the morphology of the galaxies in our sample. 
Catalogs available publicly, based on a visual classification of galaxy morphology, only include galaxies in releases up to DR15, which comprise $\sim$4700 objects, less than half the galaxies available in the final sample.
For the purposes of this article, we do not need a highly accurate determination of the morphology for each individual galaxy. It is enough to have a consistent classification in terms of statistical properties.
As such, we use the accurate visual morphology determinations from the SDSS Value Added Catalog (VAC) by \citet{vazquez22}, which includes $\sim 6000$ galaxies and is an extension of publicly available SDSS VAC\footnote{\url{https://data.sdss.org/sas/dr16/manga/morphology/manga_visual_morpho/1.0.1/}}, to obtain training and testing samples for a machine learning algorithm to classify the rest of the galaxies in the full sample. We selected features that are expected to be informative of the morphological class, namely: the Sèrsic index, the NSA stellar mass, the line-of-sight velocity to velocity dispersion ratio at the effective radius, the ellipticity, the concentration index and the $u'g'r'i'z'$ colors (extracted from the NSA catalog). For such a task, we implemented a Gradient Tree Boosting (GTB) algorithm, which is a type of ensemble method that successively trains a predefined number of decision trees, each improving upon its predecessor errors.
% clustering algorithm capable of giving an estimate of the morphology of all galaxies in our sample.
We found the resulting classification to be satisfactory for our purposes: the deviation of the estimated morphology and that of the VAC for the testing sample is consistent with the deviation observed between this VAC and other morphological classifications (discussed below). We will detail this method in a forthcoming publication (Mej\'ia-Narv\'aez et al., in prep.).

%%%%%%%%%%%%%%%%%%%%%%%%%%%%%%%%%%%%%%%%%%%%%%%%%%%%%%%%%%%%%%%%%%%%%%%5
%\begin{figure}
% \minipage{0.99\textwidth}
% \includegraphics[width=8.5cm]{figures/manga-7495-12704.map.png}%\includegraphics[width=5cm]{figures/manga-74%95-12704.Mass_t.png}\includegraphics[width=5cm]{figures/manga-7495-12704.ZH_t.png}
% \endminipage
% \caption{TBW}
% \label{fig:map}
%\end{figure}
%%%%%%%%%%%%%%%%%%%%%%%%%%%%%%%%%%%%%%%%%%%%%%%%%%%%%%%%%%%%%%%%%%%%%%%

% --------------- QC Entries ------------------------%
\begin{table*}
\begin{center}
\caption{Quality Control Flags.}
\begin{tabular}{cll}\hline\hline
QCFLAG &  Level & Meaning\\
\hline
  0 & OK            & All QC steps passed           \\
  1 & BAD           & Wrong redshift\\
  2 & BAD           & Low signal-to-noise or empty field \\
  3 & WARNING       & Possible issue with the fitting and/or presence of a strong AGN \\
%   &               & and/or presence of a strong AGN \\
  4 & WARNING       & Mass based on \pyp\ does not match NSA reported value \\
%   &               & NSA reported value \\
  5 & WARNING       & Redshift based on \pyp\ does not match NSA reported value \\
%   &               & NSA reported value \\
  6 & WARNING       & Bright foreground field star \\
  7 & WARNING       & Evident Merging system \\
\hline
\end{tabular}\label{tab:qc} 
\end{center}
\end{table*}
% --------------- QC Entries ------------------------%

\subsection{Photometric and structural properties}
\label{sec:phot}

We estimate a set of photometric and structural properties directly {\it from the MaNGA datacubes} in addition to the different parameters derived by \pyp. Among them, we compute the broad-band photometry in the Gunn $u$, $g$, $r$ and $i$ and the Johnson $B$, $V$ and $R$ filters, adopting the Vega photometric system
and using the filter parameters provided by \citet{fukugita95}, redshifted to the rest-frame of each object. Thus, no additional corrections, such as the $K$ or $E$ corrections have to be considered. From this photometry we derive the galaxy observed and absolute magnitudes, using again the redshift to estimate its cosmological distance\footnote{A standard $\Lambda$CMD cosmology with
$H_O$=73 km/s, $\Omega_M$=0.3 and $\Omega_\Lambda$=0.7 is adopted throughout this work.}. In addition, we estimate the radii R50 and R90 which include, respectively, 50\% and 90\% of the integrated flux in the $V$-band inside the MaNGA FoV, and the concentration index R90/R50. In the literature there are several estimates of similar properties derived for the MaNGA galaxies using direct imaging, like the ones provided by the SDSS \citep{blanton+2017,fisher19} or the DESI survey \citep{arora21}. However, they present some disadvantages and obvious intrinsic differences: (i) none of them can derive a direct rest-frame estimate of the photometry, requiring a $K$-correction based on modelling the spectral energy distribution to obtain them; (ii) the same limitation affects the structural properties, such as R50 or the concentration index; (iii) these photometric values are usually derived for the full (optical) extension of the galaxies, not limited to the FoV of the IFS data. This is very useful to derive the global integrated galaxy properties (e.g., stellar mass or absolute magnitude). However, they are not appropriate for comparison with aperture limited quantities, as derived by the \pyp\ analysis.% (for instance, if colors wants to be compared with $\Upsilon_\star$).

From our photometry, we estimate each galaxy stellar mass from its $M/L$-ratio assuming the relation
\begin{equation}\label{eq:Mphot}
{\rm log}(M_{\star,phot}/M_\odot) = -0.95+1.58(B-V)+0.43*(4.82-V_{\rm abs})
\end{equation}
by \citet{bell00}, valid for the \citet{chab03} IMF, using ($\Upsilon_{\rm B-V}$) as the $B-V$ color and our $V$-band absolute magnitude.
These stellar mass estimates are considered to be less accurate but most probably more precise than estimates based on the spectroscopic analysis performed by \pyp. 
We acknowledge that there are more recent $\Upsilon_{\rm color}$ derivations \citep[e.g.][]{zibetti09,taylor:2011,zhang:2017,rgb18}, some of them based on IFS data, and more precise estimates of the stellar mass that take into account multi-band photometry \citep[e.g.][]{arora21}. However, we adopt Eq.~(\ref{eq:Mphot}) for easy comparison with previous calculations \citep{pipe3d_ii}. Nevertheless, since we provide colors and absolute magnitudes, it should be easy to define other $\Upsilon_{\rm color}$ stellar mass estimators.

Finally, we derive each galaxy $V$-band surface-brightness at its effective radii R50 and Re, encircling half of the light within the FoV and half of the total integrated light of the galaxy, respectively, by averaging the flux values in an elliptical ring (using the known position angle and ellipticity of the object) of width 0.15 (in units of Re or R50). Again, the surface brightness is measured in the galaxy rest-frame, which is not directly accessible when using broad-band imaging. 
As mentioned above, parameter errors are estimated from MC iterations, perturbing the observed spectra using the error spectra included in the MaNGA datacubes.

\subsection{Quality Control}
\label{sec:qc}

The analysis by \pyp\ runs in an automatic way through the entire MaNGA dataset, without human supervision. Therefore, we must perform a quality control of the results to identify and flag any possible issue on either the data or the analysis itself. This procedure comprises different steps. First an automatic exploration is done to check that the analysis has produced all the required files, that they have the required format and the expected values (i.e., no map or table is filled with NaN, Inf or zeros). 
%Of the 10,245 analyzed datacubes, 21 present evident problems with the data and have automatically flagged as BAD in the final dataset.
Then, a second automatic exploration is done by comparing two basic parameters derived from our analysis (the redshift and the integrated stellar mass), with those provided by the NSA catalog. If significant differences ($\gtrsim$30\%) are found in any of these quantities, the corresponding datacube is flagged with a WARNING. 

%%%%%%%%%%%%%%%%%%%%%%%%%%%%%%%%%%%%%%%%%%%%%%%%%%%%%%%%%%%%%%%%%%%%%%%5
\begin{figure*}
%\captionsetup[subfigure]{labelformat=empty}
 \minipage{0.99\textwidth}
% \subfloat[]{
%\includegraphics[clip,trim=10 0 40 10,width=17.5cm]{figures/ana_spec_auto_ssp.manga-7495-12704.int.out.png}
 \begin{tabular}{cc}
 \multirow[t]{2}{*}[-3.75cm]{\includegraphics[width=11cm]{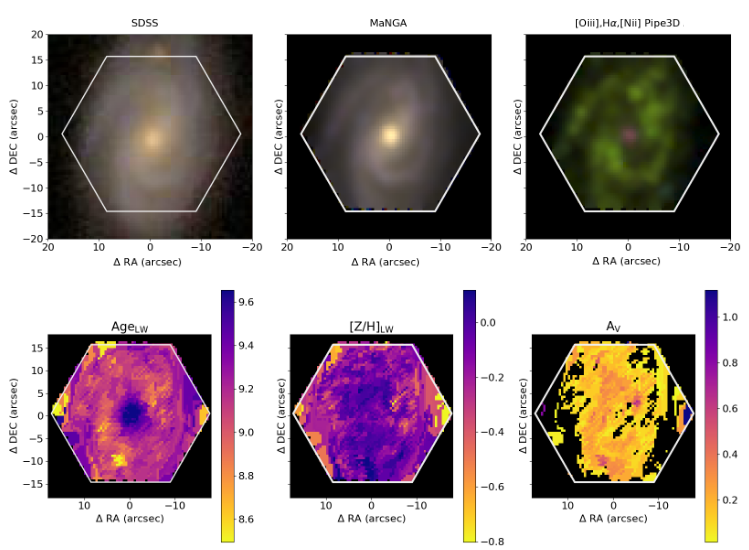}} & \includegraphics[width=5.5cm]{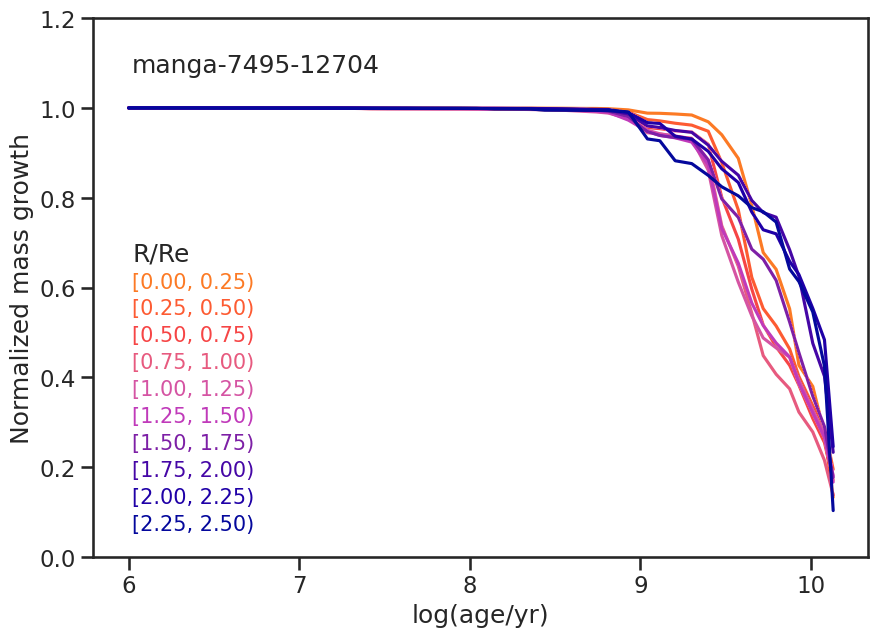}\\
 &  \includegraphics[width=5.5cm]{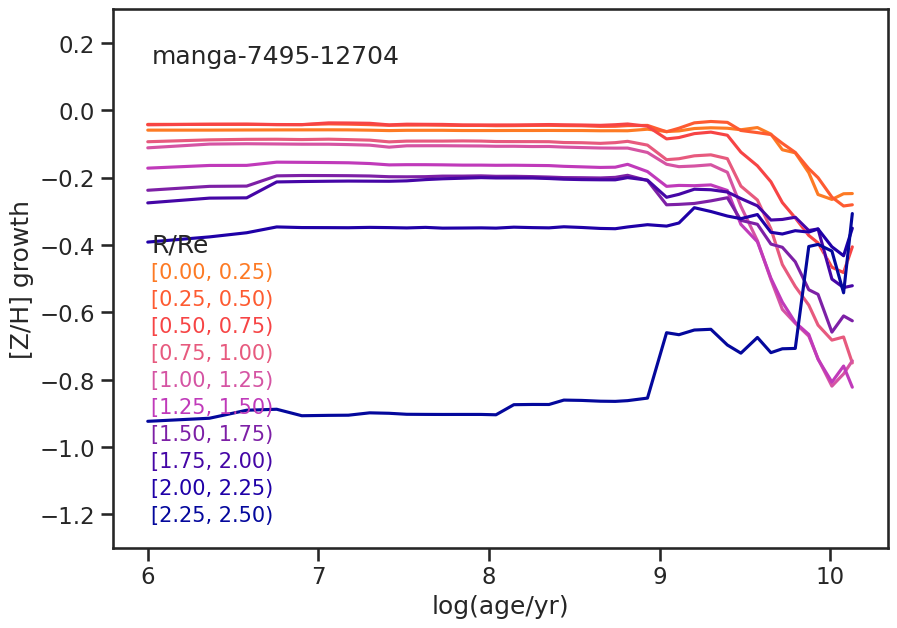} \\
% \vfill
\end{tabular}
   \endminipage
 \caption{Example of the information explored in the quality control process for the galaxy/cube manga-7495-12703. Left-panels comprise, from top-left to bottom-right, (i) a true-color image created using the SDSS u-, g- and r-band images, (ii) the same image created using images in the same bands synthesized from the MaNGA datacube, (iii) a similar true-color image created using the [OIII]5007 (blue), H$\alpha$ (green) and [NII]6584 (red) flux intensity maps, together with the (iv) \ageLW, (v) \metLW and (vi) A$_{\rm V,\star}$ maps, all of them derived as part of the \pyp\ process. Right-panels show the normalized mass-assembly (top) and metallicity enrichment history (bottom) for this galaxy at different galactocentric distances (indicated with different colors), estimated from the \pyp\ analysis following the procedures described in \citet{camps20,camps21}. { We remark that no signal-to-noise mask has been applied to the data and, therefore, the results corresponding to the outer regions in all panels are highly unreliable (e.g., the apparent increase of \ageLW and \metLW at the edge of the FoV and/or the metallicity history beyond two effective radius).}}
 \label{fig:qc_map}
\end{figure*}
%%%%%%%%%%%%%%%%%%%%%%%%%%%%%%%%%%%%%%%%%%%%%%%%%%%%%%%%%%%%%%%%%%%%%%%
In addition, a more detailed by-eye exploration is performed to determine the quality of the data. For doing so we created an interactive webpage in which we can show for each galaxy/datacube: (i) its central spectrum, corresponding to an aperture of 2.5$\arcsec$, together with the best fitted model provided by \pyp; (ii) the mass-assembly and chemical enrichment curves \citep[e.g.][]{eperez13,ibarra16,camps20}; (iii) three true-color images generated using (a) the original SDSS u-,g- and r-band images, (b) the same image generated using synthetic broad-band images extracted from the original MaNGA datacubes for the same filters, and (c) the emission line maps for [OIII]5007 (blue), H$\alpha$ (green) and [NII]6584 (red); (iv) maps of the luminosity-weighted age (\ageLW) and metallicity (\metLW), and dust extinction, derived as part of the \pyp\ process (described in detail in Sec. \ref{sec:ssp_cube}); and (v) a set of spatially resolved emission-line diagnostic diagrams described in \citep{carlos20}, including (a) the WHAN diagram \citep{cid-fernandes10}, (b) the classical BPT diagrams involving the [OIII]/H$\beta$ line ratio as a function of the [NII]/H$\alpha$, [SII]/H$\alpha$, and [OI]/H$\alpha$ ones, (c) the spatial distribution of the emission line velocity dispersion, the [NII]/H$\alpha$ ratio and the distribution of one as a function of the other, and (d) the gas and stellar velocity maps and their difference.

Figures~\ref{fig:qc_spec},~\ref{fig:qc_map} and~\ref{fig:BPT_single} (Appendix \ref{app:qc}) illustrate an example of the quality control analysis for a show-case datacube (manga-7495-12703).
We search for (i) evident problems in the quality of the explored spectra or issues in the fitting process; (ii) issues with the observations themselves (large masked areas and/or strong contamination by foreground stars); and (iii) presence of strong AGN or clear merging systems that may affect the stellar population and/or the kinematic analysis (just to warn the user). 
The visual quality control was performed by seven different people that explored a sub-sample of the full dataset, with an average overlapping of three different inspections per target/datacube. When there was a different appreciation of the quality between different inspectors, we adopted the worst reported quality control flag.

The reported quality control flags (QCFLAG) are listed in Table \ref{tab:qc}, indicating its level of importance (OK, BAD or WARNING). Of the 10,245 cubes corresponding to galaxies analyzed by \pyp, the code fails to analyze 25 cases due to different issues with the data (very low S/N in most cases and evident empty fields). Of the remaining 10,220 cubes, in 6 cases we flagged our results as BAD (for the reasons described before) and should be discarded. In addition, 386 cubes have different issues, and we strongly recommend not to use them without visual inspection of the original spectra, the results from the fitting and, in particular, the nature of the warning. In summary, we provide good quality analysis for 9,828 datacubes (corresponding 9,784 individual galaxies).

\section{Results}
\label{sec:res}

All the results from the analysis described in the previous sections are delivered through the network webpages (\url{http://ifs.astroscu.unam.mx/MaNGA/Pipe3D_v3_1_1/} and \url{https://data.sdss.org/sas/dr17/manga/spectro/pipe3d/v3_1_1/3.1.1/}). For each individual MaNGA cube/galaxy, we release a single FITS file using the Pipe3D data model described below. In addition, a catalog including individual properties of each galaxy is available too. Its format is described in the upcoming sections.

%SDSS17Pipe3D_v3_1_1.
%\Com{Description of Pipe3D. What is new. SSPs adopted. Emission lines.}

%%%%%%%%%%%%%%%%%%%%%%%%%%%%%%%%%%%%%%%%%%%%%%%%%%%%%%%%%%%%%%%%%%%%%%%5
\begin{figure*}
 \minipage{0.99\textwidth}
 \includegraphics[width=17.5cm]{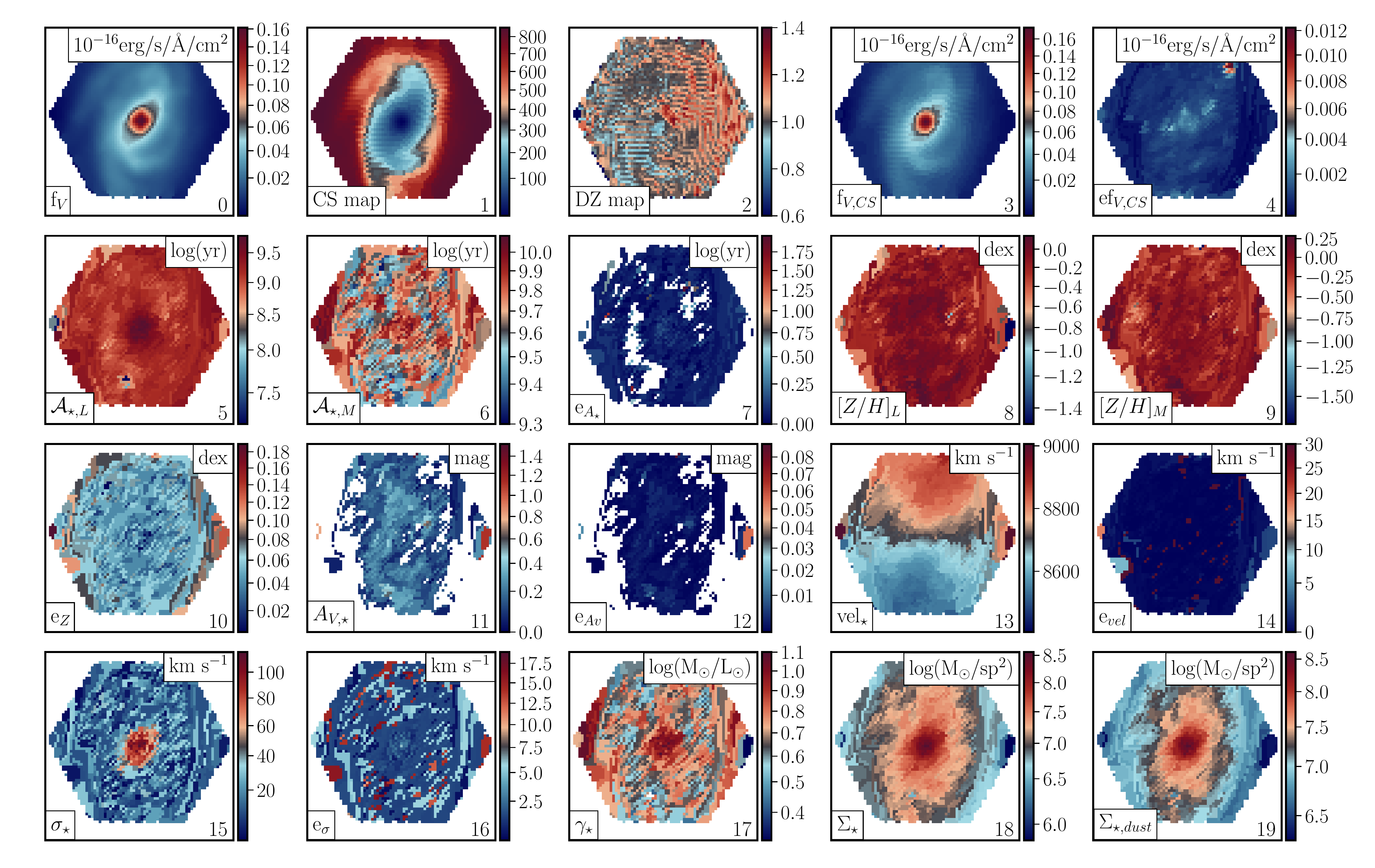}
 \endminipage
 \caption{Example of the content of the SSP extension in the Pipe3D fitsfile, corresponding to the MaNGA datacube (galaxy) manga-7495-12704. Each panel shows a colour image with the property stored in the corresponding channel of the datacube, 
 as listed in Table \ref{tab:ssp}. The index, corresponding property, and units, are indicated in each panel as a label at the bottom-right, bottom-left and top-right of the figure.}
 \label{fig:SSP_map}
\end{figure*}
%%%%%%%%%%%%%%%%%%%%%%%%%%%%%%%%%%%%%%%%%%%%%%%%%%%%%%%%%%%%%%%%%%%%%%%5

\subsection{Pipe3D data model}
\label{sec:cubes}

The result from the analysis described above for each individual spectrum is a set of parameters derived for each particular spaxel. Each of these parameters is stored as a 2D array or a map of the given quantity that preserves the World Coordinate System (WCS)
of the original data. These maps are the the final product of the analysis. For an optimal (compact and easily accessible) distribution of the products, we pack them in datacubes, i.e., 3D arrays in which each channel (slide in the z-axis) corresponds to one of the maps described before, and therefore, to one of the derived parameters. Instead of storing all the products in a single datacube, they are packed in a set of datacubes based on a physically motivated association of parameters. In the current implementation of Pipe3D there are five datacubes: (i) SSP, contains maps of the spatial distribution of the main properties of the stellar populations; (ii) SFH, comprises the distribution of the light fraction derived by the stellar decomposition for each SSP in our template library; (iii) INDICES, comprises the maps of the stellar indices; (iv) ELINES, includes properties of a set of strong emission lines derived by the Gaussian fitting procedure, and (v) FLUX\_ELINES, includes properties of a much larger set of emission lines based on the weighted moment analysis. In the current release, we distribute two versions of the ELINE analysis, based on two different sets of emission lines, with the new version labelled FLUX\_ELINES\_LONG. Finally, all the datacubes are stored as individual extensions of a single FITs file (the Pipe3D file). In this release we include, in addition to these 6 extensions: (i) a primary extension with the header of the original MaNGA cube (ORG\_HDR); (ii) a mask of the brightest field stars recovered from the Gaia survey (GAIA\_MASK), and (iii) a mask of the spaxels within the hexagon with enough signal-to-noise to provide with reliable estimates of the stellar population properties (SELECT\_REG).

% --------------- Pipe3D.cube ------------------------%
\begin{table}
\begin{center}
\caption{Description of the Pipe3D file.}
\begin{tabular}{cll}\hline\hline
HDU	&  EXTENSION & Dimensions\\
\hline
  0 & ORG\_HDR            &()            \\
  1 & SSP                 & (72, 72, 21)  \\
  2 & SFH                 & (72, 72, 319) \\
  3 & INDICES             & (72, 72, 18)  \\
  4 & ELINES              & (72, 72, 11)  \\
  5 & FLUX\_ELINES        & (72, 72, 456) \\
  6 & FLUX\_ELINES\_LONG  & (72, 72, 1536)\\
  7 & GAIA\_MASK          & (72, 72)      \\
  8 & SELECT\_REG         & (72, 72)      \\
\hline
\end{tabular}\label{tab:hdu} 
\end{center}
\end{table}
% --------------- Pipe3D.cube ------------------------%

Table~\ref{tab:hdu} summarizes the structure of the Pipe3D FITs file. 
Each file can be easily associated with its corresponding MaNGA cube from the file naming convention, i.e., {\tt manga-plate-ifudsg.Pipe3D.cube.fits}, where {\tt plate} and {\tt ifudsg} correspond, respectively, to the number of the physical plate used during the pointing and the IFU bundle design and number. These two numbers define uniquely a fixed pointing to a certain galaxy, as indicated before. In this way, {\tt manga-7495-12704.Pipe3D.cube.fits}, corresponds to the manga observation of a particular galaxy, observed using the plate number {\tt 7495}, and the 4th IFU ({\tt 04}) comprising {\tt 127} fibers. It is important to note that one galaxy may be observed using different plates and IFUs, corresponding to repeated observations that have been treated or analyzed individually, as if they were different objects. In the final MaNGA distribution there are 90 of these repeated galaxies (i.e., 2$\times$45 cubes). Since they are different observations, taken under different atmospheric conditions, and calibrated using different stars, they offer a unique opportunity to evaluate the quality of the data. We discuss them in detail in Appendix \ref{app:rep}.

%%%%%%%%%%%%%%%%%%%%%%%%%%%%%%%%%%%%%%%%%%%%%%%%%%%%%%%%%%%%%%%%%%%%%%%5
\begin{figure*}
 \minipage{0.99\textwidth}
 \includegraphics[width=17.5cm]{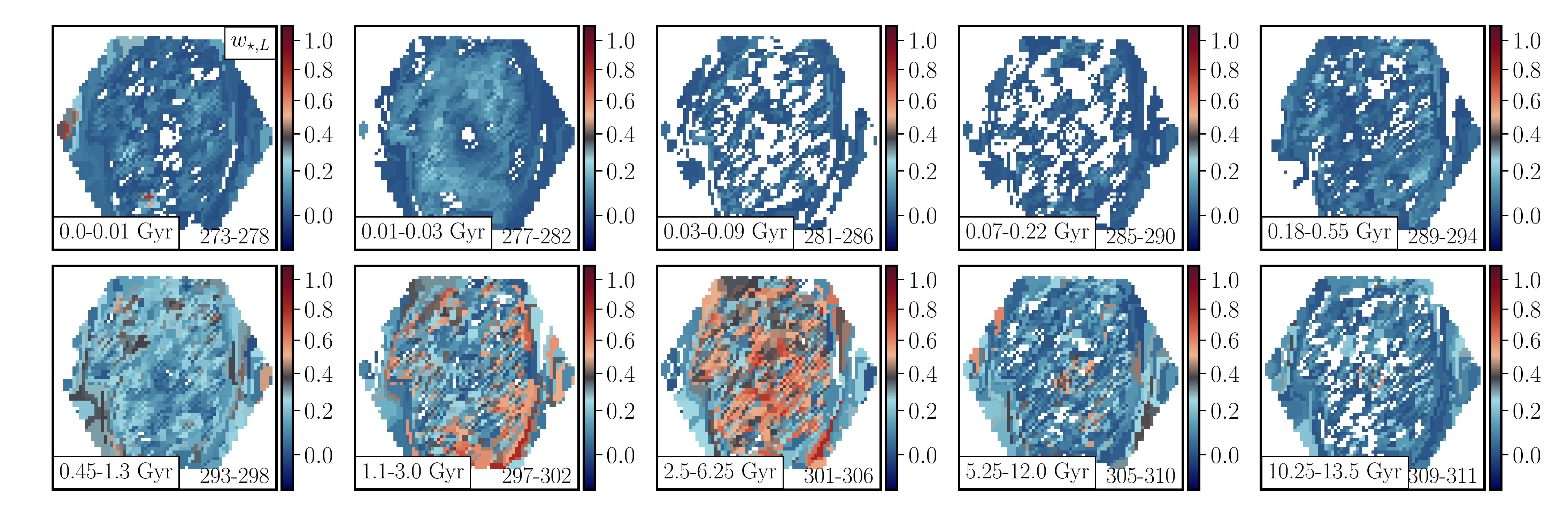}
 \endminipage
 \caption{Example of the content of the SFH extension in the Pipe3D fitsfile, corresponding to the MaNGA datacube (galaxy) manga-7495-12704. Each panel shows a colour image with the fraction of light at the normalization wavelength ($w_{\star,L}$) for different ranges of age included in the SSP library. The range of age is indicated in the bottom-left inbox, and the corresponding indices of the co-added channels in the SFH extension are shown in the bottom-right legend. For clarity we do not show the individual $w_{\star,L}$ maps included in the SFH extension.}
 \label{fig:SFH_map}
\end{figure*}
%%%%%%%%%%%%%%%%%%%%%%%%%%%%%%%%%%%%%%%%%%%%%%%%%%%%%%%%%%%%%%%%%%%%%%%5

%\Com{What is the content in each datacube }
%%%%%%%%%%%%%%%%%%%%%%%%%%%%%%%%%%%%%%%%%%%%%%%%%%%%%%%%%%%%%%%%%%%%%%%5

% --------------- SSP ------------------------%
\begin{table}
\begin{center}
\caption{Description of the SSP extension.}
\begin{tabular}{cll}\hline\hline
Channel	& Units	& Stellar index map\\
\hline
0	&	10$^{-16}$ \flux	& Unbinned flux intensity at $\sim$5500\AA, f$_{V}$ \\
1	&	none	                & Continuum segmentation index, $CS$\\
2	&	none	                & Dezonification parameter, $DZ$\\
3	&	10$^{-16}$ \flux	& Binned flux intensity at $\sim$5500\AA, f$_{V,CS}$\\
4	&	10$^{-16}$ \flux	& StdDev of the flux at $\sim$5500\AA, ef$_{V,CS}$ \\
5	&	{ log$_{10}$(yr)}& Lum. Weighted age, $\mathcal{A}_{\star,L}$, { (log scale)}\\
%        &                                     & { in logarithmic scale}\\
6	&	{ log$_{10}$(yr)}& Mass Weighted age, $\mathcal{A}_{\star,M}$, { (log scale)}\\
%        &                                     & { in logarithmic scale}\\
7	&	{ log$_{10}$(yr)}& Error of both $\mathcal{A}_{\star}$, { (log scale)}\\
8	&	dex	                & Lum. Weighted metallicity, $Z_{\star,L}$\\
        &                       & in logarithmic scale, normalized to \\
        &                       & the solar value { ($Z_{\odot}=$0.017)}\\
9	&	dex	                & Lum. Weighted metallicity, $Z_{\star,M}$\\
        &                       & in logarithmic scale, normalized to \\
        &                       & the solar value { ($Z_{\odot}=$0.017)}\\
10	&	dex	                & Error of both $Z_{\star}$\\
11	&	mag	                & Dust extinction of the st. pop., A$_{V,\star}$\\
12	&	mag	                & Error of A$_{V,\star}$, e$_{{\rm A}_V}$\\
13	&	km/s	                & Velocity of the st. pop., vel$_{\star}$\\
14	&	km/s	                & Error of the velocity, $e_{\rm vel}$ \\
15	&	km/s	                & Velocity dispersion of the st. pop., $\sigma_{\star}$\\
16	&	km/s	                & Error of $\sigma_{\star}$, $e_\sigma$\\
17	&	log$_{10}$(M$_\odot$/L$_\odot$)	& Mass-to-light ratio of the st. pop., $\Upsilon_{\star}$\\
18	&	log$_{10}$(M$_\odot$/sp$^2$)	& Stellar Mass density per spaxel., $\Sigma_{\star}$\\
19	&	log$_{10}$(M$_\odot$/sp$^2$)	& Dust corrected $\Sigma_{\star}$, $\Sigma_{\star,dust}$\\
20	&	log$_{10}$(M$_\odot$/sp$^2$)	& error of $\Sigma_{\star}$\\
\hline
\end{tabular}\label{tab:ssp}
\end{center}
Channel indicates the Z-axis of the datacube starting from 0.			
$(1)$ measured along the entire wavelength range covered by the
spectroscopic data.
\end{table}
% --------------- SSP ------------------------%

\subsubsection{SSP extension}
\label{sec:ssp_cube}

As indicated before, this extension comprises a datacube with the average properties of the stellar populations derived from the multi-SSP fitting procedure outlined in Sec.~\ref{sec:pipe3d}.  The content of each channel with index $N$ is described in the header entry named {\tt DESC\_N}, starting with $N=0$ for the 1st channel, corresponding to the properties listed in Table~\ref{tab:ssp}. An example of the content of this extension, for datacube manga-7495-12704, is shown in Fig.~\ref{fig:SSP_map}.

The first channel comprises the median flux intensity in the natural units of the Pipe3D pre-processed MaNGA datacube (Sec.~\ref{sec:data}) at the wavelength range 5450-5550 \AA. The second channel includes the index of the tessella in which each spaxel is grouped after the binning performed to increase the S/N. The third channel comprises the dezonification map, as described in Sec.~\ref{sec:pipe3d}. The average flux intensity and the standard deviation within the same wavelength range as for the unbinned flux in the 1st channel, after applying the spatial binning, are included in the following two channels. Up to now, the stored properties correspond to parameters derived directly from the original datacubes, before performing the spectral fit. The remaining channels correspond to properties extracted directly from our fits, including the luminosity- and mass-weighted age (\ageLW and \ageMW) and metallicity (\metLW and \metMW), respectively, also the dust extinction, the stellar velocity, the stellar velocity dispersion, the stellar mass-to-light ratio and the stellar mass density (with and without correction for dust extinction). 

A detailed description on how these parameters are derived has been presented in \citet[][Sec. 2.3]{pipe3d} and \citet[][Sec.~3.1, Eq.~2-4]{sanchez20}. The kinematic parameters ($vel_\star$ and $\sigma_\star$), and the dust extinction ($A_{\rm V,\star}$) are derived directly from the first step of the spectral decomposition analysis described in Sec~\ref{sec:pipe3d}, and expressed in Eq.~\ref{eq:dec}. The luminosity weighted parameters ($P_{L}$) are derived from
\begin{equation}
{\rm log} P_{L}  = \Sigma_{ssp} w_{ssp,\star,L} {\rm log} P_{ssp},
\label{eq:LW}    
\end{equation}
\noindent and the mass-weigthed parameters ($P_M$) from:
\begin{equation}
{\rm log} P_{M}  = \frac{\Sigma_{ssp} w_{ssp,\star,L} \Upsilon_{ssp,\star} {\rm log}P_{ssp}}{\Sigma_{ssp} w_{ssp,\star,L} \Upsilon_{ssp,\star}},
\label{eq:MW}    
\end{equation}
\noindent where (i) $w_{ssp,\star,L}$ is the light-fraction/weight of each SSP in the library at the normalization wavelength derived from the spectral fit according to
Eq.~\ref{eq:dec}, (ii) $P_{ssp}$ is the value of the given parameter for the corresponding SSP (e.g., age, metallicity), and (iii) $\Upsilon_{ssp,\star}$ is the stellar mass-to-light ratio. Note that both the luminosity and mass-weighted average values are indeed geometrical averages (or average of logarithmic parameters).  Thus, these maps should be interpreted as a realization of the statistical distribution of the considered parameter \citep[e.g., age or metallicity distributions, ][]{mejia20}.

%%%%%%%%%%%%%%%%%%%%%%%%%%%%%%%%%%%%%%%%%%%%%%%%%%%%%%%%%%%%%%%%%%%%%%%5
\begin{figure*}
 \minipage{0.99\textwidth}
 \includegraphics[width=17.5cm]{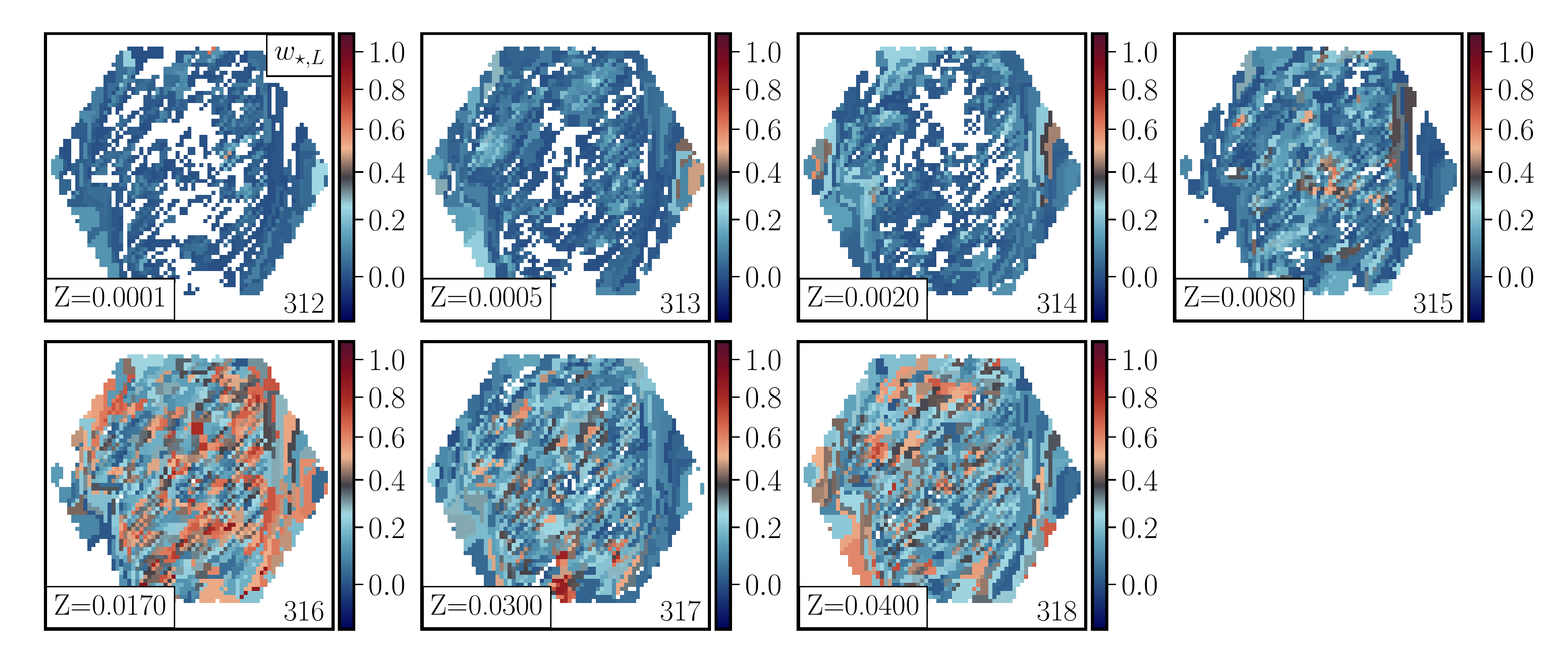}
 \endminipage
 \caption{Similar figure as Fig. \ref{fig:SFH_map}, showing the $w_{\star,L}$ for the different metallicities included in the adopted SSP library (indicated in the bottom-left inset), corresponding to the channels of the SFH extension indicated in the bottom-right legend.}
 \label{fig:SFH_Z_map}
\end{figure*}
%%%%%%%%%%%%%%%%%%%%%%%%%%%%%%%%%%%%%%%%%%%%%%%%%%%%%%%%%%%%%%%%%%%%%%%5

Finally, the mass-to-light ratio is derived from a simple weighted average:
\begin{equation}
\Upsilon_\star = \Sigma_{ssp} w_{ssp,\star,L} \Upsilon_{ssp,\star}.
\label{eq:ML}    
\end{equation}
\noindent The stellar surface density is then derived by multiplying $\Upsilon_\star$ by the observed surface density at the normalization wavelength (approximately the rest-frame $V$-band), corrected by extinction by dust and luminosity distance($D_L$), and divided by the area of each spaxel ($a_{sp}$):
\begin{equation}\label{eq:mu}
\begin{aligned}
\Sigma_* &= \mu_V \Upsilon_\star,\\
\mu_V &= \frac{4\pi D_L^2 f_{V}}{a_{sp}} 10^{0.4 A_{\rm V,*}}.
\end{aligned}
\end{equation}

% --------------- SFH ------------------------%
\begin{table}
\begin{center}
\caption{Description of the SFH extension.}
\begin{tabular}{cl}\hline\hline
Channel	&  Description of the map\\
\hline
0 &  $w_{\star,L}$ for ($\mathcal{A}_\star$,$Z_\star$)~=~(0.001 Gyr, 0.0001) \\
... &  ... \\
272 &  $w_{\star,L}$ for ($\mathcal{A}_\star$,$Z_\star$)~=~(13.5 Gyr, 0.04) \\
\hline  
273 &  $w_{\star,L}$ for $\mathcal{A}_\star$~=~0.001 Gyr\\
  ... &  ... \\
311 &  $w_{\star,L}$ for $\mathcal{A}_\star$~=~13.5 Gyr\\
\hline  
312 &  $w_{\star,L}$ for $Z_\star$~=~0.0001 \\
... &  ... \\
318 &  $w_{\star,L}$ for $Z_\star$~=~0.04 \\
\hline
\end{tabular}\label{tab:sfh}
\end{center}
Channel indicates the Z-axis of the datacube starting from 0, and $w_{\star,L}$ indicates the fraction (weight) of light at 5500\AA\ corresponding to: (i) an SSP of a certain age ($\mathcal{A}_\star$) and metallicity ($Z$)
(channels 0 to 272), 
(ii) a certain age, i.e., co-adding all $w_{\star,L}$ %the fractions of light 
corresponding to SSPs with the same age but different metallicity (channels 273 to 311), and 
(iii) a certain metallicity, i.e., co-adding all $w_{\star,L}$ %the fractions of light
corresponding to SSPs with the same metallicity but different age (channels 312 to 318). 
The adopted {\tt MaStar\_sLOG} SSP library comprises 
39 ages, $\mathcal{A}_\star$/Gyr~=~(0.001, 0.0023, 0.0038, 0.0057, 0.008, 0.0115, 0.015, 0.02, 0.026, 0.033, 0.0425, 0.0535, 0.07, 0.09, 0.11, 0.14, 0.18, 0.225, 0.275, 0.35, 0.45, 0.55, 0.65, 0.85, 1.1, 1.3, 1.6, 2, 2.5, 3, 3.75, 4.5, 5.25, 6.25, 7.5, 8.5, 10.25, 12, 13.5), and 7 metallicities, $Z_\star$~=~(0.0001, 0.0005, 0.002, 0.008, 0.017, 0.03, 0.04).
\end{table}
% --------------- SFH ------------------------%

\subsubsection{SFH extension}
\label{sec:sfh_cube}

%The SSP-library is distributed as a raw-stacked spectra fitsfile extension together with the average/characteristic properties of the galaxies, in the  SDSS17Pipe3D\_v3\_1\_1.fits, that we will describe in detail in Sec. \ref{sec:TBW}. 

This extension comprises a datacube that in each slice/channel includes the spatial distribution of the light fraction (weights $w_{\star,L}$ in Eq.~\ref{eq:dec}) of each SSP within the adopted library at the normalization wavelength (5500\AA), following the WCS of the original MaNGA cube. Since the decomposition of the stellar population is done for the spatial binning described in Sec.~\ref{sec:pipe3d}, the distributed $w_{\star,L}$ have the same value for all the spaxels within the same tessella. As indicated in Sec.~\ref{sec:ssp},
the adopted {\tt MaStar\_sLOG} library comprises 273 SSP spectra, covering 39 ages and 7 metallicities (Table~\ref{tab:sfh}). The spatial distribution of the $w_{\star,L}$ for each SSP is then stored in the first 273 channels of the SFH extension. Then, 39 additional channels (from $N$~=~273 to $N$~=~311, with the 1st channel corresponding to index $N$~=~0) comprise the weights for each individual age (i.e., the values derived co-adding the seven $w_{\star,L}$ for the same age but different metallicity). An example of the spatial distribution of the latter weights for the same datacube shown in Fig. \ref{fig:SSP_map}, manga-7495-12704, is shown in Figure \ref{fig:SFH_map}. Since the weights $w_{\star,L}$ are not weighted by the flux intensity, i.e., they are just relative quantities, the original shape of the galaxy is not evident in almost none of the panels. Only for the youngest ages ($t~<~$30Myr) it is possible to trace the location of the disk of this galaxy (somehow). Curiously, most of the light corresponds to stars with age between $\sim$1 and $\sim$6 Gyr, more homogeneously distributed than younger stars. Finally, $w_{\star,L}$ increases slightly for older stars towards the center up to $\sim$12 Gyrs. These weights/light-fractions correspond to the spatially resolved age distribution function (ADF), that is usually explored in the analysis of resolved stellar populations \citep[e.g.,][]{hass20}. 

In addition, seven more channels are included in the SFH extension comprising the weights for each individual metallicity (i.e., the values derived by co-adding the 39 $w_{\star,L}$ for the same metallicity but different age). Figure \ref{fig:SFH_Z_map} shows the spatial distribution of these light fractions for each of the metallicities included in the SSP library. Despite of the clumpy distribution which is a consequence of the local uncertainties in the derivation of $w_{\star,L}$, there are clear patterns that emerge. For instance, the higher light fractions are more distributed in the outer regions of the galaxy for the lowest metallicities, while the distributions are more homogeneous for the highest metallicities. This is a clear consequence of the spatial variation of the metallicity distribution function \citep[MDF, as reported by][]{mejia20}. Indeed, these weights are the basic ingredients to derive the MDFs using our analysis.

Table \ref{tab:sfh} summarizes the information included in each channel of the SFH cube. This information is included in the header with a set of keywords named DESC\_N, indicating the content of channel N, with N running from 0 to 318. For completeness, the original file generated by \pyp from which the information in channel N was obtained, 
has been listed in the header as FILE\_N.

%The content of each channel with index N is included in the header entry named DESC_N, where starts with 0 for the 1st channel.

%%%%%%%%%%%%%%%%%%%%%%%%%%%%%%%%%%%%%%%%%%%%%%%%%%%%%%%%%%%%%%%%%%%%%%%5
\begin{figure*}
 \minipage{0.99\textwidth}
 \includegraphics[width=17.5cm]{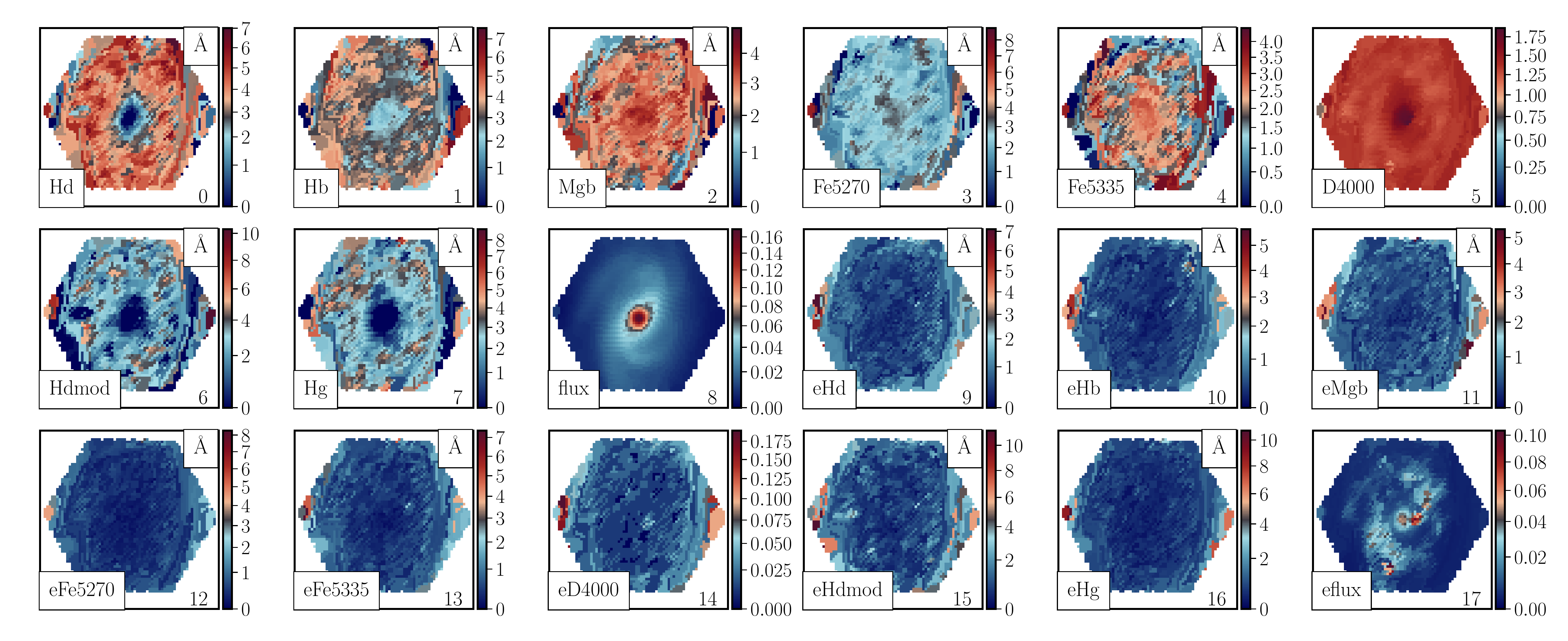}
 \endminipage
 \caption{Example of the content of the INDICES extension in the Pipe3D file, corresponding to the MaNGA dartacube (galaxy) manga-7495-12704. Each panel shows a color image with the content of a channel in this datacube. The actual content is indicated for each panel in the lower-left, the channel in the lower-right and the units of the represented quantity in the upper-right legend{. For the flux and flux error the units are \funitsI}.}
 \label{fig:IND_map}
\end{figure*}
%%%%%%%%%%%%%%%%%%%%%%%%%%%%%%%%%%%%%%%%%%%%%%%%%%%%%%%%%%%%%%%%%%%%%%%5

\subsubsection{INDICES extension}
\label{sec:indices_cube}

This extension comprises a datacube in which each channel/slice corresponds to the spatial distribution of the stellar indices derived for the emission-line subtracted spectra, following the procedure outlines in Sec.~\ref{sec:pipe3d} and described in detail in \citet{pypipe3d,pipe3d}. Like in the previous cases, this analysis is done for the averaged spectra within each spatial bin, and therefore the reported values are the same for all the spaxels within the same tessella. The content of each channel/slice is listed in Table \ref{tab:index}, indicating the explored stellar index and its error. The adopted spectral range to measure each index, together with the ranges adopted to estimate the adjacent continuum (at bluer and redder wavelength ranges with respect to the bandwidth that defines the index) are included in the table. For completeness, we include the average flux within the full wavelength range covered by the data and its standard deviation, as a proxy for the signal and noise within each tessella. The corresponding information is stored in the header keywords named INDEX{\tt N}, where {\tt N} corresponds to each channel. Figure \ref{fig:IND_map} shows an example of the content of this extension, corresponding to the same datacube shown in previous figures (i.e., manga-7495-12704, e.g., Fig. \ref{fig:SSP_map}), including both the measurements and the estimated errors. As expected for a spiral galaxy, there are clear radial patterns in the different stellar indices that follow the light distribution: (i) D4000 presents a radial gradient, with the highest values found in the center of the galaxy; (ii) all the explored Balmer indices (H$\beta$, H$\gamma$ and H$\delta$), show a decline in the inner regions, most probably associated with the location of the bulge of this galaxy, and a rise, spatially coincident with the disk and the spiral arms; and (iii) the indices more sensible to the stellar metallicity, in particular Mg$b$ and Fe5335, present a mild decline from the center to the outer regions. The error maps show a near constant distribution for all the indices, suggesting that the adopted spatial binning has indeed increased the S/N of the outer regions by adding the required spectra. Only in the outermost regions, at both the east and west edges of the FoV, the errors rise-up, suggesting that the surface brightness of the galaxy has dropped beyond the ability of the binning procedure to produce reliable (high enough S/N) spectra. Finally, the map of the standard deviation of the flux intensity presents higher values at the location of the spiral arms, as expected, since in these regions the emission-line subtracted spectra (used for the analysis of the stellar indices) present stronger variations and residuals. Similar patterns are appreciated for all the explored galaxies, with their own peculiarities and signatures.

%This extension comprises a datacube with the absorption stellar indices derived from the continuum segmentated and emission line subtracted spectra in the original cube. A set of absorption lines (plus D4000) are analyzed, using the procedure described in Sanchez et al. (2016b). The provided indices are not corrected by the effect of the velocity dispersion, so, they must be corrected using the stellar velocity dispersion provided in the SSP extension. The details of its content are included in the following table.

%--------------- INDICES ------------------------%
\begin{table*}
\begin{center}
\caption{Description of the INDICES extension.}
\begin{tabular}{llllccc}\hline\hline
ID & Channel	& Units	& Channel content & Index $\lambda$ range (\AA) & blue $\lambda$ range (\AA) & red $\lambda$ range (\AA) \\
\hline
Hd     & 0/9	&	\AA\     &H$\delta$ index/error& 4083.500-4122.250 &4041.600-4079.750 &4128.500-4161.000\\
H$\beta$    & 1/10	&	\AA	&H$\beta$ index/error& 4847.875-4876.625 &4827.875-4847.875 &4876.625-4891.625\\
Mgb    & 2/11	&	\AA	&Mg$b$ index/error  & 5160.125-5192.625 &5142.625-5161.375 &5191.375-5206.375\\ 
Fe5270 &3/12	&	\AA	&Fe5270 index/error  & 5245.650-5285.650 &5233.150-5248.150 &5285.650-5318.150\\
Fe5335 &4/13	&	\AA	&Fe5335 index/error & 5312.125-5352.125 &5304.625-5315.875 &5353.375-5363.375\\
D4000  & 5/14	&	\AA	&D4000 index/error  & 4050.000-4250.000 &3750.000-3950.000 & \\
Hdmod  &6/15	&	\AA	&H$\delta_{\rm mod}$/error & 4083.500-4122.250 &4079.000-4083.000 &4128.500-4161.000\\
Hg     &7/16	&	\AA	&H$\Upsilon$/error  & 4319.750-4363.500  &4283.500-4319.750  &4367.250-4419.750\\
Flux   &8/17	&	10$^{-16}$ \flux&median flux/standard deviation$^{1}$ & & & \\
%9	&	\AA	&error of H$\delta$\\
%10	&	\AA	&error of H$\beta$\\
%11	&	\AA	&error of Mg$b$ \\
%12	&	\AA	&error of Fe5270\\
%13	&	\AA	&error of Fe5335\\
%14	&	\AA	&error of D4000\\
%15	&	\AA\     &error of H$\delta_{\rm mod}$\\
%16	&	\AA	&error of H$\Upsilon$\\
%17	&	10$^{-16}$ cgs & flux standard deviation$^{1}$\\
\hline
\end{tabular}\label{tab:index}
\end{center}
Channel indicates the Z-axis of the datacube starting from 0.			
$(1)$ measured along the entire wavelength range covered by the
spectroscopic data. 
\end{table*}
% --------------- ELINES ------------------------%

%The definition of the wavelength range of each index is given in the following table, including the name of the index, the blue and red range at which the continuumm is estimated, and the actual range at which the index is derived, following the procedure described in Sanchez et al. 2016b:

%%%%%%%%%%%%%%%%%%%%%%%%%%%%%%%%%%%%%%%%%%%%%%%%%%%%%%%%%%%%%%%%%%%%%%%5
\begin{figure}
 \minipage{0.99\textwidth}
 \includegraphics[width=8.5cm]{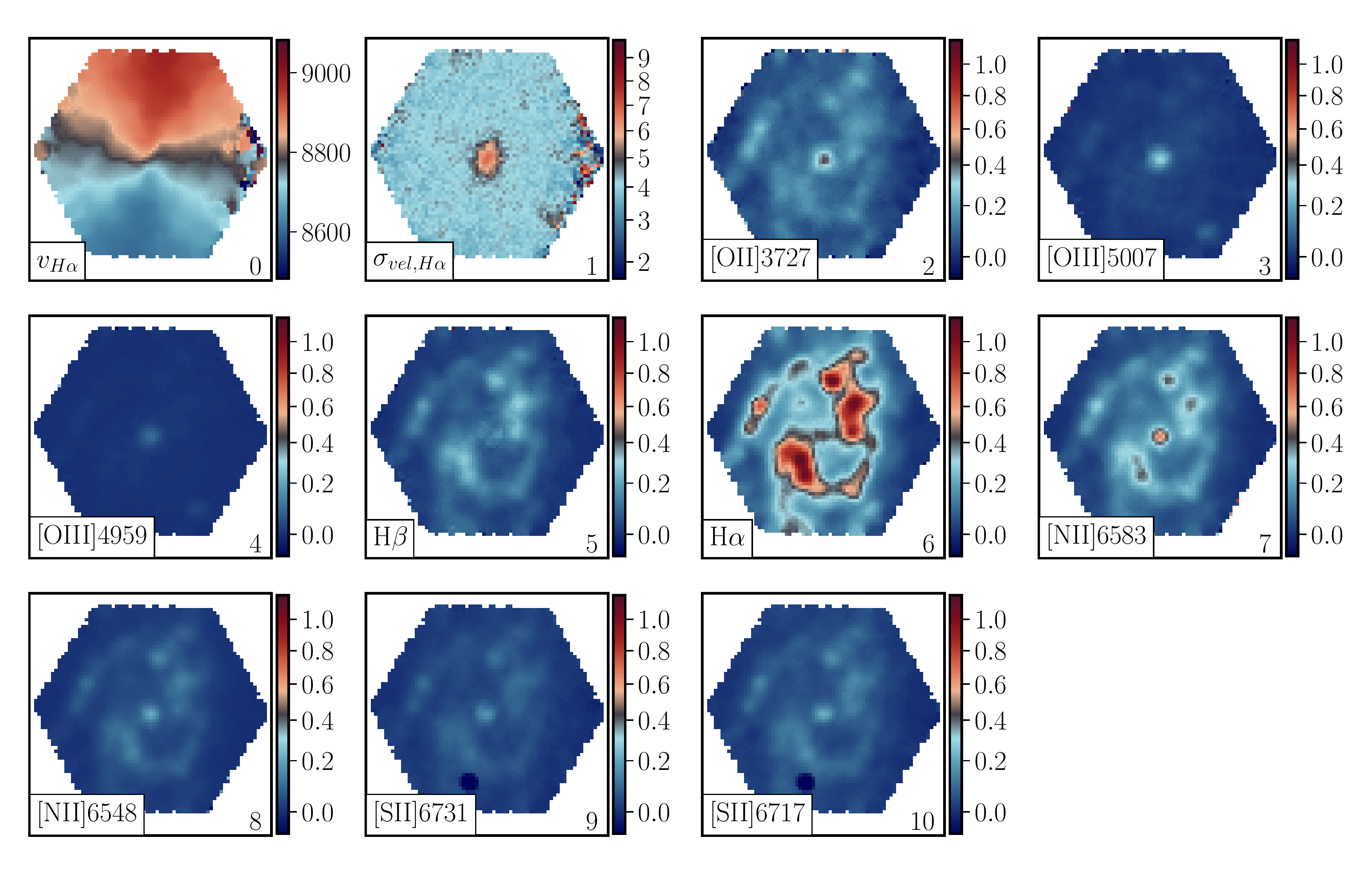}
 \endminipage
 \caption{Example of the content of the ELINES extension in the Pipe3D file, corresponding to the MaNGA dartacube (galaxy) manga-7495-12704. Each panel shows a color image with the content of a channel in this datacube. { The 1st panel corresponds to the velocity, in km/s, and the 2nd panel to \EWa, in \AA. The remaining panels represent the distribution of the flux intensities for the different analyzed emission lines (lower-left legend), in units of \funitsI}}
 \label{fig:ELINES_map}
\end{figure}
%%%%%%%%%%%%%%%%%%%%%%%%%%%%%%%%%%%%%%%%%%%%%%%%%%%%%%%%%%%%%%%%%%%%%%%5

\subsubsection{ELINES extension}
\label{sec:elines_cube}

As outlined in Sec. \ref{sec:pipe3d} and described in detail in \citet{pypipe3d} and \citet{pipe3d_ii}, the pipeline explores the properties of the emission lines performing two different analysis. In a first analysis, a set comprising the strongest emission lines in the optical range ([OII]3727, [OIII]4959,5007, H$\beta$, [NII]6548,83, H$\alpha$ and [SII]6717,31) is fitted with a single Gaussian function for each line at each spaxel within the pure-GAS cube (once the best model spectra for the stellar populations are subtracted ). The fitting procedure recovers  the flux intensity, the velocity and the velocity dispersion for each emission line. The pipeline stores the results of this analysis in the ELINES extension of the Pipe3D FITs file. Like in the previous cases, this extension comprises a datacube in which each channel/slice includes the spatial distribution of a different property derived from this fitting procedure, following the scheme described in Table \ref{tab:e}. The corresponding information is stored in the header keywords DESC\_{\tt N}, where {\tt N} corresponds to the channel index (starting with 0). An example of the content of this extension for datacube (galaxy) manga-7495-12704 is included in Figure \ref{fig:ELINES_map}. For this particular galaxy, the H$\alpha$ velocity map included in the 1st channel, shows a clear rotational pattern, with an S-shaped distortion in the central regions, suggesting the possible presence of a bar. The ionized gas presents a low velocity dispersion in all the disk, near to the instrumental dispersion, rising only in the central regions at the location of the bulge. The intensity maps of all the emission lines show a clumpy structure, as expected from ionized gas dominated by \HII\ regions and associations (at the spatial resolution of these data), with most of these regions distributed along the spiral arms of this galaxy. This is the expected ionized gas distribution for a low-inclination spiral galaxy \citep[e.g., ][]{ARAA,sanchez20}.

%For the strongest emission lines in the optical range ([OII]3727, [OIII]4959,5007, Hβ, [NII]6548,83, H&alpH$\alpha$and [SII]6717,31), derived by fitting a single Gaussian function to each of the considered emission lines at each spaxel in the datacube once subtracted the underlying stellar population. Each channel in the datacube included in this extension include one of the derived parameters for each emission line.

%--------------- ELINES ------------------------%
\begin{table}
\begin{center}
\caption{Description of the ELINES extensions.}
\begin{tabular}{cll}\hline\hline
Channel	& Units	& Description of the map\\
\hline
0	& km/s	& H$\alpha$ velocity\\
1	& $\AA$	& H$\alpha$ velocity dispersion$^a$\\
2	& 10$^{-16}$ \flux &	[OII]3727 flux intensity \\
3	& 10$^{-16}$ \flux &	[OIII]5007 flux intensity \\
4	& 10$^{-16}$ \flux &	[OIII]4959 flux intensity \\
5	& 10$^{-16}$ \flux &	H$\beta$ flux intensity \\
6	& 10$^{-16}$ \flux &	H$\alpha$ flux intensity \\
7	& 10$^{-16}$ \flux &	[NII]6583 flux intensity \\
8	& 10$^{-16}$ \flux &	[NII]6548 flux intensity \\
9	& 10$^{-16}$ \flux &	[SII]6731 flux intensity \\
10	& 10$^{-16}$ \flux &	[SII]6717 flux intensity  \\
\hline
\end{tabular}\label{tab:e}
\end{center}
Channel indicates the Z-axis of the datacube starting from 0. 
$^a$\,FWHM, i.e., 2.354$\sigma$. The instrumental velocity dispersion
has not been removed. 

\end{table}
%--------------- ELINES ------------------------%

%%%%%%%%%%%%%%%%%%%%%%%%%%%%%%%%%%%%%%%%%%%%%%%%%%%%%%%%%%%%%%%%%%%%%%%5
\begin{figure*}
 \minipage{0.99\textwidth}
 \includegraphics[width=17.5cm]{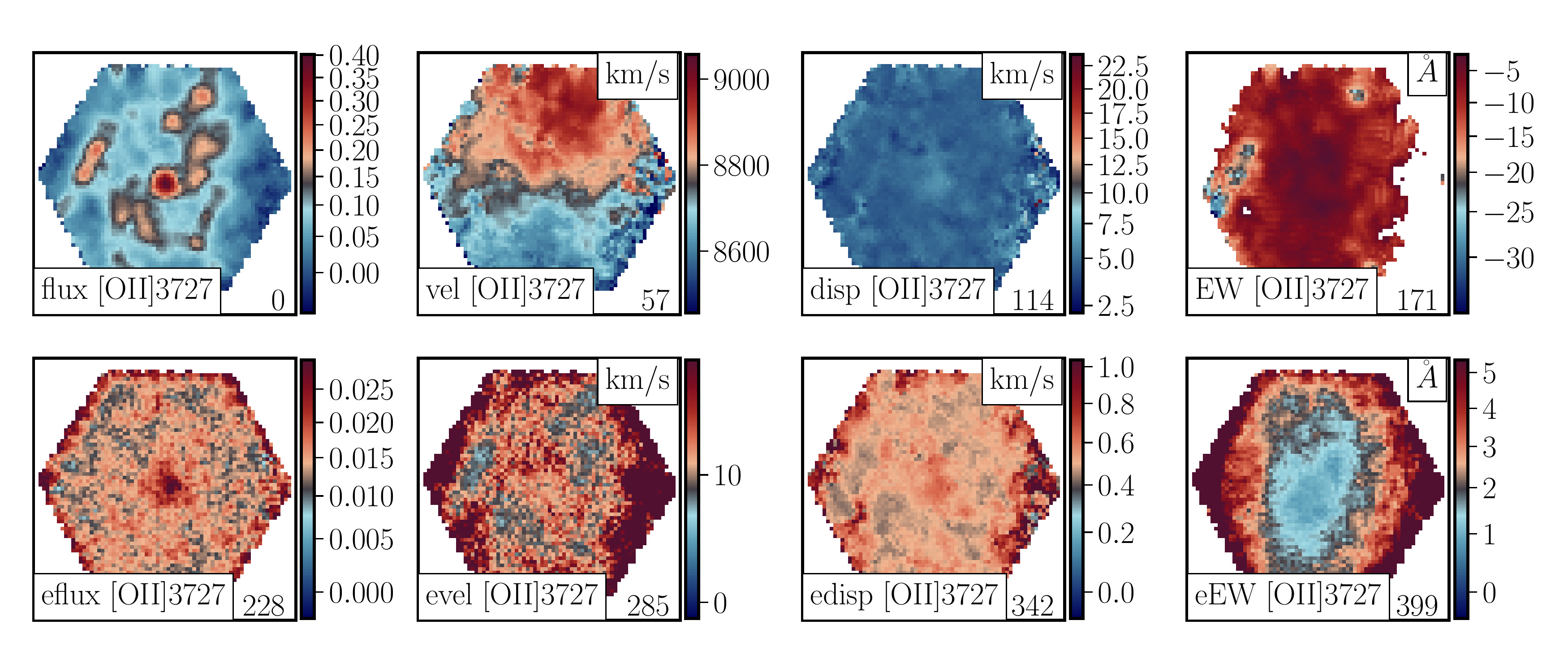}
 \endminipage
 \caption{Example of the content of the FLUX\_ELINES extension in the Pipe3D file, corresponding to the MaNGA dartacube (galaxy) manga-7495-12704. Each panel shows a color image with the content of a channel in this datacube. For each panel, the actual content is indicated in the lower-left, the channel number in the lower-right, and the units of the represented quantity in the upper-right legend. { For the flux and flux error the units are \funitsI}.}
 \label{fig:FELINES_map}
\end{figure*}
%%%%%%%%%%%%%%%%%%%%%%%%%%%%%%%%%%%%%%%%%%%%%%%%%%%%%%%%%%%%%%%%%%%%%%%5

\subsubsection{FLUX\_ELINES and FLUX\_ELINES\_LONG extensions}
\label{sec:flux_elines_cube}

The second analysis performed by \pyp to extract the main parameters of the emission lines is based on a weighted moment analysis, summarized in Sec. \ref{sec:pipe3d} and described in detail in \citet{pypipe3d} and \citet{pipe3d_ii}. This analysis is performed over a larger list of emission lines than the Gaussian fitting procedure discussed in the previous section. In former distributions of the MaNGA Pipe3D analysis \citep[DR14 and DR15,][]{sanchez18} we adopted the list of emission lines included in \citet{pipe3d_ii}, that was mostly focused on emission lines detectable in the wavelength range bluer than $\lambda$~=~7000\AA, since it was originally compiled to explore the ionized gas in the CALIFA dataset \citep{califa}. This list comprises a total of 56 emission lines. In this new distribution we decided to update the list, enlarging the number of emission lines for a better coverage of the redder wavelength range of the MaNGA dataset, and using a single/homogeneous set of wavelengths (and not a compilation). We adopted the list of emission lines in \citet{snr_elines}, selecting those lines within the formal wavelength range covered by the MaNGA dataset. The final list comprises 192 emission lines. For simpler compatibility and comparison with previous versions of the dataset, we include in the Pipe3D file two extensions, one for the analysis of the former list of emission lines (FLUX\_ELINES), and another one for the new updated list (FLUX\_ELINES\_LONG). Both extensions follow the same basic scheme, described in Table \ref{tab:fe}, comprising eight different parameters for each analyzed emission line: flux intensity, velocity, velocity dispersion and equivalent width (and their corresponding errors). Details on the two adopted sets of emission lines are presented in Appendix \ref{app:elines}, Tables~\ref{tab:fe_list} and \ref{tab:fe_long_list}. We note that in the ELINES and the two FLUX\_ELINES extensions, the velocity dispersion is given as a FWHM in \AA, without subtracting the instrumental dispersion. The following transformation should be applied to derive the velocity dispersion in km/s :
\begin{equation}
    \sigma_{\AA}={\rm FWHM}/2.354, \ \ \ \sigma_{\rm km s^{-1}} = \frac{c}{\lambda}\sqrt{\sigma_{\AA}^2-\sigma_{\rm inst}^2},
\end{equation}
where $c$ is the speed of light in km s$^{-1}$, $\lambda$ is the wavelength of the emission line and $\sigma_{\rm inst}$ is the instrumental resolution ($\sim$1.6\AA).

As for the previous extensions, we present in Figure \ref{fig:FELINES_map} an example of the content of the FLUX\_ELINES extension, corresponding to the properties derived for the [OII]3727 emission line for the datacube (galaxy) manga-7495-12704. As expected, we observe a similar distribution in the flux intensity reported for this emission line based on the moment-analysis than the one extracted using the Gaussian fitting (shown in Fig. \ref{fig:ELINES_map}). The patterns in the velocity and velocity dispersion are also similar to the ones found for H$\alpha$ using the Gaussian fitting, although they look noisier, as expected, since the [OII]3727 emission line is weaker through the FoV in this galaxy, and it is a blended doublet. Finally, EW(H$\alpha$) presents clear negative values, with an absolute value larger than 3\AA\ through most of the FoV of this dataset, in agreement with an scenario in which the ionization is dominated by young massive OB stars associated with recent star-formation activity \citep[e.g.,][]{sanchez20}.

%---------------------- FLUX_ELINES -------------------------- %
\begin{table}
\begin{center}
\caption{Description of the FLUX\_ELINES extensions.}
\begin{tabular}{cll}\hline\hline
Channel	& Units	& Description of the map\\
\hline
{\tt I}	    & 10$^{-16}$ \flux	& Flux intensity\\
{\tt I}+N	    &km/s	            & Velocity \\
{\tt I}+2N	&\AA	                & Velocity dispersion$^a$\\
{\tt I}+3N	&\AA	                & Equivalent width$^b$\\
{\tt I}+4N	&10$^{-16}$ \flux	& Flux error\\
{\tt I}+5N	&km/s	            & Velocity error\\
{\tt I}+6N	&\AA	            & Velocity dispersion error \\
{\tt I}+7N	&\AA	                & Equivalent width error\\
\hline
\hline
\end{tabular}\label{tab:fe}
\end{center}
{\tt I} is a running index over the set of emission lines
listed in Appendix~\ref{app:elines}, and $N$ is the number of analysed
lines. $N$~=~57 for FLUX\_ELINES and $N$~=~192 for FLUX\_ELINES\_LONG. 
$^a$FWHM, i.e., ~2.354$\sigma$. The instrumental velocity dispersion
has not been removed. 
$^b$We follow the convention in which the EW for
an emission line is negative and positive for an absorption line.
\end{table}
%---------------------- FLUX_ELINES ---------------------------- %

\subsubsection{GAIA\_MASK extension}
\label{sec:gaia}
%can use this data to check astrometry of MaNGA, result is generally 0.1"?? SFS:CHECK
%naturally possess GAIA-MaNGA cross-catalog, could this be useful? Easy to obtain GAIA-MaStar I assume, some study could use this to compare MaStar stellar properties to GAIA positional and kinetic properties of local stars?? Do we/could we distribute this?
%Alfredo should give it a read to see if I say anything stupid about GAIA data

Some data cubes have foreground field stars in their FoV, which need to be masked in order to prevent artifacts from appearing in the estimation of the maps of the  physical properties.
The Gaia DR3 catalog was used to find stars located within the FoV of each cube and mask them out. The Gaia DR3 catalog\footnote{https://www.cosmos.esa.int/web/gaia/dr3} \citep{gaia1,gaia3} is the most complete photometric and astrometric survey of the full sky with around 1.46 billion sources with a full astrometric solution. The astrometric information is useful because it provides a way to identify sources belonging to the Milky Way. The catalog was refined to include only sources with a measured parallax at least five times higher than its uncertainty. No further criteria in the selection of the Gaia stars were applied.
%We also include a limit on the magnitude of the stars such that objects fainter than 20 mag would not be considered.

With this information, each cube has its coordinate range cross-checked with the Gaia catalog to find field star candidates. If there are any, the algorithm checks for the presence of the star by fitting a PSF around the star coordinates. The fitted coordinates are then used to produce a mask image, a circular aperture of 2.5$\arcsec$ diameter is masked at the location of each star. This mask is stored in the GAIA\_MASK extension of the Pipe3D FITs file. This procedure allows for a further check on the precision of the astrometry of the MaNGA data cubes by comparing the Gaia coordinates to the fitted coordinates for the cube. We generally find good agreement, with typical discrepancies of $\sim$0.1". %\textbf{SFS:Check this number please, I have only my test results}
% ToDo

\subsubsection{SELECT\_REG extension}
\label{sec:reg}

%2D image including a mask of the region covered by the FoV (hexagon), and with a S/N>1sigma in the continuum emission.
This extension comprises an additional mask-image, masking all the spaxels within the FoV with a median S/N lower than 1 in the continuum emission within the wavelength range 5589-5680 \AA. To do so we use the \texttt{get\_SN\_cube} routine of the \pyf package \citep{pypipe3d}, that estimates the noise as the standard deviation of the flux intensity inside a given wavelength range. The flux standard deviation is derived for a datacube resulting from subtracting to the observed spectra a smoothed version of themselves, using a median filter of size { 6 spectral pixels} (i.e., 9\,\AA\ for the MaNGA data). This procedure removes the contribution of the shape of the spectra to the standard deviation, but not the contribution of the metallic features. Thus, the estimated error is an upper limit to the real one. Finally the S/N map is calculated dividing the mean value of the flux intensity by the estimated noise in the same wavelength for each spaxel. 

%\subsubsection{FLUX\_ELINES\_LONG cube}
%\label{sec:flux_elines_cube}

\subsection{A practical use of the Pipe3D resolved dataproducts}
\label{sec:ex_cube}

%%%%%%%%%%%%%%%%%%%%%%%%%%%%%%%%%%%%%%%%%%%%%%%%%%%%%%%%%%%%%%%%%%%%%%%5
\begin{figure*}
 \minipage{0.99\textwidth}
 \includegraphics[width=17.5cm]{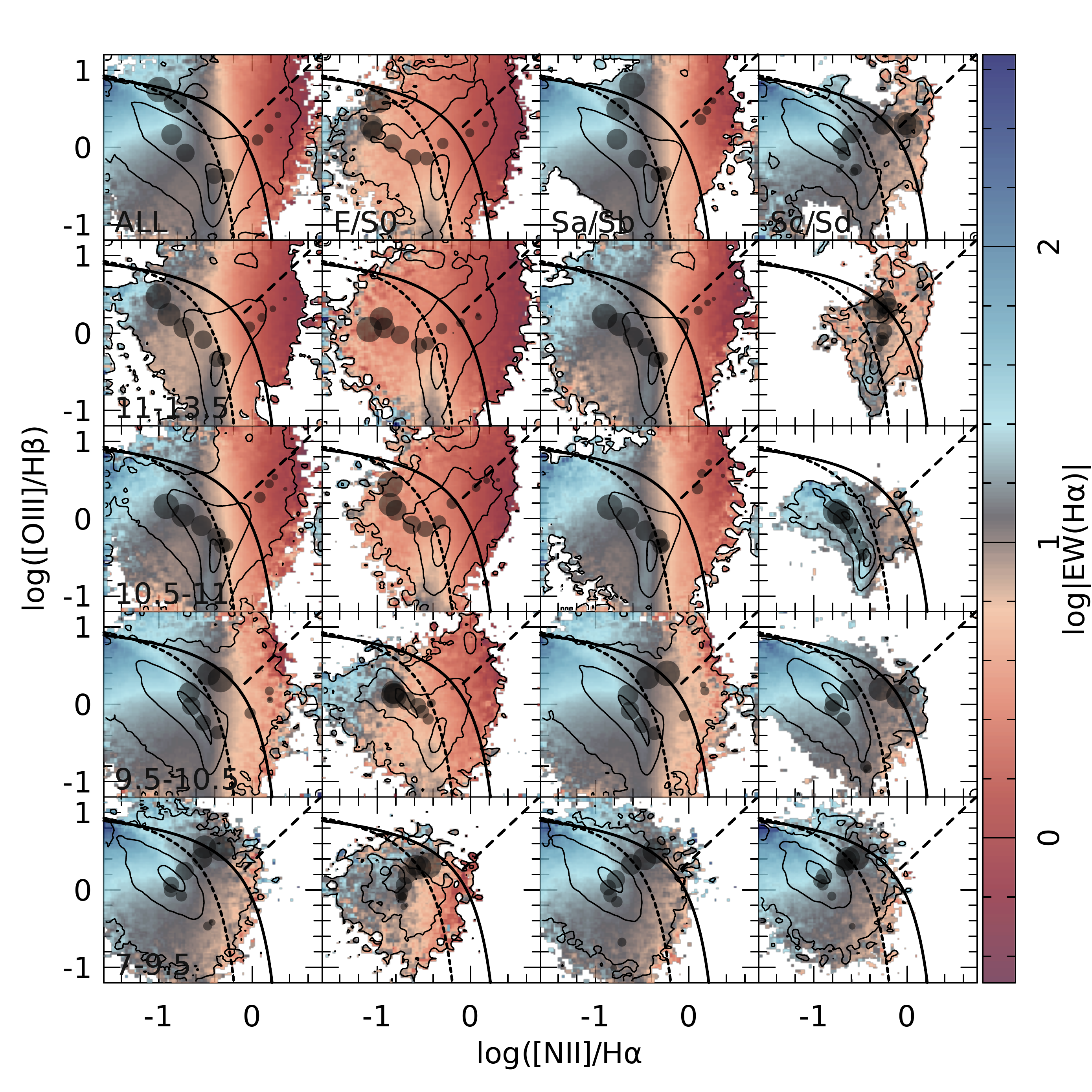}
 \endminipage
 \caption{Spatially resolved distribution of the sub-sample of 5624 well-resolved galaxies in the [OIII/H$\beta$]~vs.~[NII]/H$\alpha$ diagnostic diagram segregated by mass (rows) and morphology (columns), color-coded by the average EW(H$\alpha$) at each location (color distribution). Contours indicate the density of spaxels, with each consecutive contour encircling 90, 50 and 10\% of the points. The grey solid-circles correspond to the average location in this diagram of spaxels at different galactocentric distances, with the size of the circles being proportional to the distance (ranging from 0.1~Re to 2.1~Re in steps of 0.2~Re). In each panel the solid and short-dashed lines correspond to the \citet{kewley01} and \citet{kauff03} demarcation lines, frequently used to separate between SF- and AGN-like ionization. The long-dashed line corresponds to the separation between Seyferts and LINERs proposed by \citet{kewley01}}
 \label{fig:ex_BPT}
\end{figure*}
%%%%%%%%%%%%%%%%%%%%%%%%%%%%%%%%%%%%%%%%%%%%%%%%%%%%%%%%%%%%%%%%%%%%%%%5

The described set of dataproducts included in the current distribution comprises a vast dataset that can be used to explore a large number of spatially resolved properties of galaxies. As a simple showcase, we use this dataset to explore how the ionization conditions change across the optical extension of galaxies, as well as their dependence on galaxy properties. To do so, we emulate Figure 5 of \citet{ARAA}, a review aimed at characterizing the main spatially resolved properties of galaxies in the nearby Universe. This figure shows the distribution across the classical [OIII]/H$\beta$ vs. [NII]/H$\alpha$ diagnostic diagram of the full set of spatially resolved regions extracted from a compilation of data from four different IFS-GS (CALIFA, SAMI, MaNGA and AMUSSING++). This compilation comprises $\sim$8000 galaxies (i.e., it is of the same order of the final MaNGA dataset), from which we selected a sub-sample of $\sim$2000 well-resolved and well-sampled objects. Despite the size of this sample and the efforts to homogenize the dataproducts \citep[for more details, see][]{ARAA}, no compilation has the advantages of a well defined sample, like the one offered by MaNGA, considering that all the data were observed, reduced, and analyzed using the same procedures and tools. Following a similar philosophy, we select a subset of MaNGA datacubes with the best characteristics to explore the spatial distribution of the ionized gas. From the full sample, we retain only the galaxies/cubes observed with the largest IFUs (i.e., those comprising 127, 91 and 61 fibers), whose FoV covers at least 2 effective radii of the galaxy. This guarantees good sampling and coverage of the spatially resolved properties, based on the simulations presented by \citet{ibarra19}. Furthermore, we keep only those galaxies that are clearly resolved, i.e., whose Re is larger than the PSF FWHM of the MaNGA observations, following \citet{belf17}. Finally, we exclude highly inclined galaxies, i.e., the ones whose ellipticity is larger than 0.75. This latter condition is adopted to guarantee that the observed patterns result from changes with galactocentric distance, and are not an effect of a change of the ionization with vertical distance, due to the presence of shocks associated with galactic outflows \citep[e.g.][]{bland95,heckman90,carlos20} or extra-planar diffuse ionized gas \citep[e.g.][]{floresfajardo11,levy18}. The final sub-sample fulfilling all these criteria comprises $\sim$5,500 well-resolved/sampled galaxies/cubes.

Figure \ref{fig:ex_BPT} shows the distribution along the BPT diagram of the full set of spatially resolved regions (spaxels) included in the final sample of well-resolved/sampled galaxies, segregated by mass and morphology, color-coded by the average EW(H$\alpha$). The average location within this diagram for different galactocentric distances is indicated by a solid circle (whose size grows with radius). To generate this figure, the content of the {\tt FLUX\_ELINES} extensions in the \pyp\ FITs file was used, extracting the flux intensity maps of [OIII]$\lambda$5007, H$\beta$, [NII]$\lambda$6584 and H$\alpha$, and the map of EW(H$\alpha$). Then, for each galaxy two maps were derived, each one comprising the two line ratios involved in the diagram. We use these maps to derive (i) the azimuthal-average radial distribution of both line ratios, with the galactocentric distance normalized to the effective radius; (ii) the density distribution of spaxels across the BPT diagram; and (iii) the distribution of EW(H$\alpha$) across the same diagram. Finally, for each morphological and stellar mass sub-sample (and for the complete sample), we derive the average of all these properties. In this way, each galaxy contributes equally to the final distribution, irrespectively of its number of ionized regions (represented by the individual density distribution). 

The emerging patterns seeing in Figure~\ref{fig:ex_BPT} are very similar to those discussed already in \citet{ARAA}, although there are some mild but evident differences that are clearly related to the differences in the explored samples. For the complete dataset (upper-left panel), the peak of the density distribution traces the classical location of \HII\ regions \citep{osterbrock89}, associated with recent SF processes. The highest densities are found in the lower-right end of the distribution. There is a clear segregation of the location in the BPT diagram with EW(H$\alpha$), with most of the high(low)-EW regions being located at the left(right)-side of the diagram. This bimodality is a direct consequence of LINER-like regions (less frequent) being located preferentially in the central regions of galaxies while SF-like regions (more frequent) being located much further away. The above is reflected in the distribution of the average values at different galactocentric distances, being clearly above the \citet{kewley01} demarcation line for $R<$~0.6Re (the region dominated by the bulges), and below the \citet{kauff03} demarcation line for $R>$~Re (the region dominated by the disk, in late-type galaxies). The most plausible explanation for this pattern is that LINER-like ionization is dominated by the contribution of Hot-Evoled Low-Mass Stars \citep[HOLMES][]{floresfajardo11}, also known as post-AGB ionization \citep{binette94}. The above would be the reason why this ionization is ubiquitous in galaxies \citep{singh13}, and more frequently found in the central regions, dominated by old stellar populations \citep{belfiore17a,lacerda18}.

This pattern is modulated by both the stellar mass and the morphology. As galaxies are more massive and earlier, the less frequent is the presence of SF-like regions. The peak of the density distribution is shifted towards the LINER-like region, that is more relevant for this kind of galaxies. If there is still any SF activity, it is mostly located in the outer-regions (at least for M$_\star>$10$^{9.5}$M$_\odot$). On the contrary, as galaxies are less massive and later, the LINER-like component becomes less relevant and the dominance of the ionization by SF processes becomes more evident (to the extend that in the latest morphological bins (Sc/Sd), and in particular for M$_\star<$10$^{11}$M$_\odot$, there is almost no LINER-like component in the average distribution). An additional pattern is observed in the radial trend traced by the grey-solid circles in the SF-dominated regions. For galaxies more massive than M$_\star>$10$^{9.5}$M$_\odot$ there is a shift from the bottom-right end of the classical location of \HII\ region towards its upper-left end from the inner (0.6-0.8 Re) to the outer regions (1.8-2.1 Re) of the disks. This radial trend is directly associated with the oxygen abundance gradient observed in galaxies \citep[e.g.][]{searle73,vila92,sanchez14}. On the contrary, this trend is absent or even reverted for low mass galaxies (M$_\star<$10$^{9.5}$M$_\odot$). This suggests a possible flatenning of the oxygen abundance gradient in this mass bin \citep[e.g.,][]{belf17}. These results are in agreement with those presented in \citet{ARAA} and discussed in detail in \citet{sanchez20}, highlighting the clear connection between the ionization processes and the properties of the underlying stellar populations, that are the dominant ionizing sources in galaxies (on average). This connection is the main driver for the observed patterns and trends.

In this example we made use of a tiny fraction of the information distributed in the current dataset. Similar analysis using different stellar properties instead of the EW(H$\alpha$), would reveal the outlined connection. Explorations using different line ratios or other physical properties, like the stellar or the gas velocity dispersion \citep[e.g.][]{dagos19,law21} are easily obtained introducing simple modifications to the outlined procedure.

\subsection{Integrated and Characteristic parameters}
\label{sec:int}

%\subsection{Physical properties}
%\label{sec:phy}

From the results of the analysis performed by \pyp, we derive a set of spatially resolved properties of the stellar populations and the emission lines in the sample galaxies, following \citet{sanchez18}, \citet{ARAA} and \citet{sanchez20}. We do not distribute the entire set of resolved properties in the current release. However, we provide for each galaxy property either its integrated value (for extensive parameters, such as the stellar mass), or a characteristic value (for intensive
parameters, such as the oxygen abundance), and/or a value at a certain aperture. 
For the integrated extensive quantities we just co-add the corresponding values spaxel-by-spaxel, excluding those spaxels for which the quantity cannot be derived (we will include some examples below). For intensive properties that presents clear gradients along the galactocentric distance we derive their azimuthally averaged radial distribution. For doing so, we use the position angle, ellipticity, and effective radius (Re) provided by the NSA catalog\footnote{Petrosian $r-$band R50} for each galaxy, to create elliptical apertures of 0.15 Re width covering the galactocentric distance from 0 to 3.5 Re. Then, we estimate the average value for each parameter (and its standard deviation). From these radial distributions we derive the value at the effective radius and the slope of the average gradient based on a linear regression of the considered parameter along the radius (normalized to Re). This fitting is restricted to the galactocentric distance range between 0.5 and 2.0 Re. When the FoV does not reach this galactocentric distance, the regression is restricted to the largest distance covered by the FoV. We acknowledge that fitting the radial gradients { of the explored physical properties} to a linear relation is an oversimplification in many cases{ , as clearly demonstrated by results based on previous IFS-GS \citep[e.g., CALIFA,][]{rosa14,rosa15}}. For a more detailed description of the radial gradients and the validity of this approximation, we refer the reader to Barrera-Ballesteros et al. (in prep.).  It is worth noticing that for most of the explored quantities in galaxies, the value at the effective radius is usually considered a good proxy of the average one \citep[e.g.][]{mous10,pipe3d_ii}, being a better representation than the mean value for data with variable FoV (with respect to the optical extension of the galaxy). 

For some parameters we include also the central value, derived from a fixed aperture of 2.5$\arcsec$ diameter (i.e., one FWHM of the spatial PSF). These values reveal the effects of physical processes and structures present in the central regions of galaxies, such as AGNs, outflows, or strong bulges, without being necessarily representative of their average properties \citep[for a deeper discussion on this topic, see][]{ARAA,sanchez20}. For some particular parameters we also provide the values within 1Re aperture (i.e., not the value at the effective radius but the value within one effective radius). 

Table \ref{tab:cat} lists all the integrated, characteristic and aperture limited properties derived for each cube/galaxy. For completeness, we have included additional parameters extracted from other catalogs, together with the morphological classification, photometric/structural parameters, and quality-control flags described in Sec.~\ref{sec:morp}, \ref{sec:phot} and \ref{sec:qc}. In general, the error estimated for each quantity is labeled with the same name but with the prefix {\tt e\_}
(e.g., {\tt e\_log\_Mass} corresponds to the error of parameter {\tt log\_Mass}). 
The final catalog is included as an extension of the SDSS17Pipe3D\_v3\_1\_1.fits FITs file distributed in the following webpages: \url{http://ifs.astroscu.unam.mx/MaNGA/Pipe3D_v3_1_1/tables/} and \url{https://data.sdss.org/sas/dr17/manga/spectro/pipe3d/v3_1_1/3.1.1/}. This FITs file comprises three extensions. The first extension, HDU[0], is empty, just to fulfill the standard format adopted by the SDSS collaboration. The second extension, HDU[1], includes the described catalog, comprising 535 entries for each galaxy. The third extension, HDU[2], consists of row-stacked spectra comprising the individual spectra included in the {\tt MaStars\_sLOG} SSP library (Sec.~\ref{sec:ssp}), in the format required to run \pyf\ and \pyp\ \citep[][]{pypipe3d}.

%The final catalog is included as the 1st extension of SDSS17Pipe3D_v3_1_1.fits
%by averaging the values within radial bins of 0.15 R/Re width that follows position angle and ellipticity of the each object
\medskip
The derivation of the delivered parameters listed in Tab. \ref{tab:cat} has been described in detail in many previous articles. Below we provide a summary of the adopted procedures and some details for those parameters that require extra information.

%%%%%%%%%%%%%%%%%%%%%%%%%%%%%%%%%%%%%%%%%%%%%%%%%%%%%%%%%%%%%%%%
% Explanation of the individual parameters
%%%%%%%%%%%%%%%%%%%%%%%%%%%%%%%%%%%%%%%%%%%%%%%%%%%%%%%%%%%%%%%%

\subsubsection{Parameters inherited from the DRP}
\label{sec:cat_drpall}

The MaNGA Data Reduction Pipeline \citep[DRP,][]{law15} provides with a single file that comprises some basic information for each of the observed target. This information was either extracted from the NSA-catalog (see Sec. \ref{sec:sample}) or directly from the header information. We replicate some of this information in the current catalog to identify more easily the observed target, both in the MaNGA dataset and in the sky, and to facilitate comparisons with the properties derived in our analysis. The inherited entries in the catalog comprise the first seven columns of the table, including the object {\tt name}, observing {\tt plate} and IFU ({\tt ifudsgn}), as described already in Sec. \ref{sec:cubes}. It also includes the {\tt plateifu}, a combination of {\tt plate-ifudsgn} parameters, and the unique {\tt mangaid}, quantities frequently adopted within the community to identify the MaNGA targets. The right accession and declination of the objects, not necessarily corresponding to the center of the IFU, are also included ({\tt objra} and {\tt objdec} parameters). Most parameters extracted directly from the NSA catalog are explicitly labelled with the {\tt nsa} prefix (e.g., redshift, stellar mass, inclination, sersic index and the NSA unique identification number: {\tt nsa\_redshift}, {\tt nsa\_mstar}, {\tt nsa\_inclination}, {\tt nsa\_sersic\_n\_morph} and {\tt nsa\_nsaid}). We also list the effective radius of the galaxy adopted in this analysis, that corresponds to the Petrosian $r-$band R50 parameter ({\tt Re\_arc}), and its position angle ({\tt PA}). From these parameters we derive the luminosity ({\tt DL}) and angular ({\tt DA}) distances, the effective radius in kpc ({\tt Re\_kpc}), the relative size of the FoV with respect to Re ({\tt FoV}, defined as the radius of the circumscribed circle around the IFU in units of the effective radius, i.e. $FoV=\frac{r_{circ}}{Re}$) and
the ellipticity\footnote{For this parameters we use the eccentricity, defined as $\sqrt{1-\frac{b}{a}^2}$} ({\tt ellip}).
We correct the NSA stellar mass to our cosmology (i.e., $h=0.73$), but we do not apply any correction for the IMF \citep[they adopted the ][ IMF]{chab03}.

\subsubsection{Photometric and structural properties}
\label{sec:cat_phot}

In Sec.~\ref{sec:phot} we describe a set of photometric and structural properties that are included in our catalog (columns 468 to 511). For each photometric band ({\tt FILTER}: u, g, r, i, B, V, R\footnote{The $R-$band is denoted $RJ$ to avoid confusion with a radius}), we include (i) the rest-frame observed magnitude ({\tt FILTER\_band\_mag}) and its corresponding absolute magnitude ({\tt FILTER\_band\_abs\_mag}); (ii) the {\tt R50}, {\tt R90} and {\tt C} parameters; (iii) the {\tt B-V} and {\tt B-R} colors; (iv) the photometric stellar mass ({\tt log\_Mass\_phot}); and (v) the surface brightness in the $V-$band at the effective radius and at R50 ({\tt V-band\_SB\_at\_Re} and {\tt V-band\_SB\_at\_R50}, respectively).
For these quantities we include the corresponding errors, estimated from a MC simulation using the error extension in the MaNGA datacubes to perturb the individual spectra. Errors are indicated with a prefix, either {\tt e\_} or {\tt error\_}, or the suffix {\tt \_error}, as listed in Tab.~\ref{tab:cat}. All quantities were estimated from the datacube, except M$_{\star,phot}$, derived from Eq.~\ref{eq:Mphot}.

\subsubsection{Morphological properties}
\label{sec:cat_morp}

The morphology of each galaxy was derived as described in Sec. \ref{sec:morp}. We include in the catalog all the parameters required and derived by that procedure (rows 512 to 532), including (i) the colors ({\tt u-g}, {\tt g-r}, {\tt r-i} and {\tt i-z}) and the Sersic index extracted from the NSA catalog, (ii) the probability {\tt P(MORPH)} that the galaxy is classified in each of the thirteen {\tt MORPH} groups (cD, E, S0, Sa, Sab, Sb, Sbc, Sc, Scd, Sd, Sdm, Sm and Irr), and (iii) the best morphological type estimated by the analysis, as an integer ({\tt best\_type\_n}) and as an alphanumeric code ({\tt best\_type}). The parameter {\tt best\_type\_n} runs from -2 for cD to 10 for Irr galaxies.
%
% BUG: Best_type contain just nans!
%
%. For each morphological type it provides with a certain probability
%To derive those morphology it was used the colors and sersic indices provided by the NSA catalog

%%%%%%%%%%%%%%%%%%%%%%%%%% Mass-Comparison %%%%%%%%%%%%%%%%%%%%%%%%%%%%%%%%%%%%%%%%%
%%%%%%%%%%%%%%%%%%%%%%%%%%%%%%%%%%%%%%%%%%%%%%%%%%%%%%%%%%%%%%%%%%%%%%%5
\begin{figure}
 \minipage{0.49\textwidth}
 \includegraphics[width=4.5cm,clip,trim=0 10 0 10]{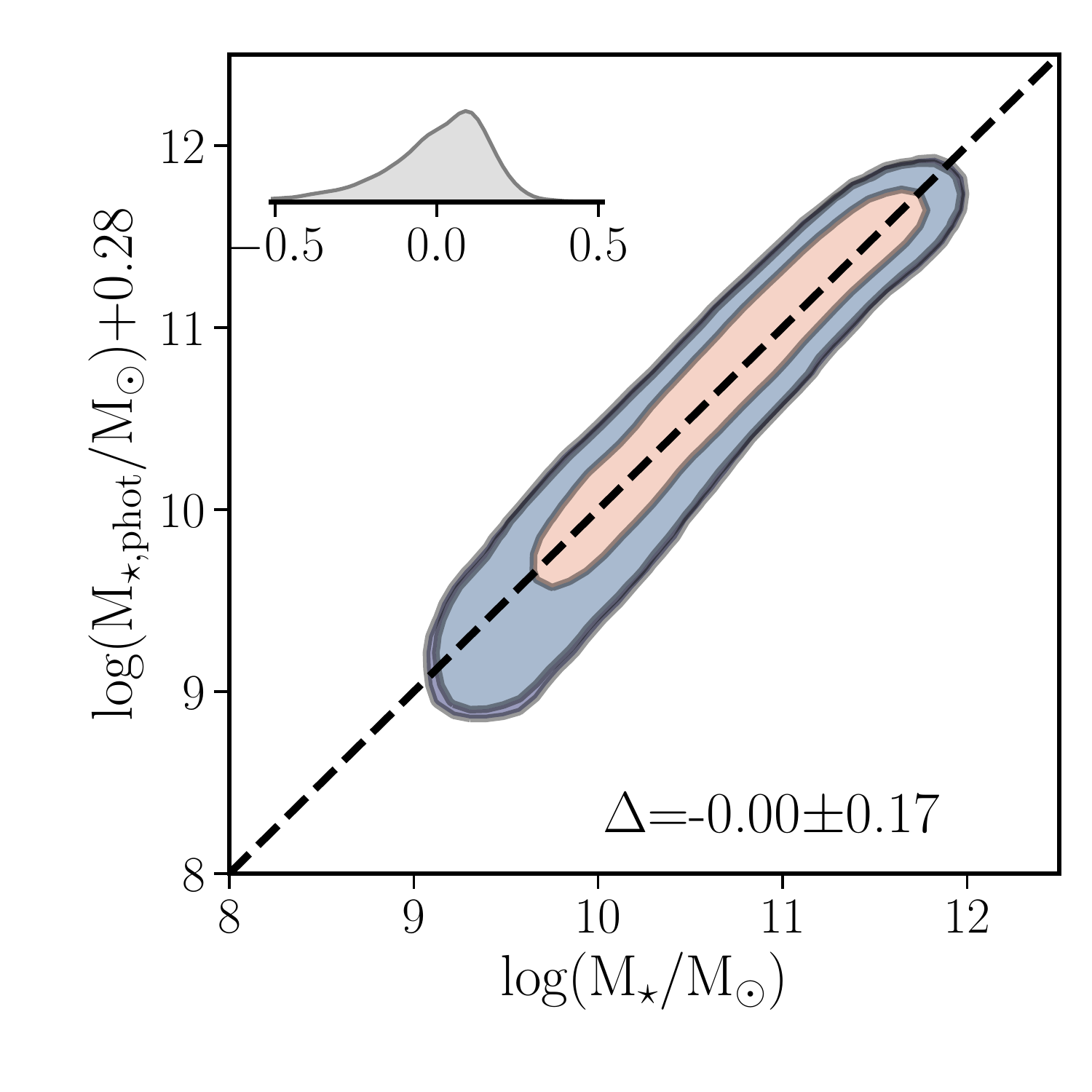}
 \includegraphics[width=4.5cm,clip,trim=0 10 0 10]{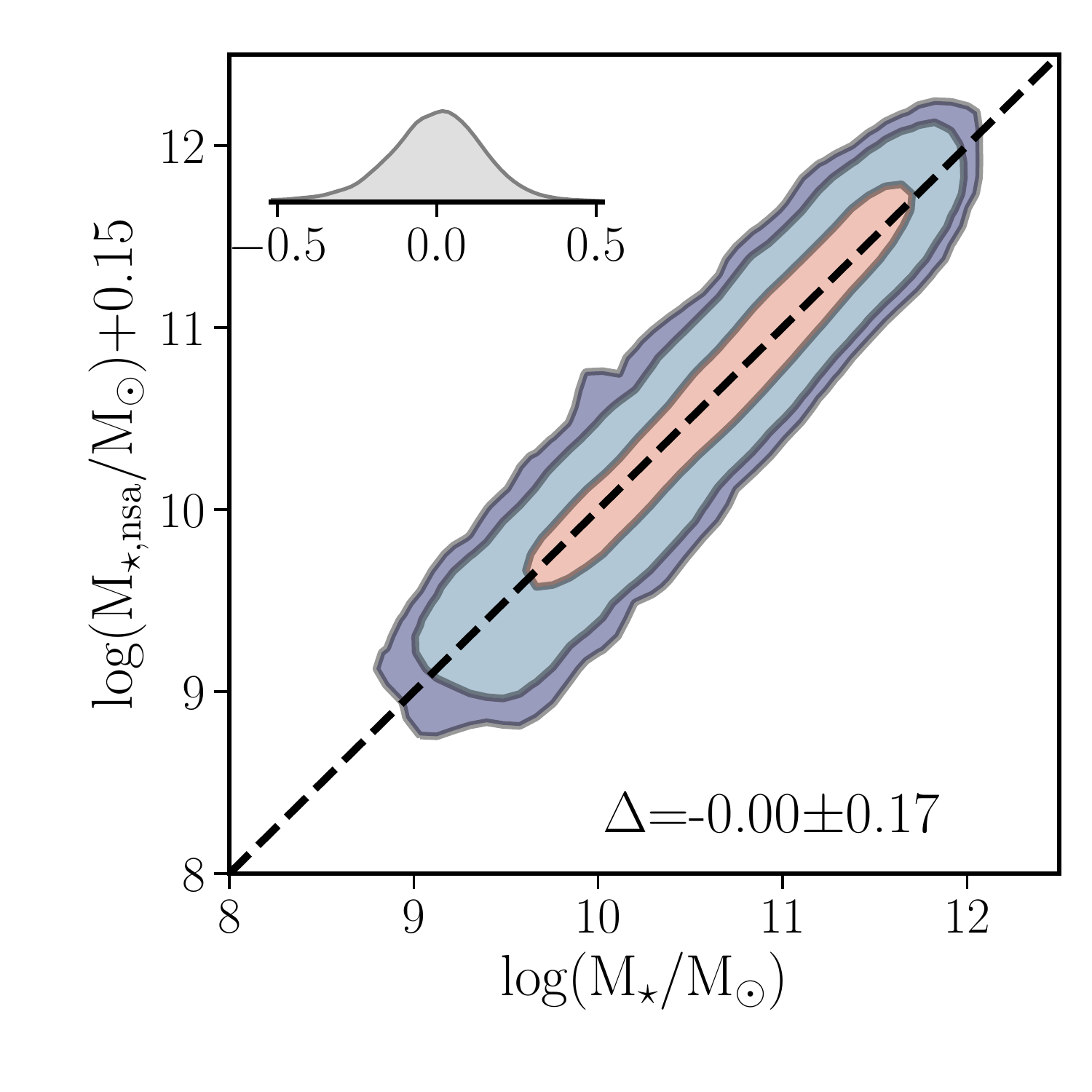}
 \endminipage
 \caption{Comparison between the stellar masses derived from the stellar population decomposition performed as part of the \pyp\ analysis (M$_\star$) and the stellar masses derived by ourselves using only photometric information (M$_{\star,\rm phot}$, left panel) and the stellar mass included in the NSA catalog (M$_{\star,\rm NSA}$, right panel), both of them included in our catalog. An offset of 0.28 dex and 0.15 dex has been applied to M$_{\star,\rm phot}$ and M$_{\star,\rm NSA}$ to correct for the different adopted IMF \citep[][in both cases]{chab03} and the different cosmology for the NSA values. We adopt the same format of Fig.~\ref{fig:comp_ssp_DR15}.}
 \label{fig:comp_Mass}
\end{figure}
%%%%%%%%%%%%%%%%%%%%%%%%%%%%%%%%%%%%%%%%%%%%%%%%%%%%%%%%%%%%%%%%%%%%%%%5
%

%%%%%%%%%%%%%%%%%%%%%%%%%%%%%%%%%%%%%%%%%%%%%%%%%%%%%%%%%%%%%%%%%%%%%%%5
\begin{figure*}
 \minipage{0.99\textwidth}
 \includegraphics[width=17.5cm]{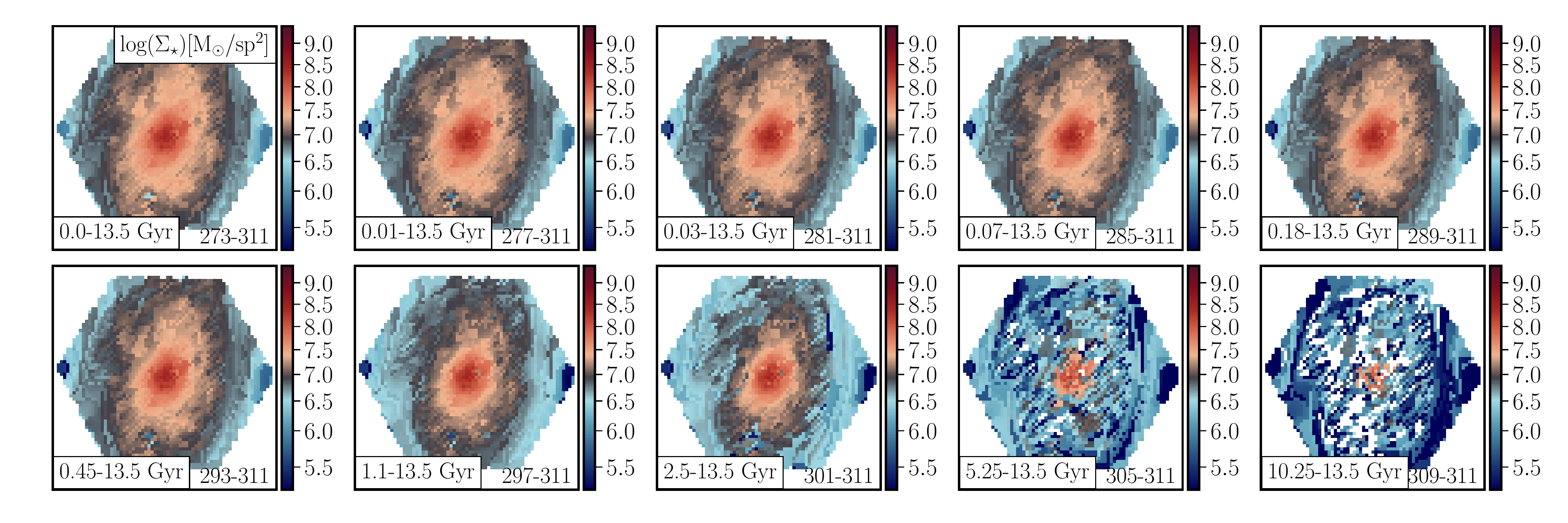}
 \endminipage
 \caption{$\Sigma_\star$ derived from the weights included in the SFH extension using the mass-to-light ratio of each SSP and the flux intensity in each tessella (spatial bin) for different age ranges, indicated in the bottom-left inset, corresponding to the channels in the bottom-right legend. The first panel shows the actual $\Sigma_\star$, corresponding to integration along all look-back times. The subsequent panels show the cumulative $\Sigma_{\star,t}$ at the look-back time $t$ corresponding to the lower age of the considered range, without considering the redshift of the target. The age (\age) of the stellar population and the look-back time of the universe at which it was formed ($\tau_{lbt,\star}$) are related via the look-back time corresponding to the redshift at which it was observed ($\tau_{lbt,z}$): $\tau_{lbt,\star}$=$\tau_{lbt,z}$+\age. For simplicity, in this plot we consider $\tau_{lbt,z}\sim$0.}
 \label{fig:SFH_CM_map}
\end{figure*}
%%%%%%%%%%%%%%%%%%%%%%%%%%%%%%%%%%%%%%%%%%%%%%%%%%%%%%%%%%%%%%%%%%%%%%%5

\subsubsection{Stellar population related quantities}
\label{sec:cat_star}
%\subsubsection{Stellar Mass related quantities}
%\label{sec:cat_mass}

{\bf Stellar Masses:} Most of the stellar parameters from which we derive the integrated and characteristic properties of each galaxy result from the decomposition of the stellar spectra in the set of SSPs described in Sec.~\ref{sec:pipe3d} and \ref{sec:ssp_cube}. We estimate the integrated stellar mass (M$_\star$, labelled as {\tt log\_Mass}) by co-adding the stellar surface-density ($\Sigma_\star$, Eq.~\ref{eq:mu}) through the unmasked region (Sec.~\ref{sec:reg}) within the FoV of the data. { Regions contaminated by foreground stars included in the {GAIA\_MASK} extension of the Pipe3D file have been masked in the derivation of any integrated quantity, including M$_\star$}. From $\Sigma_\star$ we derive not only M$_\star$ but also (i) the mass within R50 ({\tt log\_Mass\_corr\_in\_R50\_V}) using the parameters described in Sec. \ref{sec:phot}, and the mass within one Re ({\tt log\_Mass\_in\_Re}); (ii) the radius enclosing 50\%\ of the stellar mass in kpc ({\tt R50\_kpc\_Mass}, not to be confused with {\tt R50\_kpc\_V}, the radius enclosing 50\%\ of the V light in kpc, i.e., {\tt R50} transformed to kpc); (iii) the stellar mass surface density in the central aperture ({\tt Sigma\_Mass\_cen}), at one effective radius ({\tt Sigma\_Mass\_Re}), and averaged within all the FoV ({\tt Sigma\_Mass\_ALL}). The derivation of $\Sigma_\star$ is based on the estimates of the mass-to-light ratio ($\Upsilon_\star$, Eq.~\ref{eq:ML}) spaxel-by-spaxel, from which we also derive its average value across the FoV ({\tt ML\_avg}). A different method to estimate the average $\Upsilon_\star$ across the entire galaxy is to divide the integrated stellar mass by the integrated luminosity ({\tt ML\_int}). Both quantities are listed in the catalog. 
%
% Write the units! Missing in table!
%

%Among the properties delivered in the final catalog we have included different estimates of the same parameter derived using different procedures. Of them, the most relevant ones are the stellar mass, the SFR, and the different estimates of the oxygen and nitrogen abundances. 

Figure~\ref{fig:comp_Mass} shows the comparison between the stellar mass derived by us from our photometry (M$_{\star,\rm phot}$, Sec.~\ref{sec:cat_phot}),
the stellar mass provided by the NSA catalog based on multiband photometry (M$_{\star,\rm NSA}$, Sec. \ref{sec:cat_drpall}), and the stellar mass derived by spectral fitting using \pyp (M$_{\star}$, Sec.~\ref{sec:cat_star}). 
We applied a systematic offset to the photometric masses. In the case of M$_{\star,\rm phot}$, the offset is a pure consequence of the different adopted IMF \citep[0.28 dex, according to][]{rosa16a}. However, in the case of M$_{\star,\rm NSA}$, applying a similar offset produces an over-correction, with NSA masses being systematically larger than masses derived by \pyp. Aperture issues may be behind this difference. It is known that MaNGA apertures miss $\sim$22\%\ of the total flux \citep[e.g.][]{pace19b}. On average, we find that an offset of 0.15 dex is a better correction to match the NSA and \pyp\ stellar masses. Once corrected, the photometric stellar masses agree with those derived using the stellar synthesis code within $\sim$0.17 dex, which is similar to the differences found by other groups \citep[e.g.,][]{pace19b}. Furthermore, this difference is of the order of the expected error in the stellar mass due to uncertainties in the $\Upsilon_\star$-color relation, estimated to be $\sim$0.09-0.20 dex \citep[e.g.][]{rgb18,fraser19}.

{\bf Age and metallicity:} We derive from our analysis the LW and MW stellar age and metallicity, spaxel-wise as described in Sec.~\ref{sec:ssp_cube} following Eqs~\ref{eq:LW} and \ref{eq:MW}. We report their characteristic values at the effective radius, {\tt PAR\_LW\_Re\_fit} and {\tt PAR\_MW\_Re\_fit}, the slope of their radial gradients, {\tt alpha\_PAR\_LW\_Re\_fit} and {\tt alpha\_PAR\_MW\_Re\_fit}, and the corresponding errors, labelled with an {\tt e\_} prefix. {\tt PAR} corresponds to either the {\tt Age} or the metallicity {\tt ZH}. These parameters are listed in the catalog in rows 24 to 39. Additionally, we include the dust extinction at the effective radius derived from the analysis of the stellar population, {\tt Av\_ssp\_Re} (row 173) and its error {\tt e\_Av\_ssp\_Re} (row 174).

{\bf Star-formation and chemical enrichment histories:} $\Sigma_\star$, M$_\star$, \age and \met can be estimated at any look-back time ($\tau_{lbt}$) by integrating the corresponding equations (e.g., Eq. \ref{eq:LW}) from the oldest age in the SSP decomposition down to the age matching $\tau_{lbt}$. This is the standard procedure broadly applied in the literature to derive, for instance, the mass assembly history \citep[MAH,][]{eperez13,ibarra16,ibarra19}, the star-formation history {\citep[SFH,][]{panter07,rosa17,lopfer18,sanchez18b}}, or the chemical enrichment history \citep[ChEH,][]{vale09,camps20,camps22}. This derivation can be applied to integrated, characteristic, and/or spatially resolved properties. In Figure ~\ref{fig:qc_map}, left-panels, we show the normalized MAH and the ChEH at different galactocentric distances for the example galaxy/cube manga-7495-12704. Fig.~\ref{fig:SFH_CM_map} illustrates the derivation of a spatially resolved parameter for different $\tau_{lbt}$, by showing the cumulative $\Sigma_{\star,t}$ for the same prototype/example galaxy, derived making use of the content of the SFH extension, and applying Eqs.~\ref{eq:ML} and \ref{eq:mu} for a range of ages, labelled in each panel of the figure, together with the corresponding slices of the SFH cube. Integrating $\Sigma_{\star,t}$ through the FoV of the IFU vs. $\tau_{lbt}$ we derive the stellar mass vs. time, the MAH. A correction to account for mass in stars that are  dead at the observing time, but were still shining at $\tau_{lbt}$, should be applied \citep[e.g.,][]{court13}. The SFH (integrated or resolved) can be obtained from the MAH (or $\Sigma_{\star,t}$) dividing the differential mass accumulated at two consecutive $\tau_{\star,t}$ by the corresponding time interval, namely,
\begin{equation}\label{eq:sfh_int}
  {\rm SFR}_{\rm ssp,t} = \frac{\Delta {\rm M}_{\star,{\rm t}}}{\Delta {\rm t}},\ \ \ \ \ \ \ \   {\rm \Sigma}_{\rm SFR,ssp,t} = \frac{\Delta {\rm \Sigma}_{\star,\rm t}}{\Delta {\rm t}},
\end{equation}
where SFR$_{\rm ssp,t}$ ($\Sigma_{\rm SFR,ssp,t}$) is the star-formation (density) at a particular look-back time $t$, $\Delta t$ is the time interval between two consecutive $\tau_{lbt}$ (ages in the SSP library), and $\Delta$M$_{\star,{\rm t}}$ ($\Delta \Sigma_{\star,\rm t}$) is the differential mass (density) assembled during $\Delta t$. When $\tau_{lbt}$ is short enough, the estimated SFR correspond to the current one \citep[e.g.,][]{rosa16}. In our catalog we include three estimates of SFR based on the analysis of the stellar population. One corresponding to the last 32 Myr ({\tt log\_SFR\_ssp}), adopting the time interval proposed by \citet{rosa16} and already tested in \citet{sanchez18b}, and two additional values corresponding to 10 and 100 Myr, {\tt log\_SFR\_ssp\_10Myr} and {\tt log\_SFR\_ssp\_100Myr}, respectively. The latter intervals correspond to the maximum time at which an OB star can still produce some ionizing photons before it fades out (10 Myr), and the typical time interval associated with the SFR measured using far infrared emission \citep[100 Myr, e.g.,][]{kennicutt98,catalan15}.

{\bf SFH Time-scales:} Following \citet{eperez13} and \citet{rgb17}, we estimate the age at which a fraction {\tt X} (from 30 to 99\%) of the current stellar mass was formed, based on the integrated MAH derived for each galaxy. Then, {\tt T30} 
({\tt T80}) indicate the time at which 30\% (80\%) of the stellar mass was formed. In addition, making use of the integrated and spatially resolved ChEH, we derive (i) the average stellar metallicity, normalized to the solar value, in logarithmic-scale at the same look-back times, e.g., {\tt ZH\_T30} corresponds to the metallicity at the time at which 30\%\ of the current mass was formed, (ii) the metallicity at the effective radius ({\tt ZH\_Re\_T\%}), and (iii) the radial gradient of the metallicity at the same time ({\tt a\_ZH\_T\%}). These quantities are listed in the catalog in rows 107 to 142.

%%%%%%%%%%%%%%%%%%%%%%%%%%%%%%%%%%%%%%%%%%%%%%%%%%%%%%%%%%%%%%%%%%%%%%%5
\begin{figure}
 \minipage{0.49\textwidth}
\includegraphics[width=4.5cm,clip,trim=0 0 0 0]{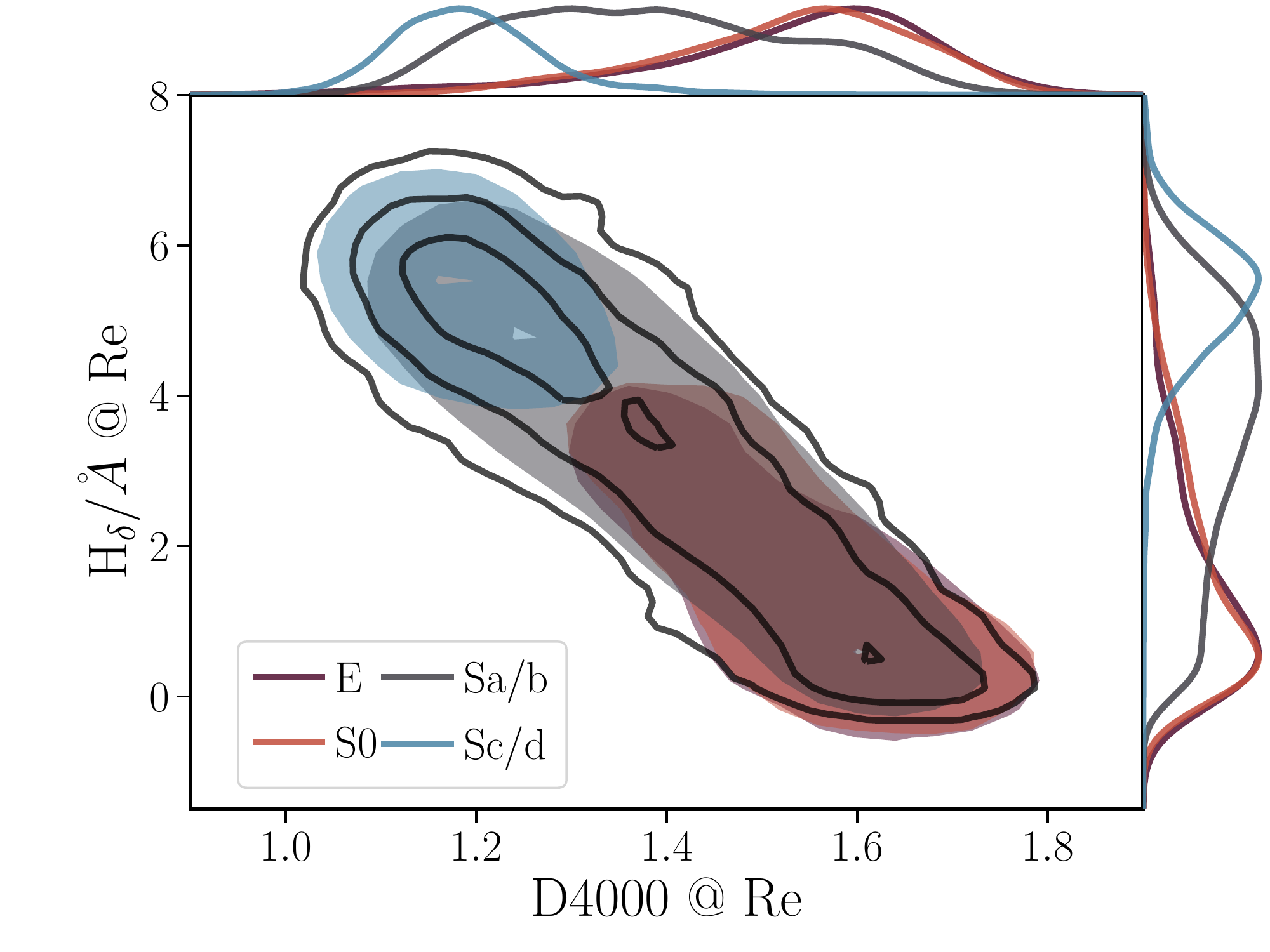}
\includegraphics[width=4.5cm,clip,trim=0 0 0 0]{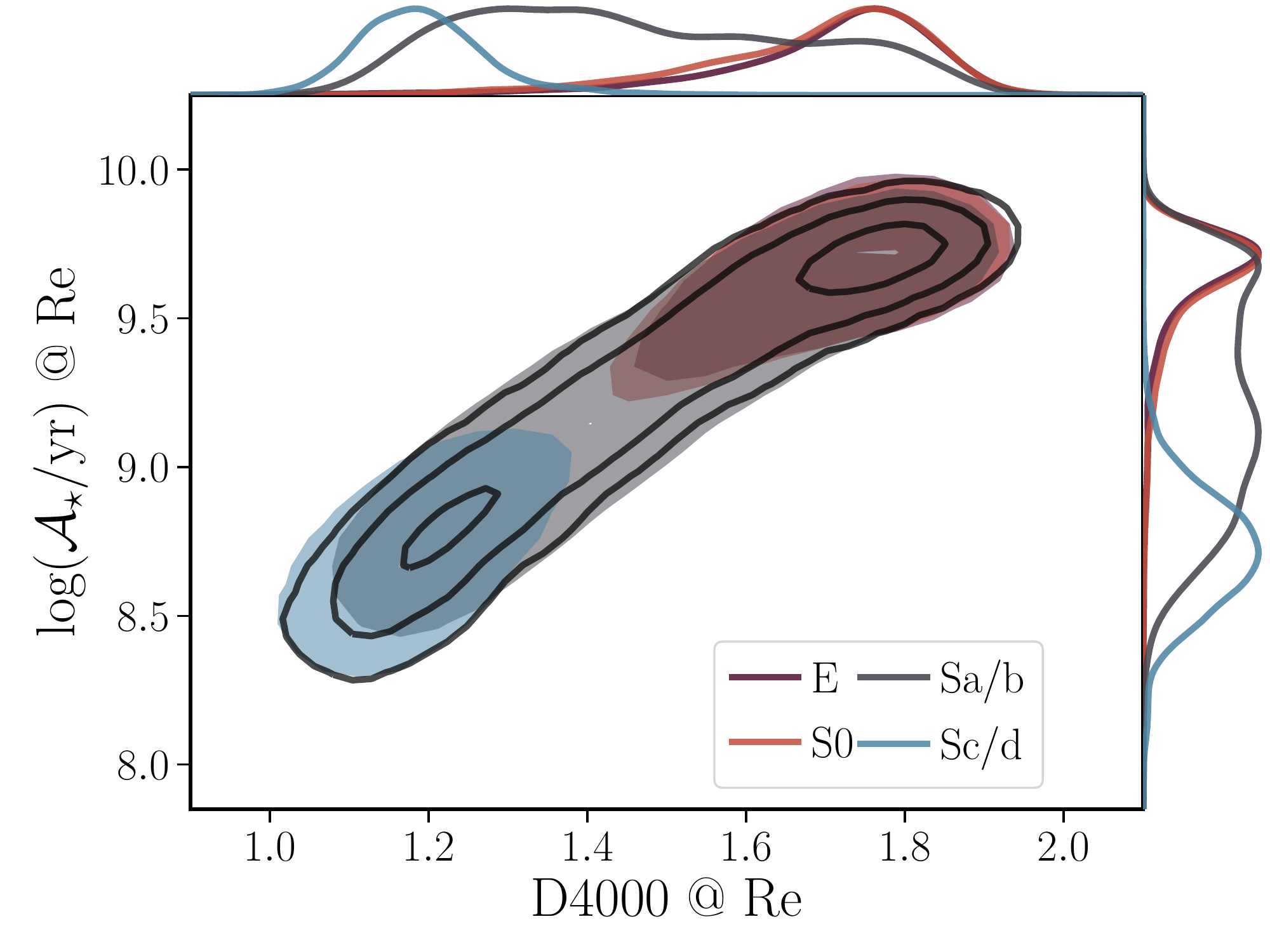}
\includegraphics[width=4.5cm,clip,trim=0 0 0 0]{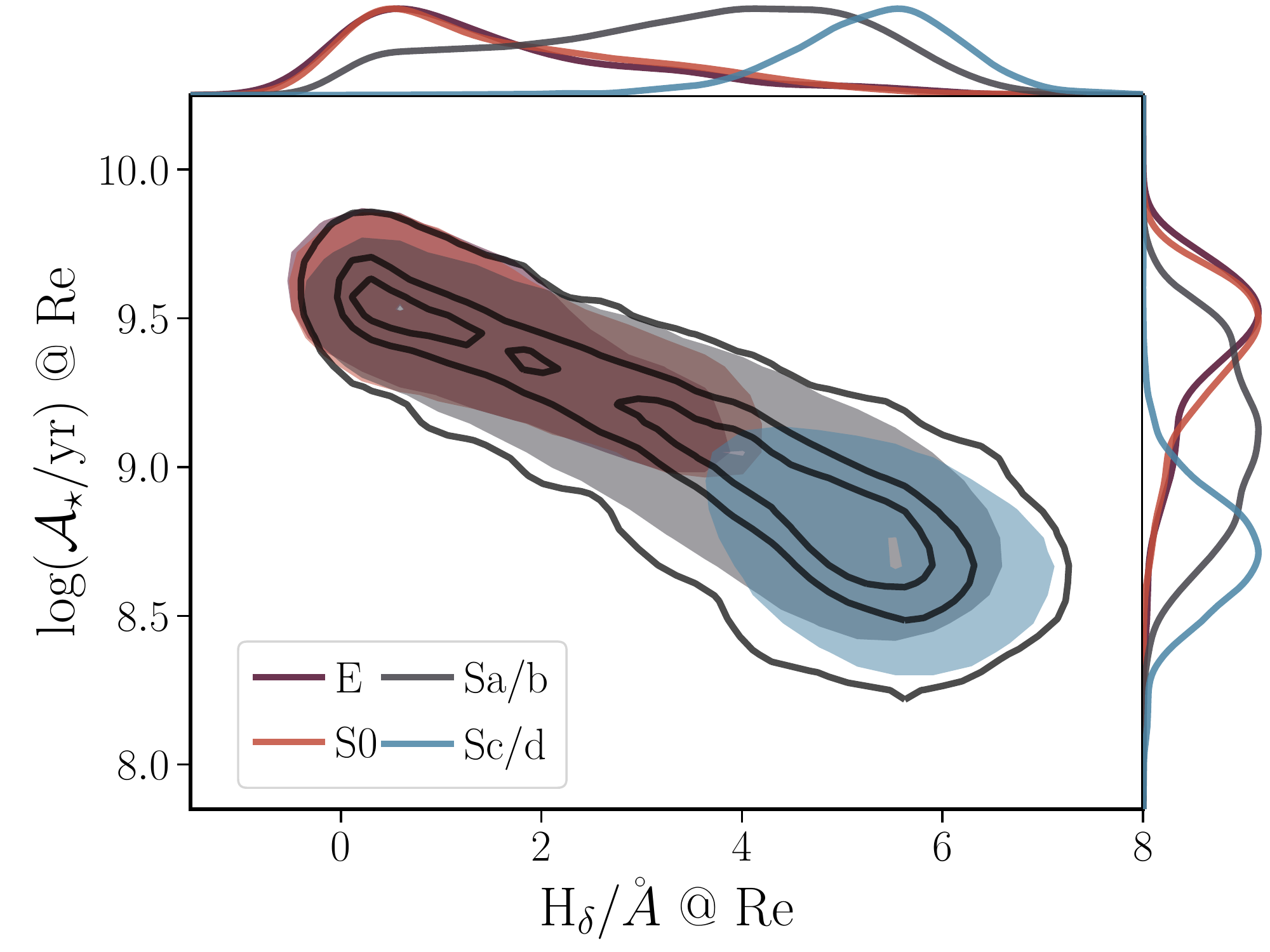}
\includegraphics[width=4.5cm,clip,trim=0 0 0 0]{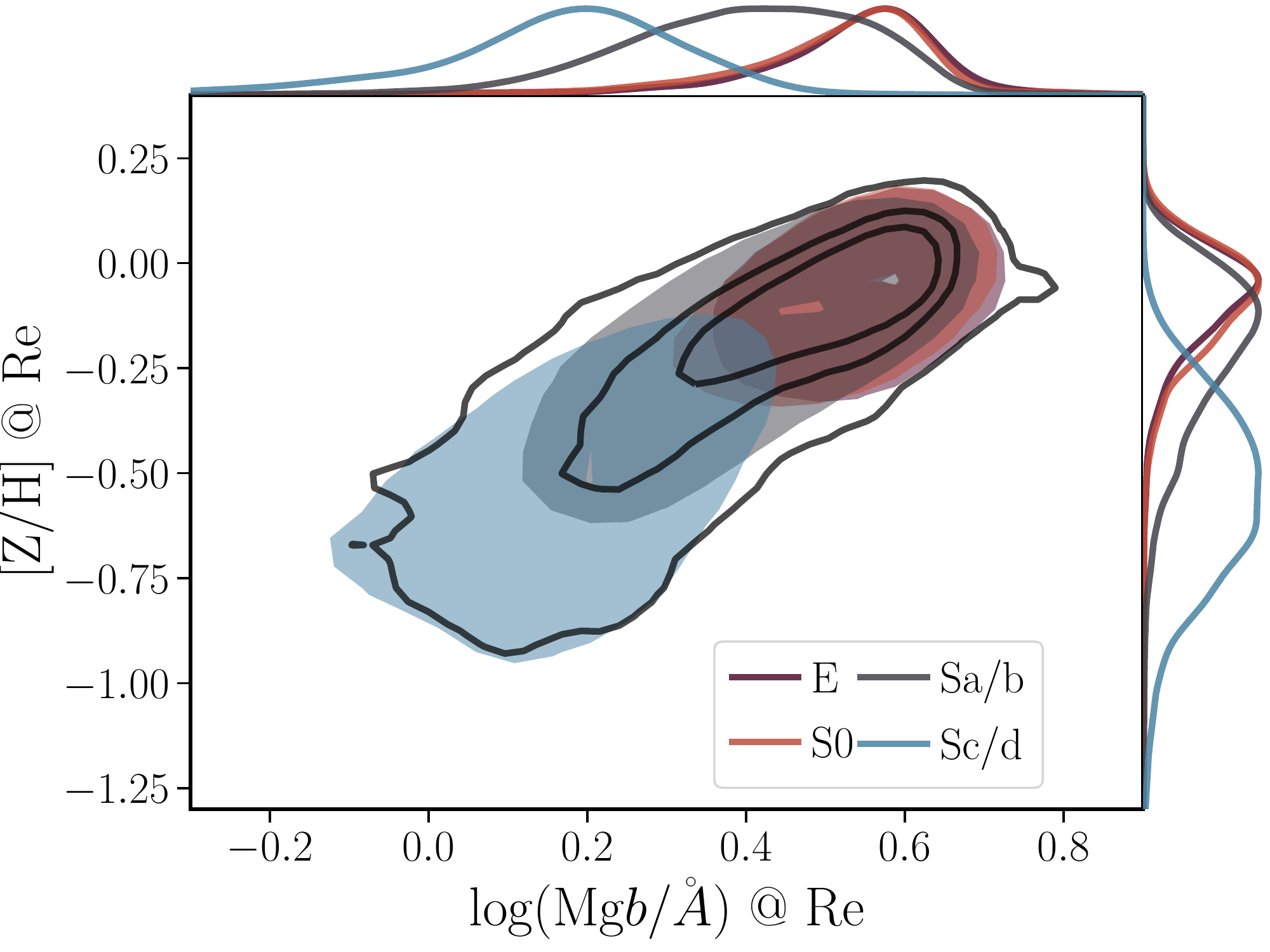}
 \endminipage
 \caption{{\it Top-left panel:} Distribution of the H$\delta$ spectral index vs. D4000 measured at the effective radius for the full sample of galaxies (contours encircling 95, 65 and 40\% of the objects, successively). The location of different morphological types (E, S0, Sa/b, Sc/d) is color-coded in the shaded areas. The upper and right sub-panels represent histograms of the parameters segregated by morphological type. {\it Top-right panel:} Distribution of \ageLW vs. D4000. {\it Bottom-left panel:} Distribution of \ageLW\ vs. H$\delta$. {\it Bottom-right panel:} Distribution of \metLW\ vs. Mg$b$.}
 \label{fig:ind}
\end{figure}
%%%%%%%%%%%%%%%%%%%%%%%%%%%%%%%%%%%%%%%%%%%%%%%%%%%%%%%%%%%%%%%%%%%%%%%5

%\subsubsection{Stellar-indices related quantities}
%\label{sec:cat_index}

{\bf Stellar Spectral Indices:} The \pyp\ analysis provides us with the spatial distribution of a set of stellar spectral indices for each galaxy/cube, as described in 
Sec.~\ref{sec:indices_cube}. For each index we estimate its value at the effective radius ({\tt ID\_Re\_fit}), the slope of its radial gradient ({\tt ID\_alpha\_fit}), and corresponding errors (labelled with the {\tt e\_} prefix), where {\tt ID} is the index identification in Table~\ref{tab:index}. The indices listed in rows 436 to 467 of the final catalog (Table~\ref{tab:cat}). 

Spectral indices are frequently used to derive average properties of stellar populations, such as age and metallicity \citep[e.g.,][]{gallazzi05}. Some indices, such as D4000, are sensitive to the age of the stellar population, tracing the fraction of young to intermediate-to-old stars, or H$\delta$, which is more sensitive to the presence of young stars. Other indices, like Mg$b$ are sensitive to metallicity. Since indices permit an alternative analysis to the stellar decomposition at the core of \pyp, it is worth comparing the results from both methods. Figure \ref{fig:ind} shows (i) the distribution of H$\delta$ as a function of D4000; (ii) the distribution of \ageLW as a function of these two indices, and (iii) the distribution of \metLW as a function of Mg$b$, where all the index values are estimated at the effective radius. As additional information, we indicate the location of different morphological types in these diagrams. The H$\delta$-D4000 diagram shows the typical distribution observed for the bulk population of nearby galaxies \citep[e.g.,][]{kauff03b}, a clear anti-correlation in the sense that galaxies with higher H$\delta$ (young stellar populations) have a lower D4000 (a lower fraction of old stars with respect to young ones). Late-type (early-type) galaxies are found at the upper-left (lower-right) area of the distribution. \ageLW\ shows a well defined positive (negative) relation with D4000 (H$\delta$), with different morphological types following the expected behavior. Finally, \metLW\ presents a well defined positive relation with Mg$b$, that seems to be less tight than that tracing \ageLW with D4000 or H$\delta$. As expected, early-type (late-type) galaxies are found in the regime of high (low) \metLW and Mg$b$ values. This comparison suggests that our estimated stellar indices are indeed good tracers of the stellar content.

%We include in the catalog the look-back time at which it is formed a 

%All stellar parameters derived using our analysis are the result of co-adding (e.g., stellar mass) or averaging (e.g., \age or $\Upsilon_\star$) the corresponding value for the different SSP in the adopted library (e.g. Eq. \ref{eq:LW} or Eq. \ref{eq:ML}. The same calculation could be derived for a certain range of ages.
%Finally, following the procedure illustrated in Fig. \ref{fig:SFH_CM_map}

\subsubsection{Emission line related quantities}
\label{sec:cat_elines}

We derive a wide variety of parameters based on the emission line analysis performed by \pyp. First, we determine which are the most frequently detected emission lines. For doing so we explore how often the different emission lines included in the {\tt FLUX\_ELINES\_LONG} extension (Sec.~\ref{sec:flux_elines_cube}, Table~\ref{tab:fe_long_list}) are detected in all the analyzed datacubes. We select emission lines that are detected with a S/N ratio larger than 3 in at least 5\%\ of all galaxies/cubes (i.e., in at least $\sim$500 galaxies). For these emission lines we estimate the flux intensity at the effective radius and the slope of the radial gradient ({\tt flux\_ELINE\_Re\_fit} and {\tt flux\_ELINE\_alpha\_fit}), and their corresponding errors (labelled with the {\tt e\_} prefix) following the procedure described above (Sec. \ref{sec:int}). For each parameter, {\tt ELINE} corresponds to the Id+wavelength listed in Tab. \ref{tab:fe_long_list}, where the most frequently detected emission lines are labelled with an $^*$ symbol. All those parameters and errors comprise the rows between 180 and 307 of the final catalog, as listed in Tab. \ref{app:cat}. 

%Fig. \ref{fig:comp_Mass} shows the comparison between the stellar masses derived using the 
%%%%%%%%%%%%%%%%%%%%%%%%%%%%%%%%%%%%%%%%%%%%%%%%%%%%%%%%%%%%%%%%%%%%%%%5
\begin{figure}
 \minipage{0.49\textwidth}
\includegraphics[width=4.5cm,clip,trim=0 10 0 10]{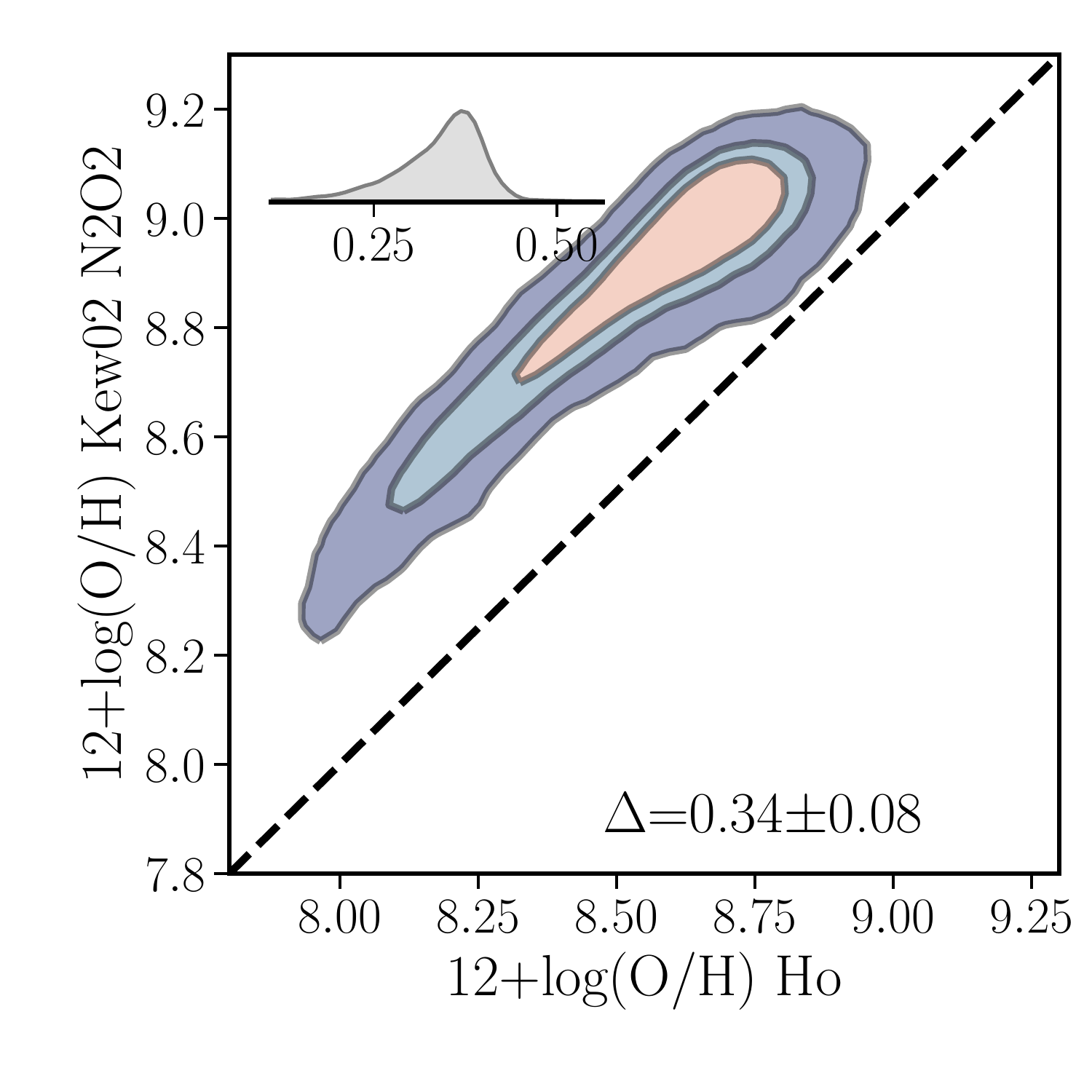}
\includegraphics[width=4.5cm,clip,trim=0 10 0 10]{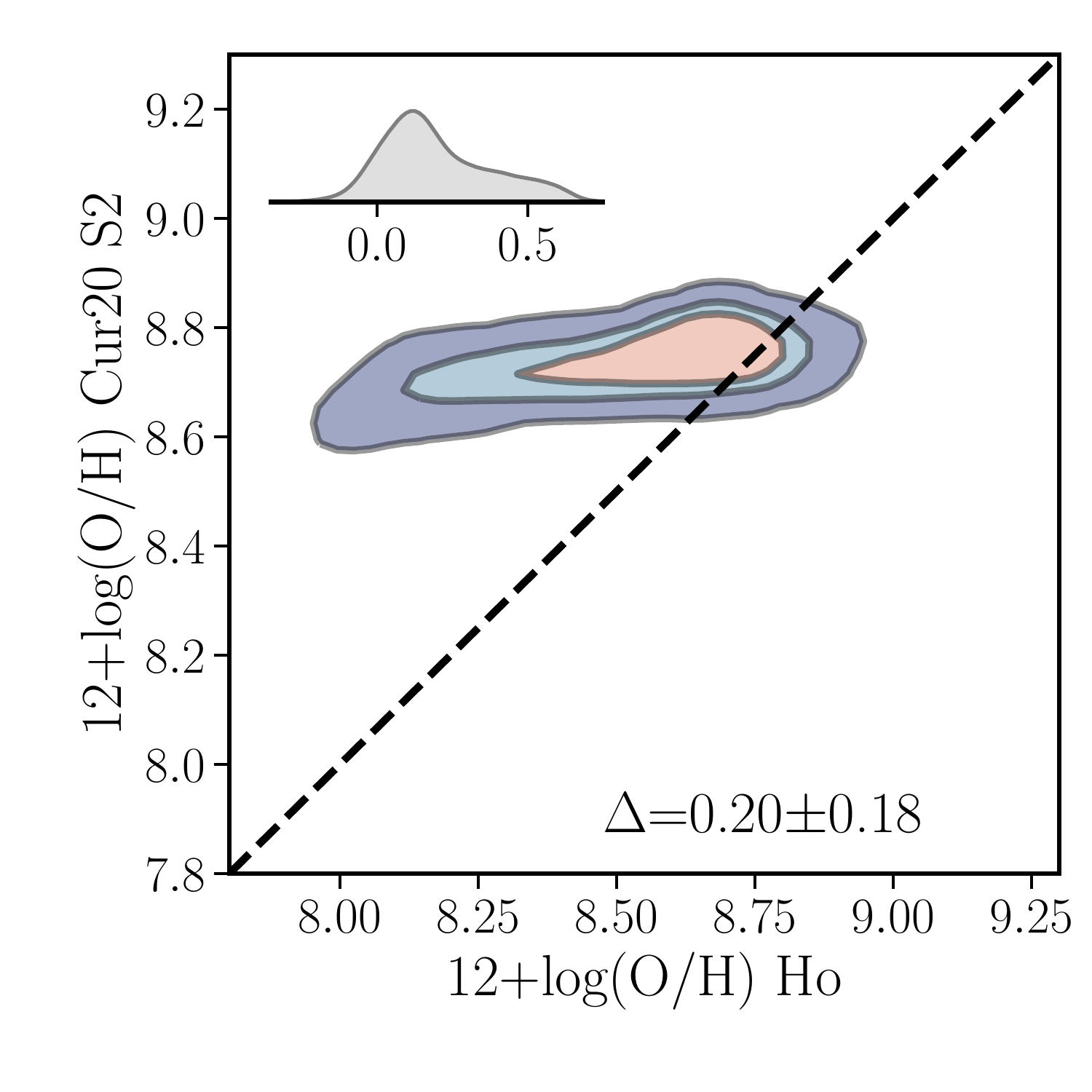}
\includegraphics[width=4.5cm,clip,trim=0 10 0 10]{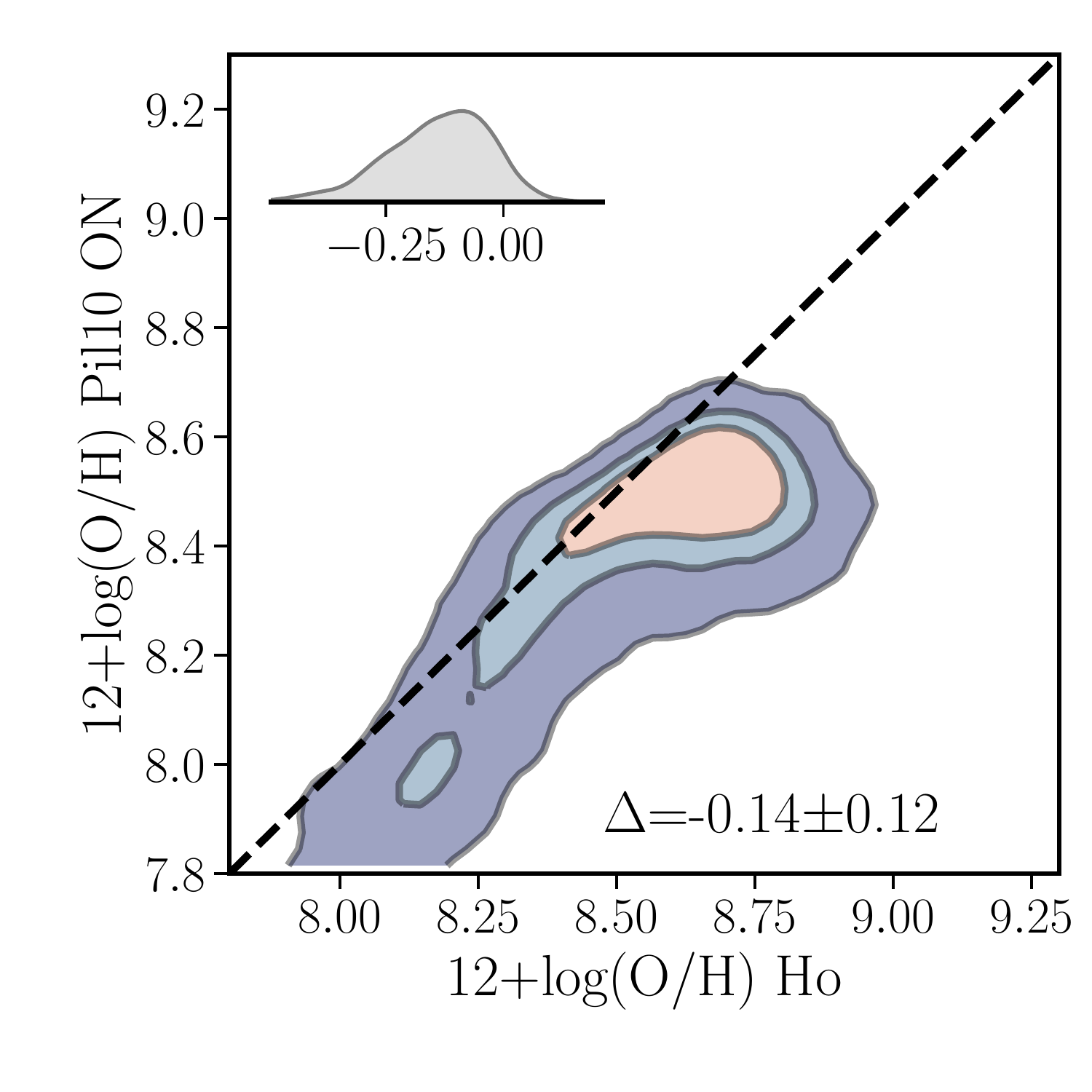}
\includegraphics[width=4.5cm,clip,trim=0 10 0 10]{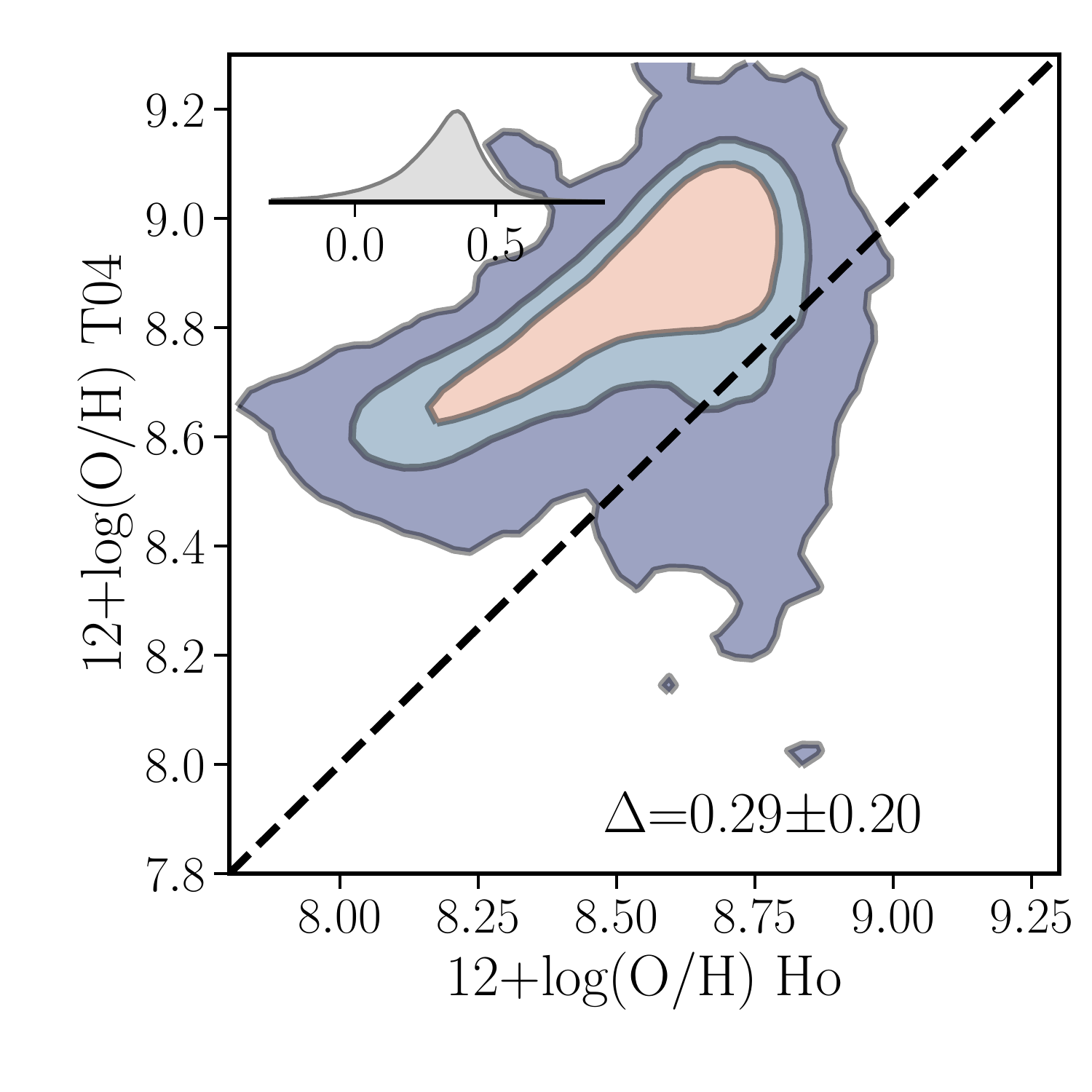}
 \endminipage
 \caption{Comparison between the oxygen abundances, 12+log(O/H), derived at the effective radius using four different calibrators (one for each panel) listed in the final catalog as a function of the values derived using the \citet{ho19} one (adopted as fiducial one in an arbitrary way). We adopt the same format as the one described in Fig. \ref{fig:comp_ssp_DR15} for the different panels in the figure. }
 \label{fig:comp_fewOH}
\end{figure}
%%%%%%%%%%%%%%%%%%%%%%%%%%%%%%%%%%%%%%%%%%%%%%%%%%%%%%%%%%%%%%%%%%%%%%%5

{\bf Line ratios:} For those emission lines more frequently used in the classical diagnostic diagrams (i.e., [OII]$\lambda$3727, [OIII]$\lambda$5007, [OI]$\lambda$6300, [NII]$\lambda$6584 and [SII]$\lambda$6717,31), we derive their ratios with respect to H$\beta$ (in the case of [OII] and [OIII]) and H$\alpha$ (for the remaining lines) using the values included in the {\tt FLUX\_ELINES} extension at (i) the central aperture ({\tt \_cen} suffix), (ii) the effective radius ({\tt \_Re} suffix), and (iii) average across the full FoV of the datacube ({\tt \_ALL} suffix). In addition we estimate the equivalent width of H$\alpha$ ({\tt EW\_Ha\_}), and the H$\alpha$ to H$\beta$ ratio ({\tt Ha\_Hb}) at the three locations. The corresponding errors have been estimated too. Rows between 13 and 21, and between 70 and 100 comprise all those parameters. A simple nomenclature including the two involved line ratios is adopted to label those quantities (e.g., {\tt log\_OI\_Ha\_cen} corresponds to the line ratio between [OI] and H$\alpha$ in the central aperture in logarithm scale, including an {e\_} prefix it will
corresponds to its error). The H$\alpha$ flux in the central aperture is also included for comparison purposes with SDSS single aperture fiber data (rows 162-163). A practical example of the use of those line ratios is included in Sec. \ref{sec:agns}.
%
% e_Ha_Hb_ALL is missing in the table!
%

{\bf Dust extinction:} We use the H$\alpha$ to H$\beta$ ratio to derive the dust extinction of the ionized gas (A$_{\rm V,gas}$) by assuming a nominal ratio of 2.86. This value corresponds to a nebulae fulfilling the case-B recombination scenario with an electron density of n$_e$=100 cm$^{-3}$ and an electron temperature of T$_e$=10$^4$ K \citep{osterbrock89}. We adopted a MW-like dust extinction \citep{cardelli89}, with a total-to-selective extinction R$_V=$3.1. By doing this derivation spaxel-by-spaxel we are able to derive a dust extinction map for each galaxy/cube (A$_V$). From those maps we estimate the dust extinction at the effective radius ({\tt Av\_gas\_Re}) and its error, included in the final catalog in rows 171-172. These A$_V$ maps are used to correct all the observed emission lines by their dust extinction, providing with dust-corrected emission lines that can be used in the derivation of additional parameters.

{\bf Oxygen and nitrogen abundances:} The emission line maps can be used to examine the possible ionizing sources spaxel-by-spaxel. In particular, following the criteria described in detail in \citet{ARAA} and \citet{sanchez20}, we classify as SF-areas those regions below the \citet{kewley01} demarcation line in the classical BPT diagram involving [OIII]/H$\beta$ and [NII]/H$\alpha$ with an EW(H$\alpha$)$>3$\AA. For those regions, and only for them, it is possible to estimate the oxygen and nitrogen abundance, and the ionization parameter, based on strong-line indicators (since those are calibrated only for ionization due to young OB stars, associated with recent SF activity). The dust-corrected emission line maps are used in the derivation of the oxygen abundances using the different calibrators adopted along this exploration. For the central aperture we estimate the oxygen abundance using the calibrators adopted in \citet{sanchez19}, that comprise a mixed set including calibrators anchored to measurements based on the direct method and calibrators derived using photoionizaiton models. Among them, we include in here the O3N2- and N2-based calibrators proposed by \citet{marino13} ({\tt OH\_O3N2\_cen} and {\tt OH\_N2\_cen}), the ONS calibrator by \citet{pilyugin10} ({\tt OH\_ONS\_cen}), the R23-based calibrator by \citet{kobu04} ({\tt OH\_R23\_cen}), the {\tt pyqz} calibrator \citet{vogt15} ({\tt OH\_pyqz\_cen}), the calibrators by \citet{maio08} ({\tt OH\_M08\_cen}), \citet{tremonti04} ({\tt OH\_T04\_cen}), \citet{dopita1996} ({\tt OH\_dop\_cen}), \citet{epm09} ({\tt OH\_O3N2\_EPM09\_cen}), and the t2-corrected calibrator \citet{2012apj...756l..14p} as derived by \citet{sanchez19} ({\tt OH\_t2\_cen}). All those oxygen abundances and their errors are included in rows between 50 and 69 of the final catalog.

%%%%%%%%%%%%%%%%%%%%%%%%%%%%%%%%%%%%%%%%%%%%%%%%%%%%%%%%%%%%%%%%%%%%%%%5
\begin{figure}
 \minipage{0.49\textwidth}
 \includegraphics[width=4.5cm,clip,trim=0 10 0 10]{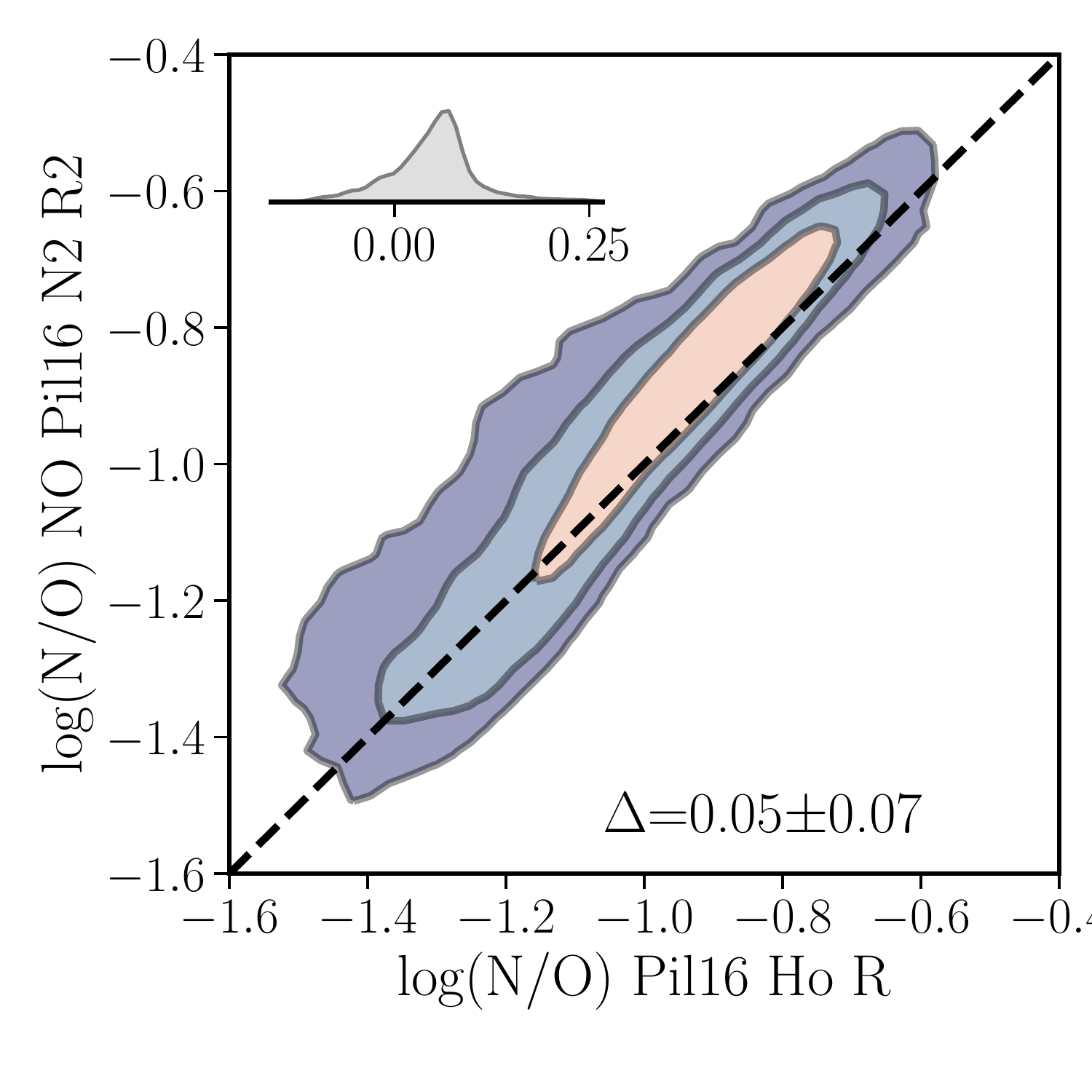}
 \includegraphics[width=4.5cm,clip,trim=0 10 0 10]{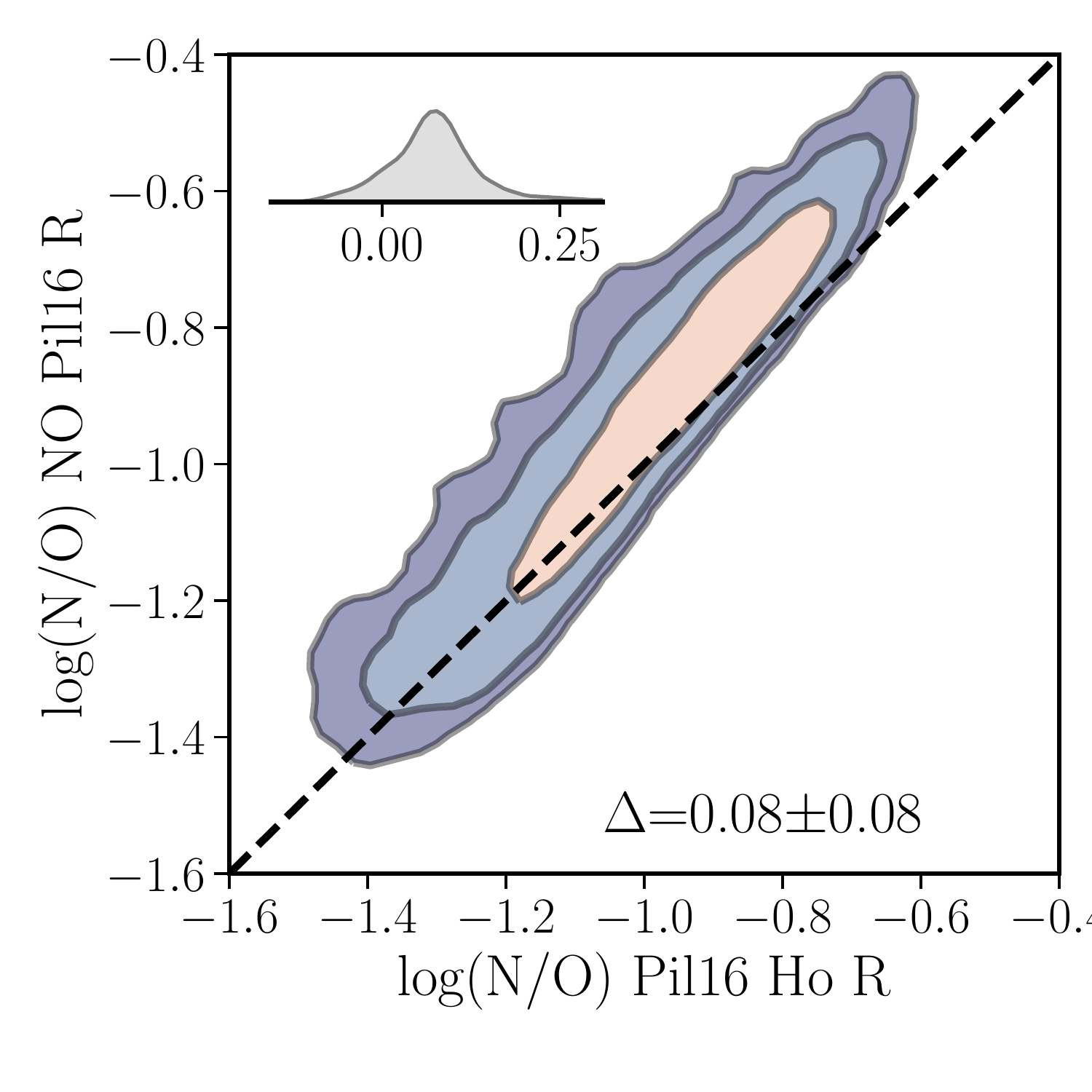}
 \endminipage
 \caption{Comparison between the three values for N/O relative abundance measured at the effective radius included in the final catalog. We adopt the same format as the one described in Fig. \ref{fig:comp_ssp_DR15}. Details on the compared quantities are given in Sec. \ref{sec:int}.}
 \label{fig:comp_NO}
\end{figure}
%%%%%%%%%%%%%%%%%%%%%%%%%%%%%%%%%%%%%%%%%%%%%%%%%%%%%%%%%%%%%%%%%%%%%%%5
%

A larger and more complete set of calibrators was recently used by \citet{espi22} in their exploration of the properties of \HII\ regions. We adopt most of those calibrators to derive the oxygen and nitrogen abundances (or the nitrogen-to-oxygen abundance ratio), together with the ionization parameter, for those spaxels which ionization is compatible with young-massive OB stars (i.e., associated with recent SF). The complete list of adopted calibrators is included in Tab. \ref{tab:OH_calibrators} of Sec. \ref{app:OH}.
Once derived those parameters spatially resolved (i.e., spaxel-by-spaxel), we estimate their corresponding values at the effect radius ({\tt OH\_CAL\_Re\_fit}, where {\tt CAL} corresponds to the ID of each calibrator), the slope of their radial gradient ({\tt OH\_CAL\_alpha\_fit}), and their errors (labelled with a prefix {\tt e\_}). All these values are included in the final catalog, covering the rows between 308 and 431. They comprise a total of 24 oxygen abundance estimates, 4 different nitrogen (and nitrogen-to-oxygen) estimates, and 4 estimates of the ionization parameter.

Figure \ref{fig:comp_fewOH} shows the comparison of a sub-set of the oxygen abundance calibrators included in the final catalog as a function of the values derived using the \citet{ho19} calibrator, selected as an arbitrary fiducial one following \citet{espi22}. This calibrator implements a state-of-the-art machine learning routine to provide with a calibrator that anchors the oxygen abundance to direct measurements using the direct method. It uses a large number of emission line ratios, as listed in Tab. \ref{tab:OH_calibrators}, providing with a quite accurate estimation of the oxygen abundance. A similar comparison for the full set of calibrators is included in App. \ref{app:OH} (Figure \ref{fig:comp_OH}). It is beyond the scope of the current study to discuss in detail the well-known discrepancies among different O/H calibrators and the possible transformations between them \citep[e.g., ][]{kewley08,angel12,2020Curti_MNRAS491,espi22}. The main aim of this comparison is to show that indeed there is a correspondence between the oxygen abundances derived using different calibrators included in our catalog. This correspondence is far from being a one-to-one relation in most of the cases. In a few cases the oxygen abundances present just a constant offset for all the considered dynamical range (e.g., Kew02 N2O2, Pil16 R and S calibrators). For a large number of calibrators the relation is well characterized by an almost linear relation with a slope different than one (e.g., Cur20 N2, Mar N2, Pet04 N2). In some of them the dynamical range is considerably different calibrator to calibrator (e.g., Cur20 R2 or RS32), what would translate into a large variety of slopes for the linear relation that would match them. Finally, there are cases in which there are clear deviations from the linear relation, with a change in the slope or even a plateau found at high abundances (e.g., Pil10 ON and NS), or even large discrepancies in the same regime (e.g., T04). In summary, any analysis using the oxygen abundances included in the delivered catalog should acknowledge the differences described among the different calibrators, and the impact on the results. For instance, explorations of the shape of the mass-metallicity relation (MZR) or the oxygen abundance gradient are deeply affected by the adopted calibrator, both qualitatively \citep[e.g., the presence of plateaus in the distribution of the MZR,][]{sanchez19,paola22} and quantitatively \citep[e.g., the actual value of radial gradients,][]{belf17,laura18}.

A less relevant difference is found among the three estimates of the nitrogen-to-oxygen relative abundance included in our catalog, as seen in Figure \ref{fig:comp_NO}. They present clear linear relations among them, with a slope near to one, and small offsets ($\Delta$log(N/O)$\sim$0.03-0.08 dex) and scatters ($\sigma_{\rm \Delta log(N/O)}\sim$0.03-0.07 dex). In this case the use of a different calibrator would produce little quantitative and essentially no qualitative differences.

%%%%%%%%%%%%%%%%%%%%%%%%%% SFR-Comparison %%%%%%%%%%%%%%%%%%%%%%%%%%%%%%%%%%%%%%%%%
%%%%%%%%%%%%%%%%%%%%%%%%%%%%%%%%%%%%%%%%%%%%%%%%%%%%%%%%%%%%%%%%%%%%%%%5
\begin{figure}
 \minipage{0.49\textwidth}
 \includegraphics[width=4.5cm,clip,trim=0 10 0 10]{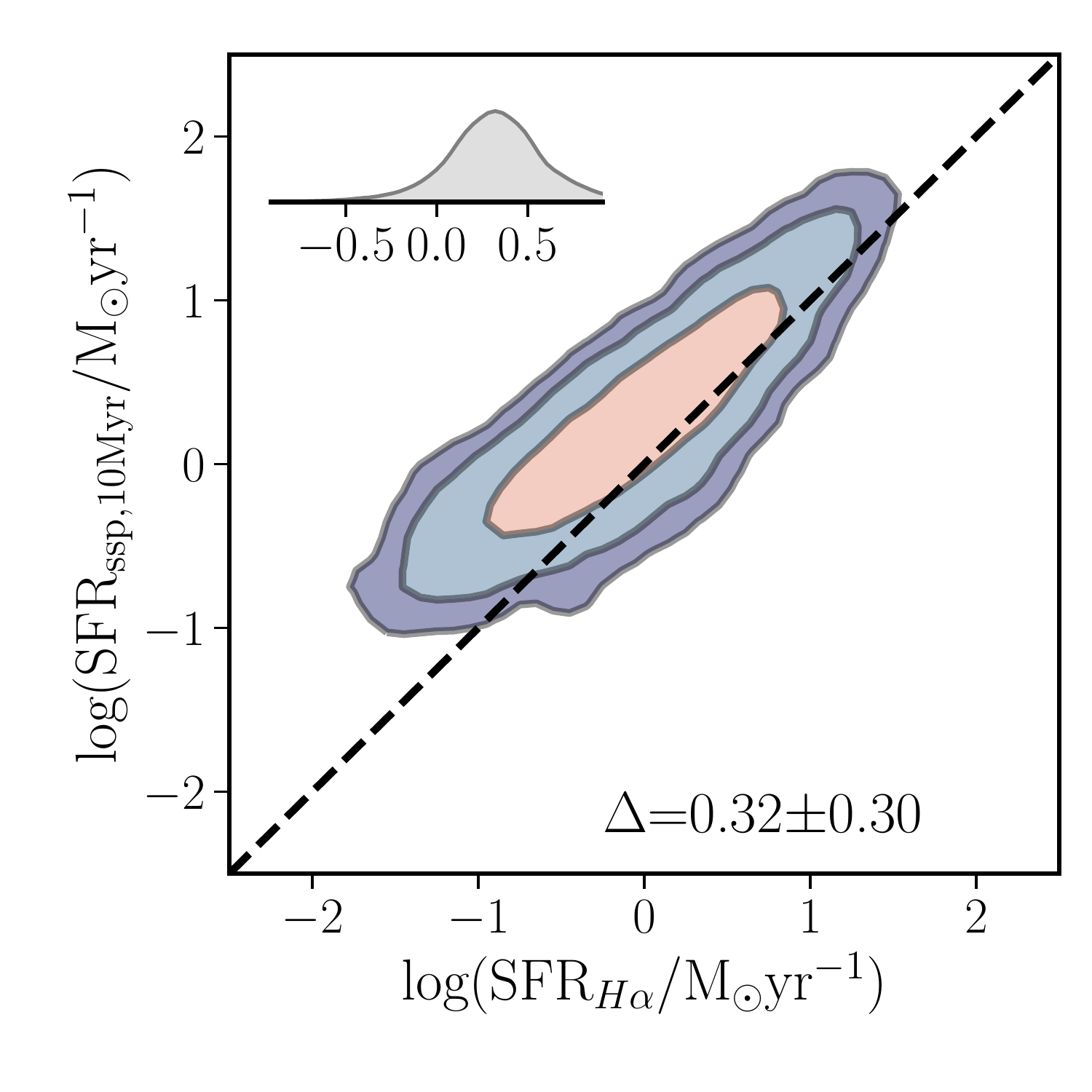}
 \includegraphics[width=4.5cm,clip,trim=0 10 0 10]{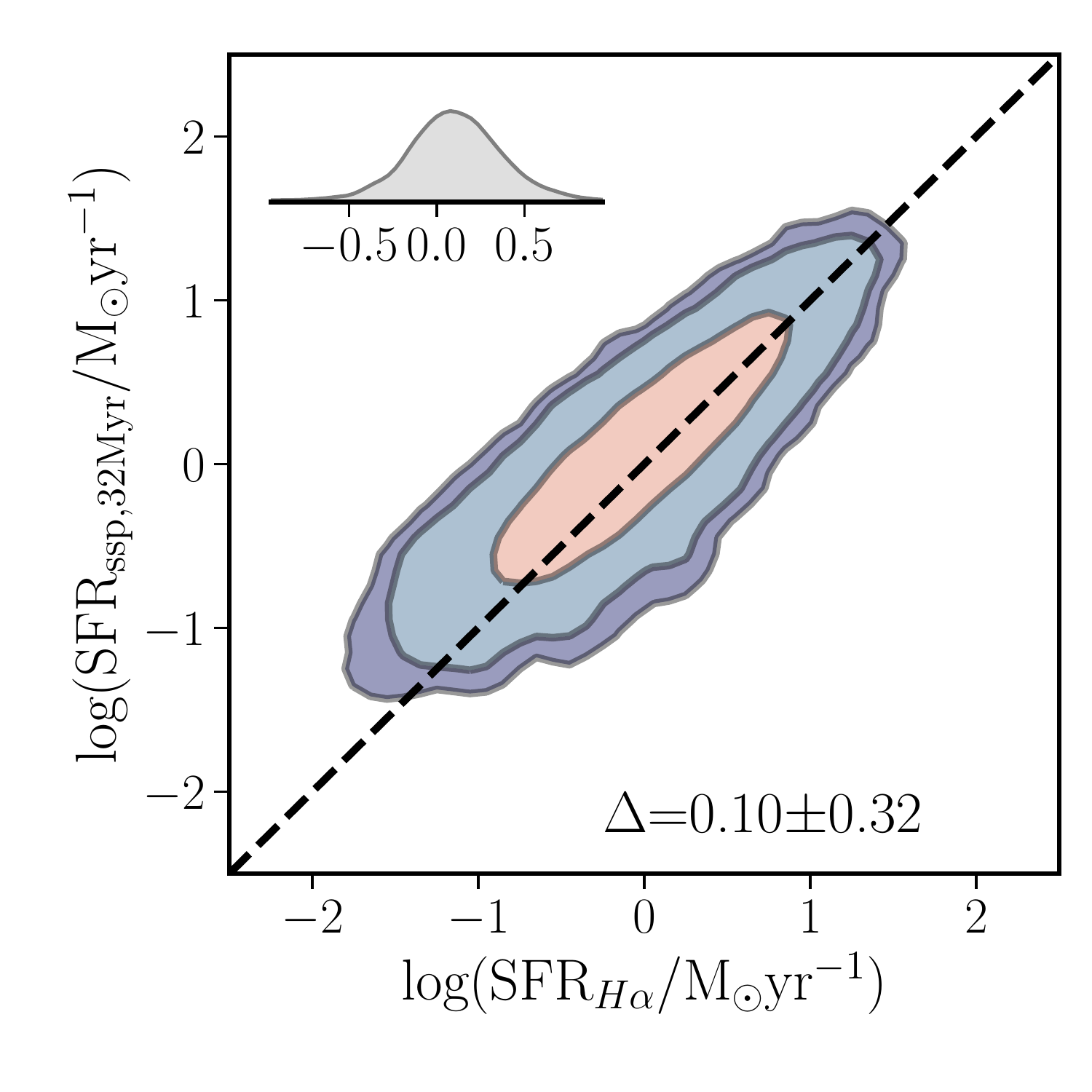}
 \includegraphics[width=4.5cm,clip,trim=0 10 0 10]{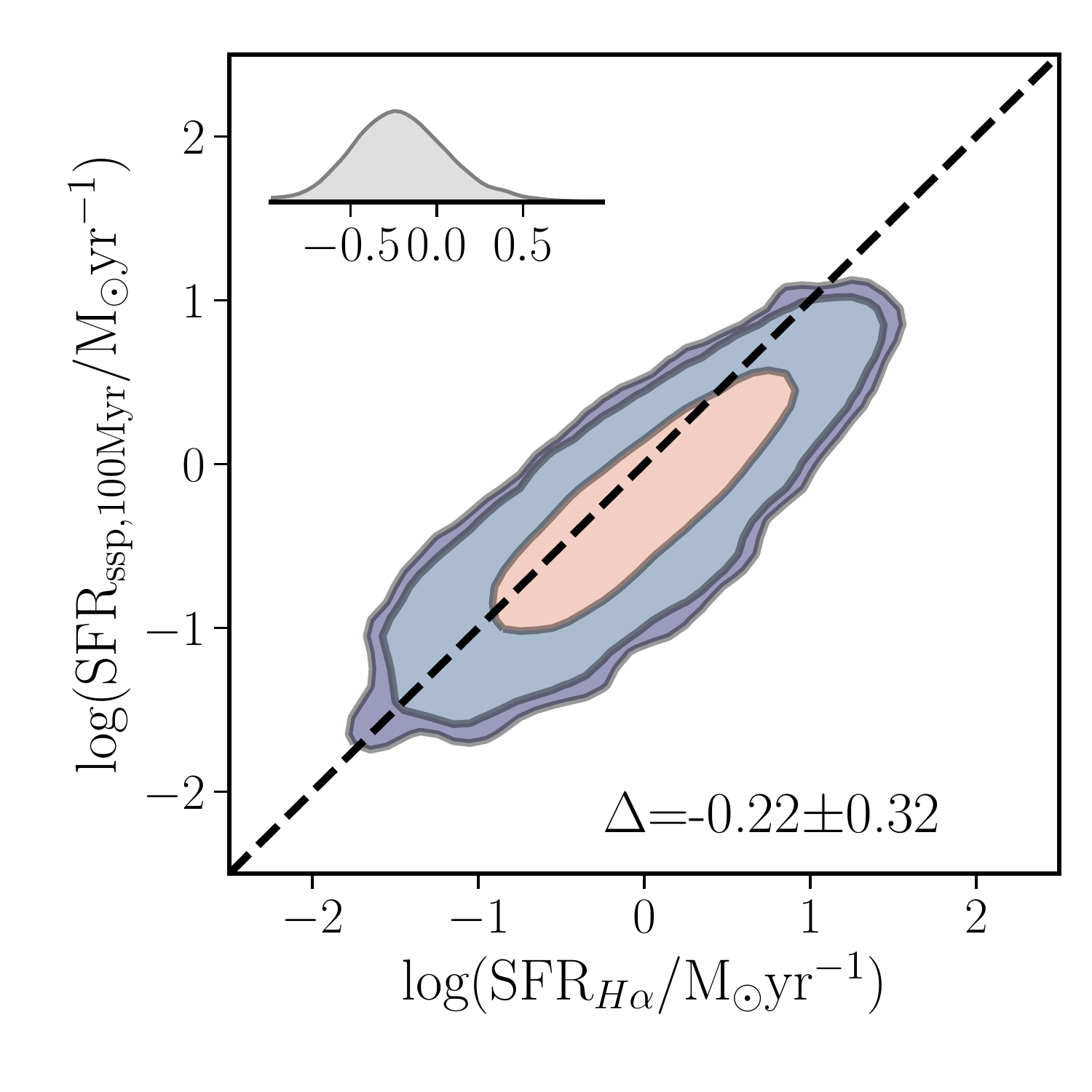}
  \includegraphics[width=4.5cm,clip,trim=0 10 0 10]{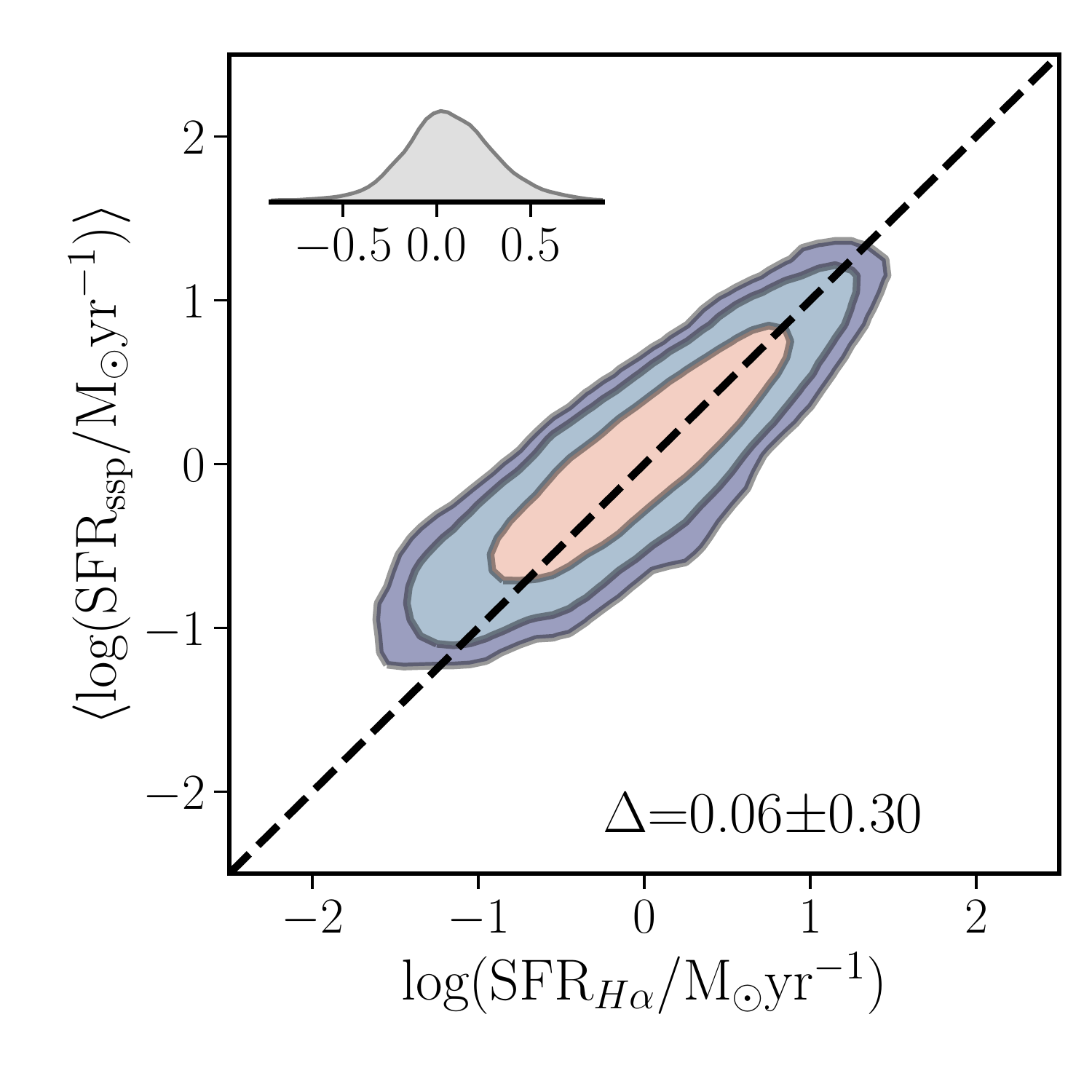}
 \endminipage
 \caption{Comparison between the SFR derived based on the dust-corrected H$\alpha$ luminosity (SFR$_{H\alpha}$) and { four different estimates based on the stellar population decomposition performed as part of the \pyp\ analysis (SFR$_{\rm ssp}$), for three time ranges: 10 Myr (top-left panel), 32 Myr (top-right panel) and 100 Myr (bottom-left panel), and the average of the three (bottom-right panel)}. We exclude those galaxies without evidence of SF activity from this comparison by excluding those ones for which the $\big |$EW(H$\alpha$)$\big | <$3 \AA\  at the effective radius. We adopt the same format as the one described in Fig. \ref{fig:comp_ssp_DR15}. Details on the compared quantities are given in Sec. \ref{sec:int}.}
 \label{fig:comp_SFR}
\end{figure}
%%%%%%%%%%%%%%%%%%%%%%%%%%%%%%%%%%%%%%%%%%%%%%%%%%%%%%%%%%%%%%%%%%%%%%%5
%
% Select Only SFGs?
%

{\bf Star-formation rate:} The dust corrected H$\alpha$ luminosity is used to estimate the star-formation rate (SFR) spaxel-wise by applying the relation proposed by \citet{kennicutt98}, for the \citet{salpeter55} IMF:
\begin{equation}\label{eq:sfr}
  {\rm SFR}\ (M_\odot\ yr^{-1})\ =\ 0.79\ 10^{-41}\ {\rm L}_{\rm H\alpha}\ (erg/s)
\end{equation}
From this distribution it is possible to derive the SFR surface density ($\Sigma_{\rm SFR}$). Like in case of the stellar-mass, we can derive the integrated SFR just co-adding the spaxel-by-spaxel values across the FoV of the IFU data. As already discussed in previous articles \citep{mariana16,sanchez18}, this SFR is an upper-limit to the real one, since in this derivation all H$\alpha$ flux is integrated irrespectively of the nature of the detected ionization. In other words, this calculation would derive an SFR even for those galaxies in which there is no ionization that can be directly associated with a recent star-formation event (i.e., in the case of retired galaxies, such as elliptical ones). However, it is still useful as an upper-limit in case that a fraction of this ionization is still due to SF but it is so weak that it is over-shaded by ionization due to other sources \citep[i.e., shocks, ionization by old stars, like post-AGBs and HOLMES, or AGNs, see][and references therein]{sanchez20}. We include this integrated SFR in the catalog ({\tt log\_SFR\_Ha}), together with its error ({\tt e\_log\_SFR\_Ha}), in rows 7 and 11. However, for completeness, we also derive the SFR just co-adding those regions in each galaxy which ionization is compatible with recent SF, using the combination of the BPT diagram plus the EW(H$\alpha$), as described above ({\tt log\_SFR\_SF}). Furthermore, we estimate the SFR once considered the possible contamination by a diffuse ionization gas (DIG) due to old-stars \citep[e.g][]{binette94,flor11}, assuming a constant EW(H$\alpha$)=1 \AA\ for this component ({\tt log\_SFR\_D\_C}).

Figure \ref{fig:comp_SFR} shows the comparison between the SFR derived using the dust-corrected H$\alpha$ luminosity (SFR$_{\rm H\alpha}$) and the values derived based on the stellar synthesis analysis (SFR$_{\rm ssp}$) for three different time scales (10, 32 and 100 Myr), and for the average of the three of them. We find a clear correspondence between the four SFR$_{\rm ssp}$ with SFR$_{H\alpha}$, with the average of the three time-scales providing the lower offset with respect to the one-to-one relation ($\Delta$log($SFR$)=0.06$\pm$0.30 dex), followed by the value for 32 Myr. In average the SFR derived for 10Myr (100Myr) is higher (lower) than then one estimated using the H$\alpha$ luminosity. These results agree with previous explorations using different IFS datasets \citep[e.g.][]{rosa16,jkbb21a}. In summary, the use of the SFR$_{\rm ssp}$ instead of SFR$_{\rm H\alpha}$ would provide with similar statistical results, for instance when exploring global relations such as the star-formation main sequence \citep[e.g.][]{sanchez18,sanchez18b}. In principal this quantity should be more accurate, since it does not involve any assumed SFH and ChEH as it is required to derive the scaling between the H$\alpha$ luminosity and the SFR \citep{kennicutt98}. However, it is most probably less precise, due to the additional uncertainties introduced by the stellar synthesis analysis to derive this quantity. As a consequence, the derived relations may present a larger scatter.

%%%%%%%%%%%%%%%%%%%%%%%%%%%%%%%%%%%%%%%%%%%%%%%%%%%%%%%%%%%%%%%%%%%%%%%%%

{\bf Molecular gas estimation:} The dust extinction is a tracer of the molecular gas content, via the dust-to-gas relation \citep[e.g.][and reference there in]{brin04}. We adopted the recent calibrator by \citet{jkbb21b} to estimate the molecular gas surface density ($\Sigma_{\rm mol}$) through the spaxel-by-spaxel A$_{V,gas}$ parameter, using the formula:
\begin{equation}\label{eq:mol}
  \Sigma_{\rm mol}\ (M_\odot\ pc^{-2})\ =\ 1.06\ A_{\rm V,gas}^{2.58}\ (mag)
\end{equation}
then, co-adding throughout the FoV of each IFU we estimated the integrated molecular gas ({\tt log\_Mass\_gas\_Av\_gas\_log\_log}, row 170). For completeness we provide with the same parameter estimated based on the linear calibrator proposed by \citet{jkbb20}:
\begin{equation}\label{eq:mol_old}
  \Sigma_{\rm mol}\ (M_\odot\ pc^{-2})\ =\ 23\ A_{\rm V,gas}\ (mag)
\end{equation}
included in row 40 ({\tt log\_Mass\_gas}). We do not provide with the formal error of those parameters since the error budget is dominated by the calibrator itself ($\sim$0.2 dex), rather than by the uncertainties in the derivation of the dust extinction. 

Two additional estimates of the integrated molecular gas mass has been included in the catalog. One of them is derived by applying a possible correction taking into account the dependence of the dust-to-gas ratio with the oxygen abundance ({\tt log\_Mass\_gas\_Av\_gas\_OH}), and the other is derived adopting this correction but using the stellar dust extinction instead of the ionized gas one in the estimation of the molecular gas surface density ({\tt log\_Mass\_gas\_Av\_ssp\_OH}). To derive those estimates we followed the main procedures described in \citet{jkbb21b}. However, the dynamical range of the metallicity explored in that study were too narrow to provide a clear conclusion regarding the improvement of introducing these two new calibrators. We include them here in order to allow future explorations to compare them with other estimates of the molecular gas.
%These too later proxies of the molecular gas mass have to be confronted with direct observations so far.

{\bf Electron density:} Finally, we use the [SII]$\lambda$6717,31 line ratio to derive the electron density spaxel-by-spaxel, solving the equation:
\begin{equation}\label{eq:ne}
  \frac{[{\rm SII}]\lambda 6717}{[{\rm SII}]\lambda 6731} = 1.49 \frac{1+3.77x}{1+12.8x},
\end{equation}
where $x=10^{-4}n_{e}t^{-1/2}$ and $t$ is the electron temperature in units of $10^{4}$ K \citep{McCa85}, and $n_e$ is the electron density in units of cm$^{-3}$. Like in the case of the dust-extinction we adopted a typical electron temperature of  T$_e=10^{4}$ K in this derivation. We should note that the dependence of the electron on this parameter is weak in the adopted formula. Like in the case of other parameters we deliver the value at the effective radius ({\tt Ne\_Oster\_S\_Re\_fit}), the slope of its radial gradient ({\tt Ne\_Oster\_S\_alpha\_fit}), and their corresponding errors for each galaxy/cube (rows 432 to 435).

%\subsubsection{Emission line related quantities}
%\label{sec:cat_elines}

%\subsubsection{Molecular gas Mass related quantities}
%\label{sec:cat_mass_gas}

\subsubsection{Kinematics related quantities}
\label{sec:cat_kin}

The analysis performed to the data provides with spatial distribution of the stellar and ionized gas velocity and velocity dispersion. From those quantities we derive different characteristic parameters for each galaxy/cube that have been included in the final catalog: (i) the average velocity to velocity dispersion ratio within an aperture of 1 Re ($\frac{v}{\sigma}$, labelled as {\tt vel\_sigma\_Re}, row 45), and its corresponding error ({\tt e\_vel\_sigma\_Re}, row 46), estimated as:
\begin{equation}\label{eq:v_s}
\frac{v}{\sigma} = \sqrt{\frac{\sum_{r<1Re} f_\star v_\star^2}{\sum_{r<1Re} f_\star \sigma_\star^2}},
\end{equation}
where $f_\star$, $v_\star$ and $\sigma_\star$ correspond to the stellar flux-intensity in the V-band at any position (x,y) within the FoV, the stellar velocity and the stellar velocity dispersion in each spaxel within the considered apertures; (ii) The stellar and H$\alpha$ ionized gas velocities ({\tt vel\_ssp\_R} and {\tt vel\_Ha\_R}), and their corresponding errors (labelled with an {\tt e\_} prefix), derived in an elliptical ring (following the PA and ellipticity of the galaxy) at one and two effective radius ({\tt R}=1 or 2), listed in rows 145 to 152; (iii) The stellar and H$\alpha$ ionized gas velocity dispersion ({\tt vel\_disp\_ssp\_R} and {\tt vel\_disp\_Ha\_R} in the central aperture ({\tt R}={\tt cen}) and at one effective radius ({\tt R}={\tt 1Re}), listed in rows 155 to 158; (iv) The apparent stellar angular momentum parameter at the effective radius \citep{emsellem07}, $\lambda_{\rm Re}$ (labelled as {\tt Lambda\_Re}, row 175), and its corresponding error ({\tt e\_Lambda\_Re}, row 176), defined as:
\begin{equation}\label{eq:lambda}
\lambda_{\rm R} = \frac{\sum_{r<1.15Re} f_\star\ r\ |v_\star|}{\sum_{r<1.15Re} f_\star\ r\ \sqrt{v_\star^2+\sigma_\star^2}},
\end{equation}
where $f_\star$, $v_\star$ and $\sigma_\star$ correspond to the same parameters as in Eq. \ref{eq:v_s}, and $r$ is the deprojected galactocentric distance.

For the current dataset we introduce an inclination correction on $\lambda$ that was not included in the calculations for DR14 and DR15. This correction was described in the Appendix of \citet{emsellem11}:
\begin{equation}\label{eq:l_cor}
\begin{array}{l}
\lambda^\prime = \frac{\lambda}{C_1}\\
C_1 = C_0^2-\lambda^2(C_0^2-1)\\
C_0 = \frac{{\rm sin}(i)}{\sqrt{1-0.3*{\rm cos}^2(i)}}\\
\end{array}
\end{equation}
where $\lambda$ corresponds to the apparent angular momentum described in Eq. \ref{eq:lambda}, and $i$ is the inclination angle of the galaxy.

\subsubsection{Volume correction}
\label{sec:VM}
%%%%%%%%%%%%%%%%%%%%%%%%%%%%%%%%%%%%%%%%%%%%%%%%%%%%%%%%%%%%%%%%%%%%%%%5
\begin{figure}
 \minipage{0.99\textwidth}
 \includegraphics[width=8.5cm,clip]{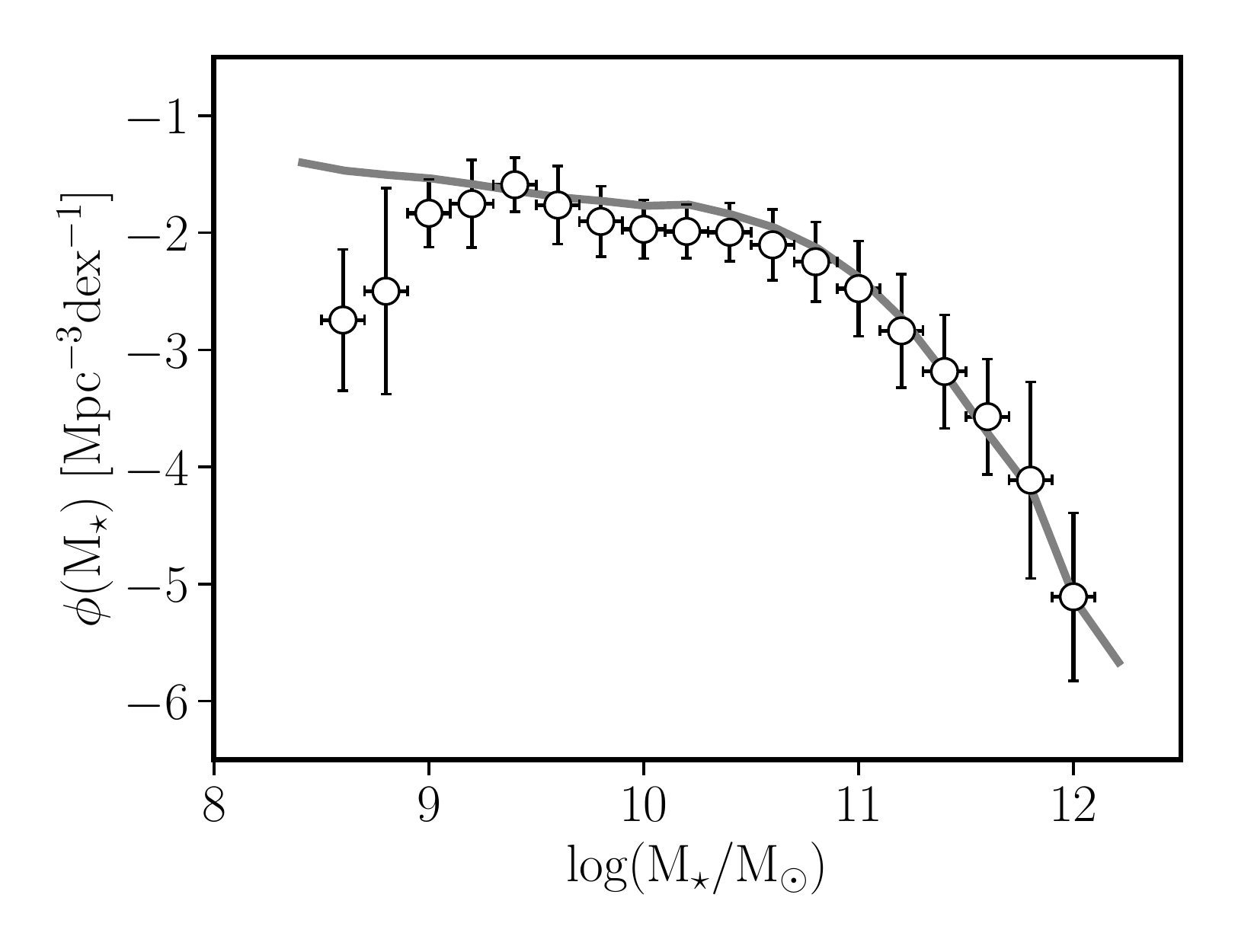}
 \endminipage
 \caption{Stellar Mass function derived using the the volume corrections included in our final catalog (white solid-circles) compared with the one derived using a volume-complete sample extracted from the NSA catalog \citep{blanton+2017}, shifted in mass to correct for the different adopted IMFs, and in density to match our adopted cosmology (i.e., $h=0.73$). The agreement is particularly good for the mass range above 10$^{9}$ M$_\odot$.}
 \label{fig:VC}
\end{figure}
%%%%%%%%%%%%%%%%%%%%%%%%%%%%%%%%%%%%%%%%%%%%%%%%%%%%%%%%%%%%%%%%%%%%%%%5

The MaNGA sample has a complicated selection function. As already outlined in Sec. \ref{sec:sample}, the final sample was built from a set of different sub-samples, each of them adopting different selection criteria. Despite of this complicated construction, in principal, it is possible to perform a volume correction and derive representative quantities from this sample. \citet{wake17} proposed a volume correction in which the volume accessible for each sub-sample is explored separately, and then the full volume accessible for each individual object is evaluated. In \citet{sanchez18b}, Appendix E, we proposed a different approach in which the individual volume accesible for each target in the final sample is estimated {\it a posteriori}. The procedure is described in detail in \citet{rodriguez-puebla20}. In summary, we adopted the SDSS galaxies stellar-mass function \citep{blanton+2017} to estimate the volume accessible in a set of bins in stellar mass, redshift and color for this parent sample (as all MaNGA galaxies are extracted from the SDSS spectroscopic sample). Then we estimate the fraction of MaNGA galaxies in each of those bins with respect to the total number of SDSS galaxies. Considering this fraction and the previously estimated volume it is possible to estimate, for each galaxy, its accessible volume. We repeat this calculation for the final MaNGA sample deriving, for each galaxy, the weight to correct for its accessible volume ({\tt Vmax\_w}) and number ({\tt Num\_w}),  included in the final catalog as rows 533 and 534, respectively. 
Figure \ref{fig:VC} shows the comparison between the original stellar mass function adopted to estimate the accessible volume for each galaxy (i.e., the SDSS one) and the one derived from our data using these estimated volume corrections. There is a considerable agreement between both distributions in the regime in which the sample may be considered representative and statistically significant, i.e., for M$_\star>$10$^{9}$M$_\odot$. Below this stellar mass the current same is clearly incomplete. 

%\Com{M* SFR SFH ChEH O/H N/O Lambda****}
%\Com{Description of what ``proc\_elines.pl'' does}

%\Com{Volume correction}

\section{Comparison with previous results}
\label{sec:comp}

The current (and previous) versions of the MaNGA datasets have been extensively analyzed with earlier versions of our pipeline and with different tools and procedures developed within the community. The results of these analyses are publicly accessible in many cases. Here we present a brief comparison with published datasets in order to determine which scientific results may change (or not) when using different analysis, anchoring our results to the published ones when possible.
%understand their possible differences (or similarities) derived when using them and narrow-down
%As indicated before the current results comprise the analysis using \pyp\ of the last MaNGA Data Relase, included in the SDSS DR17 {\bf REF!}. For previous releases (DR14 and DR15, {\bf REFs}), we applied the former version of our data analysis pipeline ({\tt Pipe3D}), and adopted a different SSP template library, as already described in Sec. \ref{sec:pipe3d} and \ref{sec:ssp}. Since the dataproducts of those analyses are still accessible for the community \footnote{\url{https://www.sdss.org/dr14/manga/manga-data/manga-pipe3d-value-added-catalog/} and \url{https://www.sdss.org/dr15/manga/manga-data/manga-pipe3d-value-added-catalog/}}, despite that they comprise a lower number of analyzed objects and include a lower number of dataproducts, we consider that it is relevant to present some basic comparisons between the values derived in both datasets. 

\subsection{Morphological classification}
\label{sec:comp_morph}
%%%%%%%%%%%%%%%%%%%%%%%%%%%%%%%%%%%%%%%%%%%%%%%%%%%%%%%%%%%%%%%%%%%%%%%5
\begin{figure}
 \minipage{0.49\textwidth}
 \includegraphics[width=8.5cm,clip,trim=0 10 20 0]{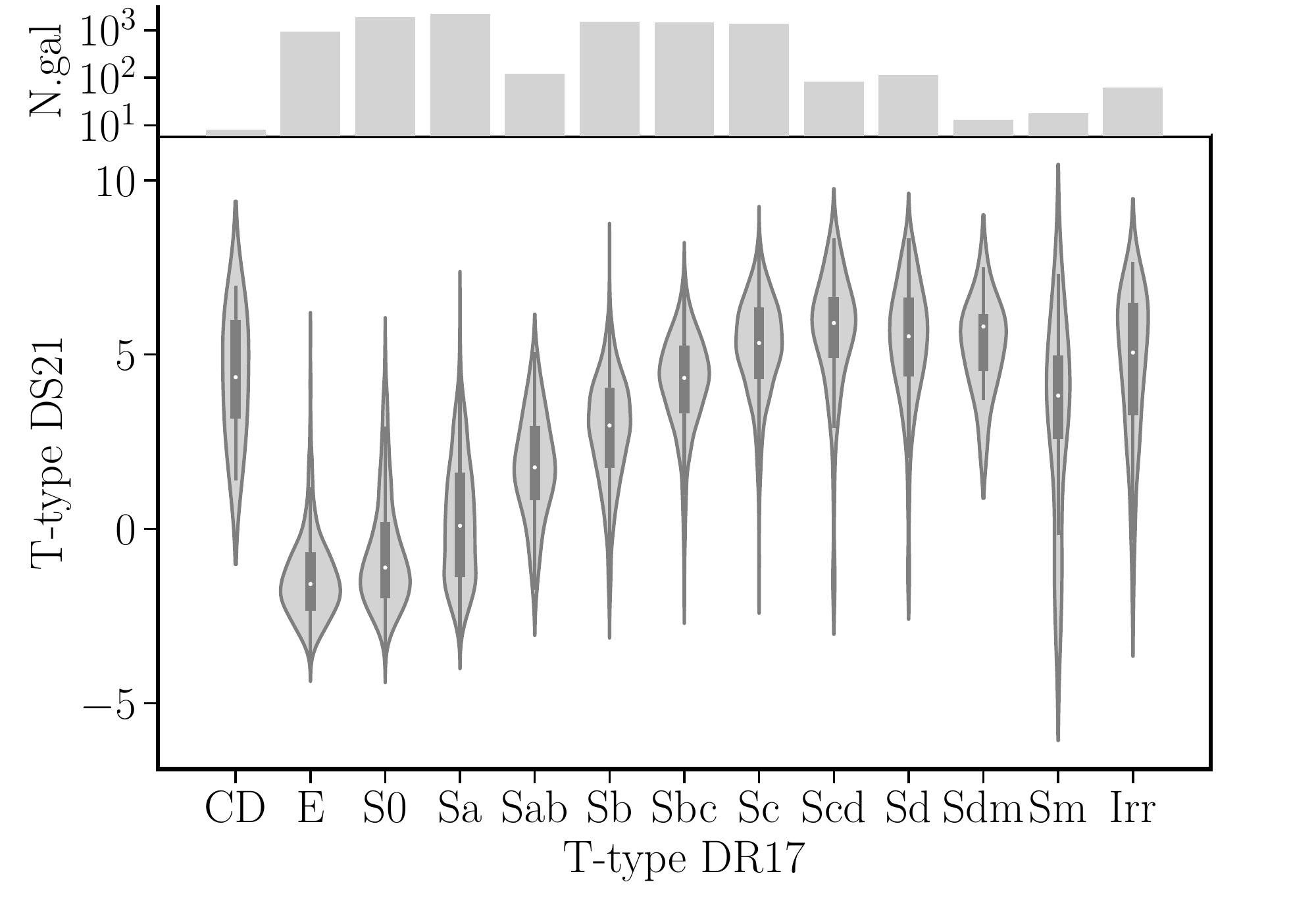}
 \endminipage
 \caption{Comparison between the morphological classification included in our final catalog (Sec.~\ref{sec:morp}) and the values reported by DS21. The violin plots show the distribution of the DS21 T-type parameter for each of our morphological classes. The upper-panel shows a histogram of the number of galaxies in each morphological bin in a logarithmic scale.}
 \label{fig:comp_Morph}
\end{figure}
%%%%%%%%%%%%%%%%%%%%%%%%%%%%%%%%%%%%%%%%%%%%%%%%%%%%%%%%%%%%%%%%%%%%%%%5
%

Different groups have addressed the morphological classification of the MaNGA galaxies. Among them, we highlight the visual classification presented by Vazquez-Mata et al. (in prep.) as a MaNGA VAC\footnote{\url{ https://www.sdss.org/dr17/data_access/value-added-catalogs/?vac_id=manga-visual-morphologies-from-sdss-and-desi-images}}, that has been the basic training dataset for our own classification. \citet[][DS21 hereafter]{DS21} present a detailed state-of-the-art morphological classification based on supervised Deep Learning models applied directly to SDSS images\footnote{\url{https://www.sdss.org/dr17/data_access/value-added-catalogs/?vac_id=manga-morphology-deep-learning-dr17-catalog}}. They use two different visual classification catalogues to train their method \citep{nair2010,willett:2013aa}. This analysis provides a T-Type for each galaxy, together with an automatic identification of edge-on and barred galaxies.
%based on the parameters provided by the {\tt pyMorph} analysis of the SDSS images of the MaNGA galaxies \citep{fisher19}. This classification is derived by adopting a state-of-the-art deep learning scheme that is supposed to provide an optimal segregation between different morphological types. 
Our morphological classification is expected to provide a statistical segregation of different morphological types, matching the visual classifications mentioned above. In Fig.~\ref{fig:comp_Morph} we compare the results from both methods as a violin plot of the DS21 T-types vs. our morphological classification ({\tt best\_type\_n} parameter). Despite of the significant differences between both methods, the agreement between both morphological classifications for most types is remarkable. The largest differences are found at the extremes ({\tt best\_type\_n}~$<$~1 and {\tt best\_type\_n}~$>$~8). On one side, mild but clear differences are found in the merging spiral (Sm) and irregular (Irr) galaxy types. 
%This is expected as our classification is primarily anchored to a visual inspection of the data, although it uses photometric and structural parameters to perform the classification. On the contrary, DS21 entirely relies on the structural parameters derived by {\tt pyMorph}.
This is expected since both schemes rely on different visual classifications of the galaxies. The more {\it regular} the galaxy, the better the expected agreement between both methods, as is apparent in Fig.~\ref{fig:comp_Morph}. On the other side, E and S0 galaxies in our catalog present very similar distributions of T-type in the DS21 catalog. We stress that DS21 consider that their T-type number is not enough to distinguish between these two morphological types, proposing a more complex decision tree based on additional parameters \footnote{\url{https://data.sdss.org/datamodel/files/MANGA_MORPHOLOGY/deep_learning/DL_VER/manga-morphology-dl.html}}. Our comparison indeed supports their results in this regards.

The largest differences are found in the cD-class, included in our classification scheme as {\tt best\_type\_n}$=-$2, but absent in DS21. Based on this result we consider that our classification for these galaxies is dubious. In some cases, like manga-11968-6103, from visual inspection it may indeed be a cD galaxy. However, there are other galaxies, like manga-8144-3703, which we consider that  should be classified as Irregular (it was classified as Scd/Sd by DS21). In summary, the cD morphological type should be excluded from any further analysis.
In any case, its removal has a negligible impact from a statistical point of view, since it comprises only 8 galaxies. For the remaining galaxies, our morphological classification is at least as good as the published ones for any statistical analysis.
%Based on this results we consider that the classification for these galaxies is dubious, what does not introduce a significant bias in the results sin   

\subsection{Values reported by {\tt Pipe3D} in previous DRs}
\label{sec:comp_DR15}

%%%%%%%%%%%%%%%%%%%%%%%%%%%%%%%%%%%%%%%%%%%%%%%%%%%%%%%%%%%%%%%%%%%%%%%5
\begin{figure*}
 \minipage{0.99\textwidth}
 \includegraphics[width=6.25cm,clip,trim=0 10 0 10]{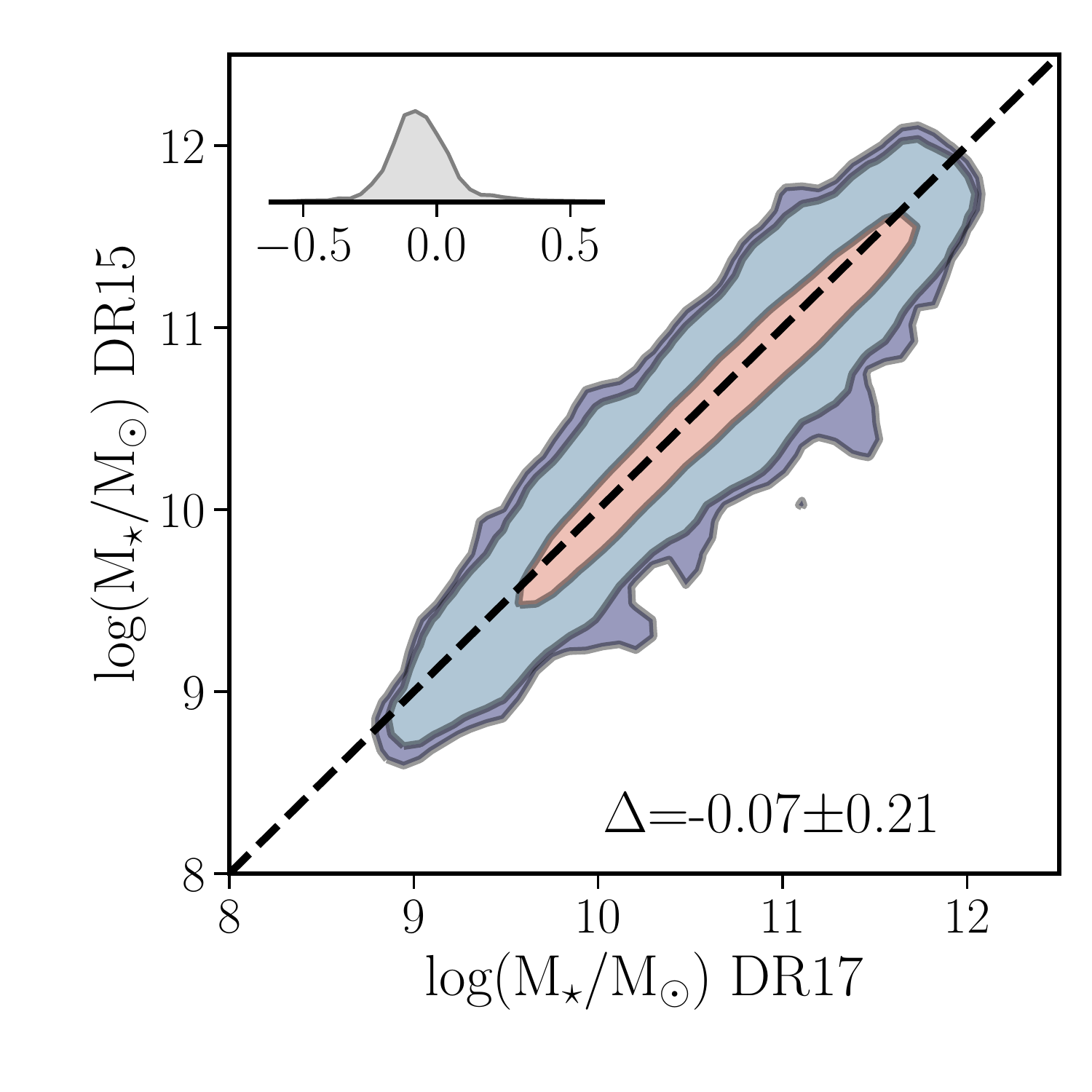}\includegraphics[width=6.25cm,clip,trim=0 10 0 10]{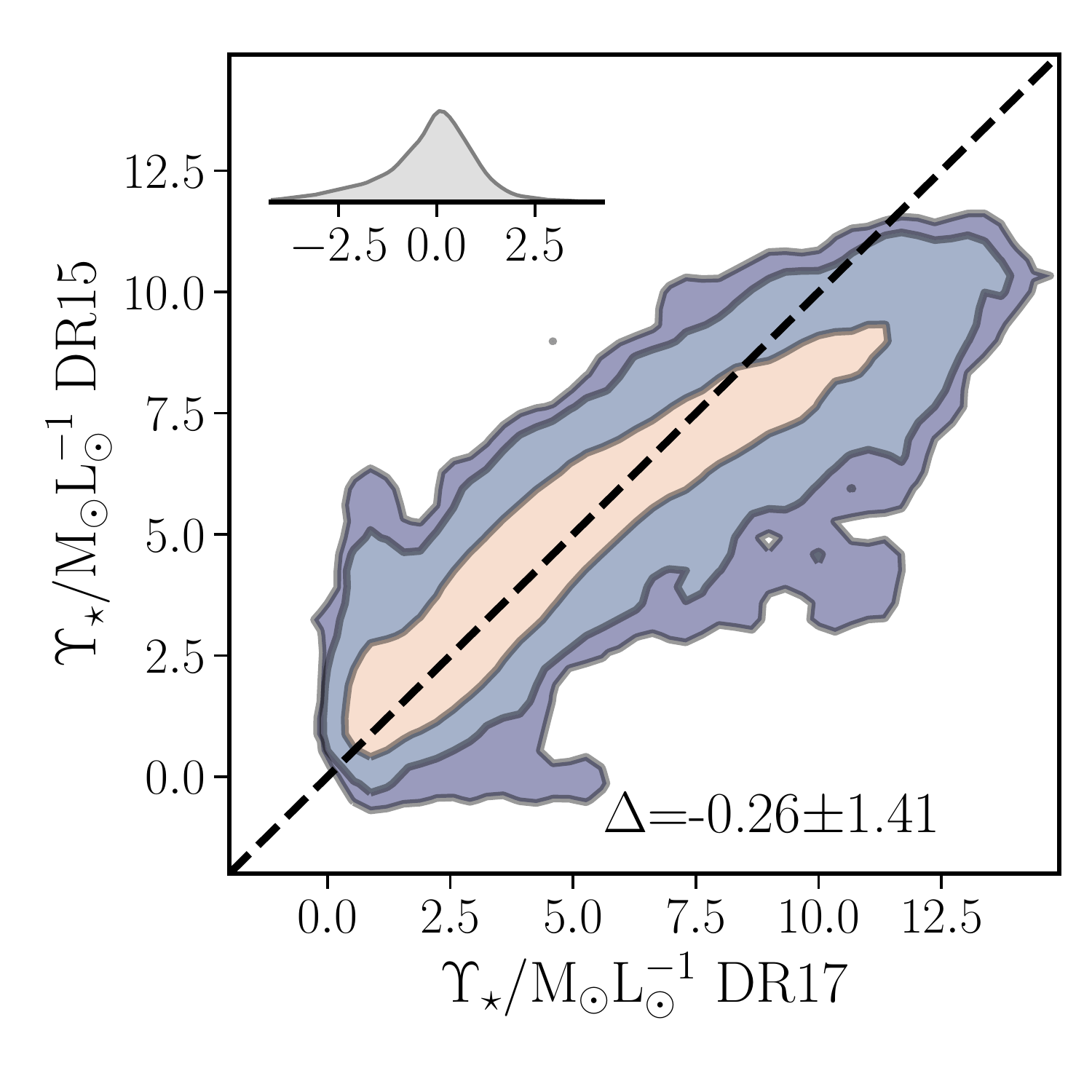}\includegraphics[width=6.25cm,clip,trim=0 10 0 10]{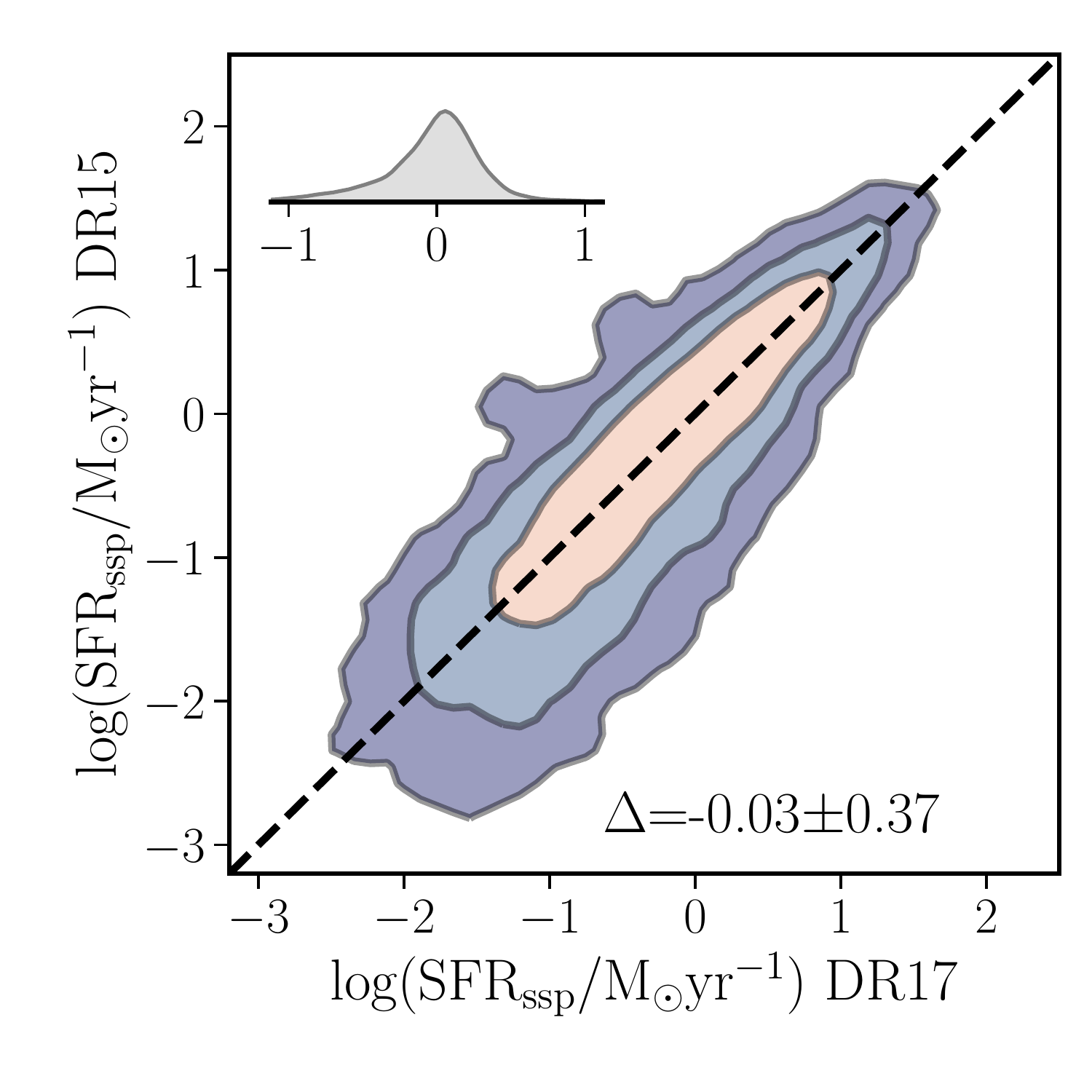}
  \includegraphics[width=6.25cm,clip,trim=0 10 0 10]{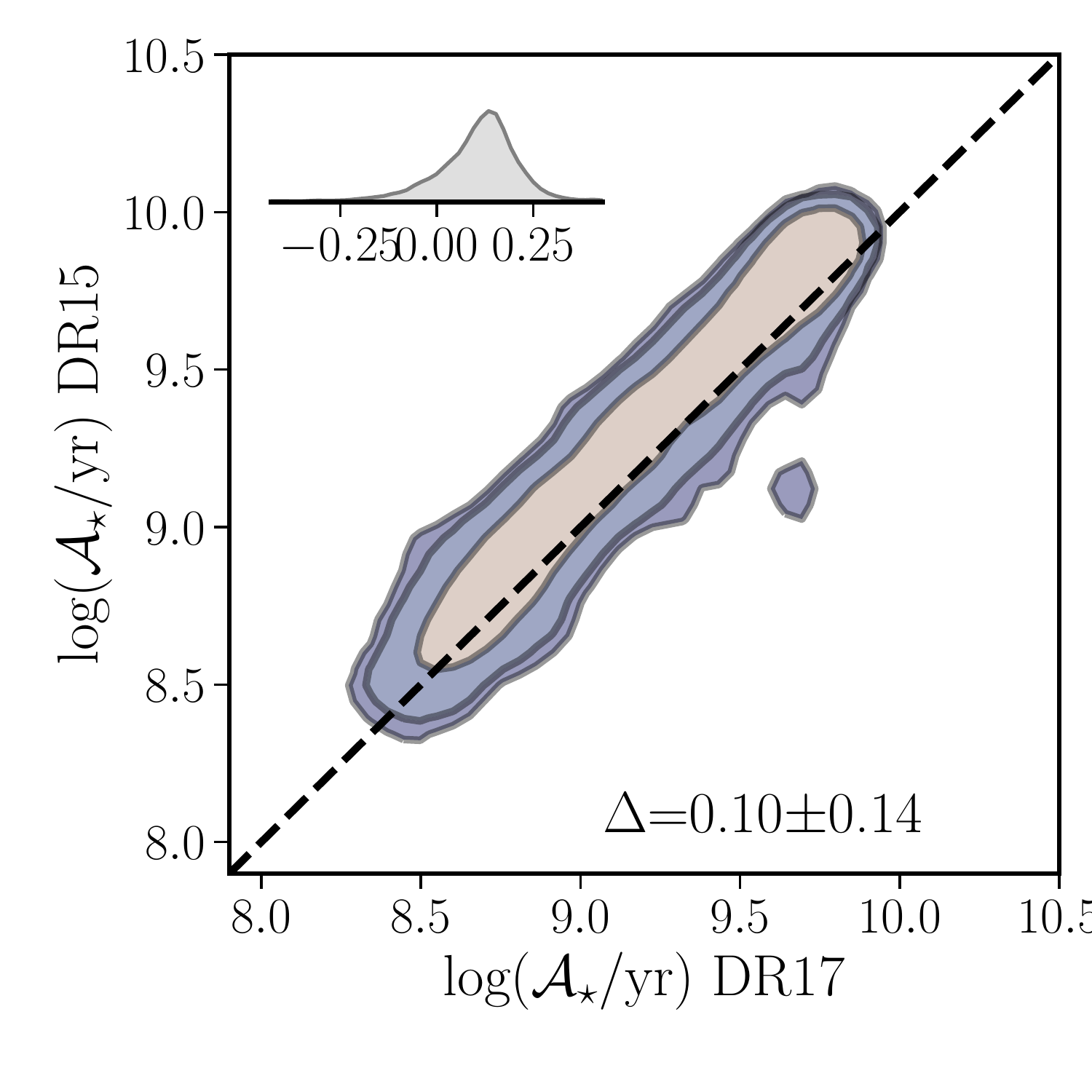}\includegraphics[width=6.25cm,clip,trim=0 10 0 10]{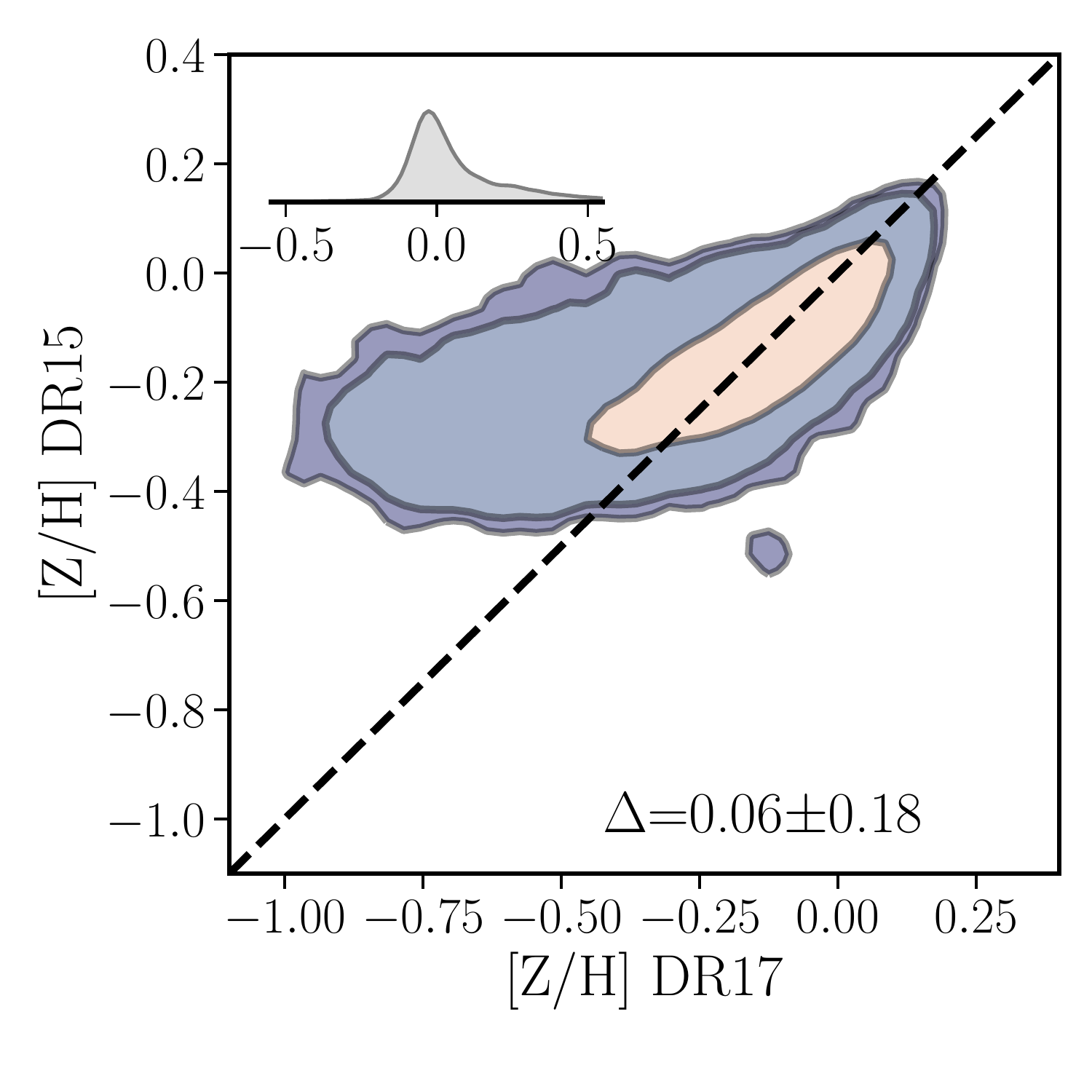}\includegraphics[width=6.25cm,clip,trim=0 10 0 10]{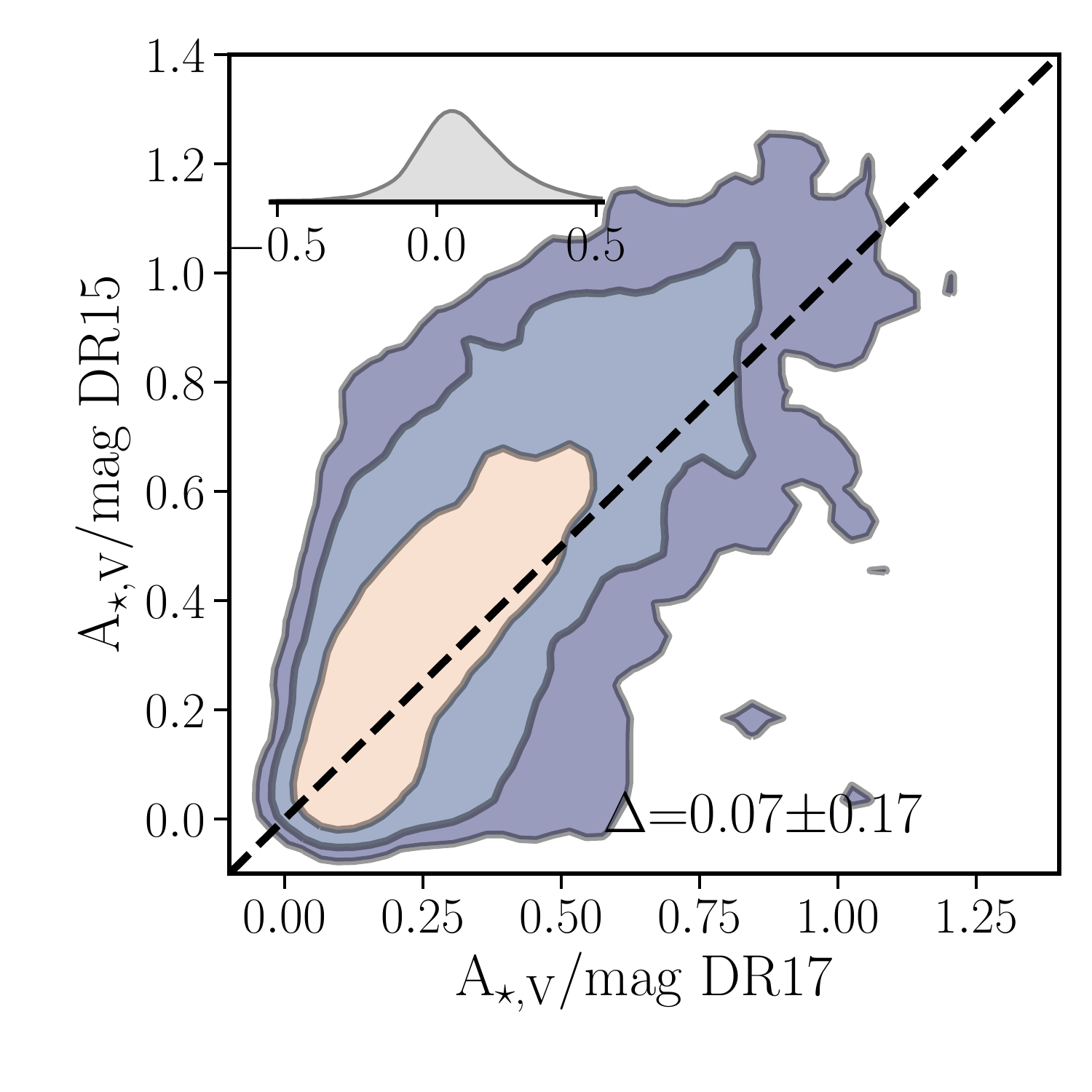}
  \includegraphics[width=6.25cm,clip,trim=0 10 0 10]{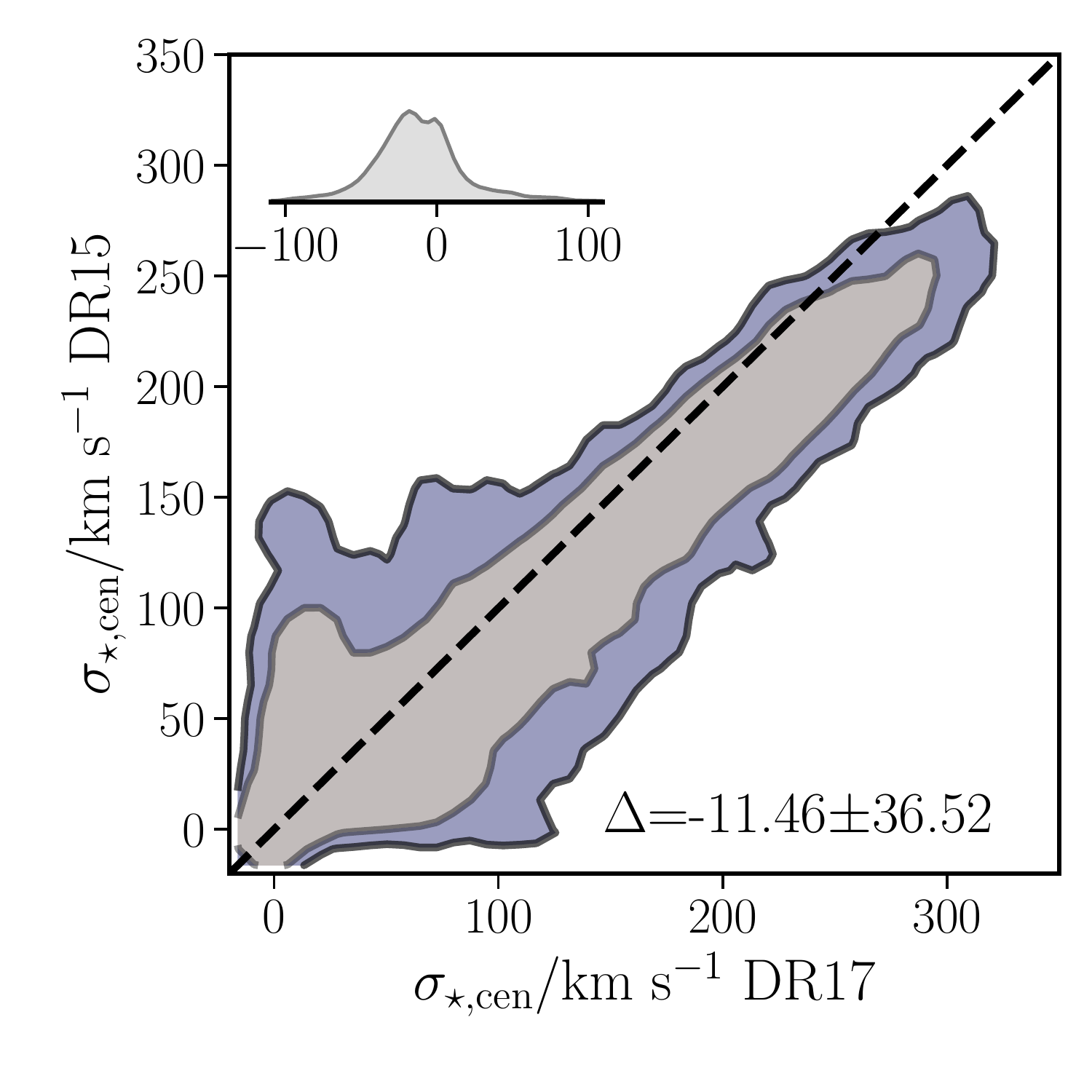}\includegraphics[width=6.25cm,clip,trim=0 10 0 10]{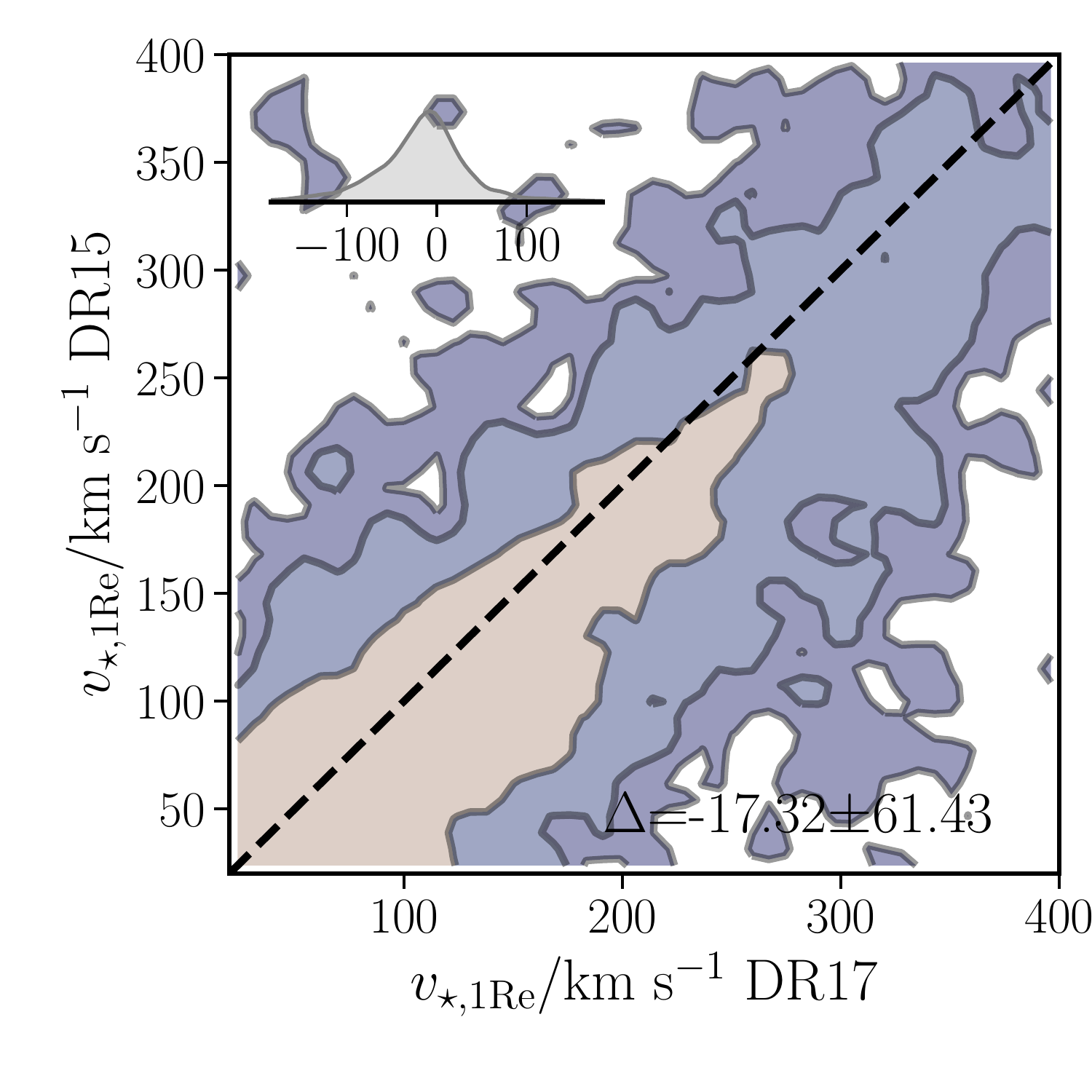}\includegraphics[width=6.25cm,clip,trim=0 10 0 10]{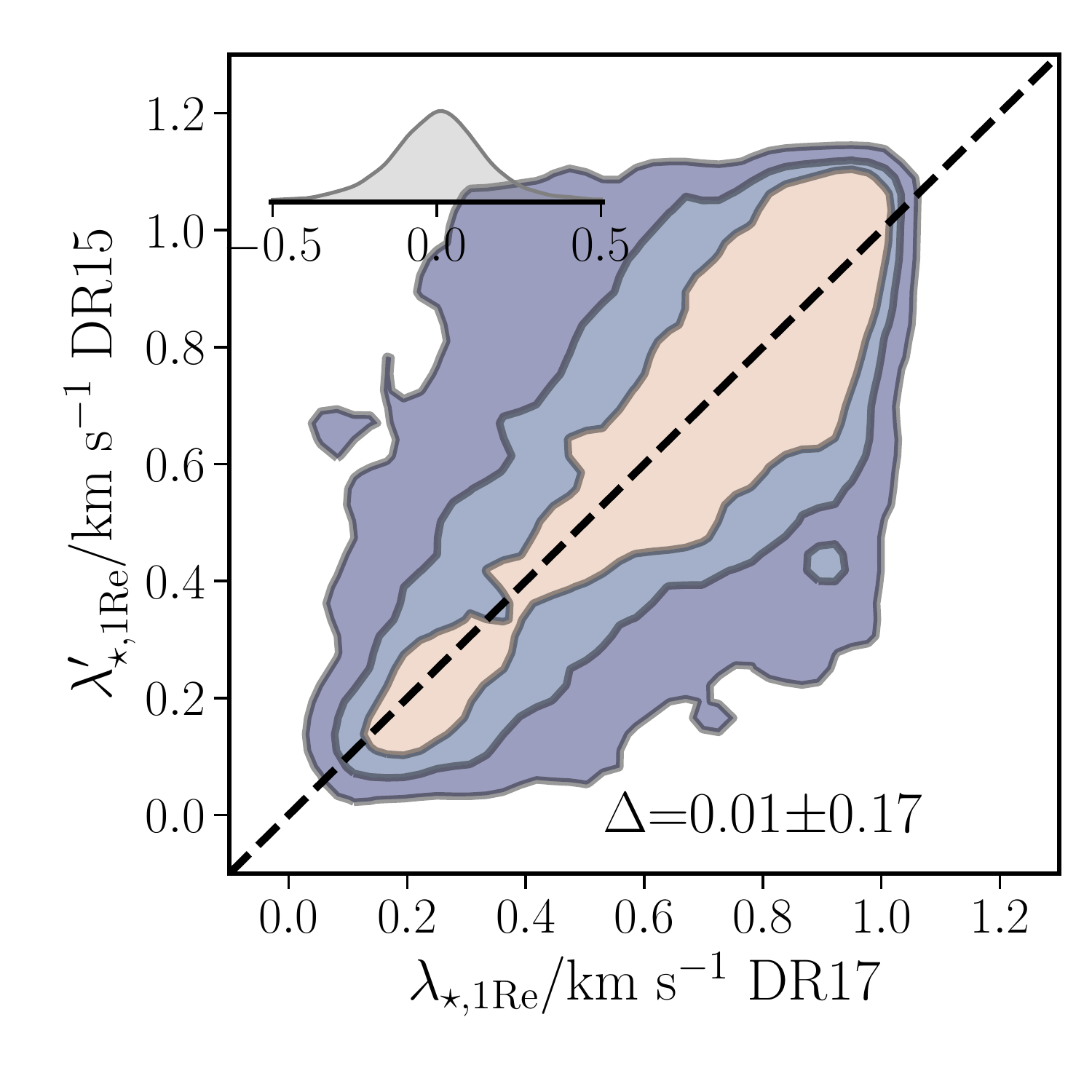}
 \endminipage
 \caption{Comparison of a set of stellar properties derived using {\tt Pipe3D} for the $\sim$4500 galaxies in MaNGA DR15 and \pyp\ for the same galaxies in MaNGA DR17. We show only the galaxies/cubes with good quality reported in both analyses. Each panel shows a density distribution in the DR15 vs. DR17 parameter diagram as a set of filled contours, with each consecutive contour encircling 99, 95 and 65\% of the points (i.e., 3, 2 and 1$\sigma$). The dashed-line shows the one-to-one relation. The upper-left inset shows the density distribution of the difference between the DR15 and DR17 values for the corresponding parameter. The mean value of this difference ($\Delta$) and its standard deviation are shown in the lower-right legend in each panel. From left-to-right and from top-to-bottom we show this comparison for: stellar mass (M$_\star$), average mass-to-light ratio ($\Upsilon_\star$), SFR  derived from the analysis of the stellar population (SFR$_{\rm ssp}$), LW age at the effective radius (\ageLW), LW metallicity at the effective radius (\metLW), average stellar dust extinction (A$_{\star,V}$), stellar velocity dispersion in the central aperture ($\sigma_{\star,\rm cen}$), stellar velocity at the effective radius ($v_{\star,\rm 1Re}$) and apparent angular momentum ($\lambda_{\star,\rm 1Re}$). For details on the derivation of these quantities see Sec.~\ref{sec:int}.}
 \label{fig:comp_ssp_DR15}
\end{figure*}
%%%%%%%%%%%%%%%%%%%%%%%%%%%%%%%%%%%%%%%%%%%%%%%%%%%%%%%%%%%%%%%%%%%%%%%5
%
%  Check Figures! Borders/Labels!
%  Check colors/contours!

As already mentioned, we have published previous versions of the analysis presented here using an out-dated version of the {\tt Pipe3D} pipeline and a different set of SSP templates for the stellar decomposition analysis. In \citet{pipe3d_ii} we presented this analysis for the IFS data of the CALIFA DR2 dataset \citep[][]{dr2}. This distribution is a very limited catalog with just 12 stellar and ionized gas characteristic/integrated properties, using a previous version of the {\tt Pipe3D} dataproducts for each of the 200 analyzed cubes (including only the {\tt SSP}, {\tt SFH}, {\tt FLUX\_ELINES} and {\tt INDICES} extensions). A similar analysis was presented in \citet{sanchez18} for the MaNGA DR14 dataset for $\sim$2800 galaxies. In this case we distributed a catalog\footnote{\url{https://www.sdss.org/dr14/manga/manga-data/manga-pipe3d-value-added-catalog}} comprising almost one hundred integrated/characteristic properties, in addition to the corresponding set of {\tt Pipe3D} files (one per galaxy). Finally, we distributed exactly the same set of properties and dataproducts for the MaNGA DR15 dataset\footnote{\url{https://www.sdss.org/dr15/manga/manga-data/manga-pipe3d-value-added-catalog}} including $\sim$4500 galaxies. 

Here we present a comparison of a set of selected properties derived for DR17 (current analysis) with their DR15 counterparts for the objects in common.
Fig.~\ref{fig:comp_ssp_DR15} shows this comparison for frequently used stellar parameters, % (Sec.~\ref{sec:cat_star}), 
including: (i) integrated stellar mass (M$_\star$); (ii) average mass-to-light ratio ($\Upsilon_\star$); (iii) star-formation rate derived from the analysis of the stellar population (SFR$_{\rm ssp}$); (iv) LW age at the effective radius (\ageLW); (v) LW metallicity at the effective radius (\metLW); (vi) average stellar dust extinction (A$_{\star,V}$); (vii) stellar velocity dispersion in the central aperture ($\sigma_{\star,\rm cen}$); (viii) stellar velocity at the effective radius ($v_{\star,\rm 1Re}$), and (ix) apparent angular momentum ($\lambda_{\star,\rm 1Re}$). 

The stellar mass, shown in the upper-left panel, presents a small systematic offsets of $\sim$~-0.07 dex between the two distributions, with DR15 masses being slightly larger than DR17 ones. Considering that the two datasets have been reduced using a different version of the MaNGA DRP (v2\_4\_3 vs. v3\_1\_1), and that we use a completely different SSP library, some systematic differences are expected.
This difference is considerably smaller than the scatter between both values, characterized by a standard deviation $\sigma_{\rm M\star}\sim$0.21 dex. Most of this scatter cannot be attributed to photometric differences as the expected spectrophotometric precision and accuracy of the MaNGA data is of the order of $\sim$5\% \citep{renbin16,renbin16b}. Indeed, a simple comparison of the $V$-band photometry measured directly on the DR15 and DR17 datacubes yields a differences of $\sim$0.02$\pm$0.05 dex. We consider that the scatter observed in M$_\star$ is introduced by differences in the estimated $\Upsilon_\star$, which are a combination of the use of a different stellar population template and, to a lower extent, to differences in the fitting procedure. This is evident from the comparison of the results for $\Upsilon_\star$ shown in the upper-middle panel of Fig.~\ref{fig:comp_ssp_DR15}. 
Although there is a clear correspondence between the two values, near to a one-to-one relation for $\Upsilon_\star<$~7.5 { M$_\odot$/L$_\odot$}, there is a deviation, with DR15 values being slightly lower than DR17 values for $\Upsilon_\star>$~7.5 { M$_\odot$/L$_\odot$}. This introduces a bias of -0.26 M$_\odot$L$^{-1}_\odot$ in $\Delta\Upsilon_\star$. 
The considerable scatter in the distribution, with $\sigma_{\Upsilon\star}$=1.41 M$_\odot$L$^{-1}_\odot$ in $\Delta\Upsilon_\star$, corresponding to $\sim$0.15 dex on average, is clearly the main driver for the observed differences in the stellar masses. This scatter is propagated to any other quantity that requires the use of $\Upsilon_\star$, like the SFR derived from the stellar population spectral decomposition (Sec.~\ref{sec:cat_star}), shown in the upper-right panel
of Fig.~\ref{fig:comp_ssp_DR15}. In this case we compare the SFR$_{\rm ssp}$ estimated at 32 Myr. There is clear agreement between the DR17 and DR15 SFR values, following on average a one-to-one relation. This is reflected in the small average difference ($\sim-$0.03 dex). However, the dispersion around the one-to-one relation, $\sim$0.37 dex, is larger than the dispersion reported for both $\Upsilon_\star$ and M$_\star$. This is expected, since the derivation of SFR depends directly on the derivation of the two previous parameters, with the additional uncertainty of estimating the stellar mass for just a tiny fraction of the total stellar component (younger than 32 Myr, according to Eq.\ref{eq:sfh_int}). 

%%%%%%%%%%%%%%%%%%%%%%%%%%%%%%%%%%%%%%%%%%%%%%%%%%%%%%%%%%%%%%%%%%%%%%%5
\begin{figure*}
 \minipage{0.99\textwidth}
 \includegraphics[width=6.25cm,clip,trim=0 10 0 10]{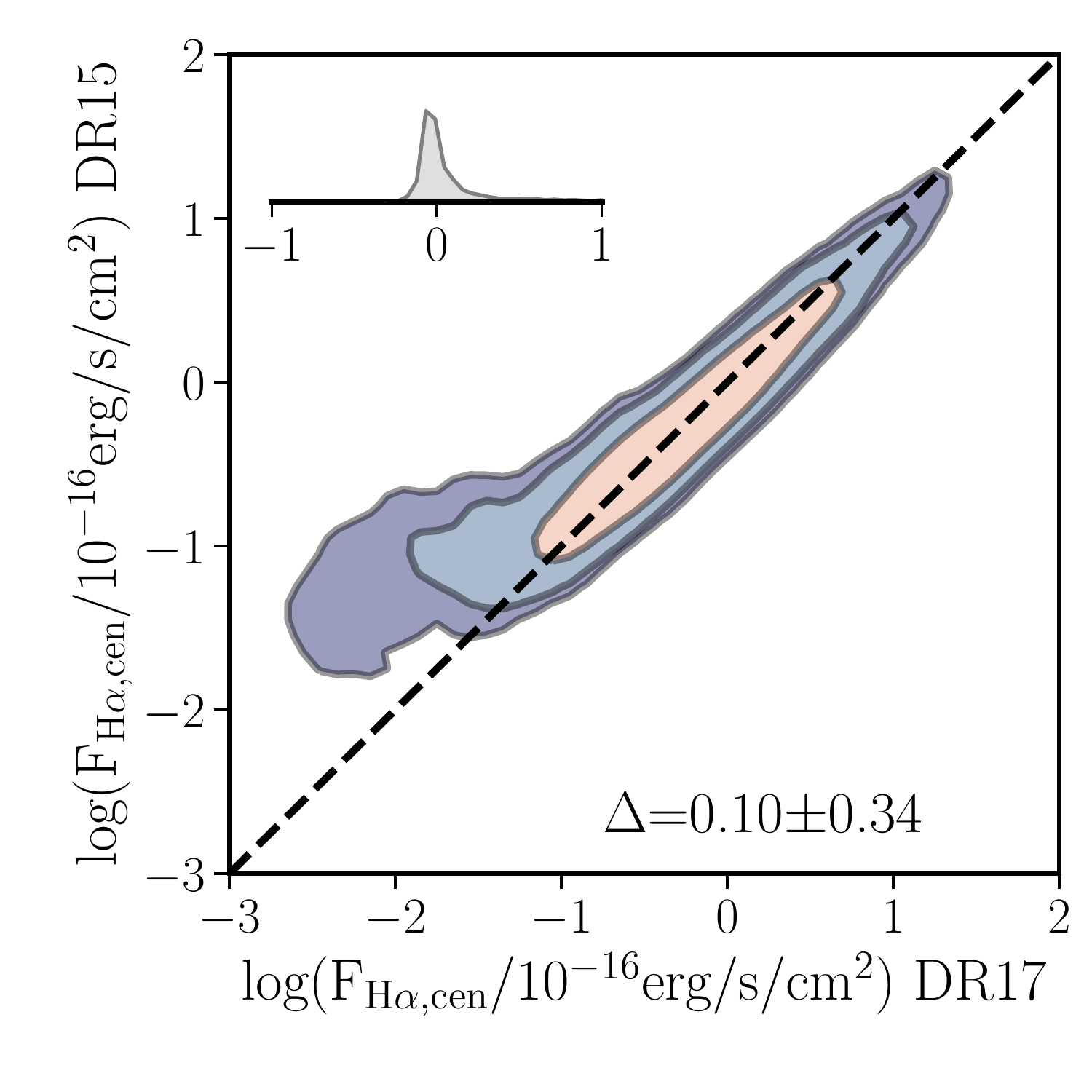}\includegraphics[width=6.25cm,clip,trim=0 10 0 10]{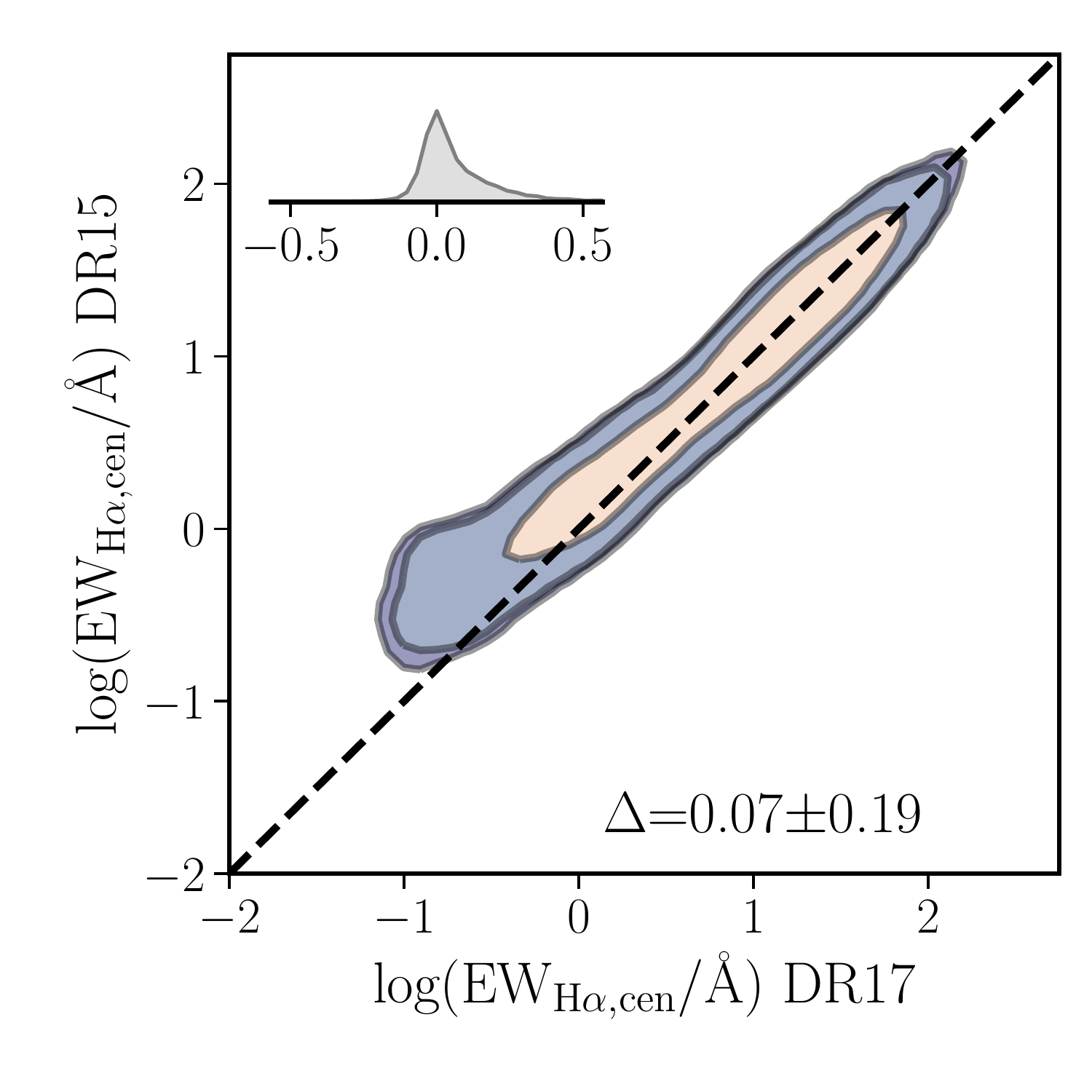}\includegraphics[width=6.25cm,clip,trim=0 10 0 10]{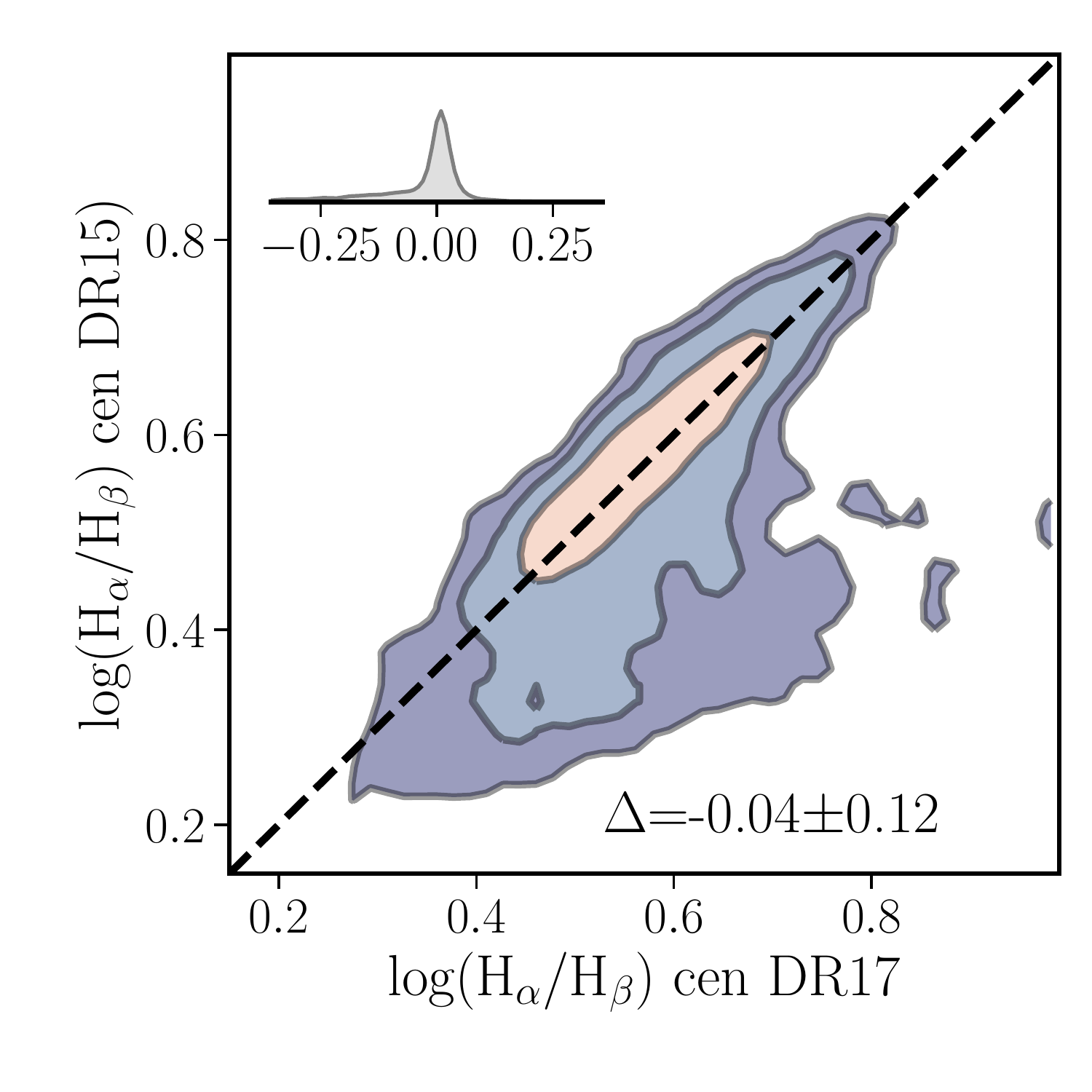}
  \includegraphics[width=6.25cm,clip,trim=0 10 0 10]{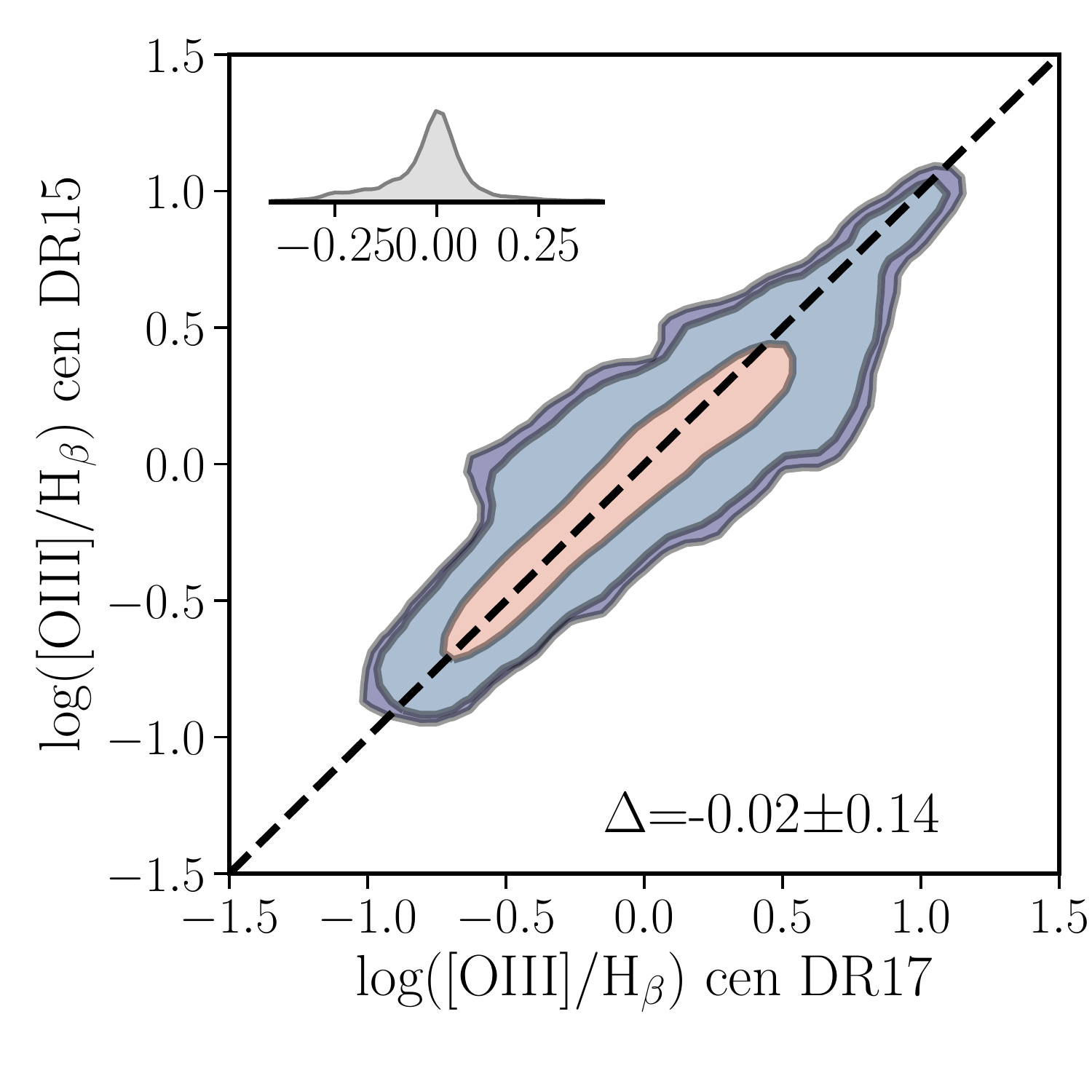}\includegraphics[width=6.25cm,clip,trim=0 10 0 10]{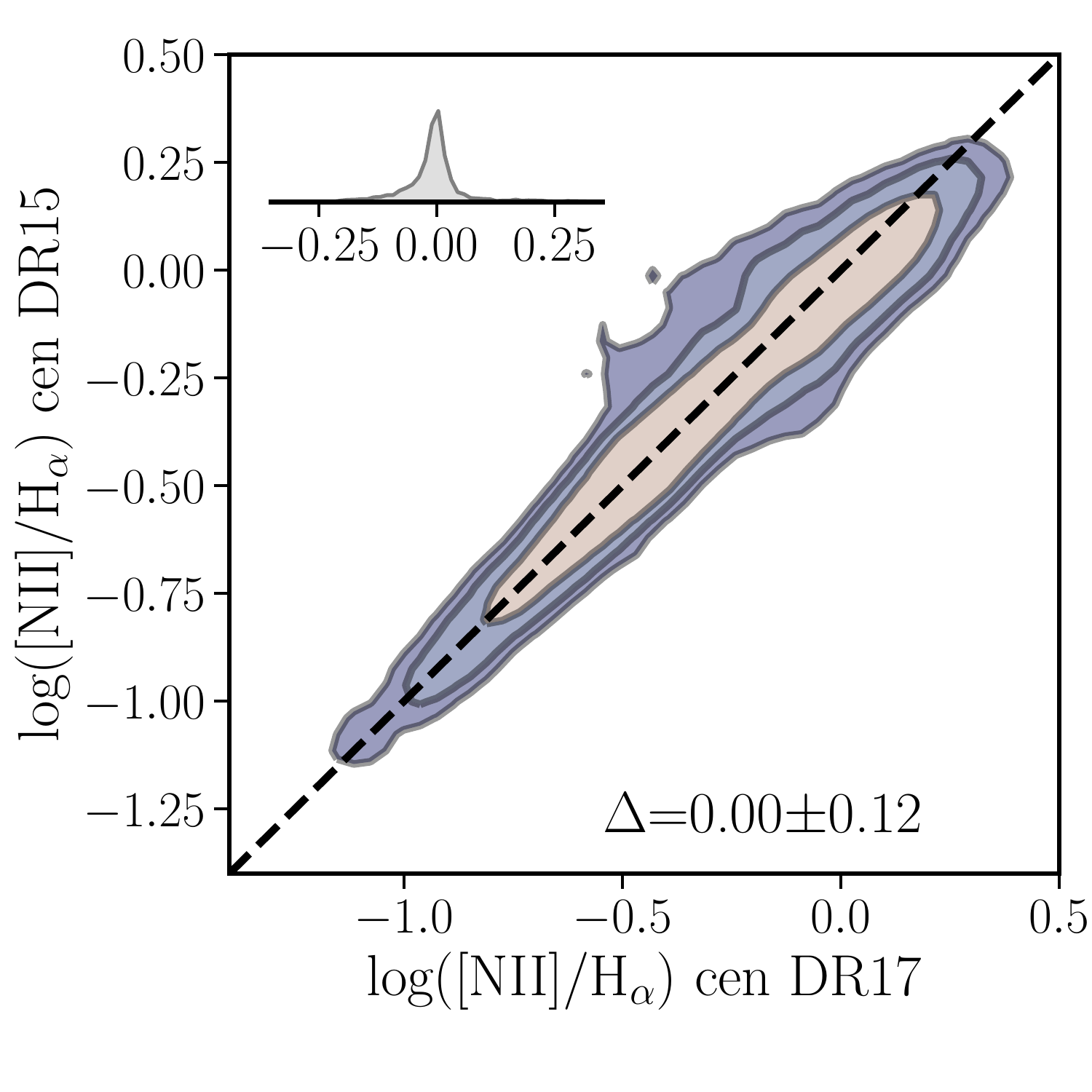}\includegraphics[width=6.25cm,clip,trim=0 10 0 10]{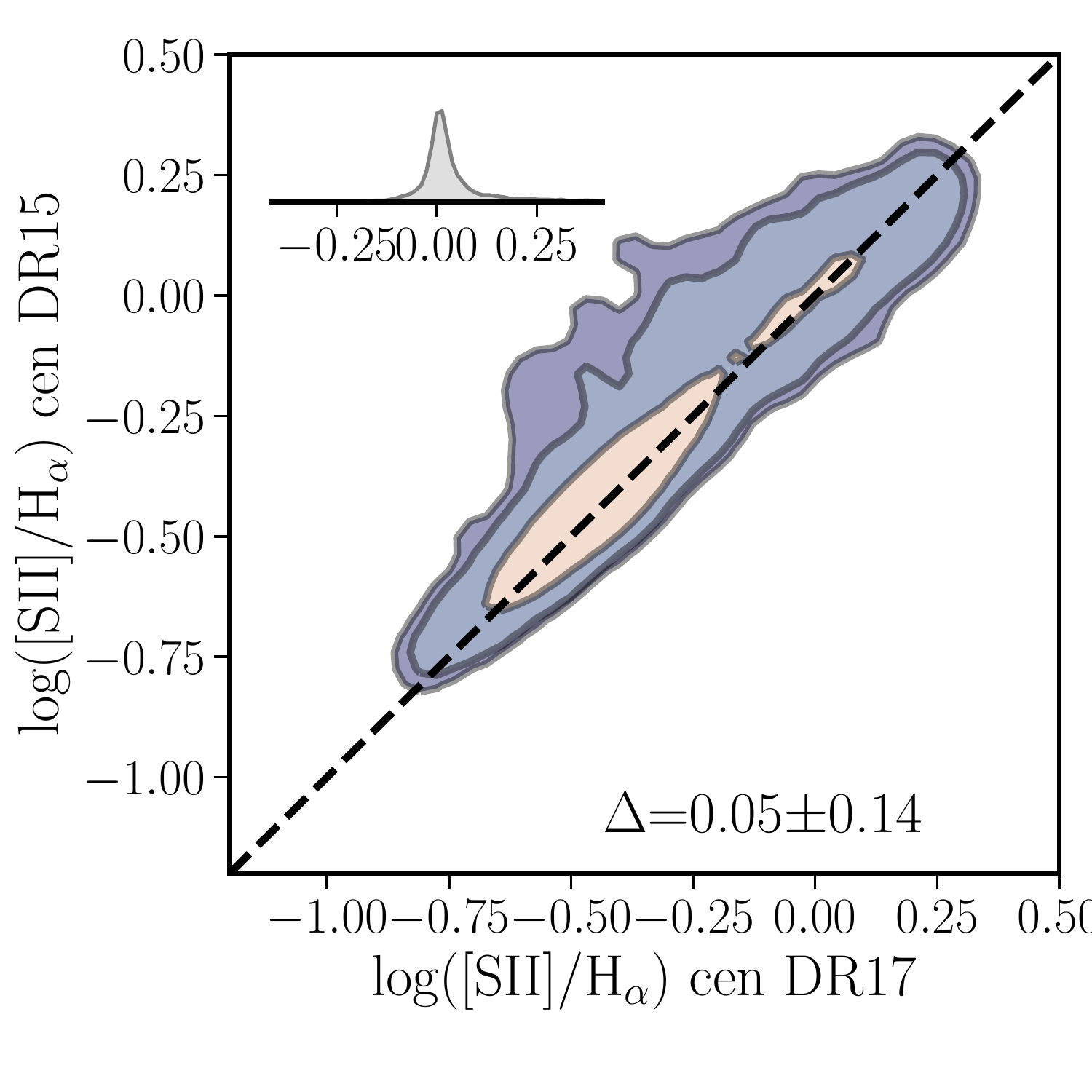}
  \includegraphics[width=6.25cm,clip,trim=0 10 0 10]{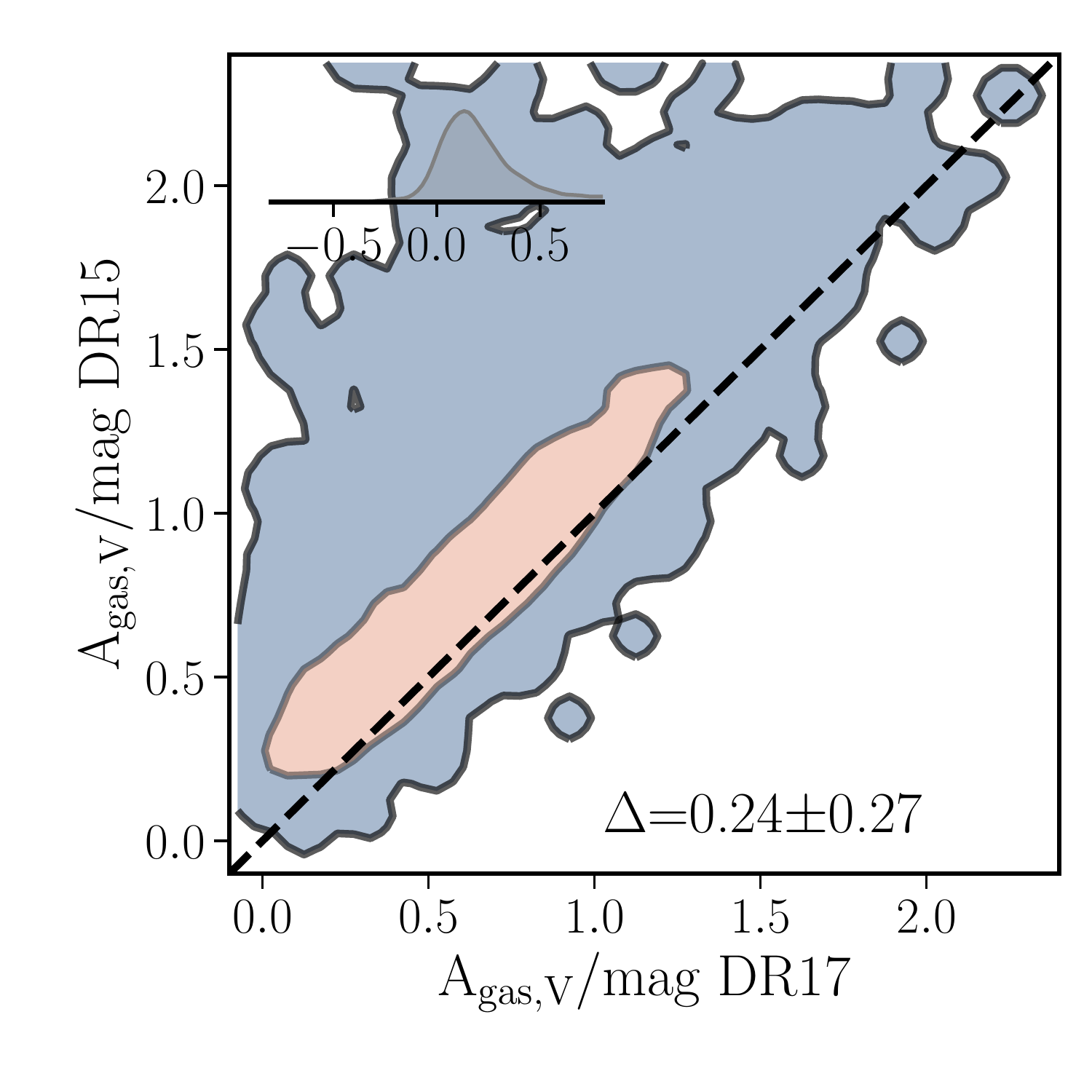}
  \includegraphics[width=6.25cm,clip,trim=0 10 0 10]{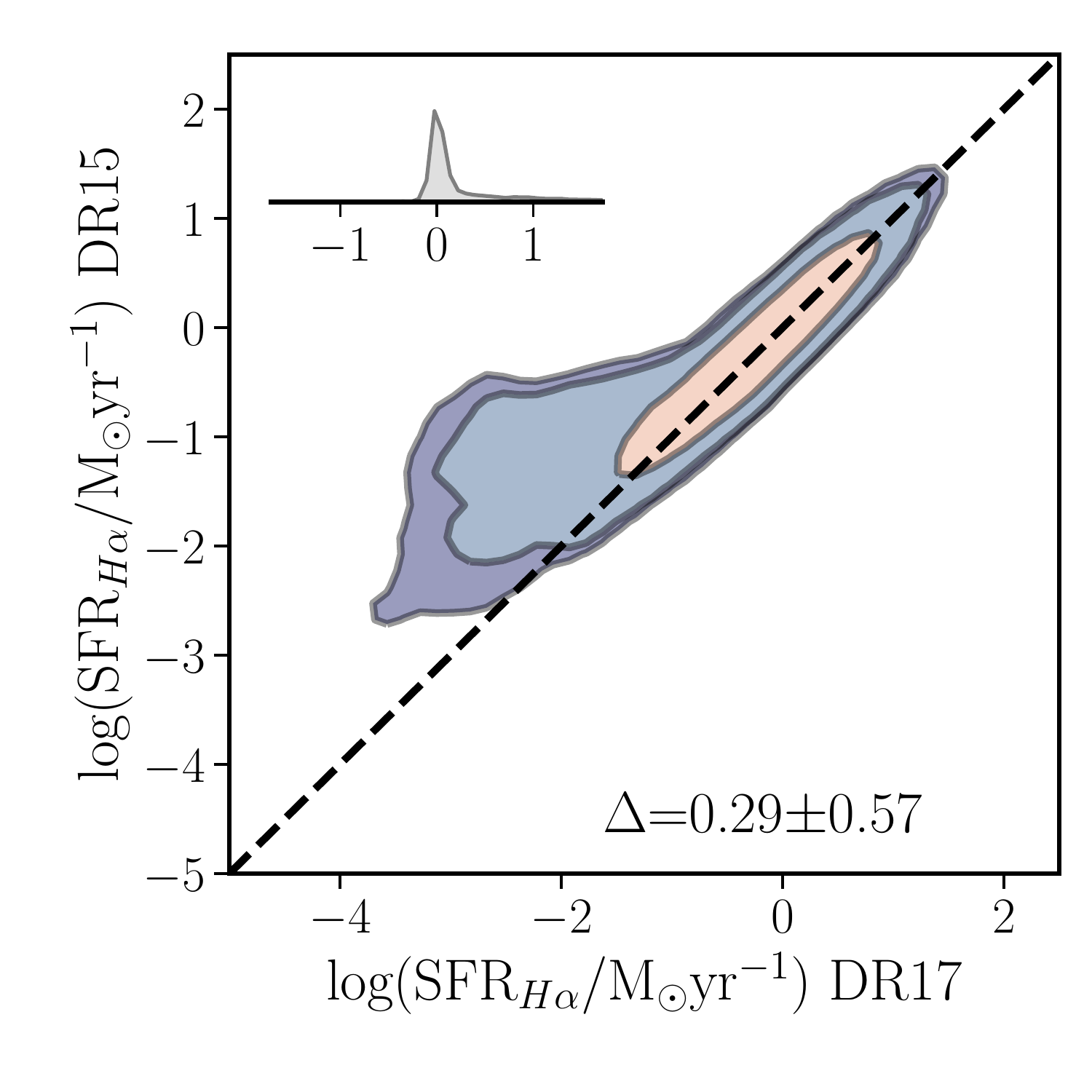}\includegraphics[width=6.25cm,clip,trim=0 10 0 10]{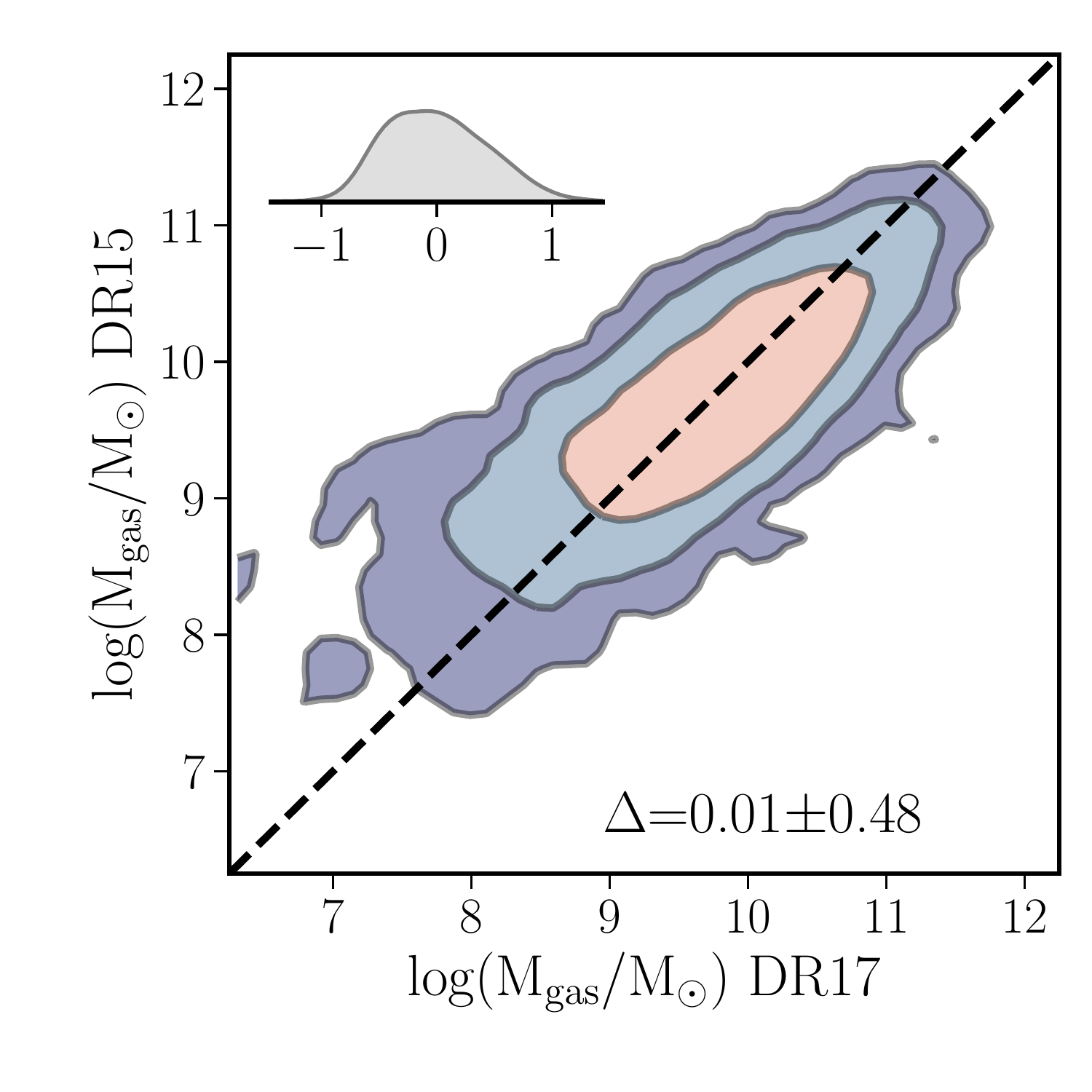}
 \endminipage
 \caption{Comparison of a set of ionized gas properties derived using {\tt Pipe3D} for the $\sim$4500 galaxies in MaNGA DR15 and \pyp\ for the same galaxies in MaNGA DR17. We show only the galaxies/cubes with good quality reported in both analyses, 
 using the same format adopted in Fig.~\ref{fig:comp_ssp_DR15}. From left-to-right and from top-to-bottom we compare: H$\alpha$ flux (F$_{H\alpha,\rm cen}$), equivalent width of H$\alpha$ (EW$_{H\alpha,\rm cen}$), H$\alpha$ to H$\beta$ line ratio, [OIII] to H$\alpha$ line ratio, [NII] to H$\alpha$ line ratio, and [SII] to H$\alpha$ line ratio, all of them at the central aperture; the average dust extinction across the optical extension of the galaxy derived from the H$\alpha$ to H$\beta$ ratio, the integrated SFR derived from the H$\alpha$ luminosity, and the molecular gas integrated mass derived from the dust-to-gas ratio. For details on the derivation of these quantities see Sec. \ref{sec:int}. }
 \label{fig:comp_gas_DR15}
\end{figure*}
%%%%%%%%%%%%%%%%%%%%%%%%%%%%%%%%%%%%%%%%%%%%%%%%%%%%%%%%%%%%%%%%%%%%%%%5
%
%  Check Figures! Borders/Labels!
%  Check colors/contours!

%So far no significant systematic difference is appreciated in the comparison of the stellar population parameters derived for the DR15 and DR17 datasets. This is not the case for the next parameters to be compared. 
The central-left panel of Fig.~\ref{fig:comp_ssp_DR15} shows the comparison of the LW stellar age. Again there is a correspondence between the DR15 and DR17 values. However, the relation is not just on top of the one-to-one line. There is a clear systematic offset of $\sim$0.1 dex, with ages derived for DR15 being older than for DR17. The difference is small compared to  the dynamical range of the parameter ($\sim$2 dex), but it is well determined, even considering the scatter around the central relation, $\sigma_{\age}$=0.14 dex. We built the age sampling of the new SSP template following a similar philosophy as the one that was adopted to generate the {\tt GSD156} template (a pseudo logarithmic sampling of the time steps). However, neither the detailed set of ages nor the range of metallicities (and their sampling) is equal between the two sets of SSPs. In addition, the two templates use a different stellar library (MILES vs. MaStar), a different set of isochrones, and were computed by two different synthesis codes. All together, this can easily explain the observed age differences, that, in any case, are rather small. 

Larger differences are found for the luminosity-weighted stellar metallicity (central-middle panel of Fig.~\ref{fig:comp_ssp_DR15}). For this parameter we find a one-to-one correspondence between the two datasets for the stellar \metLW$_{,\rm DR17}>$-0.5 dex. Below this value the distribution bends towards a plateau for DR15 metallicities at $\sim-0.3$ dex. Despite of this deviation from the one-to-one relation the average difference between the two values is rather small, with a $\Delta$\metLW=0.06$\pm$0.18 dex. The reported differences are most probably due to the different sampling and coverage in stellar metallicity in the newly adopted SSP template, ranging from \met=-2.30 to 0.30 dex, sampled in seven metallicities, compared to the DR15 template, ranging from -0.73 to 0.20 dex, sampled in four metallicities. However, this alone cannot explain the observed difference, since the plateau is not reached at the minimum sampled metallicity for the {\tt GSD156}. A combination of the difference in the metallicity coverage, the different stellar libraries, the adopted isochrones and the different synthesis codes must be behind the observed behavior. As a result of this difference, similar qualitative results will be found when using the DR15 and DR17 \metLW values, but quantitative results will change. Global relations, like the stellar mass-metallicity relation, or spatially resolved trends, like radial gradients, will present a wider dynamical range for the DR17 dataset. However, we do not expect strong changes, neither in the shape of the relations, nor in the sign of the gradients.

The right-middle panel of Fig.~\ref{fig:comp_ssp_DR15} shows the comparison between the dust extinction for the two datasets. Like in the case of the two previous parameters, there is a well defined correspondence between the two values. The best agreement is found for low dust-extinction values (A$_{\rm V}<$0.3 mag), with a distribution near to the one-to-one relation. For larger values, the A$_{\rm V}$ reported in DR15 is slightly higher than for DR17, which translates into a positive $\Delta$A$_{\rm V}=~$0.07$\pm$0.17 mag. Again, despite of the lack of a one-to-one correspondence, for the three cases shown in the central panels these differences are rather small, being just two or three times larger than the expected errors for the parameters \citep[Fig. 6 of][]{pypipe3d}

In the lower panels of Fig.~\ref{fig:comp_ssp_DR15} we compare the DR15 and DR17 results for the kinematical parameters: $\sigma_{\star,\rm cen}$, vel$_{\star,\rm Re}$ and $\lambda_{\star,\rm Re}$.
For $\sigma_{\star,\rm cen}$, and vel$_{\star,\rm Re}$ the distributions follow a one-to-one relation, with offsets lower than the reported standard deviations: $\Delta\sigma_{\star,\rm cen}=$-12$\pm$37 km\ s$^{-1}$ and $\Delta$vel$_{\star,\rm cen}=$-17$\pm$61 km\ s$^{-1}$. In the case of $\sigma_{\star,\rm cen}$, the scatter of the difference is of the order of the expected errors for this parameter based on simulations, $\sim 22 $ km\ s$^{-1}$ \citep{pypipe3d}, and slightly smaller than the instrumental resolution, $\sim$75 km\ s$^{-1}$ \citep[e.g.,][]{law21}. For the stellar velocity at the effective radius, the scatter of the difference is three times larger than the expected errors, $\sim$20 km\ s$^{-1}$ \citep{pypipe3d}, being of the order of the instrumental resolution. In both cases the differences are larger for the lowest values of the parameters, in particular for the velocity. In general we do not anticipate any qualitative or quantitative difference in results derived from both datasets related to this two kinematic parameters. Larger differences between the DR15 and DR17 values are found for $\lambda_{\star,\rm Re}$ due to the inclination correction described in Sec.~\ref{sec:cat_kin} (Eq. \ref{eq:l_cor}).
Once corrected, $\lambda^\prime_{\star,\rm Re,DR15}$ follows the DR17 results along the one-to-one relation (lower-right panel of Fig.~\ref{fig:comp_ssp_DR15}). The two values agree within $\Delta\lambda^\prime_{\star,\rm Re}$=0.4$\pm$0.19, with the larger discrepancies being found for the larger values of the angular momentum. 

%with DR17 values being systematically larger and not showing a one-to-one relation with t
%The age sampling is slightly different from the two SSP templates ({\tt GSD156} vs. {\tt MaStar\_sLOG}), but 

Fig.~\ref{fig:comp_gas_DR15} shows the same type of comparison for a selected subset of parameters derived from the ionized gas emission lines included in the final catalog (Sec.~\ref{sec:cat_elines}). We compare: (i) H$\alpha$ flux (F$_{H\alpha,\rm cen}$), (ii) equivalent width of H$\alpha$ (EW$_{H\alpha,\rm cen}$), (iii) H$\alpha$ to H$\beta$ line ratio, (iv) [OIII] to H$\alpha$ line ratio, (v) [NII] to H$\alpha$ line ratio, and (vi) [SII] to H$\alpha$ line ratio, all measured in the central aperture; additionally we include: (vii) average dust extinction across the entire galaxy derived from the H$\alpha$ to H$\beta$ ratio, (viii) integrated SFR derived from the H$\alpha$ luminosity, and (ix) the molecular gas integrated mass derived from the dust-to-gas ratio. In most cases, we find quite good agreement between the two estimates of the parameter. In the case of F$_{H\alpha,\rm cen}$ there is a tight one-to-one relation with an offset lower than the standard deviation over most of the dynamical range. Most of the scatter and the deviation from the one-to-one relation is found at the very low flux intensities, where the differences are driven by the accuracy in the subtraction of the underlying stellar population rather than by the properties of the emission lines themselves (e.g., their signal-to-noise ratio). Indeed, the scatter presents a standard deviation $\sim$0.34 dex, three times larger than expected based on the emission line properties \citep[$\sim$0.1 dex, Fig. 9 of][]{pypipe3d}. For log(F$_{H\alpha,\rm cen, DR17}$)$<$-1.5 dex the distribution bends, with DR15 values reaching a plateau. The newly adopted SSP template is based on a stellar library observed with the same instrument and reduced with the same tools, and therefore its spectral resolution should match better that of the observed data. Therefore, we consider that the new derivation of the emission line fluxes is more reliable, in particular in the central regions, where the flux intensity of the emission lines is low, in general, and the stellar component is brighter and older (i.e., the worst case scenario for a proper estimate of the emission line properties). For the same reasons a similar trend is observed for the EW(H$\alpha$), although in this case the agreement with respect to the one-to-one relation seems to be slightly better, with a $\Delta$log $\big ($EW(H$\alpha$)$\big )$=0.07$\pm$0.19.

We find similar results for the four line ratios in Fig.~\ref{fig:comp_gas_DR15}. In all cases the bulk of the distribution is located around the one-to-one relation, with tiny offsets of  $\sim$0.0-0.5 dex and a scatter of $\sim$0.13 dex. The largest differences are found for the H$\alpha$/H$\beta$ ratio that shows a tail towards larger values for DR17 compared to DR15 below $<$2.8. This tail is not observed in the other line ratios, indicating that it is most probably related to an effect that affects the Balmer lines simultaneously. We consider that the differences in the SSP templates are behind this effect. In general, based on these results, none of the conclusions regarding the ionization conditions, and most probably none of the properties derived from the line ratios (e.g., oxygen abundance, ionization parameter), should be affected by the choice of either the DR15 or the DR17 dataproducts.

%%%%%%%%%%%%%%%%%%%%%%%%%%%%%%%%%%%%%%%%%%%%%%%%%%%%%%%%%%%%%%%%%%%%%%%5
\begin{figure*}
 \minipage{0.99\textwidth}
 \includegraphics[width=6.25cm,clip,trim=0 10 0 10]{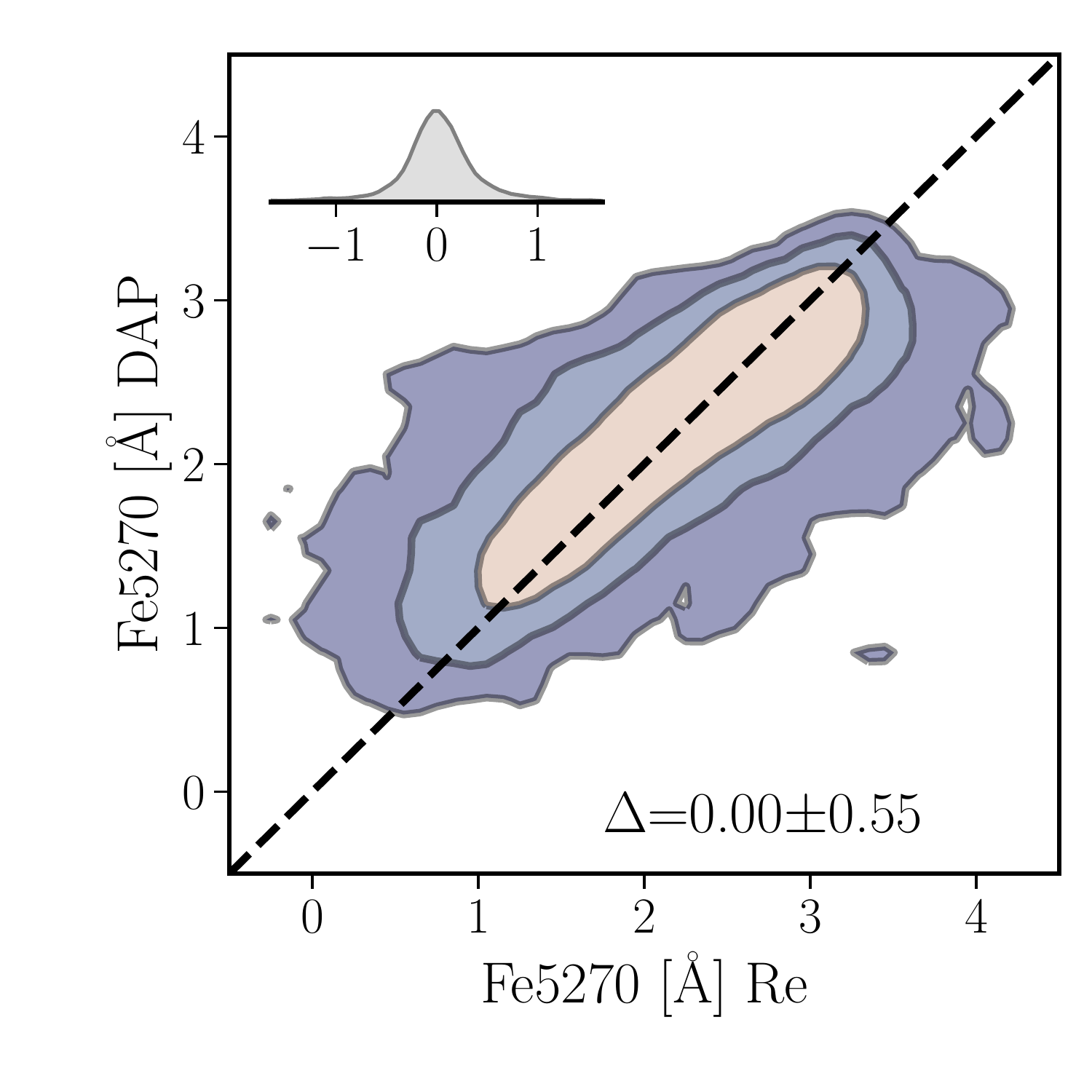}\includegraphics[width=6.25cm,clip,trim=0 10 0 10]{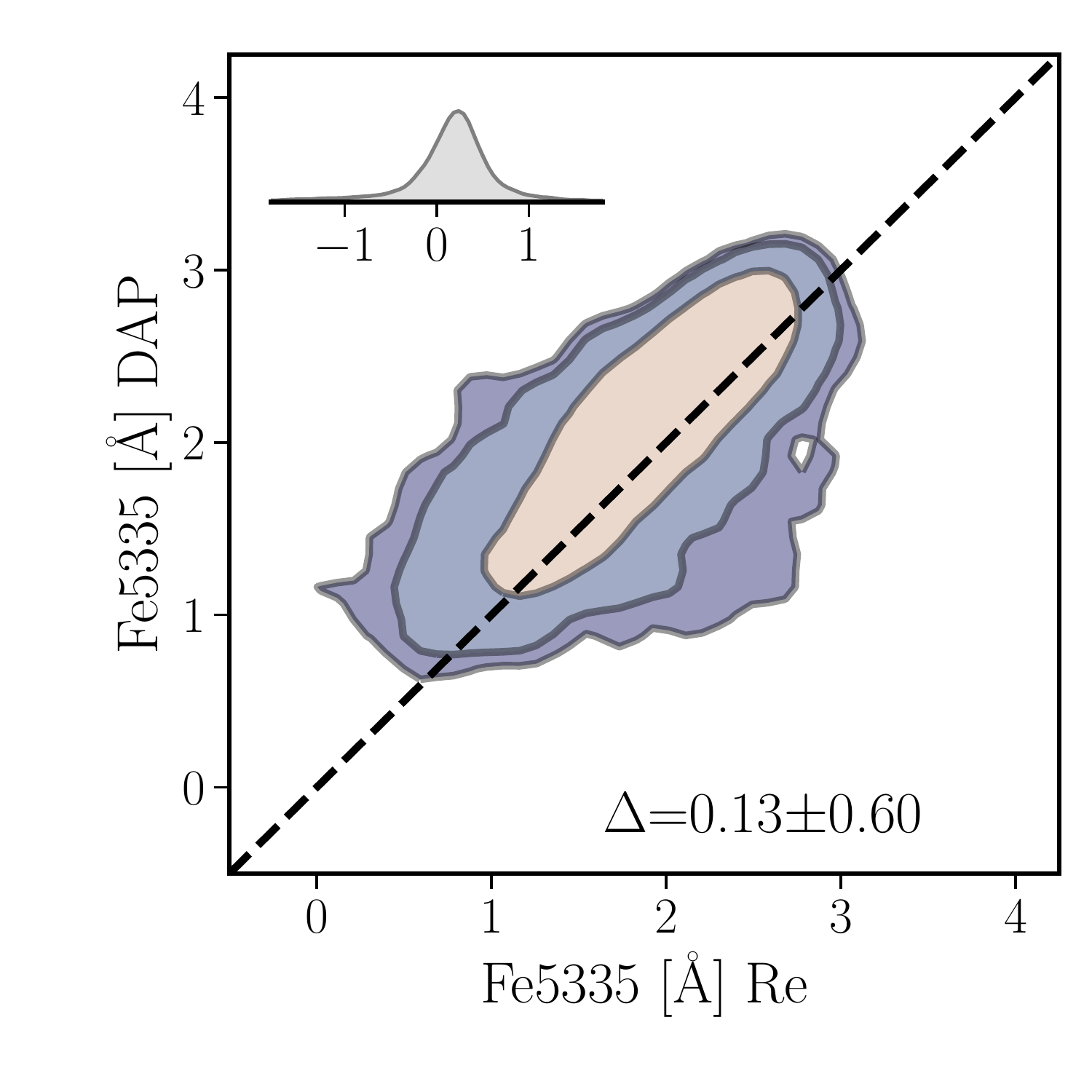}\includegraphics[width=6.25cm,clip,trim=0 10 0 10]{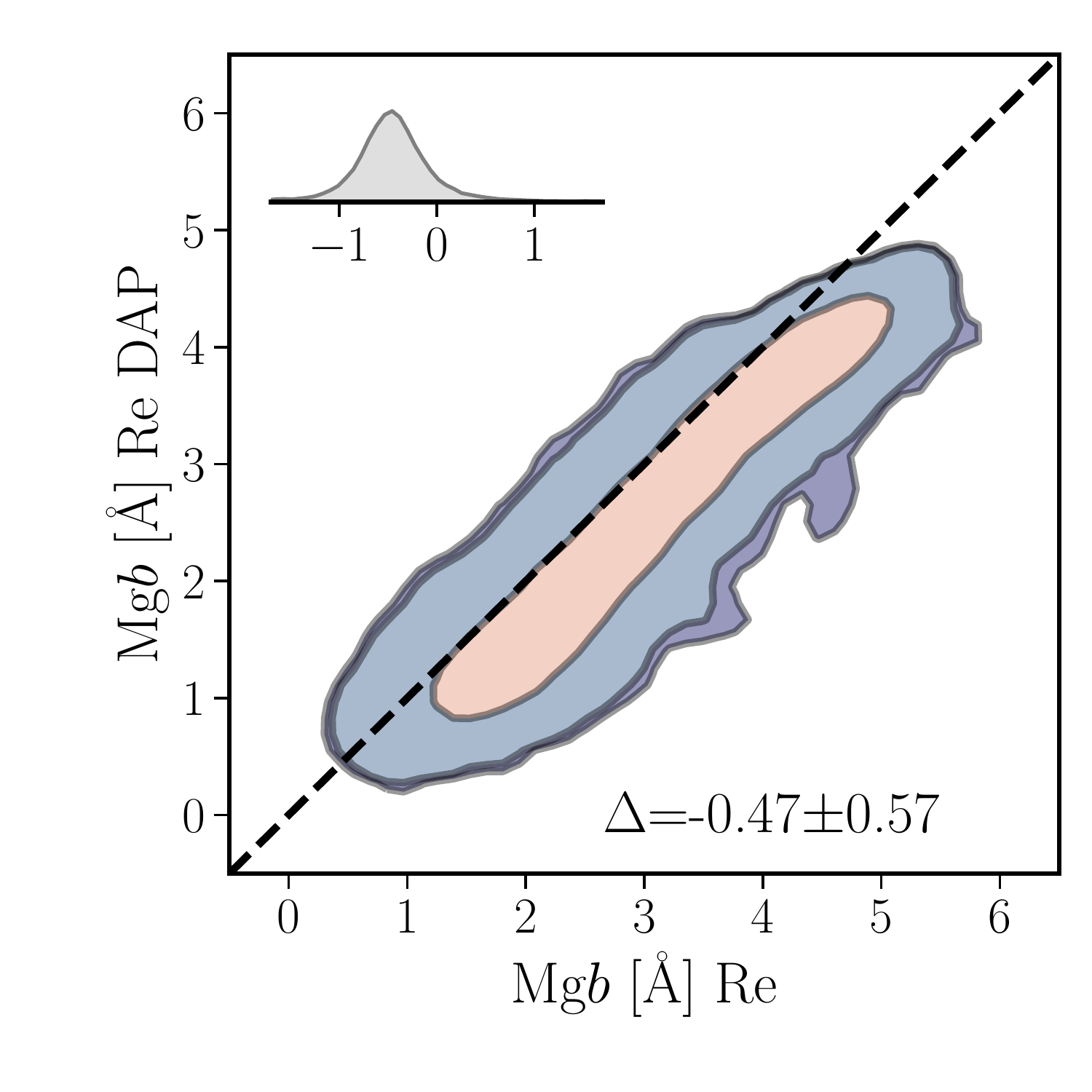}
  \includegraphics[width=6.25cm,clip,trim=0 10 0 10]{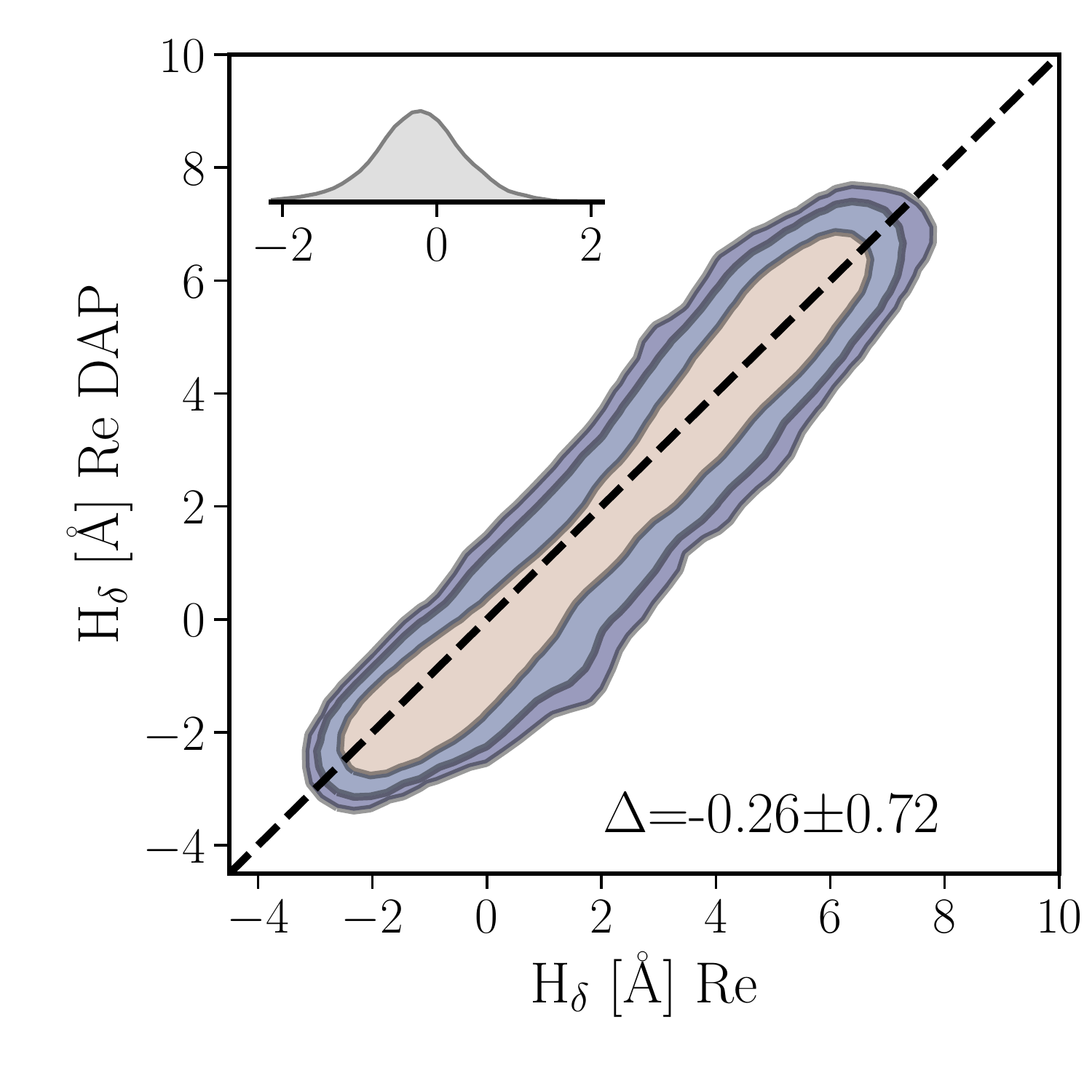}\includegraphics[width=6.25cm,clip,trim=0 10 0 10]{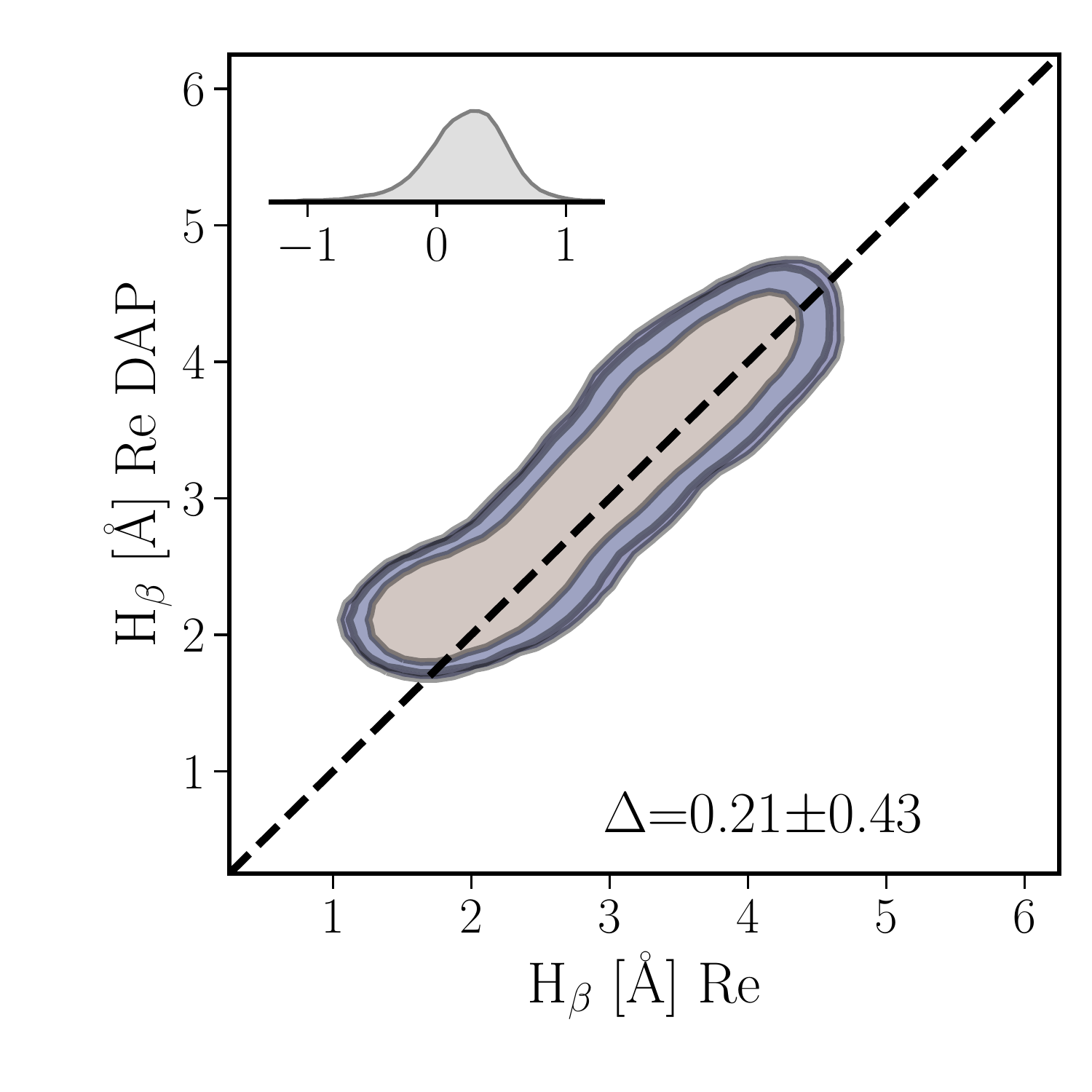}\includegraphics[width=6.25cm,clip,trim=0 10 0 10]{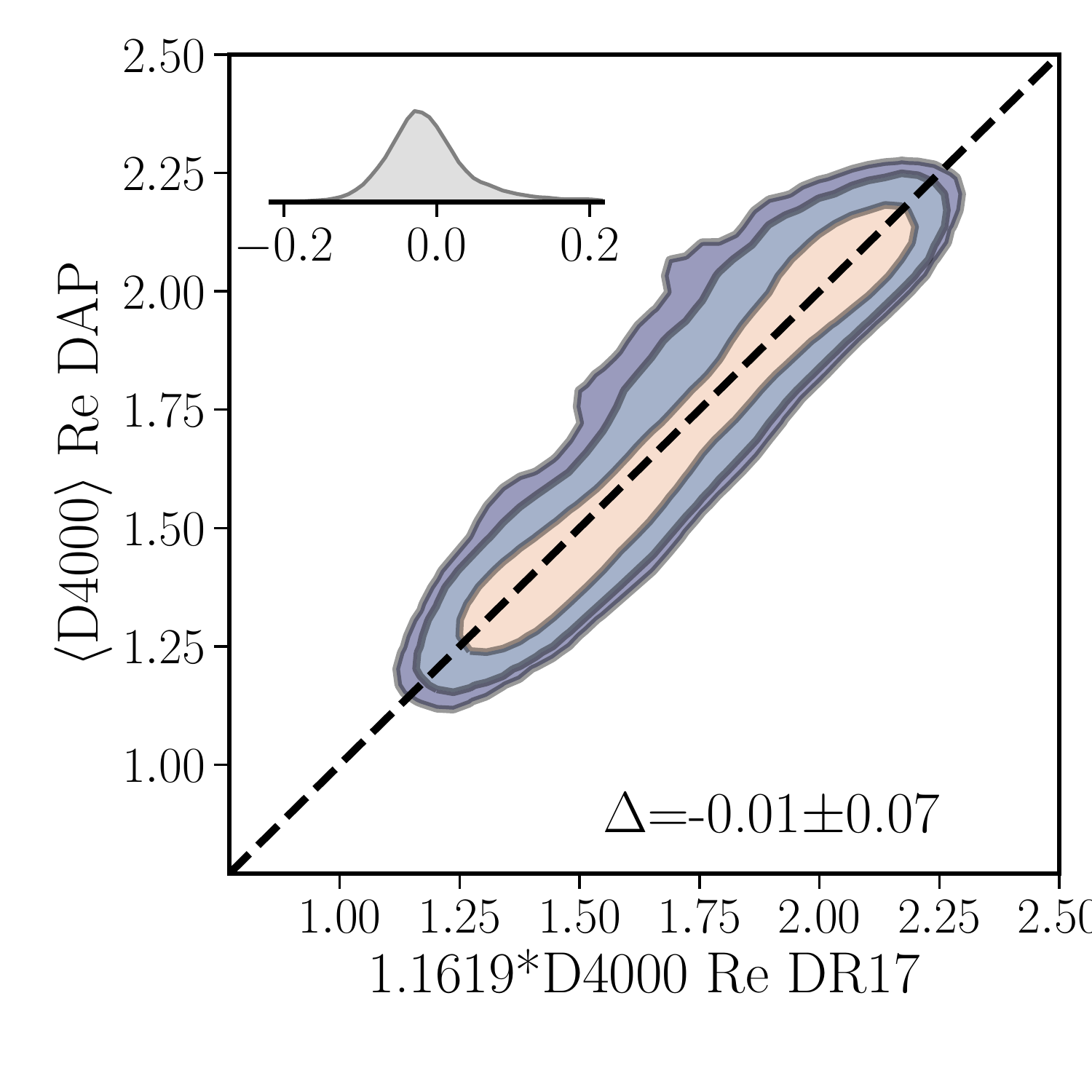}
\endminipage
 \caption{Comparison between the set of stellar indices in common in our analysis and in the MaNGA-DAP data release for the full DR17 sample. We adopt the same format as in Fig.~\ref{fig:comp_ssp_DR15}. 
 From left-to-right we compare the values derived at the effective radius for the Fe5270, Fe5335 and Mg$b$ metallic indices (top-panels), and the
 H$\delta$, H$\beta$ and D4000 age-sensitive indices (bottom-panels). For H$\delta$ we adopt our {\tt Hd\_Re\_fit} parameter and the {\tt HDeltaA} parameter in the {\tt dapall} file. For D4000, we scale our estimate by the factor proposed by \citet{gorgas99} (Sec.~\ref{sec:comp_DAP}), and for DAP we adopt the average between ({\tt D4000} and {\tt Dn4000}). Details on the compared quantities are given in Sec.~\ref{sec:int} and in the MaNGA-DAP presentation article \citep{dap}. }
 \label{fig:comp_ind_DAP}
\end{figure*}
%%%%%%%%%%%%%%%%%%%%%%%%%%%%%%%%%%%%%%%%%%%%%%%%%%%%%%%%%%%%%%%%%%%%%%%5
%
% Check colors and correct
%

The lower row of Fig.~\ref{fig:comp_gas_DR15} shows the comparison for a set of physical properties derived from the emission line parameters. The lower-left panel shows the comparison of the dust extinction at the effective radius, where we find a systematic offset of $\Delta$A$_{gas,\rm V}$=0.24$\pm$0.27 mag. We do not have a clear explanation for the origin of this difference since, in principle, all the analysis was repeated following exactly the same steps. We can only guess that the masking associated with the minimum signal-to-noise required to compute the radial distribution of the dust extinction has introduced this change. In any case, this offset has not introduced any significant change in the integrated properties included in the two final panels of this figure. The lower-middle panel of 
Fig.~\ref{fig:comp_gas_DR15} shows the comparison between the integrated SFR, derived using the dust-corrected H$\alpha$ luminosity. Like in the rest of the properties the distribution lies around the one-to-one relation, in particular for SFR$_{\rm H\alpha, DR17}>$10$^{-2}$ M$_\odot$\ yr$^{-1}$, i.e., for most star-forming galaxies. At low SFRs the DR15 values are slightly larger, following a similar trend (and most probably for the same reasons) as the distributions for F$_{H\alpha,\rm cen}$ and EW(H$\alpha$). Finally, the lower-right panel of Fig.~\ref{fig:comp_gas_DR15} shows the comparison between the molecular gas mass estimated from the dust extinction. In this particular case, we use the {\tt log\_Mass\_gas} parameter listed in the catalog (Sec. \ref{sec:cat_elines}). The distribution is centred in the one-to-one relation. However, this parameter shows a larger scatter than the previous one ($\sim$0.48 dex), reflecting a considerable variation in the spaxel-by-spaxel dust-extinction between the two DRs. Since this parameter is derived from the H$\alpha$/H$\beta$ ratio, from the discussion regarding the top-right panel of the figure, most likely
this scatter is due to the differences in the subtraction of the underlying stellar population introduced by the new SSP library (and its effects on the estimates of the shape and depth of the H$\alpha$ and H$\beta$ stellar absorption).

\subsection{Values reported by the MaNGA Data Analysis Pipeline}
\label{sec:comp_DAP}

As already mentioned, the MaNGA DR17 data have been analyzed using other tools. In particular, all the dataset was processed using the MaNGA Data Analysis Pipeline \citep[DAP,][]{dap}.\footnote{\url{https://sdss-mangadap.readthedocs.io/en/latest/}} This tool performs a decomposition of the stellar and ionized gas components of the observed spectra using different spatial binning schemes, delivering the spatial distribution for a set of properties of the emission lines, a set of stellar indices and the stellar kinematic properties. From these dataproducts they extract a set of characteristic values for each galaxy estimated at the central regions (for the emission lines) and at the effective radius (for the stellar indices), that are then integrated into the {\tt DAPall database}. In this database they also include parameters for different analysis, comprising (i) a treatment of the stellar and ionized gas on individual spectra spaxel-by-spaxel, (ii) an analysis of both the stellar and ionized gas using a spatial binning scheme, and (iii) a hybrid analysis, in which the stellar population is analyzed adopting a binning scheme and the ionized gas is studied spaxel-by-spaxel. In addition, they use a different combination of stellar libraries to explore the stellar kinematics and to decouple the stellar and ionized gas components, and both a Gaussian fitting and a moment analysis in the exploration of the emission lines. The most similar approach to the one performed by \pyp\ is the hybrid method, using the combination of the MILES stellar library for the kinematics and the MaStar SSP library for decoupling the stellar and ionized gas components ({\tt HYB10-MILESHC-MASTARSSP} dataset in their nomenclature). We compare our DR17 results with this catalog.

%%%%%%%%%%%%%%%%%%%%%%%%%%%%%%%%%%%%%%%%%%%%%%%%%%%%%%%%%%%%%%%%%%%%%%%5
\begin{figure*}
 \minipage{0.99\textwidth}
 \includegraphics[width=6.25cm,clip,trim=0 10 0 10]{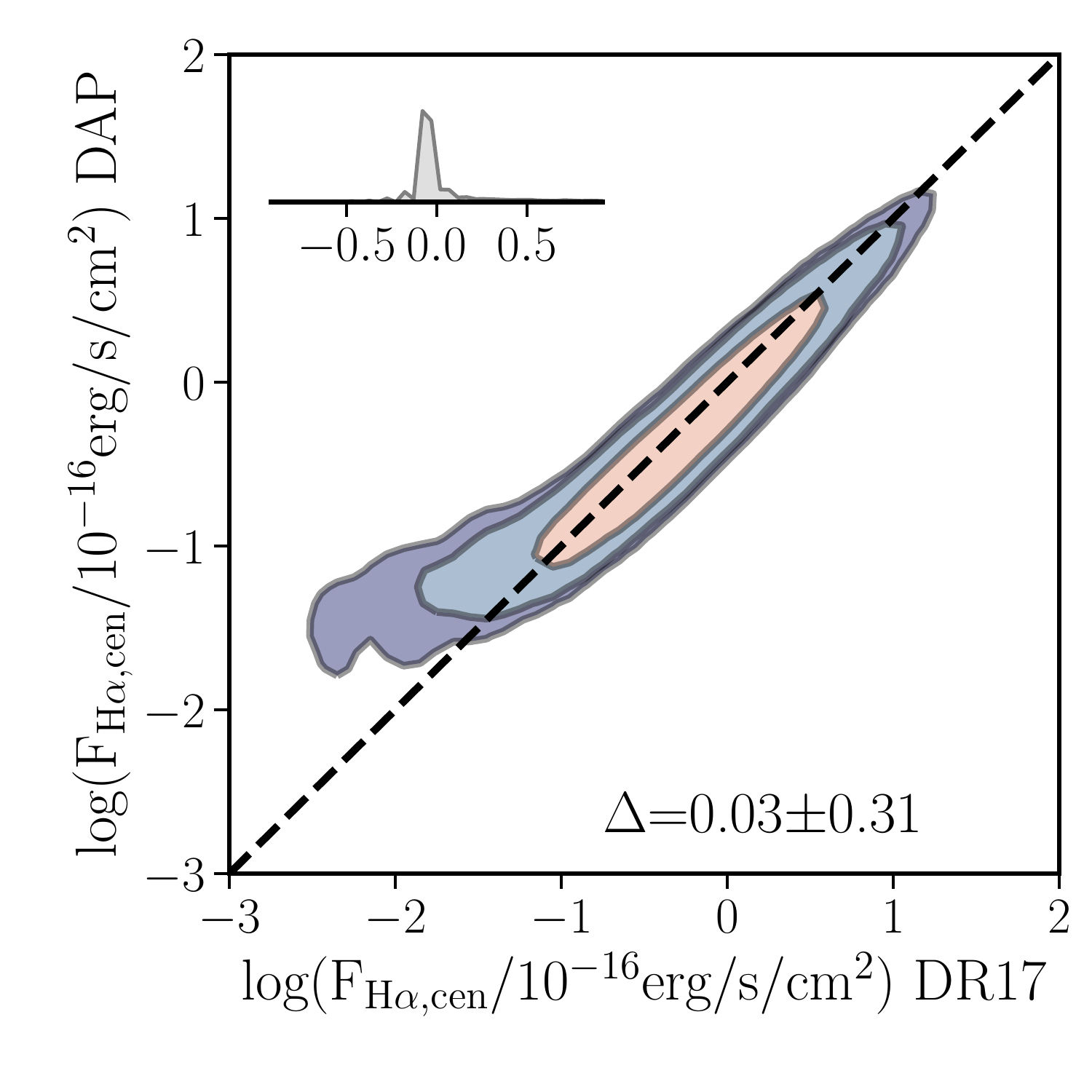}\includegraphics[width=6.25cm,clip,trim=0 10 0 10]{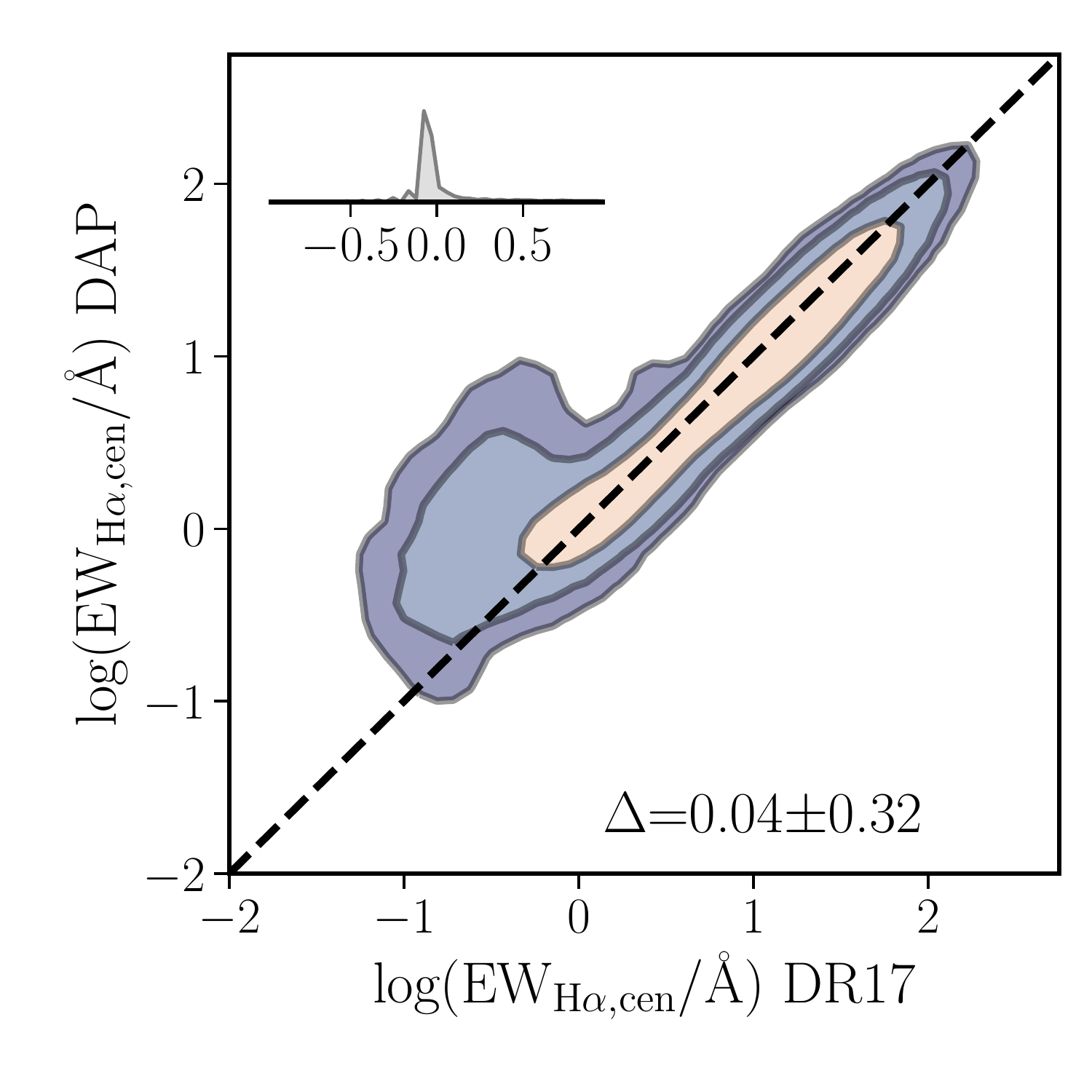}\includegraphics[width=6.25cm,clip,trim=0 10 0 10]{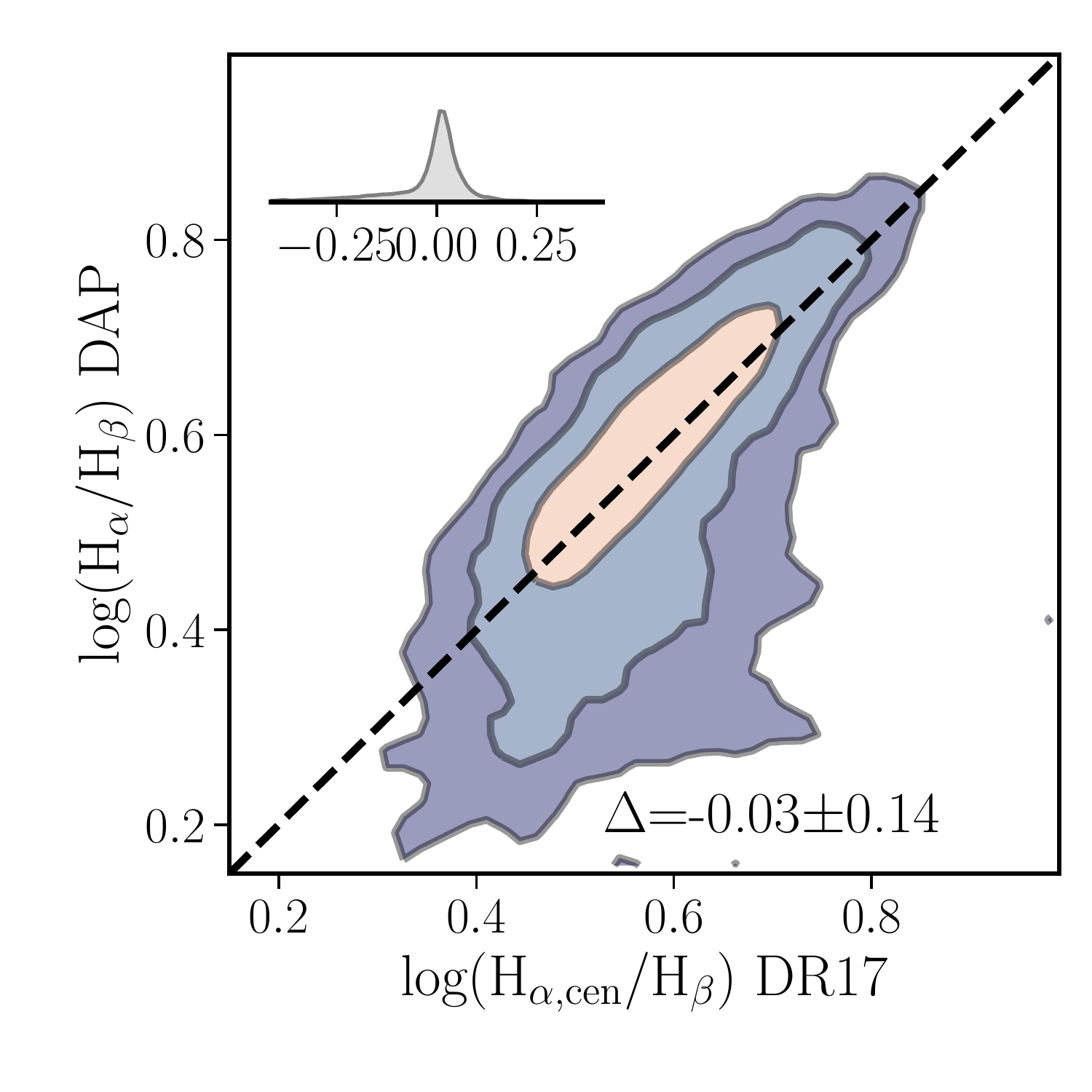}
  \includegraphics[width=6.25cm,clip,trim=0 10 0 10]{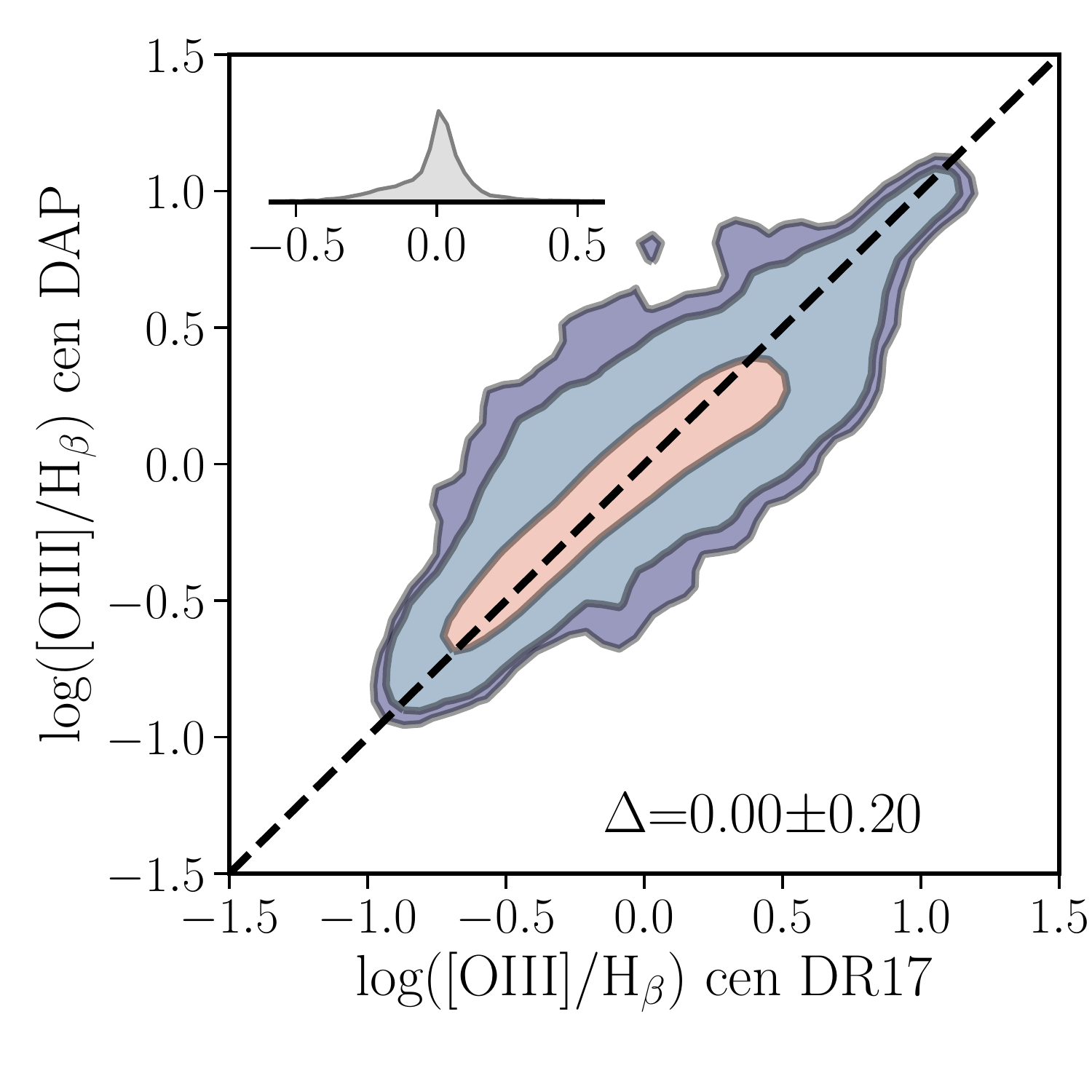}\includegraphics[width=6.25cm,clip,trim=0 0 0 10]{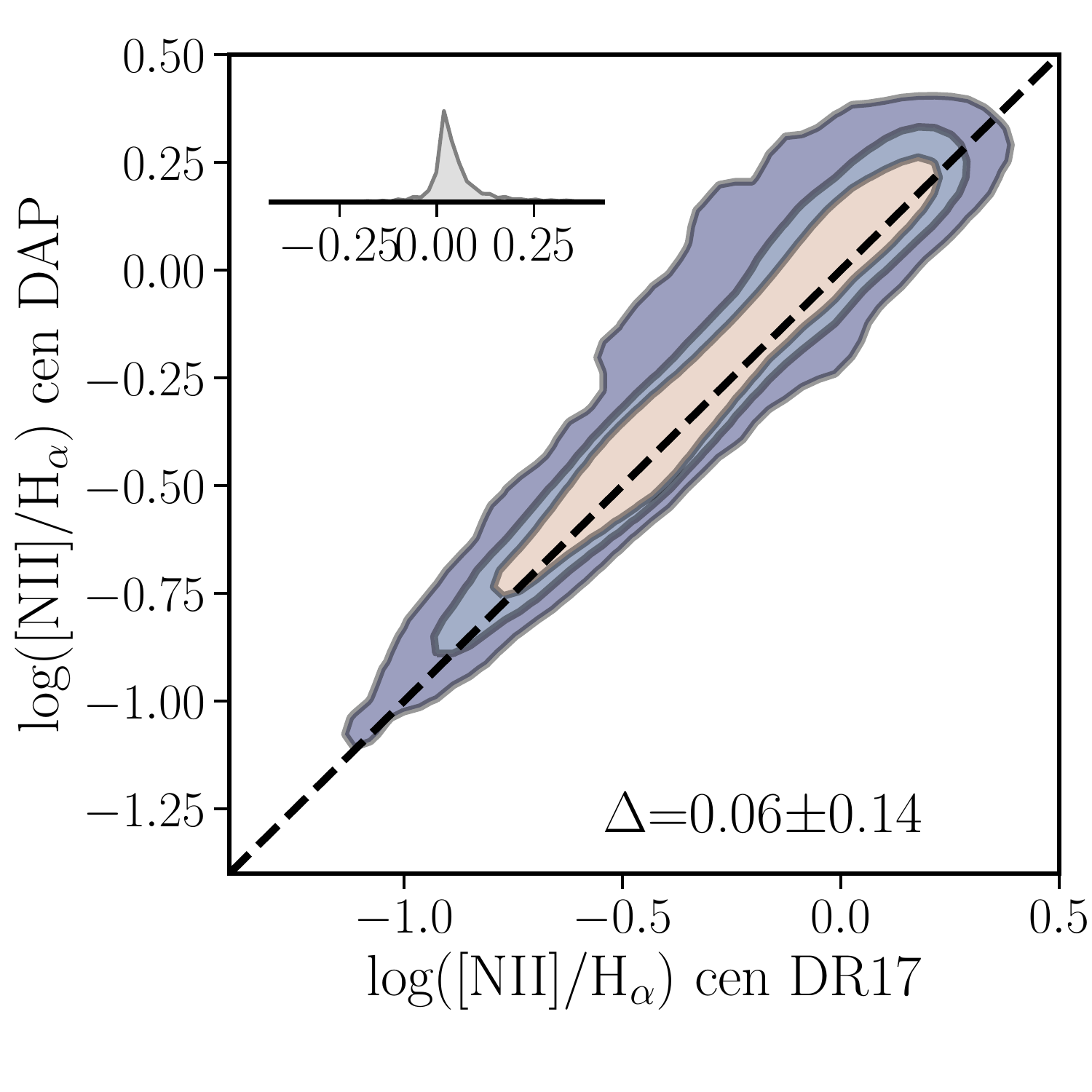}\includegraphics[width=6.25cm,clip,trim=0 10 0 10]{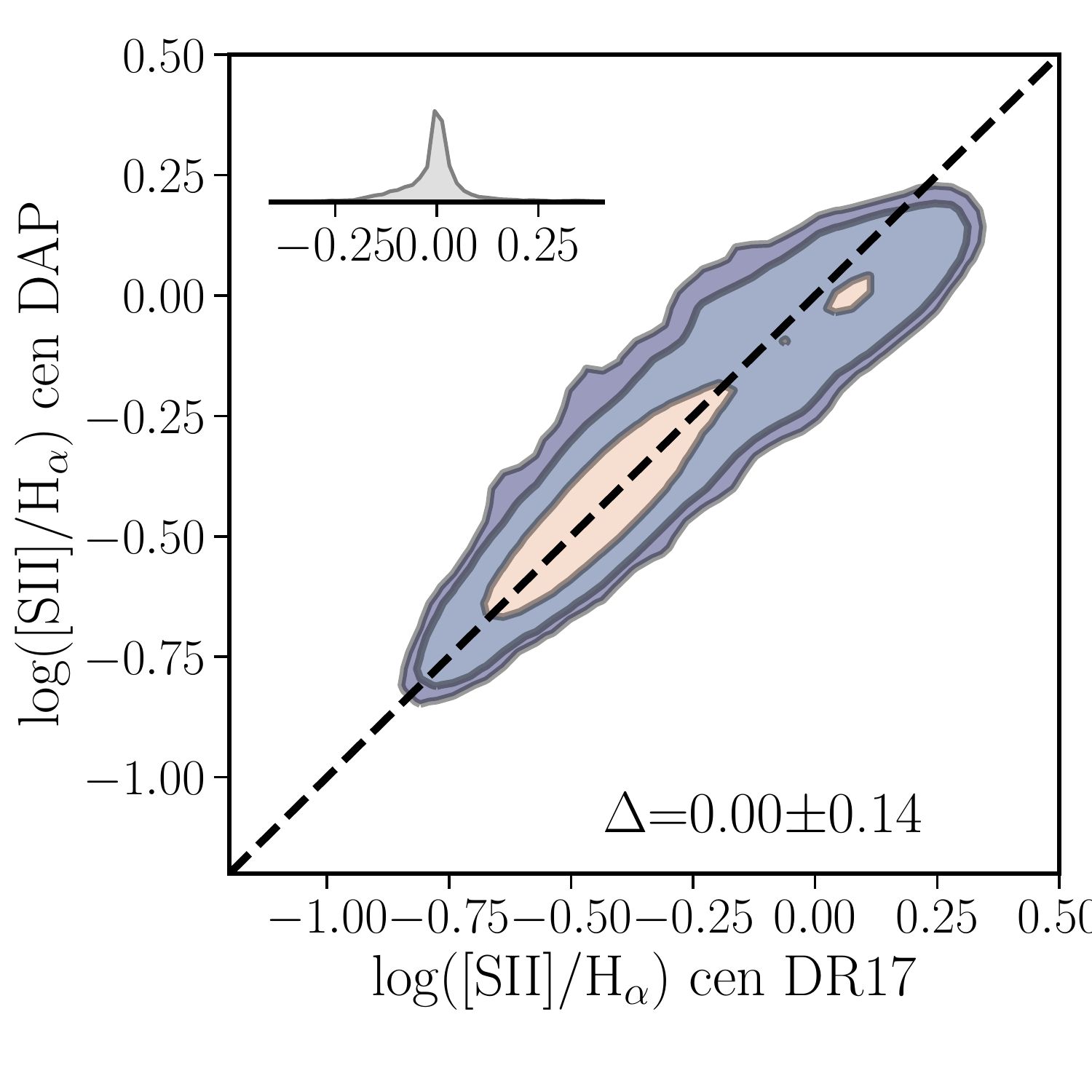}
\endminipage
 \caption{Comparison between a set of ionized gas properties derived using \pyp\ and those distributed as part of the MaNGA-DAP data release for the full DR17 sample. We adopt the same format as in Fig.~\ref{fig:comp_ssp_DR15}. From left-to-right and from top-to-bottom: H$\alpha$ flux (F$_{H\alpha,\rm cen}$), equivalent width of H$\alpha$ (EW$_{H\alpha,\rm cen}$), H$\alpha$ to H$\beta$ line ratio, [OIII] to H$\alpha$ line ratio, [NII] to H$\alpha$ line ratio, and [SII] to H$\alpha$ line ratio, all of them derived for the central aperture. For details on these quantities see Sec.~\ref{sec:int} and the MaNGA-DAP presentation article \citep{dap}.}
 \label{fig:comp_gas_DAP}
\end{figure*}
%%%%%%%%%%%%%%%%%%%%%%%%%%%%%%%%%%%%%%%%%%%%%%%%%%%%%%%%%%%%%%%%%%%%%%%5
% Check colors and labels!
%

Fig.~\ref{fig:comp_ind_DAP} shows the comparison of the stellar indices in common between the two datasets. In general, we find a good match between DAP and \pyp. The best agreement is found for D4000. For this parameter the \pyp\ value has been corrected by the scaling factor proposed by \citet{gorgas99} to transform the index  derived from a flux density in units of wavelength to a flux density in units of frequency (see Sec.~\ref{sec:pipe3d}). For the DAP value we adopt the average between the two listed parameters, {\tt D4000} and {\tt Dn4000}, since this combination provides the best comparison with our results. There is a tight one-to-one relation between the two estimates, with $\Delta$D4000=-0.01$\pm$0.07. For the remaining spectral indices there is a clear correspondence between DAP and \pyp, following a distribution on top of the one-to-one relation (e.g., Fe5270) or just showing a systematic offset with respect to that relation. In the latter cases, the offsets range between -0.47\AA\ for Mg$b$ and 0.21\AA\ for H$_\beta$. The scatter around these relations is of the order of $\sim$0.6\AA, corresponding to $\sim$20-30\% of the typical index value. 

The main driver of the observed differences is not in the definition of the pass-bands adopted to derive the indices, since a cross-check of our adopted values (Table~\ref{tab:index}) with the ones published in \citet{dap} indicates that there is no difference down to the third decimal. Therefore, the difference should be in the details of the procedure. For instance, DAP adopts a different binning scheme than the one implemented by our code. Thus, the indices are not calculated in exactly the same spectra. For indices affected by the subtraction of the emission lines, like H$\beta$, there is an additional source of discrepancy as the treatment of the emission lines is different in DAP and \pyp. However, both effects should contribute to the observed scatter, but we have found some clear systematic offsets (e.g. Mg$b$). We do not have a definitive explanation for the offsets, although we suspect that the pre-processing of the data performed as part of our analysis (Sec.~\ref{sec:data}), that involves a homogenization of the spectral resolution, may be behind the observed differences. In any case, despite of the reported offsets and discrepancies, the similarities between both sets of indices are such that no significant differences should be introduced by using either the DAP or the \pyp\ indices in a further analysis.

Fig.~\ref{fig:comp_gas_DAP} shows the comparison of the same set of emission line properties derived by DAP and \pyp for the central region of the galaxies shown in Fig.~\ref{fig:comp_gas_DR15}. In general, we find the same differences/similarities between the parameters derived using DAP and \pyp than when we compare the results of DR15 and DR17 (Sec.~\ref{sec:comp_DR15}). In some cases, like F$_{H\alpha,\rm cen}$ and [SII]/H$\alpha$, the systematic offset is smaller than the one found in the previous comparison. In other cases, like [OIII]/H$\beta$ and [NII]/H$\alpha$, they are slightly larger. However, in no case the offsets are significant compared with the standard deviation of the difference between the two sets of values. The larger differences are found for (i) EW(H$\alpha$) for values lower than 1\AA\ (measured by \pyp), a range for which DAP predicts slightly larger values; (ii) H$\alpha$/H$\beta$ ratio for values lower than 2.5 (measured by the DAP), a range for which \pyp\ derives slightly larger values; (iii) F$_{H\alpha,\rm cen}$, for values lower than 0.1 10$^{-16}$ \flux (measured by \pyp), a regime for which DAP predicts slightly larger values; and (iv) [SII]/H$\alpha$ ratio, for values larger than 0 dex (measured by \pyp), where the DAP predicts slightly lower values. For { case (i) and (ii)} the differences occur in regimes of very low intensity (and S/N) of the emission lines, where any measurement is unreliable. For { case (iii)} the difference is irrelevant in most of the calculations, since this line ratio is usually adopted to estimate the dust extinction (e.g., Sec.~\ref{sec:cat_elines}) and the considered regime corresponds to either unphysical values (H$\alpha$/H$\beta<$2.5) or a regime of very low dust extinction. Finally, for { case (iv)} the difference is so small than it should not affect any further analysis. In summary, we consider that any analysis using the emission line fluxes derived by both procedures should provide very consistent results.

%%%%%%%%%%%%%%%%%%%%%%%%%%%%%%%%%%%%%%%%%%%%%%%%%%%%%%%%%%%%%%%%%%%%%%%5
\begin{figure*}
 \minipage{0.99\textwidth}
 \includegraphics[width=6cm,clip,trim=0 10 0 10]{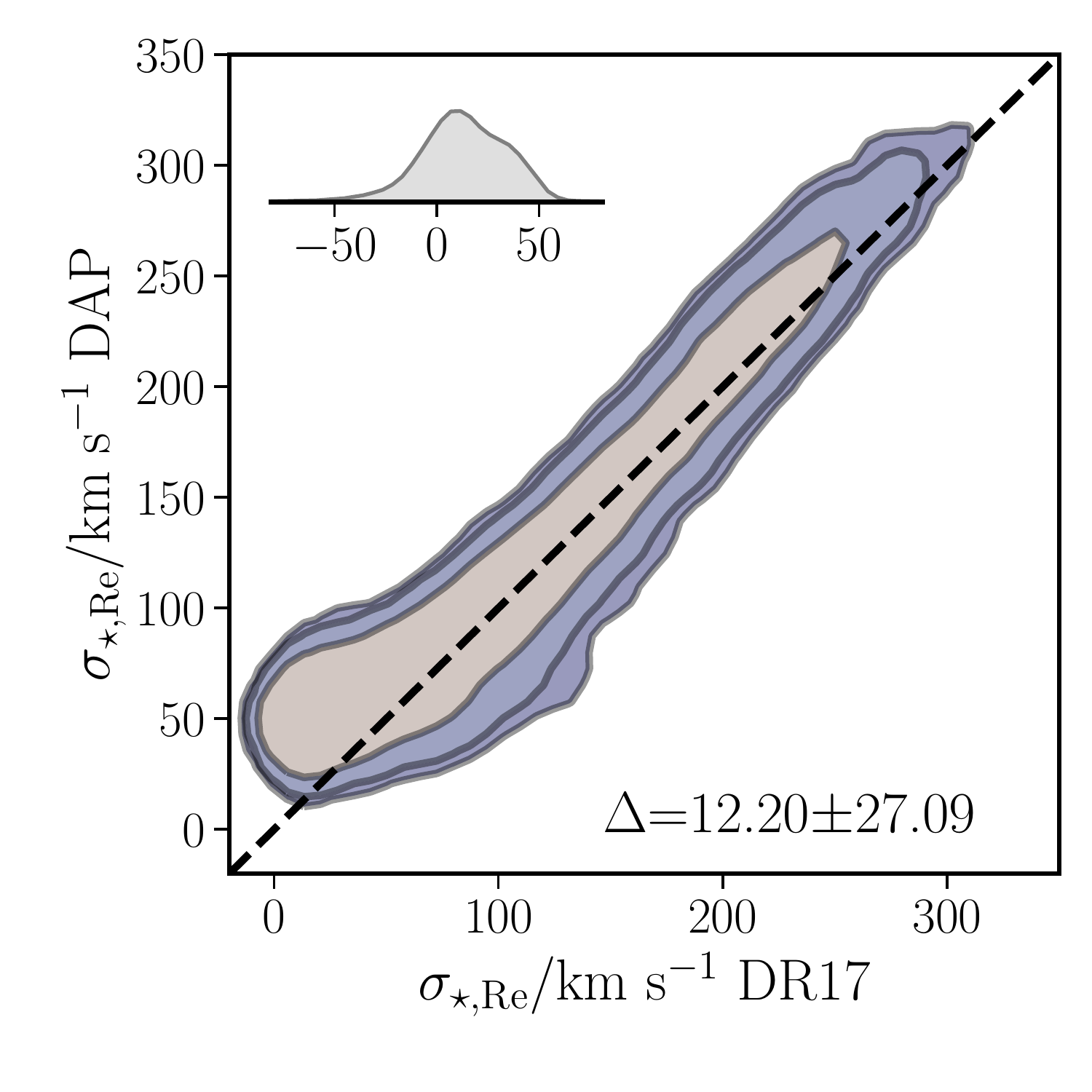}\includegraphics[width=6cm,clip,trim=0 10 0 10]{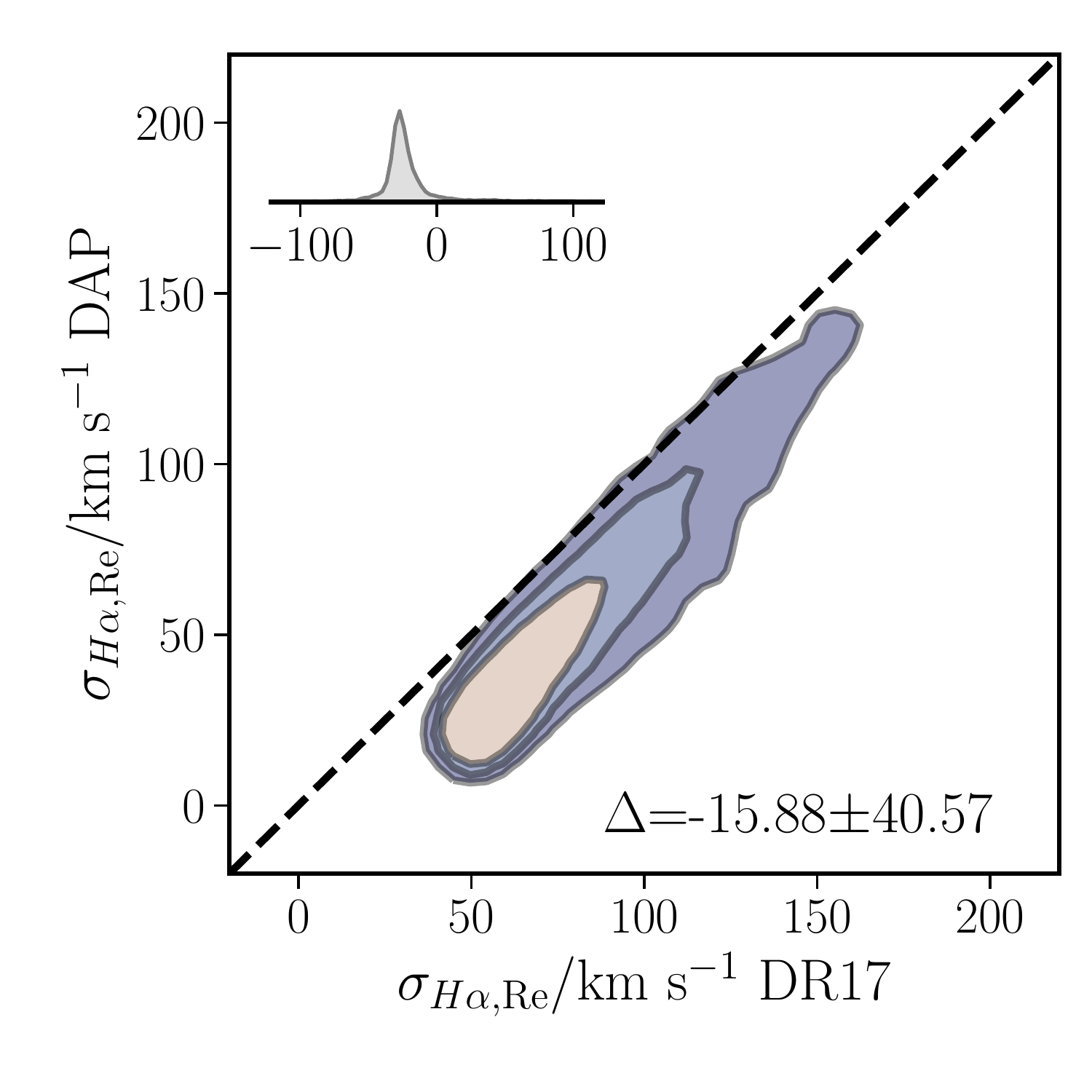}\includegraphics[width=6cm,clip,trim=0 10 0 10]{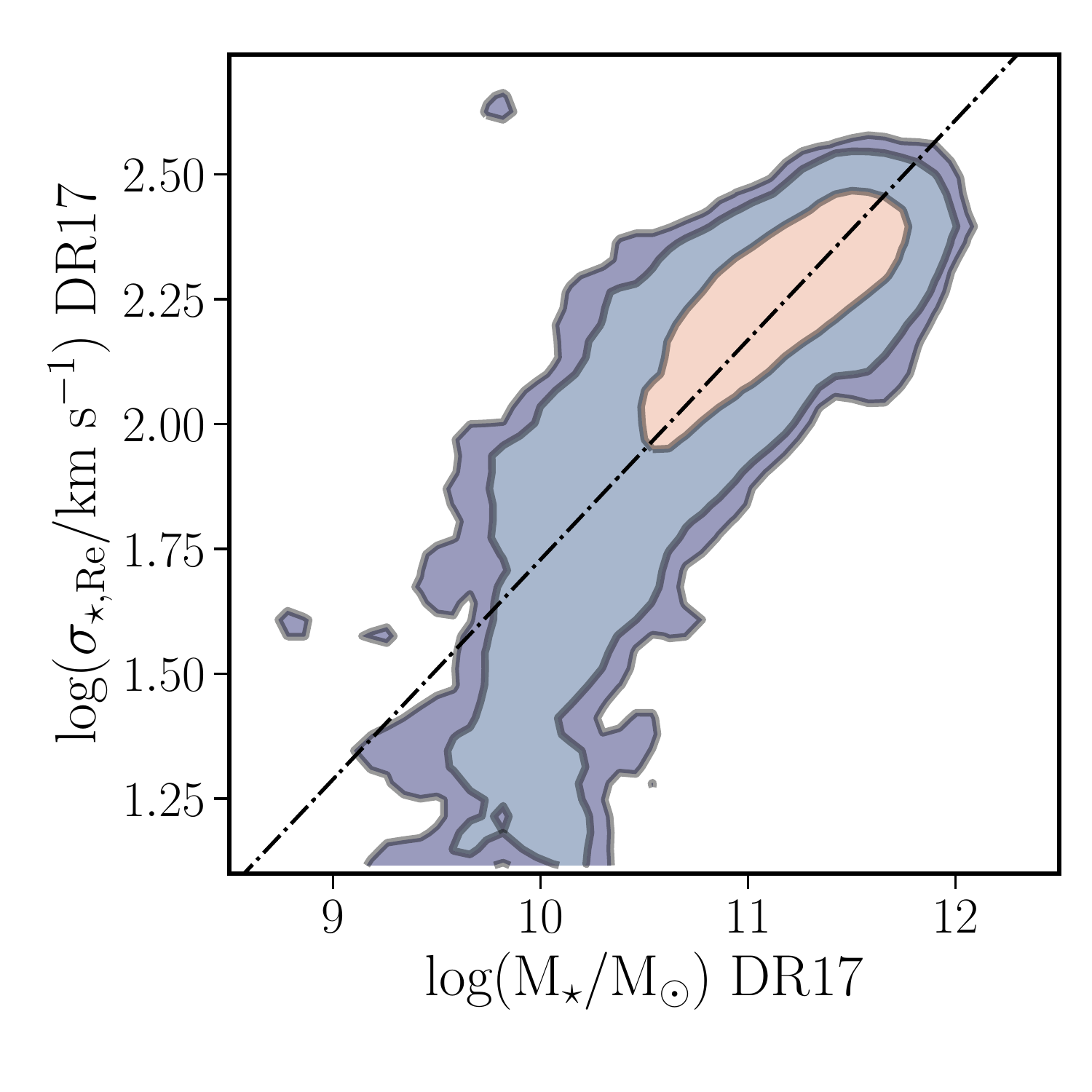}
 \endminipage
 \caption{Comparison between the stellar ($\sigma_\star$, left panel) and H$\alpha$ ($\sigma_{H\alpha}$, central panel) velocity dispersion derived using \pyp\ and the values derived by the MaNGA-DAP with one effective radius for the full DR17 sample. The right panel shows the distribution of $\sigma_\star$ as a function of M$_\star$, derived from our analysis of the retired galaxies, those with $\big |$EW(H$\alpha$)$\big | <$3\AA\ at the effective radius. We adopt the same format as in Fig.~\ref{fig:comp_ssp_DR15}. In the right panel we include the Faber-Jackson relation \citep{faber76} as published by \citet{erik18},  dotted-dashed line, for comparison purposes.}
 \label{fig:comp_disp_DAP}
\end{figure*}
%%%%%%%%%%%%%%%%%%%%%%%%%%%%%%%%%%%%%%%%%%%%%%%%%%%%%%%%%%%%%%%%%%%%%%%5

The final parameters in common between DAP and \pyp\ correspond to the kinematical properties of galaxies. Fig.~\ref{fig:comp_disp_DAP} shows the comparison of the stellar $\sigma_\star$ (left panel) and the H$\alpha$ ($\sigma_{H\alpha}$, central panel) velocity dispersion estimated within one effective radius. In both cases there is good correspondence between the DAP and DR17 estimates. However, there are noticeable differences. In the case of $\sigma_\star$ the DAP values are slightly larger ($\sim$12 km\ s$^{-1}$), showing a systematic offset that is evident even though it is smaller than the scatter ($\sim$27 km\ s$^{-1}$). Furthermore, there is a clear bend in the distribution for  $\sigma_{\star,\rm DR17}<$70 km\ s$^{-1}$, where the DAP values reach a plateau at $\sim$50 km\ s$^{-1}$, while the \pyp\ values are still decreasing. These bends occur when there is a discrepancy in the treatment of the instrumental velocity dispersion as a consequence of a miss-match between the spectral resolution of the adopted SSP or stellar templates and the observations. In our case we use SSPs based on the MaStar library, and therefore, they have a similar spectral resolution as the analyzed data. However, DAP adopts a sub-set of the MILES stellar library for the kinematics analysis. Without pre-judging which of the two approaches provides a more realistic estimate of the velocity dispersion, we just note that the difference in the adopted procedure may explain the observed differences at low $\sigma_\star$. 

In order to get an independent judgment of the quality of the \pyp\ estimates of $\sigma_\star$, we compare it with the values expected from the stellar masses of early-type galaxies, as predicted by the well known Faber-Jackson relation \citep[FJ,][]{faber76}. The right panel of Fig.~\ref{fig:comp_disp_DAP} shows the distribution of $\sigma_\star$ measured within the effective radius by \pyp, together with the FJ relation estimated by \citet{erik18} for the same aperture. We find that the DR17 results follow the FJ relation down to $\sim$60 km\ s$^{-1}$, a value that roughly corresponds to the spectral resolution of the data. Thus, we consider that our DR17 values of $\sigma_\star$ are reliable at least above this limit.
In order to provide with a direct comparison between the two estimates of the velocity dispersion, we derive a prescription to transform from one to the other, following the equation
\begin{equation}\label{eq:disp_cor_ssp}
\sigma_{\star,\rm DAP} = 1.15\sqrt{\sigma^2_{\star,\rm pyPipe3D}+75^2}-41.
\end{equation}
After applying this transformation, the two estimates of the velocity dipersion show a one-to-one relation with $\Delta\sigma_\star$=0.3$\pm$22 km\ s$^{-1}$, a scatter similar to the expected error in this quantity \citep{pypipe3d}.

Finally, for the ionized gas velocity dispersion we find a systematic offset between the values reported by DAP and \pyp\ of $\Delta\sigma_{H\alpha}$=-16$\pm$41 km\ s$^{-1}$. We recall that our estimate is based on the analysis of the emission lines included in the {\tt FLUX\_ELINES} extension of the {\tt Pipe3D} fitsfile. In this extension we include the velocity dispersion measured in \AA\ (Table~\ref{tab:fe}). We realized that for the published table we underestimated the instrumental dispersion by 45 km\ s$^{-1}$, which leads to the observed discrepancy. Therefore, for any further analysis we recommend to apply the following correction to the H$\alpha$ velocity dispersion included in the final catalog
\begin{equation}\label{eq:disp_cor_Ha}
\sigma^\prime_{H\alpha} = \sqrt{\sigma^2_{H\alpha}-45^2}.
\end{equation}
Once corrected, the two quantities agree within a standard deviation, $\sim$27 km\ s$^{-1}$.

%%%%%%%%%%%%%%%%%%%%%%%%%%%%%%%%%%%%%%%%%%%%%%%%%%%%%%%%%%%%%%%%%%%%%%%5
% Firefly!
\begin{figure}
 \minipage{0.49\textwidth}
 \includegraphics[width=4.5cm,clip,trim=0 10 0 10]{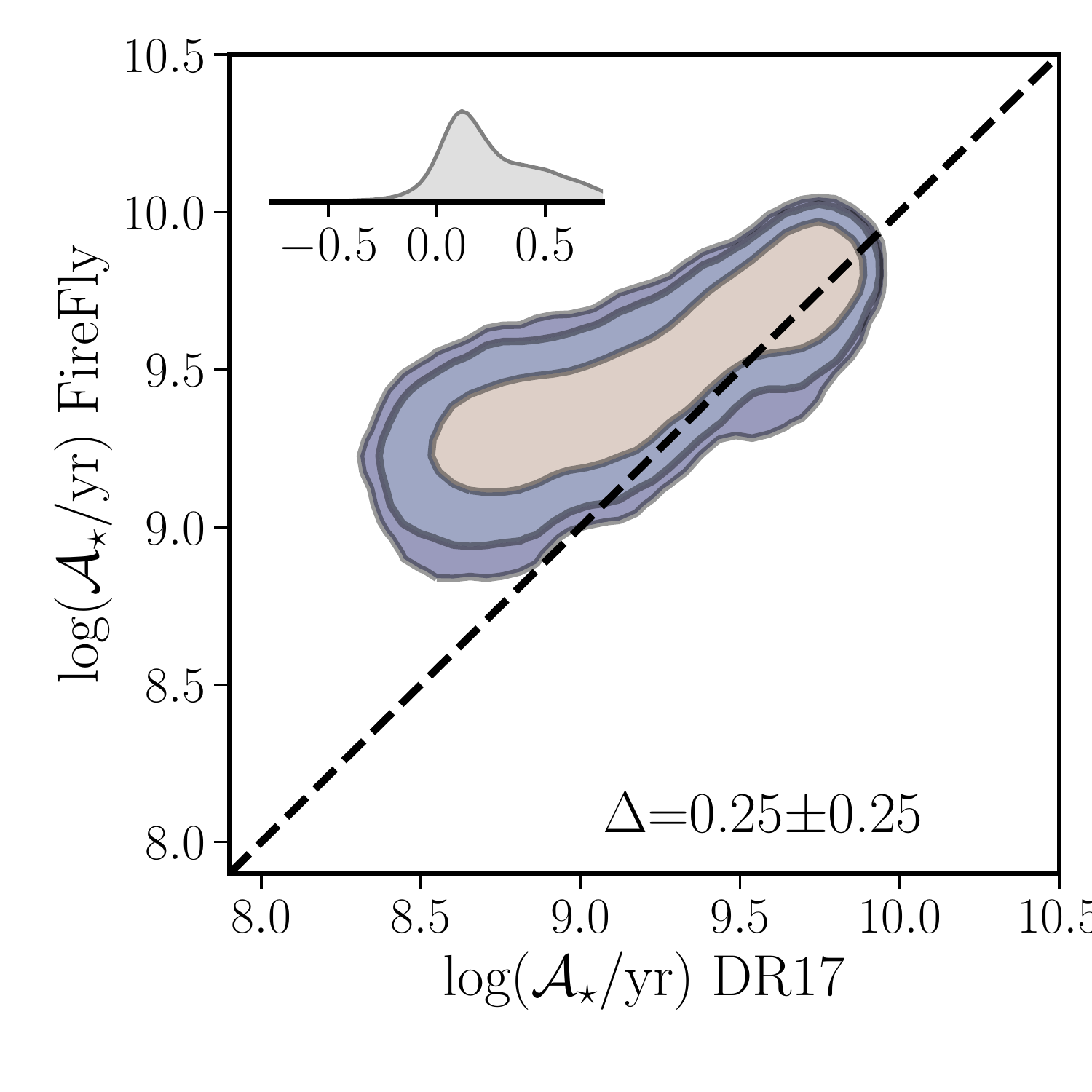}\includegraphics[width=4.5cm,clip,trim=0 10 0 10]{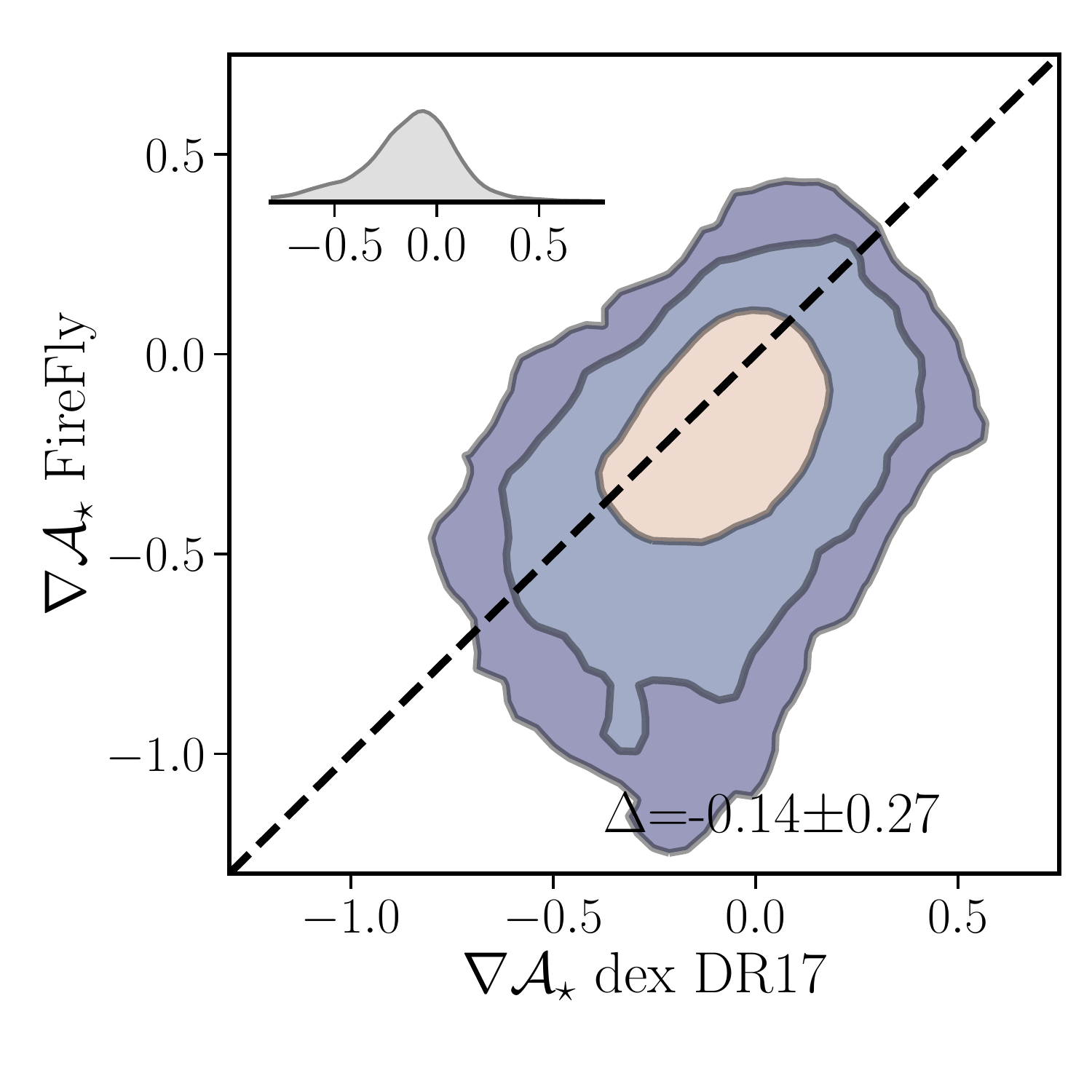}
 \includegraphics[width=4.5cm,clip,trim=0 0 0 10]{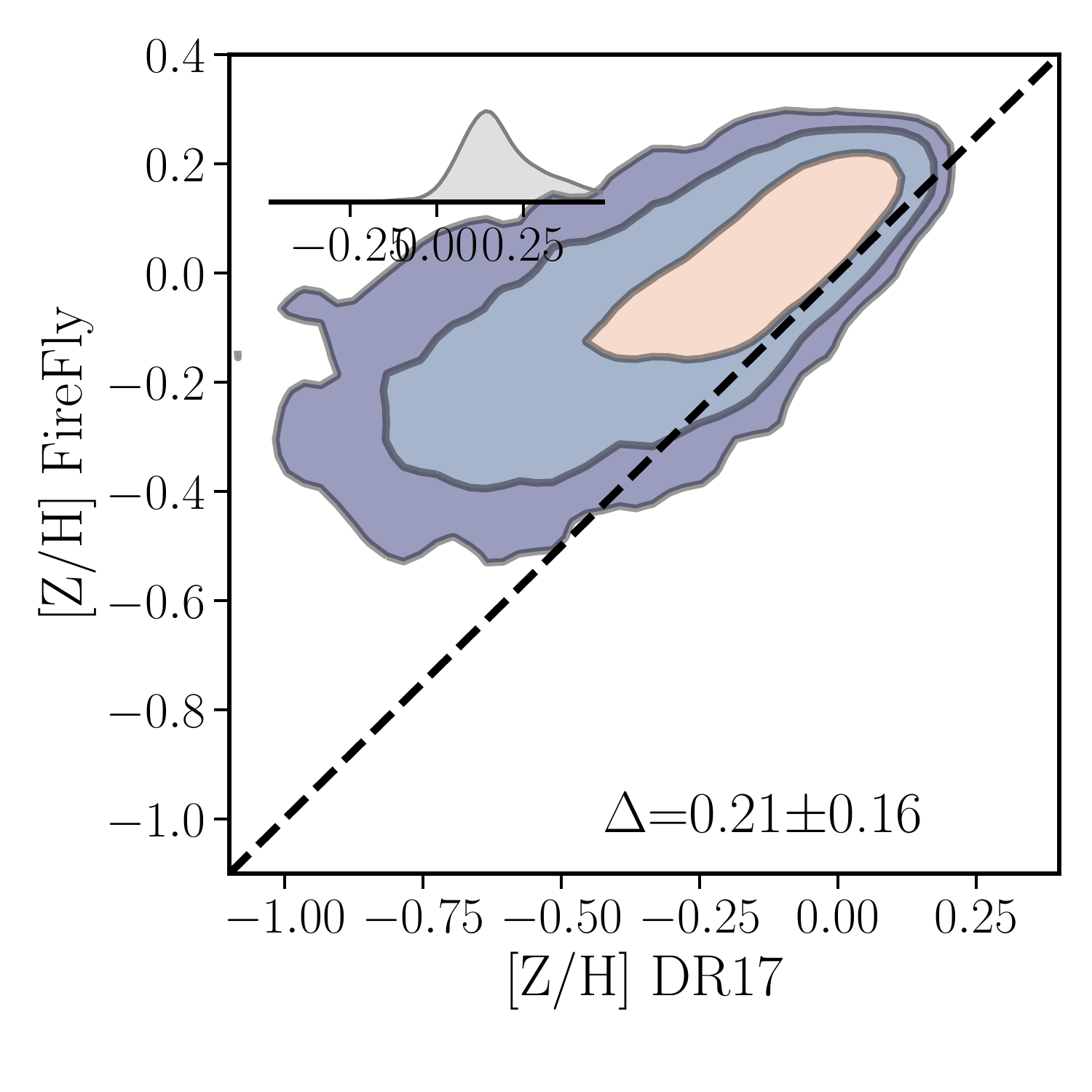}\includegraphics[width=4.5cm,clip,trim=0 10 0 10]{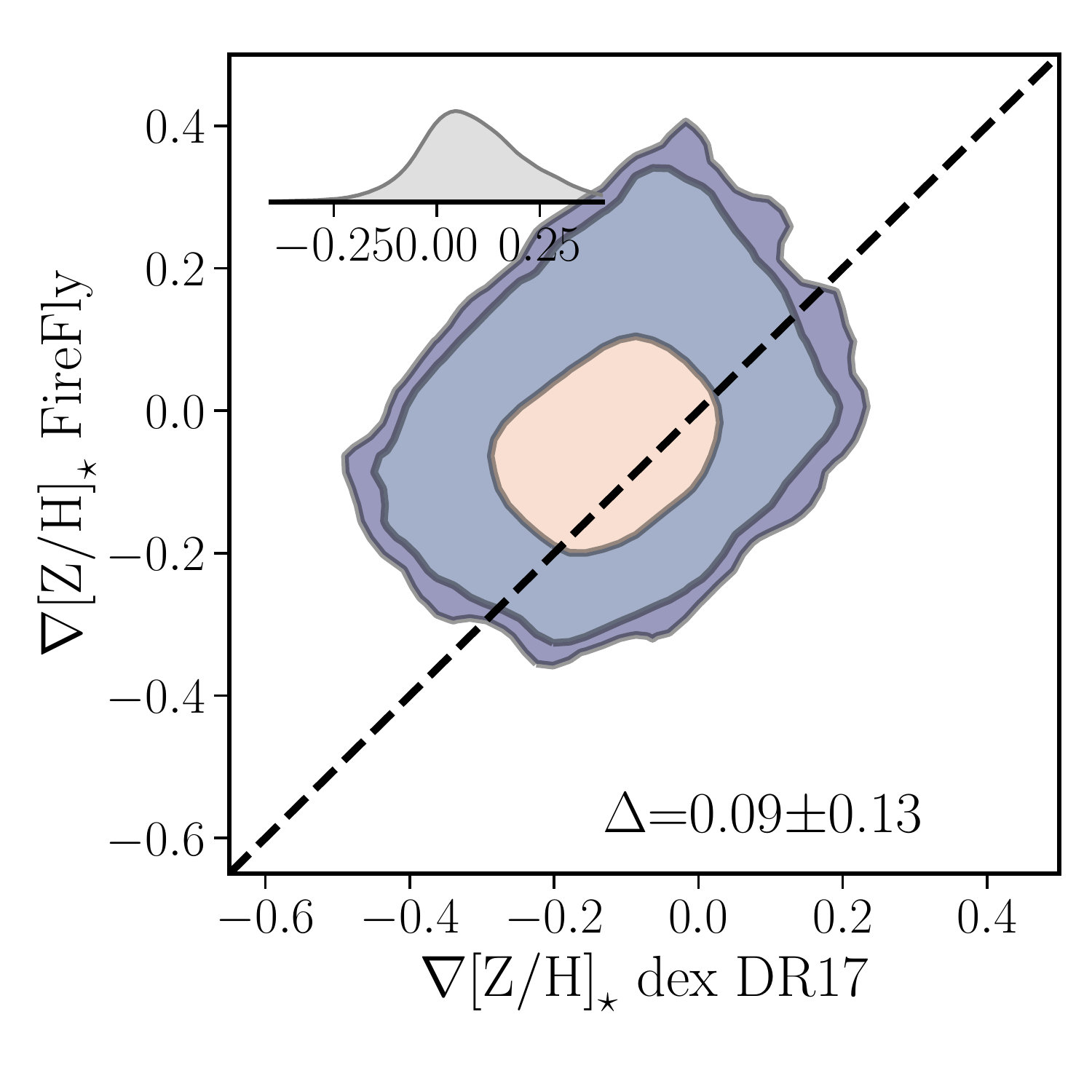}
 \endminipage
 \caption{Comparison between the LW age (\ageLW) and metallicity (\metLW) at the effective radius, together with their corresponding radial gradients ($\nabla$\ageLW and $\nabla$\metLW), derived by us and the {\tt FireFly} MaNGA DR17 VAC values. We adopt the same format as in Fig. \ref{fig:comp_ssp_DR15}. Details on the compared quantities are given in Sec.~\ref{sec:int} and in the {\tt FireFly} presentation article \citep{goddard+2017}.}
 \label{fig:comp_ff}
\end{figure}
%%%%%%%%%%%%%%%%%%%%%%%%%%%%%%%%%%%%%%%%%%%%%%%%%%%%%%%%%%%%%%%%%%%%%%%5
%
% Compare the Morphology!!!!
%

\subsection{Values reported by FIREFLY}
\label{sec:comp_FIRE}

The MaNGA DR17 dataset has been analyzed using the {\tt FIREFLY} \citep{wilki17} full spectral fitting code to derive the main spatially resolved and stellar population properties of each individual galaxy. Like \pyp, {\tt FIREFLY} implements a chi-square minimization fitting code that combines a template of SSPs to model the observed spectra. A Bayesian information criterium to select the best fitting model, without assuming any priors, is adopted. Beside this, the major difference with respect to \pyp\ is the treatment of dust extinction. Although {\tt FIREFLY} includes a method in which extinction by dust is included as part of the stellar decomposition process, the default procedure involves a pre-processing of both the observed spectra and the SSP templates that removes their shape (over a band-width of $\sim$100\AA), and from this first best-fitted model the dust extinction is derived.
This analysis applied to MaNGA DR17 provides with a full set of spatially resolved properties for the stellar populations, including the corresponding maps for LW and MW \age\ and \met,\footnote{\url{https://www.sdss.org/dr17/manga/manga-data/manga-firefly-value-added-catalog/}} from which they extracted the values at the effective radius and the slope of their radial gradients (Neumann et al. in prep\footnote{\url{https://www.sdss.org/dr17/data_access/value-added-catalogs/?vac_id=manga-firefly-stellar-populations}}). Here we compare the values derived for the LW parameters by {\tt FIREFLY} and \pyp. In this comparison we should bear in mind that the {\tt FIREFLY} dataproducts were derived using (i) a different SSP template \citep{maraston20} than the one adopted in our analysis, although we selected for this comparison the results based on the same stellar library (i.e., MaStars); (ii) the spatial binning and the results of the kinematical analysis derived by DAP (Sec.~\ref{sec:comp_DAP}); (iii) a different wavelength to weight the stellar populations, the average observed wavelength ($\sim$6600\AA), instead of the rest-frame 5500\AA\ adopted by \pyp; (iv) a different method to derive the LW values, as they adopted an arithmetic weighted average \citep[according to Neumann et al. in prep.,  Eq.2, although this was not the case in previous versions of the code, e.g.,][]{godd15}; and (v) use a different procedure to derive the radial gradients.
As indicated in Sec.~\ref{sec:int} our method excludes the central regions of the galaxies ($<$0.5 Re), since we consider that they are usually affected by PSF/beam effects. Furthermore, the gradient is derived up to 2.0 Re (or the maximum extension covered by the FoV). On the contrary, the values reported by {\tt FIREFLY} are derived for the region within 1.5 Re. In both cases an azimuthal average is performed \citep[contrary to previous derivations of this parameter by {\tt FIREFLY}, e.g., Fig.~8 and Sec.~3.2 of][]{godd15}.
%than the one described in Sec. \ref{sec:int}. 
%First, no azimuthal average is applied to the radial distribution. Second, the gradient is derived by a linear regression to the full galactocentric distances, without exclude the central regions (severely affected by PSF effects in the case of MaNGA).

Fig.~\ref{fig:comp_ff} compares the {\tt FIREFLY} and \pyp\ values of \ageLW\ and \metLW, measured at the effective radius, and the slope of their corresponding gradients.\footnote{A similar figure is included in Neumann et al. in prep, Fig. 11.}. For age and metallicity we observe a clear correspondence between the values reported by both methods, but not a one-to-one relation. On average {\tt FIREFLY} derives older stellar populations than \pyp, a bias that is stronger for the younger stellar populations (with a difference of $\sim$0.5 dex) than for the older ones (with a difference of $\sim$0.1 dex). As a result, we find a positive offset of $\Delta\ageLW\sim$0.25 dex, with a scatter of $\sim$0.25 dex. Both the offset and the scatter are larger than the values found in the comparison for the same parameter in Sec.~\ref{sec:comp_DR15}, and than in any of our previous comparisons between this tool and other stellar population analysis techniques \citep[e.g., Fig.~17 and Sec.~4 of][]{pipe3d_ii}. A similar pattern is observed for the metallicity: {\tt FIREFLY} provides larger values than \pyp, although in this case the offset seems to be similar (or at least of the same order) for low and high metallicities. On average we find $\Delta\metLW=$0.21$\pm$0.16 dex. 

The reported discrepancies are a consequence of the combined effects of the differences in the adopted procedures listed in the previous paragraph. Systematic differences are expected just due to the use of a different SSP library \citep[e.g., appendices in ][]{rosa14,rosa15,pipe3d}, differences that are enhanced by the adoption of a different binning scheme \citep[e.g.][]{ibarra19}, a different method to derive the mean values, and the normalization at a redder wavelength (which enhances the older and more metal rich stellar populations). In any case, despite of these quantitative differences, qualitatively both procedures classify the old (metal rich) and young (metal poor) populations in a similar way.

%%%%%%%%%%%%%%%%%%%%%%%%%%%%%%%%%%%%%%%%%%%%%%%%%%%%%%%%%%%%%%%%%%%%%%%5
\begin{figure*}
 \minipage{0.99\textwidth}
 \includegraphics[width=18.5cm, clip, trim= 80 0 50 0]{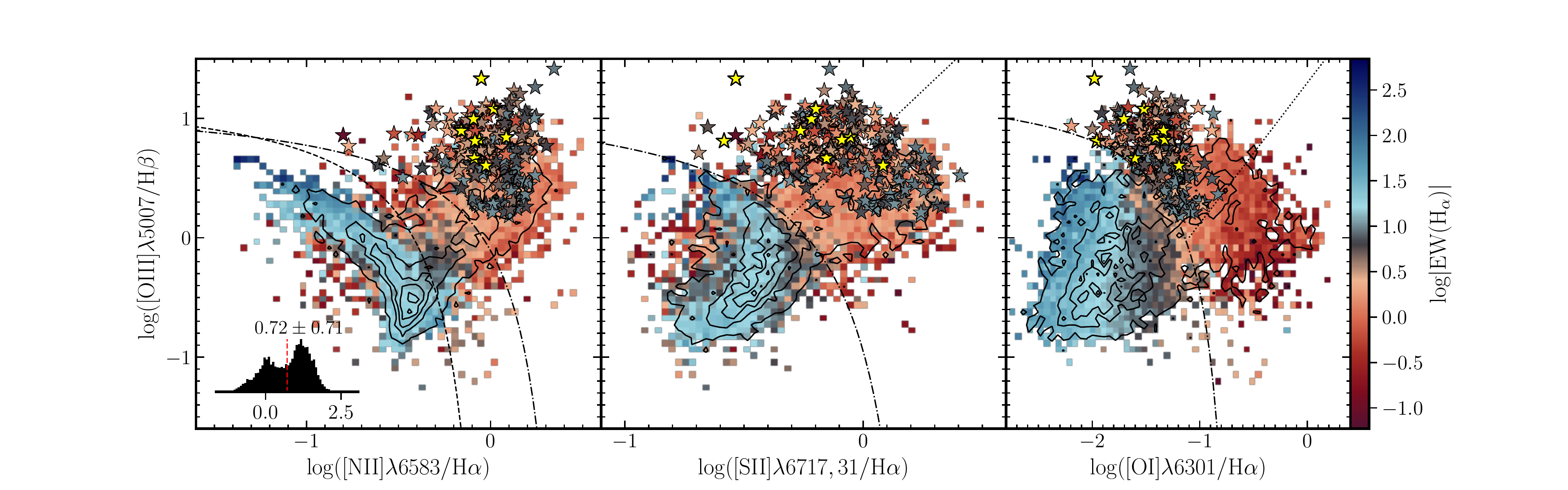}
 \endminipage

 \caption{Distribution of [OIII]/H$\beta$ vs. [NII]/H$\alpha$ (left-panel), [SII]/H$\alpha$ (central-panel) and [OI]/H$\alpha$ (right-panel) corresponding to the central aperture measurements for all the galaxies color-coded by the average EW(H$\alpha$) { in logarithm-scale} in the same aperture (with its histogram, { in logarithm-scales too}, shown as an inset in the left panel including the mean value: { $\sim5\AA$}). Contours correspond to the density of points, with the first one encircling an 80\% of the data, and each consecutive one encircling a 20\% less number of points. Stars corresponds to the AGNs selected as described in the text, color-coded by the EW(H$\alpha$) in the case of type-II and using a yellow color for type-I candidates. In each panel the dash-dotted lines correspond to the \citet{kewley01} demarcation line, and the dotted-line in the left-panel corresponds to the \citet{kauff03} demarcation line. Those lines are frequently used to separate between SF- and AGN-like ionization. The diagonal-line in the top-right quadrant corresponds to the separation between Seyferts and LINERs proposed by \citet{kewley01} too.}
 \label{fig:BPT_cen}
\end{figure*}
%%%%%%%%%%%%%%%%%%%%%%%%%%%%%%%%%%%%%%%%%%%%%%%%%%%%%%%%%%%%%%%%%%%%%%%5

Stronger differences are found for the radial gradients (right panels of Fig.~\ref{fig:comp_ff}). The gradients derived using the two methods present a less clear correspondence to one another than parameters discussed previously, with a scatter of the same order of the dynamical range covered by the gradients. The \ageLW\ gradient presents a somehow better correspondence, with {\tt FIREFLY} deriving slightly sharper gradients than \pyp: $\Delta\nabla\ageLW$=-0.14$\pm$0.27 dex/Re. On the contrary, the relation between the \metLW\ gradients is very loose, with an average difference of $\Delta\nabla\metLW$=0.09$\pm$0.13 dex/Re, with the {\tt FIREFLY} gradients tending to be shallower. 

We consider that the combination of the differences in the {\tt FIREFLY} procedure outlined above, and, in particular, the use of a different scheme to derive the gradient, can easily explain the reported discrepancies between the {\tt FIREFLY} and the \pyp\ gradients. It is worth noticing that despite these differences, both methods predict a slightly sharper (shallower) gradient for both quantities at higher (lower) stellar masses, in agreement with previous results \citep[e.g.][and references therein]{rosa15,sanchez20}. Thus, both methods provide similar qualitative results on a statistical sense, despite the large reported differences in the gradients for individual galaxies.

%Possible reasons for the observed quantitative differences could be the wavelength at which both methods normalize the weights of the individual SSPs 
%The ages derived by {\it FIREFLY} are in general older than the

%\citet{goddard+2017} 
%\Com{List of the different quantitites: Comparisons with previous results}

%\subsection{How to download the products}
%\label{sec:down}

%\section{Quantities estimated using different methods}
%\label{sec:comp_Mass_SFR}

% SFS: 05.03.2022

\section{A practical use of the catalog: AGN selection}
\label{sec:agns}

\citet{sanchez18}  illustrated the use of the analysis performed using {\tt Pipe3D} for the MaNGA DR14 dataset by selecting the candidates to AGNs and exploring the properties of their host galaxies in comparison to the bulk population of non-active galaxies. That analysis implies the exploration of many different integrated, characteristics and resolved properties, which it is beyond the scope of the present study. We present in here an update of the AGN selection using the \pyp\ analysis for the final MaNGA data release, which covers five time more galaxies, showing a few examples of the comparisons between active and inactive galaxies as shown in that article.

%%%%%%%%%%%%%%%%%%%%%%%%%%%%%%%%%%%%%%%%%%%%%%%%%%%%%%%%%%%%%%%%%%%%%%%5
\begin{figure}
 \minipage{0.95\textwidth}
 \includegraphics[width=8cm, clip, trim=0 80 0 85]{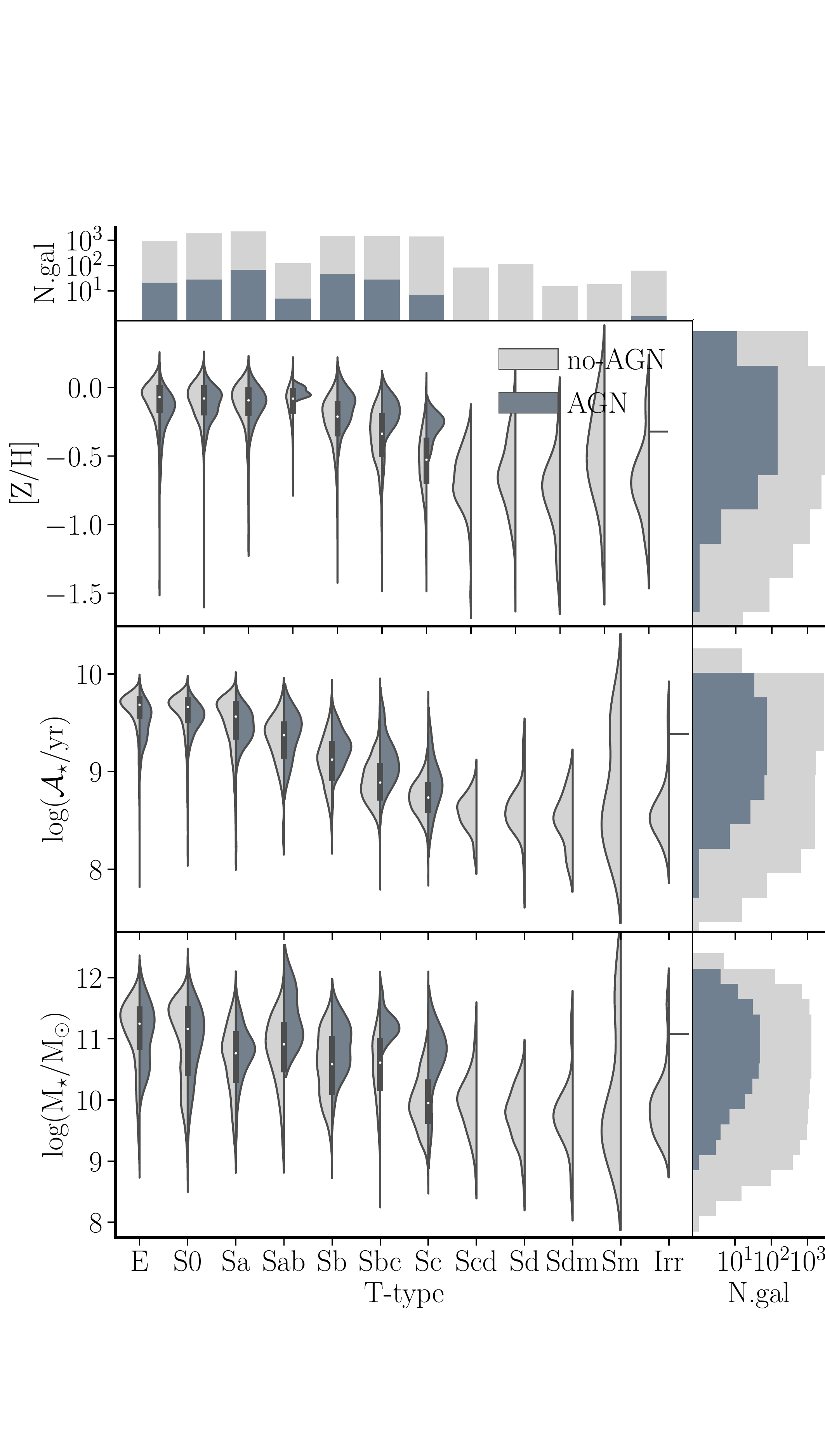}
 \endminipage
 \caption{Violin-plot of the { \metLW (top panel), \ageLW (middle panel) and stellar masses (bottom panel)} segregated by the morphology (with cD galaxies excluded, see Sec. \ref{sec:comp_morph}) for the complete sample (light-grey) and the AGN hosts (stale-grey). An histogram of the number of galaxies segregated by morphology and mass for both sub-samples have been included in the top and right panel, respectively. In both cases it was adopted a logarithmic-scale for the histograms to highlight the bins with low numbers.}
 \label{fig:morph_mass}
\end{figure}
%%%%%%%%%%%%%%%%%%%%%%%%%%%%%%%%%%%%%%%%%%%%%%%%%%%%%%%%%%%%%%%%%%%%%%%5
%the galaxies above the K01 demarcation line in the O3N2, O3S2 and O3O1 diagnostic 
%$\sim$1700 galaxies are located above the 
%#N.AGNs [OIII] vs. [NII] candidates = 1685  EW>3 404
%#N.AGNs [OIII] vs. [SII] candidates = 1525  EW>3 352
%#N.AGNs [OIII] vs. [OI] candidates = 1229
%#N.AGNs candidates (EW>3*1) = 224
%#N.AGNs strong (EW>6*1)= 142 , Very strong = 97
%#N.P-AGB = 2039 ,  y>Kewley => 1281
%#N.SF = 4717 , SF_string EW_Ha>6*1 4119

Following \citet{sanchez18} and the recent reviews on the topic \citep[e.g.][and references in there]{ARAA,sanchez20}, we select the candidates for hosting an AGN based on the combination of the use of classical diagnostic diagrams and a minimum value for the EW(H$\alpha$). We use the emission line ratios derived for the 2.5$\arcsec$/diameter central aperture described in Sec. \ref{sec:cat_elines} to explore the distribution across the [OIII]/H$\beta$ vs. [NI]/H$\alpha$ (O3N2), [SII]/H$\alpha$ (O3S2) and [OI]/H$\alpha$ (O3O1) diagnostic diagrams. Then, we select as AGN candidates those objects located above the \citet{kewley01} demarcation line (K01) in the three diagrams simultaneously and with EW(H$\alpha$)$>$3\AA. 

For the full sample of $\sim$10,000 galaxies only $\sim$180 objects present no evidence of ionized gas in the central region in at least one of the analyzed emission lines. Thus, in the vast majority of the galaxies we find evidence of ionized gas even in the central aperture, although in many cases at a very low-S/N. Selecting only those objects with a detection of H$\alpha$ in the central aperture with a S/N$>$10, and a S/N$>$1 for the remaining involved emission lines, we restrict the sample to $\sim$7000 galaxies. Of them, only $\sim$1700, $\sim$1500 and $\sim$1200 objects are located above the K01 demarcation lines in the O3N2, O3S2 and O3O1 diagnostic diagrams respectively. Limiting the sample to those objects in common in the three subsets and imposing the minimum cut in EW(H$\alpha$), we end-up with 224 AGN candidates. By construction, they corresponds to optically selected type-II AGNs. Following \citet{sanchez18}, we perform an automatic search for the presence of a broad emission line component in the emission line of H$\alpha$. This procedure allows us to recover some possible type-I AGNs that have not been selected by the previous criteria. The implemented algorithm maximize the detection of any broad component. Therefore, it is required to impose a cut in the S/N of this component for a proper type-I selection. This additional analysis recovers three more targets not included in the selection described before. In summary, we select a final sample of 227 candidates to host an AGN. From this sample, we find possible evidence of a broad component in 119, but for only 9 of them we can recover this broad component with a S/N$>$10.

Figure \ref{fig:BPT_cen} shows the distribution across the three considered diagnostic-diagrams for the values used in the AGN selection, color-coded by the EW(H$\alpha$). As expected, in average the objects with high-EWs ($>$6\AA) are found below the demarcation lines usually adopted to select the ionization due to young-OB stars that trace recent SF activity  (e.g., Fig. \ref{fig:ex_BPT}), in particular below the \citet{kauff03} demarcation line. On the contrary, the objects above those lines are those that in general present low-EW($<$3\AA), in particular those above the \citet{kewley01} demarcation line. This bimodality is evident in the histogram of the EWs shown in the inset, where the valley between the two peaks at low- and high-EWs is located at 6\AA, broadly separating the SF from the non-SF regions. These results are in agreement with the ones discussed in Sec. \ref{sec:ex_cube} and the current understanding of the dominant ionization sources in galaxies. 

The AGN candidates are distributed above the demarcation lines, having large values of the EW(H$\alpha$) by selection; thus, objects that are clearly not following the general trend. The few type-I AGNs are also distributed about the demarcation lines. However, in three cases they do not fulfill the EW criteria. This is not surprising since those EWs are derived using the moment analysis, that do not implement a detailed deblending between broad and narrow-lines.

%It is worth noticing that indeed they are objects that do not follow the general trend, 

%%%%%%%%%%%%%%%%%%%%%%%%%%%%%%%%%%%%%%%%%%%%%%%%%%%%%%%%%%%%%%%%%%%%%%%5
\begin{figure}
 \minipage{0.9\textwidth}
 \includegraphics[width=8.5cm, clip, trim= 10 0 10 0]{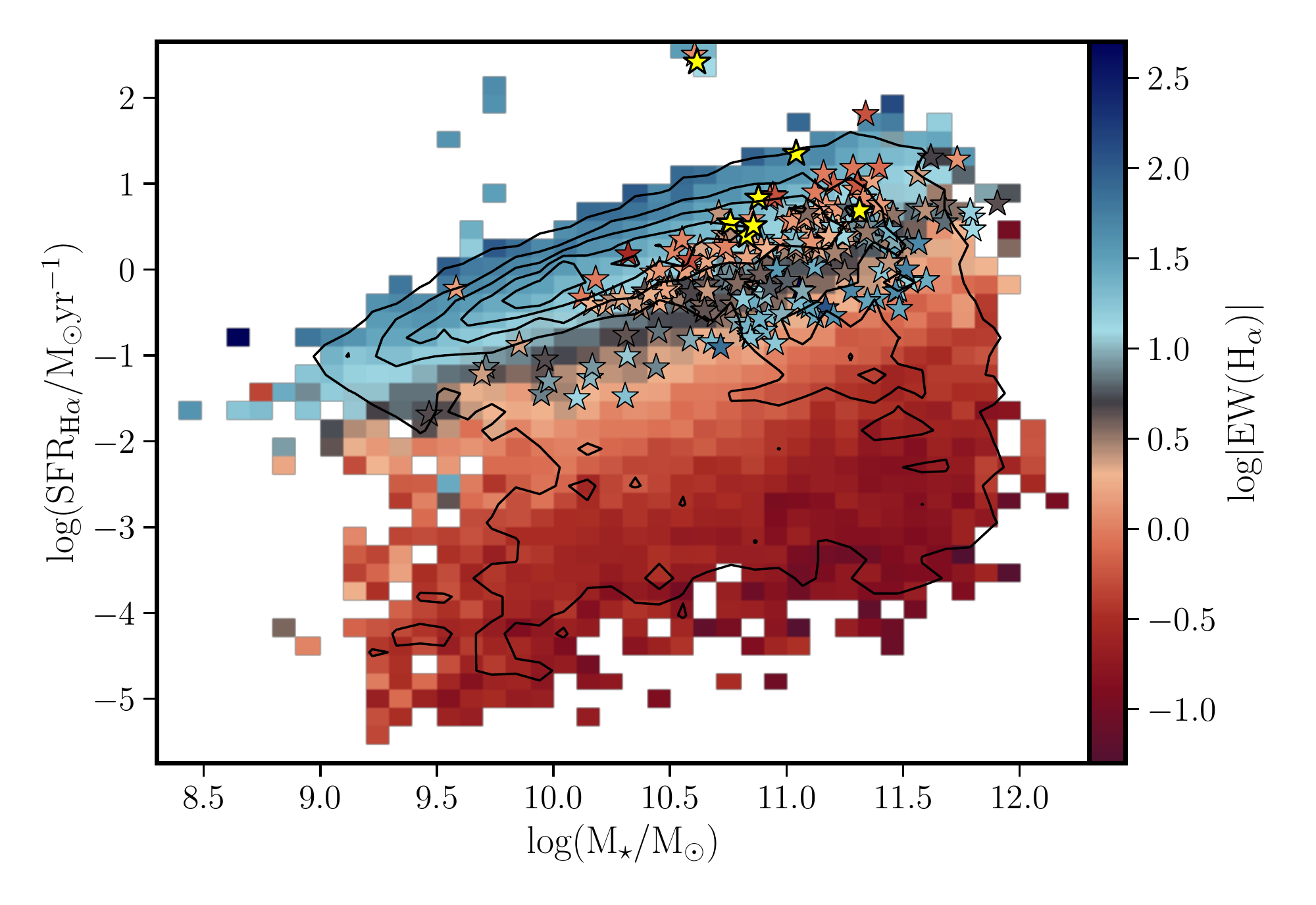}
 \endminipage
 \caption{Distribution of galaxies along the SFR-M$_\star$ diagram color coded by the average EW(H$\alpha$) at the effective radius.
 For the SFR we adopted the value derived based on the dust-corrected H$\alpha$ luminosity, and for the stellar mass we adopted the value derived from the stellar analysis performed by \pyp. Contours correspond to the density of points, with the first one encircling an 80\% of the data, and each consecutive one encircling a 20\% less number of points.  Stars corresponds to the AGNs selected as described in the text, color-coded by the EW(H$\alpha$) at the effective radius in the case of type-II and using a yellow color for type-I candidates.}
 \label{fig:SFR_mass}
\end{figure}
%%%%%%%%%%%%%%%%%%%%%%%%%%%%%%%%%%%%%%%%%%%%%%%%%%%%%%%%%%%%%%%%%%%%%%%5

Once selected the candidates to host AGNs we can compare their properties with those of non-active galaxies. Figure \ref{fig:morph_mass} shows the distribution of { \metLW, \ageLW and } stellar-masses segregated by morphology as violin-plot for both the AGN hosts and the inactive galaxies. cD galaxies have been excluded for the reasons explained in Sec. \ref{sec:comp_morph}. { For the general population of galaxies we note the well known trends between morphology, stellar masses, and age and metallicity of the stellar populations \citep[e.g.][]{rgb17}.} In agreement with the most recent results based on similar selections \citep[e.g.,][and references therein]{lacerda20}, host galaxies are found in the range of high stellar masses (with M$_\star>$10$^{10}$M$_\odot$, in most of the cases, and $\sim$10$^{11}$M$_\odot$ in average){, with metal-rich and intermediate-to-old stellar populations}, being absent in galaxies without a clear bulge (i.e., later than Sc). Indeed, the two properties, high stellar mass and the presence of a bulge, seems to be equally important in general for hosting an AGN.

Figure \ref{fig:SFR_mass} shows the distribution of the full sample of galaxies across the SFR-M$_\star$ diagram color-coded by the EW(H$\alpha$) measured at the effective radius of the galaxies, with the location of the selected AGN hosts highlighted. As expected the galaxies with high-EW, that correspond essentially to SF galaxies (SFGs), follow a well defined linear distribution across this diagram, i.e., the so-called star-formation main sequence \citep[e.g.][]{brinchmann:2004,renzini15,mariana19}. On the contrary, the galaxies with low-EW, those that are not actively forming stars \citep[retired galaxies, RGs][]{sta08}, are distributed in a lousy cloud well below this sequence. Again, this distribution reflects a well known bimodality observed in galaxies that involves several properties, including not only the star-formation stage but also the colors, morphology, dynamical stage, gas content, among several others \citep[e.g.,][]{blanton+2017}. 

In between the two peaks in density defined by SFGs and RGs there is a region with a relative dearth of galaxies (see the inset in Fig. \ref{fig:BPT_cen}), usually called the Green Valley (GV). This GV was first described for the color-magnitude diagram \citep[e.g.][]{kauff03a}, in which the RGs follow a red-sequence and the SFGs a loose blue-cloud. However, this valley is more easily identified in the SFR-M$_\star$ diagram, where the distinction between SFGs and RGs is sharper. The GV is supposed to be a transition zone between a star-formation to a retired stage of galaxies. Contrary to the main population of non-active galaxies, AGN hosts do not present a bimodal distribution in neither the color-magnitude nor the SFR-M$_\star$ diagrams. In both cases they are preferentially found at the Green Valley, with a foot-print that covers the edges of the SFMS (red-sequence) towards the red (blue cloud) for the SFR-M$_\star$ (color-magnitude) diagram \citep[e.g.][]{kauff03,sanchez04,schawinski+2014}.
\citet{sanchez18} confirmed this result by exploring the dataproducts of {\tt Pipe3D} for the MaNGA DR14 dataset, avoiding the problems introduced by the mixing of ionizing sources due to single aperture spectroscopic observations, like the ones provided by the original SDSS survey. 
Fig. \ref{fig:SFR_mass} corroborates this result using the current updated dataset. It is clearly observed that the distribution of AGN hosts in the SFR-M$_\star$ diagram peak at the Green-Valley: the vast majority of them are constrained in a region between the location of the SFMS (traced by the peak in the density distribution for SFGs), and a SFR approximately one order of magnitude below this location. We find some AGNs with extreme SFR with respect to the average population, but there are just low number of them (2 or 3 out of two hundred). %{ Finally, we notice that the AGNs with the largest \EWa\ in the central corresponds to the most retired hosts.}

In summary, for the two explored distributions we find very similar results for the current dataset than the ones already published in \citet{sanchez18} using a more limited sample (five times less number of galaxies). The updated list of AGN candidates is accessible online \footnote{\url{https://ifs.astroscu.unam.mx/MaNGA/Pipe3D_v3_1_1/tables/AGNs_candidates.csv}} for further explorations. In this table we include the complete list of AGNs. The relevant entries to select between type-I and type-II are the {\tt agn\_type} entry (1 for objects with some evidence of a broad component in H$\alpha$, 2 for objects with no evidence of such component), and {\tt sn\_Ha\_broad} (S/N of the broad H$\alpha$ component). For the current discussion we classified the AGNs as type-I if {\tt agn\_type} is equal to one and {\tt sn\_Ha\_broad} is larger than 10. On the contrary, it was classified as a type-II.

\section{Summary and Conclusions}
\label{sec:summary}

Along this article we present the dataproducts derived from the analysis performed using the \pyp\ pipeline to the full MaNGA dataset, comprising $\sim$10,000 IFS datacubes for a similar number of galaxies. We describe briefly the sample of galaxies, observing technique and data reduction. We explain in detail the implemented analysis, including a summary of the procedures that comprise the \pyp\ pipeline, highlighting the differences with the previous version of the code ({\tt Pipe3D}) when needed. We describe the new SSP library, and a set of additional analyses performed, including (i) statistical estimate of the morphological type of the galaxies, (ii) photometric and structural analysis, and (iii) the procedures adopted to estimate the quality of the data.

As a result of this analysis, we deliver to the community one of the largest datasets of fully analyzed IFS data, comprising (i) a single FITs file for each analyzed datacube in the {\tt Pipe3D} format and (ii) a catalog of $\sim$500 integrated and characteristic properties for each object. The data model of the {\tt Pipe3D} file is described in detail, including an explanation of the content of each of its different extensions, the spatially resolved dataproducts, the procedures to derive them, their units and uncertainties. An example of the content of each extension is included, using an arbitrarily selected target (the spiral galaxy manga-7495-12704) as a showcase. Particular care has been taken to describe the newly included extensions, like the {\tt ELINES} and {\tt FLUX\_ELINES\_LONG} ones. Using the complete set of {\tt Pipe3D} files, we explore the distribution across the classical [OIII]/H$\beta$ vs. [NII]/H$\alpha$ diagnostic-diagram of all the spatially resolved ionized regions, segregating them by the stellar mass and morphology of their host galaxies, finding consistent results with previous publications \citep{ARAA}. This show-case example demonstrates the scientific use of the delivered spatially resolved dataproducts.  

A detailed description of the individual parameters derived for each galaxy included in the delivered catalog, comprising both integrated and characteristic properties, is included. A clear distinction has been established between parameters inherited from previous tables (included to facilitate the identification of the targets in the sky and in other catalogs or for being part of the quality control procedure), and parameters derived as part of our analysis. We explain the derivation of parameter values at different apertures (integrated, central and at the effective radius), and of the corresponding slopes for the radial gradients (when required). We present separate descriptions of the properties derived for the stellar population, the emission lines and the kinematic properties, indicating when needed the {\tt Pipe3D} file extension to which they belong. In addition, we describe the delivered results from our morphological analysis and an update of the volume corrections presented in \citet{sanchez18b}. In all cases, we clearly state the entry in the final catalog that corresponds to each of the derived and delivered quantities.

Our set of estimated parameters and properties has been compared with similar sets that are already publicly available. The main results from this comparison are:
\begin{itemize}
    \item Our proposed morphological classification is statistically similar to the DS21 classification, with very consistent results for all morphological types in common.
    \item The comparison of our results with galaxy properties estimated using (i) a previous version of the {\tt Pipe3D} pipeline for the MaNGA DR15 dataset ($\sim$4500 galaxies in commom) and (ii) the MaNGA DAP for the DR17 dataset, show very good agreement in most of the cases. On the contrary, the comparison with the {\tt FIREFLY} results shows a poorer degree of agreement, depending on the parameter.
    \item The larger differences are found for those properties of the stellar populations that depend more strongly on the SSP template adopted for the stellar decomposition. Comparing with DR15 results, we find very good agreement for M$_\star$, $\Upsilon_\star$, SFR$_{\rm ssp}$, $\sigma_{\star,cen}$, $v_{\star,Re}$ and $\lambda_{\rm Re}$, a systematic offset for \ageLW$_{\rm Re}$, and a deviation from the one-to-one relation towards an asymptotic lower value (in the case of DR15 data) for \metLW$_{\rm Re}$, expected due to the lower dynamical range in metallicity of the DR15 SSP library. Finally, the dust extinction (A$_{\star,\rm V}$) presents slightly larger values in the DR15 dataset (although within the observed dispersion).
    \item In the case of {\tt FIREFLY} we find deviations from the one-to-one relation for both \ageLW$_{\rm Re}$ and \metLW$_{\rm Re}$, although in this case the metallicity seems to present a systematic offset and the age a linear relation (with slope different than 1). The largest discrepancies occur for the slope of the radial gradients of both parameters, that show only a mild correspondence.
    \item Considerably good agreement is found between our estimates of the spectral indices  and those reported by DAP. In some cases there is a systematic offset, but in general the degree of agreement is similar to the one found when comparing repeated observations of the same object analyzed using the same tool.
    \item The best agreement when comparing the {\tt Pipe3D}-DR15 and DAP-DR17 datasets is found for the emission line properties, especially when F$_{H\alpha,cen}>$10$^{-17}$\flux, $|$EW(H$\alpha$)$|>$0.5\AA. This is the regime in which the uncertainties (or differences) due to the subtraction of the stellar population are less relevant. The only parameter for which we find a significant deviation from the one-to-one relation is A$_{\rm V, gas}$, which shows a systematic offset, with DR15 values being $\sim$0.24 mag higher than those for DR17. 
    \item Finally, we find a systematic offset in $\sigma_{H\alpha,Re}$ and a deviation from the one-to-one relation in $\sigma_{\star,Re}$ between our estimates and those reported by DAP. We consider that an error in our estimates of the first parameter, and the differences in the adopted template of stellar spectra in the second parameter, are the most suitable cause of these differences.
\end{itemize}

As a practical example of the use of our final catalog, we update our selection of AGN-host candidates following \citet{sanchez18}. We select the candidates using the emission line properties extracted from the central apertures of the datacubes, adopting a criterium that combines the exploration of their location in three diagnostic diagrams with a cut in EW(H$\alpha$). In addition, we look for evidence of a broad component in H$\alpha$ in order to detect possible type-I AGNs. As a result of this analysis we found a total of 227 AGN-host candidates, 9 of them being candidates to host a type-I AGN. A brief comparison of the properties of this set of candidates with the properties of non-active AGNs confirms the main results from previous studies, suggesting that these objects are located in the transition regime between SFGs and RGs \citep[e.g.][]{kauff03,sanchez04,schawinski+2014,sanchez18}.

In summary, we consider that the current delivered analysis comprises a unique dataset for the exploration of the spatial, integrated and characteristic properties of galaxies in the nearby Universe.

%including 9 candidates to 
%we select a sample of 227 candidates, nine of them with clear evidence  
%A practical example of the use of the spatially resolved dataproducts included in the delivered {\tt Pipe3D} files is shown
%Problems found in previous data-releases have been corrected and the corresponding solution implemented in the analysis and the final delivered files, and described in the text when required.

%The current version of the file consists of the same extensions of previous data-releases \citep[e.g., DR15 Pipe3D VAC][]{sanchez18}, including: (i) the {\tt SSP} extension (average properties of the stellar populations); (ii) the {\tt SFH} extension (), {\tt FLUX\_ELINES} and {\tt INDICES} extensions, with the spatially resolved distribu
%We include new extensions to the {\tt Pipe3D} file not considered in previous distributions, comprising the results 
%New extensions not included in previous distributions have been 

%, together a description of the adopted stel

%\appendix

\section*{Acknowledgements}

{ We thank the referee for the suggestions and comments that have improved this manuscript and helped us to clean it.

We thanks S. Charlot for his contribution to generate the SSP templates.} We would like to thanks J. Neumann and H. Dom\'\i nguez-S\'anchez for the comments that have improved this manuscript.

SFS and J.B-B are grateful for the support of a CONACYT grant CB-285080 and FC-2016-01-1916, and funding from the PAPIIT-DGAPA-IN100519 and IG100622 (UNAM) projects. J.B-B acknowledges support from the CONACYT grant CF19-39578. GB acknowledges financial support from the National Autonomous University of M\'exico (UNAM) through grant DGAPA/PAPIIT BG100622.

This research made use of
Astropy,\footnote{http://www.astropy.org} a community-developed core
Python package for Astronomy \citep{astropy:2013, astropy:2018}.

Funding for the Sloan Digital Sky 
Survey IV has been provided by the 
Alfred P. Sloan Foundation, the U.S. 
Department of Energy Office of 
Science, and the Participating 
Institutions. 

SDSS-IV acknowledges support and 
resources from the Center for High 
Performance Computing  at the 
University of Utah. The SDSS 
website is www.sdss.org.

SDSS-IV is managed by the 
Astrophysical Research Consortium 
for the Participating Institutions 
of the SDSS Collaboration including 
the Brazilian Participation Group, 
the Carnegie Institution for Science, 
Carnegie Mellon University, Center for 
Astrophysics | Harvard \& 
Smithsonian, the Chilean Participation 
Group, the French Participation Group, 
Instituto de Astrof\'isica de 
Canarias, The Johns Hopkins 
University, Kavli Institute for the 
Physics and Mathematics of the 
Universe (IPMU) / University of 
Tokyo, the Korean Participation Group, 
Lawrence Berkeley National Laboratory, 
Leibniz Institut f\"ur Astrophysik 
Potsdam (AIP),  Max-Planck-Institut 
f\"ur Astronomie (MPIA Heidelberg), 
Max-Planck-Institut f\"ur 
Astrophysik (MPA Garching), 
Max-Planck-Institut f\"ur 
Extraterrestrische Physik (MPE), 
National Astronomical Observatories of 
China, New Mexico State University, 
New York University, University of 
Notre Dame, Observat\'ario 
Nacional / MCTI, The Ohio State 
University, Pennsylvania State 
University, Shanghai 
Astronomical Observatory, United 
Kingdom Participation Group, 
Universidad Nacional Aut\'onoma 
de M\'exico, University of Arizona, 
University of Colorado Boulder, 
University of Oxford, University of 
Portsmouth, University of Utah, 
University of Virginia, University 
of Washington, University of 
Wisconsin, Vanderbilt University, 
and Yale University.

\section*{Data Availability}

As indicated before the full dataproducts and final catalog produced by the pipeline are freely distributed to the community: \url{http://ifs.astroscu.unam.mx/MaNGA/Pipe3D_v3_1_1/}. The MaNGA DR17 reduced dataset analyzed along this article are available in the SDSS-DR17 webpage: \url{https://www.sdss.org/dr17/manga/}.

%################
%\bibliography{sample631}{}
\bibliographystyle{aasjournal}
\bibliography{my_bib} %#,carlos,extra}

%%%%%%%%%%%%%%%%%%%%%%%%%%%%%%%%%%%%%%%%%%%%%%%%%%%%%%%%%%%%%%%%%%%
%%%%%%%%%%%%%%%%%  APPENDIX START  %%%%%%%%%%%%%%%%%%%%%%%%%%%%%%%%
%%%%%%%%%%%%%%%%%%%%%%%%%%%%%%%%%%%%%%%%%%%%%%%%%%%%%%%%%%%%%%%%%%%
\appendix

{
\input AppendixSSP
}

\section{Emission lines included in the FLUX\_ELINES and FLUX\_ELINES\_LONG extensions}
\label{app:elines}

The list of emission lines which properties, derived using the moment-analysis, are included
in the FLUX\_ELINES and FLUX\_ELINES\_LONG extensions are listed in Table \ref{tab:fe_list} and 
\ref{tab:fe_long_list}. For each emission line we provide with the running index {\tt I} described
in Tab. \ref{tab:fe}, that allows to identify each emission line with each of their properties stored
in the different channels of the considered extensions. In addition we list the name of each of the adopted wavelength 
and the name that design each emission line. For the FLUX\_ELINES extension the wavelength are
based on the compilation presented in \citep{pipe3d_ii}, while for the FLUX\_ELINES\_LONG we extracted the
wavelengths from the detailed list presented by \citet{snr_elines}.

%----------------------------------------------------------------
\begin{table}
\begin{center}
\caption{Emission lines included in the FLUX\_ELINES extension}
\begin{tabular}{lll|lll}\hline\hline
{\tt I} &$\lambda$ (\AA) & Id & {\tt I}  & $\lambda$ (\AA) & Id \\
\hline
0 &  3727.4  &   [OII]3727  & 29 &  4889.62  &   [FeII]  \\ 
1 &  3750.0  &   H12  & 30 &  4905.34  &   [FeII]  \\ 
2 &  3771.0  &   H11  & 31 &  5111.6299  &   [FeII]  \\ 
3 &  3798.0  &   H10  & 32 &  5158.7798  &   [FeII]  \\ 
4 &  3819.4  &   HeI3819  & 33 &  5199.6001  &   [NI]  \\ 
5 &  3835.0  &   H9  & 34 &  5261.6201  &   [FeII]  \\ 
6 &  3869.0  &   [NeIII]  & 35 &  5517.71  &   [ClIII]  \\ 
7 &  3889.0  &   H8  & 36 &  5537.6  &   [ClIII]  \\ 
8 &  3967.0  &   [NeIII]  & 37 &  5554.94  &   OI  \\ 
9 &  3970.1  &   H$\epsilon$  & 38 &  5577.3101  &   [OI]  \\ 
10 &  4026.29  &   HeI4026  & 39 &  5754.52  &   [NII]  \\ 
11 &  4069.17  &   [SII]  & 40 &  5875.62  &   HeI5876  \\ 
12 &  4076.72  &   [SII]  & 41 &  6300.2998  &   [OI]  \\ 
13 &  4101.74  &   H$\delta$  & 42 &  6312.4  &   [SIII]  \\ 
14 &  4276.83  &   [FeII]  & 43 &  6347.28  &   SiII  \\ 
15 &  4287.4  &   [FeII]  & 44 &  6363.7798  &   [OI]  \\ 
16 &  4319.62  &   [FeII]  & 45 &  6562.68  &   H$\alpha$  \\ 
17 &  4340.47  &   H$\Upsilon$  & 46 &  6583.41  &   [NII]6584  \\ 
18 &  4363.21  &   [OIII]4363  & 47 &  6548.08  &   [NII]6548  \\ 
19 &  4413.78  &   [FeII]  & 48 &  6677.97  &   HeI6678  \\ 
20 &  4416.27  &   [FeII]  & 49 &  6716.39  &   [SII]6717  \\ 
21 &  4471.0  &   HeI4471  & 50 &  6730.74  &   [SII]6731  \\ 
22 &  4657.93  &   [FeIII]  & 51 &  7136.0  &   [ArIII]  \\ 
23 &  4686.0  &   HeII  & 52 &  7325.0  &   [OII]  \\ 
24 &  4712.98  &   HeI4713  & 53 &  7751.0  &   [ArIII]  \\ 
25 &  4922.16  &   HeI4922  & 54 &  9068.6  &   [SIII]  \\ 
26 &  5006.84  &   [OIII]5007  & 55 &  9530.6  &   [SIII]  \\ 
27 &  4958.91  &   [OIII]4959  & 56 &  ---  &   --- \\ 
28 &  4861.32  &   H$\beta$  &      &    &  \\ 
\hline
\end{tabular}\label{tab:fe_list}
\end{center}
Where {\tt I} corresponds to the index in Tab. \ref{tab:fe}, and Id is the label to each emission line
included in the FLUX\_ELINES extension.
\end{table}
%----------------------------------------------------------------

%----------------------------------------------------------------
\begin{table*}
\begin{center}
\caption{Emission lines included in the FLUX\_ELINES\_LONG extension}
\begin{tabular}{lll|lll|lll|lll|lll}\hline\hline
{\tt \#I} &$\lambda$ (\AA) & Id &
{\tt \#I} &$\lambda$ (\AA) & Id &
{\tt \#I} &$\lambda$ (\AA) & Id &
{\tt \#I} &$\lambda$ (\AA) & Id &
{\tt \#I} & $\lambda$ (\AA) & Id \\
\hline
0 &  3686.83  &   HI            & 39 &  4416.27  &   [FeII]  & 78 &  5039.1  &   [FeII]  & 117 &  6087.0  &   [FeVII]  & 156 &  8300.99  &   [NiII]  \\ 
1 &  3691.56  &   HI            & 40 &  4452.11  &   [FeII]  & 79 &  5072.4  &   [FeII]  & 118 &  6300.3  &   [OI]$^*$  & 157 &  8308.39  &   [CrII]  \\ 
2 &  3697.15  &   HI            & 41 &  4457.95  &   [FeII]  & 80 &  5107.95  &   [FeII]  & 119 &  6312.06  &   [SIII]  & 158 &  8345.55  &   HI  \\ 
3 &  3703.85  &   HI            & 42 &  4470.29  &   [FeII]  & 81 &  5111.63  &   [FeII]  & 120 &  6363.78  &   [OI]  & 159 &  8357.51  &   [CrII]  \\ 
4 &  3711.97  &   HI            & 43 &  4471.48  &   HeI     & 82 &  5145.8  &   [FeVI]  & 121 &  6374.51  &   [FeX]  & 160 &  8359.0  &   HI  \\ 
5 &  3726.03  &   [OII]$^*$     & 44 &  4474.91  &   [FeII]  & 83 &  5158.0  &   [FeII]  & 122 &  6435.1  &   [ArV]  & 161 &  8374.48  &   HI  \\ 
6 &  3728.82  &   [OII]$^*$     & 45 &  4485.21  &   [NiII]  & 84 &  5158.9  &   [FeVII]  & 123 &  6548.05  &   [NII]$^*$  & 162 &  8392.4  &   HI  \\ 
7 &  3734.37  &   HI$^*$        & 46 &  4562.48  &   [MgI]  & 85 &  5176.0  &   [FeVI]  & 124 &  6562.85  &   H$\alpha^*$  & 163 &  8446.0  &   OI  \\ 
8 &  3750.15  &   HI            & 47 &  4571.1   &   MgI]  & 86 &  5184.8  &   [FeII]  & 125 &  6583.45  &   [NII]$^*$  & 164 &  8467.25  &   HI  \\ 
9 &  3758.9  &   [FeVII]        & 48 &  4632.27  &   [FeII]  & 87 &  5191.82  &   [ArIII]  & 126 &  6678.15  &   HeI  & 165 &  8498.02  &   CaII  \\ 
10 &  3770.63  &   HI           & 49 &  4658.1   &   [FeIII]  & 88 &  5197.9  &   [NI]$^*$  & 127 &  6716.44  &   [SII]$^*$  & 166 &  8502.48  &   HI  \\ 
11 &  3797.9  &   HI$^*$        & 50 &  4685.68  &   HeII  & 89 &  5200.26  &   [NI]$^*$ & 128 &  6730.82  &   [SII]$^*$  & 167 &  8542.09  &   CaII  \\ 
12 &  3819.61  &   HeI          & 51 &  4701.62  &   [FeIII]  & 90 &  5220.06  &   [FeII]  & 129 &  6855.18  &   FeI  & 168 &  8545.38  &   HI  \\ 
13 &  3835.38  &   HI           & 52 &  4711.33  &   [ArIV]  & 91 &  5261.61  &   [FeII]  & 130 &  7005.67  &   [ArV]  & 169 &  8578.7  &   [ClII]  \\ 
14 &  3868.75  &   [NeIII]      & 53 &  4713.14  &   HeI  & 92 &  5268.88  &   [FeII]  & 131 &  7065.19  &   HeI  & 170 &  8598.39  &   HI  \\ 
15 &  3888.65  &   HeI$^*$      & 54 &  4724.17  &   [NeIV]  & 93 &  5270.3  &   [FeIII]  & 132 &  7135.8  &   [ArIII]$^*$  & 171 &  8616.96  &   [FeII]  \\ 
16 &  3889.05  &   HI$^*$       & 55 &  4733.93  &   [FeIII]  & 94 &  5273.38  &   [FeII]  & 133 &  7155.14  &   [FeII]  & 172 &  8662.14  &   CaII  \\ 
17 &  3933.66  &   CaII         & 56 &  4740.2   &   [ArIV]  & 95 &  5277.8  &   [FeVI]  & 134 &  7171.98  &   [FeII]  & 173 &  8665.02  &   HI  \\ 
18 &  3964.73  &   HeI$^*$      & 57 &  4754.83  &   [FeIII]  & 96 &  5296.84  &   [FeII]  & 135 &  7236.0  &   CII  & 174 &  8727.13  &   [CI]  \\ 
19 &  3967.46  &   [NeIII]$^*$  & 58 &  4769.6   &   [FeIII]  & 97 &  5302.86  &   [FeXIV]  & 136 &  7281.35  &   HeI  & 175 &  8750.47  &   HI  \\ 
20 &  3968.47  &   CaII$^*$     & 59 &  4774.74  &   [FeII]  & 98 &  5309.18  &   [CaV]  & 137 &  7290.42  &   [FeI]  & 176 &  8862.78  &   HI  \\ 
21 &  3970.07  &   H$\epsilon^*$ & 60 &  4777.88 &   [FeIII]  & 99 &  5333.65  &   [FeII]  & 138 &  7291.46  &   [CaII]  & 177 &  8891.88  &   [FeII]  \\ 
22 &  4026.19  &   HeI          & 61 &  4813.9  &   [FeIII]  & 100 &  5335.2  &   [FeVI]  & 139 &  7318.92  &   [OII]  & 178 &  9014.91  &   HI$^*$  \\ 
23 &  4068.6  &   [SII]         & 62 &  4814.55  &   [FeII]  & 101 &  5376.47  &   [FeII]  & 140 &  7323.88  &   [CaII]  & 179 &  9033.45  &   [FeII]  \\ 
24 &  4076.35  &   [SII]        & 63 &  4861.36  &   H$\beta^*$  & 102 &  5411.52  &   HeII  & 141 &  7329.66  &   [OII]  & 180 &  9051.92  &   [FeII]  \\ 
25 &  4101.77  &   H$\delta^*$  & 64 &  4881.11  &   [FeIII]  & 103 &  5412.64  &   [FeII]  & 142 &  7377.83  &   [NiII]  & 181 &  9069.0  &   [SIII]$^*$  \\ 
26 &  4120.81  &   HeI          & 65 &  4889.63  &   [FeII]  & 104 &  5424.2  &   [FeVI]  & 143 &  7388.16  &   [FeII]  & 182 &  9123.6  &   [ClII]  \\ 
27 &  4177.21  &   [FeII]       & 66 &  4893.4  &   [FeVII]  & 105 &  5426.6  &   [FeVI]  & 144 &  7411.61  &   [NiII]  & 183 &  9226.6  &   [FeII]  \\ 
28 &  4227.2  &   [FeV]         & 67 &  4905.35  &   [FeII]  & 106 &  5484.8  &   [FeVI]  & 145 &  7452.5  &   [FeII]  & 184 &  9229.02  &   HI  \\ 
29 &  4243.98  &   [FeII]       & 68 &  4921.93  &   HeI  & 107 &  5517.71  &   [ClIII]  & 146 &  7637.52  &   [FeII]  & 185 &  9266.0  &   OI  \\ 
30 &  4267.0  &   CII           & 69 &  4924.5  &   [FeIII]  & 108 &  5527.33  &   [FeII]  & 147 &  7686.19  &   [FeII]  & 186 &  9267.54  &   [FeII]  \\ 
31 &  4287.4  &   [FeII]        & 70 &  4930.5  &   [FeIII]  & 109 &  5577.34  &   [OI]  & 148 &  7686.9  &   [FeII]  & 187 &  9399.02  &   [FeII]  \\ 
32 &  4340.49  &   H$\Upsilon^*$  & 71 &  4942.5  &   [FeVII]  & 110 &  5631.1  &   [FeVI]  & 149 &  7751.06  &   [ArIII]  & 188 &  9470.93  &   [FeII]$^*$  \\ 
33 &  4358.1  &   [FeII]        & 72 &  4958.91  &   [OIII]$^*$  & 111 &  5677.0  &   [FeVI]  & 150 &  7774.0  &   OI  & 189 &  9531.1  &   [SIII]$^*$  \\ 
34 &  4358.37  &   [FeII]       & 73 &  4972.5  &   [FeVI]  & 112 &  5720.7  &   [FeVII]  & 151 &  7891.8  &   [FeXI]  & 190 &  9545.97  &   HI  \\ 
35 &  4359.34  &   [FeII]       & 74 &  4973.39  &   [FeII]  & 113 &  5754.59  &   [NII]  & 152 &  7999.85  &   [CrII]  & 191 &  9682.13  &   [FeII]  \\ 
36 &  4363.21  &   [OIII]       & 75 &  4985.9  &   [FeIII]  & 114 &  5876.0  &   HeI$^*$  & 153 &  8125.22  &   [CrII]  &  &    &  \\ 
37 &  4413.78  &   [FeII]       & 76 &  5006.84  &   [OIII]$^*$  & 115 &  5889.95  &   NaI$^*$  & 154 &  8229.55  &   [CrII]  & &    &  \\ 
38 &  4414.45  &   [FeII]       & 77 &  5015.68  &   HeI$^*$  & 116 &  5895.92  &   NaI$^*$  & 155 &  8236.77  &   HeII  &   &    &  \\ 
\hline
\end{tabular}\label{tab:fe_long_list}
\end{center}
Where {\tt \#I} corresponds to the index in Tab. \ref{tab:fe}, and Id is the label to each emission line
included in the FLUX\_ELINES\_LONG extension. $^*$ emission lines detected in at least a 5\%\ of the galaxies with a S/N$>$3.
\end{table*}
%----------------------------------------------------------------

\section{Additional Quality Control}
\label{app:qc}

\begin{figure}
 \minipage{0.99\textwidth}
 \includegraphics[width=0.5\textwidth]{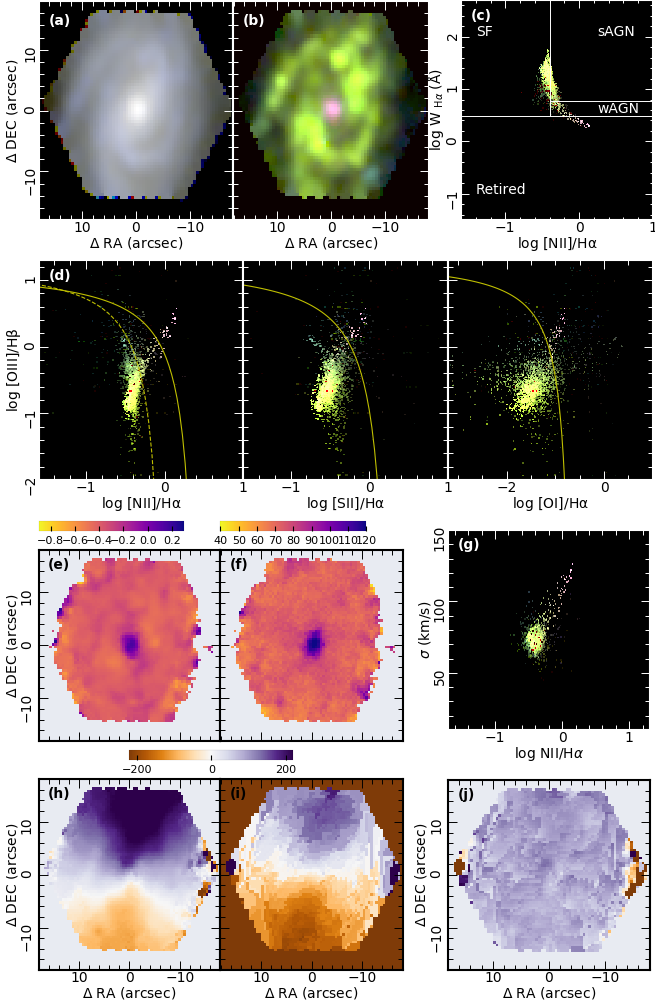}
 \endminipage
 \caption{Example of the information explored in the quality control process for the galaxy/cube manga-7495-12703, extracted from the analysis presented by \citet{carlos20}. (a) and (b) panels corresponds to the same color images created using broad-band and emission lines images shown in Fig. \ref{fig:qc_map}; (c) panel comprises the spatially resolved WHAN diagnostic diagram, with each spaxel color-coded using the scheme included in panel (b); classical BPT diagnostic diagrams showing the spatially resolved distribution of [OIII]/H$\beta$ as a function of [NII]/H$\alpha$, [SII]/H$\alpha$ and [OI]/H$\alpha$ are shown in panels (d), using a similar color scheme as panel $c$; panels (e) and (f) show the spatial distributions of the [NII]/H$\alpha$ line ratio, in logarithm scales, and the H$\alpha$ velocity dispersion, with panel (g) showing the distribution of those parameters one against each other (color coded using the same scheme of panels $c$ and $d$.); panels (h), (i) and (j), show the ionized gas velocity map (derived from H$\alpha$), the stellar velocity map, and the residual of subtracting the latter from the former, respectively.}
 \label{fig:BPT_single}
\end{figure}

Figure \ref{fig:BPT_single} shows the additional information used in the visual exploration of the data as part of the quality control process described in Sec. \ref{sec:qc}. This information was used to identify the presence of strong interacting systems, and/or AGNs, that may affect the results from the kinematic analysis (due to the presence of multiple kinematic components) and/or the stellar population exploration (due to strong contaminations by a non-stellar source).

%%%%%%%%%%%%%%%%%%%%%%%%%%%%%%%%%%%%%%%%%%%%%%%%%%%%%%%%%%%%%%%%%%%%%%%5
%%%%%%%%%%%%%%%%%%%%%%%%%%%%%%%%%%%%%%%%%%%%%%%%%%%%%%%%%%%%%%%%%%%%%%%5

\section{List of Oxygen and Nitrogen abundance calibrators}
\label{app:OH}

Table \ref{tab:OH_calibrators} comprises the list of calibrators extracted from the compilation by \citet{espi22} adopted
along this exploration. We order the calibrators as they appear in the final catalog for the values reported at the effective radius. We include
the correspondence with those calibrators adopted to estimate the oxygen abundance in the central regions when available.
Those calibrators anchored to measurements based on the direct method (usually \HII\ regions) are labelled as {\it Empirical}, while
those based on photoionization models are labelled as {\it Theoretical}. 

%%%%%%%%%%%%%%%%%%%%%%%%%%%%%%%%%%%%%%%%%%%%%%%%%%%%%%%%%%%%%%%%%%%%%%%%%%%%%%%%%%%%%%%%%%%%%%%%%%%%%%%%%%%%%%%%%%%%%%%%%%%%
% List of calibrators
\begin{table*}
    \centering
        \caption{List of oxygen, nitrogen and ionization parameter calibrators adopted to estimate the values at the effective radius.}
    \begin{tabular}{l|l|c|c|c}
    \hline
    \hline
    ID at Re & ID central aperture & Emission lines and ratios & Calibration Type & Reference  \\
    \hline
    OH\textunderscore Mar13\textunderscore N2 & OH\_N2 &   [\ION{N}{II}]$\lambda 6548$/H$\alpha$ & Empirical & \cite{marino13} \\
    OH\textunderscore Mar13\textunderscore O3N2 & OH\_O3N2 &  O3N2 & Empirical & \cite{marino13} \\
    OH\textunderscore T04 & OH\_T04& [\ION{N}{II}]$\lambda 6548$/H$\alpha$, $R_{23}$ & Empirical & \cite{tremonti04} \\
    OH\textunderscore Pet04\textunderscore N2\textunderscore lin & &  [\ION{N}{II}]$\lambda 6548$/H$\alpha$ & Empirical & \cite{pettini04} \\
    OH\textunderscore Pet04\textunderscore N2\textunderscore poly & &  [\ION{N}{II}]$\lambda 6548$/H$\alpha$ & Empirical & \cite{pettini04} \\
    OH\textunderscore Pet04\textunderscore O3N2 & & O3N2 & Empirical & \cite{pettini04} \\
    OH\textunderscore Kew02\textunderscore N2O2 & &  [\ION{N}{II}]$\lambda 6583$/[\ION{O}{II}]$\lambda 3727$ & Theoretical & \cite{kewley02} \\
    OH\textunderscore Pil10\textunderscore ONS & OH\_ONS&  [\ION{N}{II}]$\lambda 6583$/H$\beta$, $R_2$, $R_3$, $P$, \SII & Empirical & \cite{pilyugin10} \\
    OH\textunderscore Pil10\textunderscore ON & &  [\ION{N}{II}]$\lambda 6583$/H$\beta$, $R_2$, $R_3$, \SII & Empirical & \cite{pilyugin10}  \\
    OH\textunderscore Pil11\textunderscore NS & &  [\ION{N}{II}]$\lambda 6583$/H$\beta$, $R_3$, \SII & Empirical & \cite{pilyugin11} \\
    OH\textunderscore Cur20\textunderscore RS32 & & $R_3$+\SII/H$\alpha$  & Empirical & \cite{2020Curti_MNRAS491} \\
    OH\textunderscore Cur20\textunderscore R3 & &  $R_3$ & Empirical & \cite{2020Curti_MNRAS491} \\
    OH\textunderscore Cur20\textunderscore O3O2 & & [\ION{O}{III}]$\lambda 5007$, [\ION{O}{II}]$\lambda 3727+29$ & Empirical & \cite{2020Curti_MNRAS491} \\
    OH\textunderscore Cur20\textunderscore S2 & & \SII  & Empirical & \cite{2020Curti_MNRAS491} \\
    OH\textunderscore Cur20\textunderscore R2 & & $R_2$  & Empirical & \cite{2020Curti_MNRAS491} \\
    OH\textunderscore Cur20\textunderscore N2 & & [\ION{N}{II}]$\lambda 6583$/H$\alpha$  & Empirical & \cite{2020Curti_MNRAS491} \\
    OH\textunderscore Cur20\textunderscore R23 & &  $R_{23}$ & Empirical & \cite{2020Curti_MNRAS491} \\
    OH\textunderscore Cur20\textunderscore O3N2 & & $R_3$, [\ION{N}{II}]$\lambda 6583$/H$\alpha$  & Empirical & \cite{2020Curti_MNRAS491} \\
    OH\textunderscore Cur20\textunderscore O3S2 & & $R_3$/\SII/H$\alpha$  & Empirical & \cite{2020Curti_MNRAS491} \\
    OH\textunderscore KK04 & OH\_R23 & $R_{23}$, [\ION{O}{III}]/[\ION{O}{II}] & Theoretical & \cite{kobu04} \\
    OH\textunderscore Pil16\textunderscore R & &  [\ION{N}{II}]$\lambda 6583$/H$\beta$, $R_2$, $R_3$ & Empirical & \cite{pilyugin16} \\
    OH\textunderscore Pil16\textunderscore S & &  [\ION{N}{II}]$\lambda 6583$/H$\beta$, $R_3$, \SII & Empirical & \cite{pilyugin16} \\
    OH\textunderscore Ho & & $R_2$, $R_3$, [\ION{N}{II}]$\lambda 6583$/H$\beta$, \SII/H$\alpha$  & Empirical & \cite{ho19} \\
     \hline
    U\textunderscore Dors\textunderscore O32 & & [\ION{O}{III}]$\lambda 5007$, [\ION{O}{II}]$\lambda 3727+29$  & Theoretical & \cite{dors11} \\
    U\textunderscore Dors\textunderscore S2 & & [\ION{N}{II}]$\lambda 6583$/H$\alpha$, \SII  & Theoretical & \cite{dors11} \\
    U\textunderscore Mor16\textunderscore O32\textunderscore fs & & [\ION{O}{III}]$\lambda 5007$, [\ION{O}{II}]$\lambda 3727+29$  & Empirical/Theoretical & \cite{mori16} \\
    U\textunderscore Mor16\textunderscore O32\textunderscore ts & & [\ION{O}{III}]$\lambda 5007$, [\ION{O}{II}]$\lambda 3727+29$  & Empirical/Theoretical & \cite{mori16} \\
     \hline
    NH\textunderscore Pil16\textunderscore R & &  [\ION{N}{II}]$\lambda 6583$/H$\beta$, $R_2$, $R_3$ & Empirical & \cite{pilyugin16} \\
    NO\textunderscore Pil16\textunderscore R & &  [\ION{N}{II}]$\lambda 6583$/H$\beta$, $R_2$, $R_3$ & Empirical & \cite{pilyugin16} \\
    NO\textunderscore Pil16\textunderscore Ho\textunderscore R & &  [\ION{N}{II}]$\lambda 6583$/H$\beta$, $R_2$, $R_3$, \SII/H$\alpha$  & Empirical & \cite{pilyugin16} \& \cite{ho19}\\
    NO\textunderscore Pil16\textunderscore N2\textunderscore R2 & &  [\ION{N}{II}]$\lambda 6583$/H$\beta$, $R_2$, $R_3$ & Empirical & \cite{pilyugin16} \\
%
% This last calibrator???
%
%    
%    NO\textunderscore Pil16\textunderscore S &  [\ION{N}{II}]$\lambda 6583$/H$\beta$, $R_3$, \SII & Empirical & \cite{pilyugin16} \\
%    OH\textunderscore NB & \vtop{\hbox{\strut\OII, [\ION{O}{III}]$\lambda 5007$, [\ION{O}{I}]$\lambda 6100$,} \hbox{\strut[\ION{N}{II}]$\lambda 6583$, \SII, H$\alpha$, H$\beta$}} & Theoretical & \cite{2018Thomas_ApJ856} \\
%    OH\textunderscore HCm\textunderscore(interp/no\textunderscore interp) & \vtop{\hbox{\strut [\ION{O}{II}]$\lambda 3727$, [\ION{Ne}{III}]$\lambda 3868$, [\ION{O}{III}]$\lambda 4363$,} \hbox{\strut[\ION{O}{III}]$\lambda 4959$, [\ION{O}{III}]$\lambda 5007$, [\ION{N}{II}]$\lambda 6584$, [\ION{S}{II}]$\lambda 6725$ }}& Theoretical & \cite{2014PerezMontero_MNRAS441}\\
%    OH\textunderscore IZI\textunderscore value\textunderscore models &
%    \vtop{\hbox{\strut [\ION{O}{II}]$\lambda 3727$, [\ION{O}{III}]$\lambda\lambda 4959+5007$,} \hbox{\strut[\ION{N}{II}]$\lambda\lambda 6583,48$,  \SII, H$\alpha$, H$\beta$}}& Theoretical &  \cite{2020Mingozzi_AA636} \\
    \hline
    \end{tabular}
    \label{tab:OH_calibrators}
\end{table*}
%%%%%%%%%%%%%%%%%%%%%%%%%%%%%%%%%%%%%%%%%%%%%%%%%%%%%%%%%%%%%%%%%%%%%%%%%%%%%%%%%%%%%%%%%%%%%%%%%%%%%%%%%%%%%%%%

\begin{figure*}
 \minipage{0.99\textwidth}
\includegraphics[width=4.4cm,clip,trim=0 10 0 10]{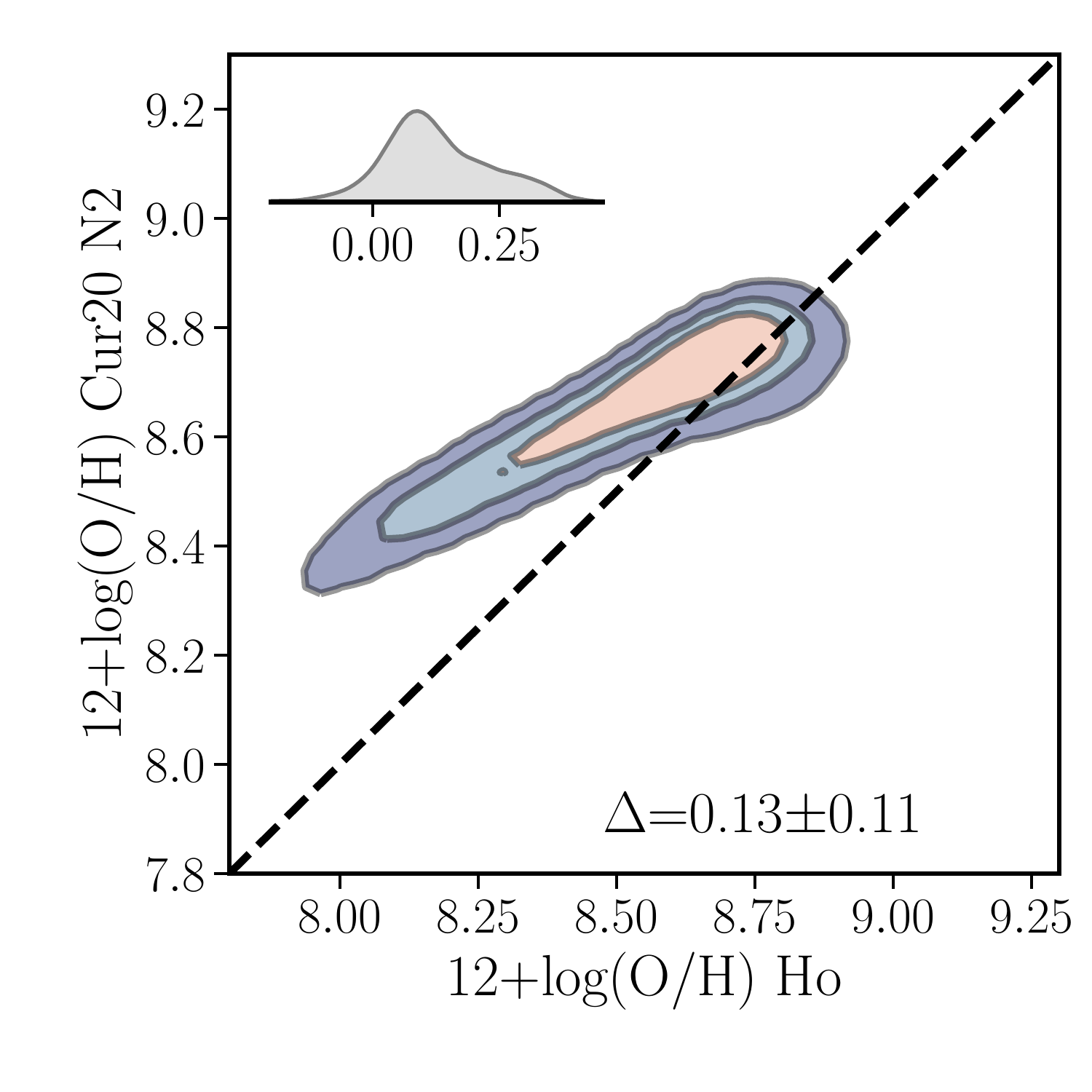}
\includegraphics[width=4.4cm,clip,trim=0 10 0 10]{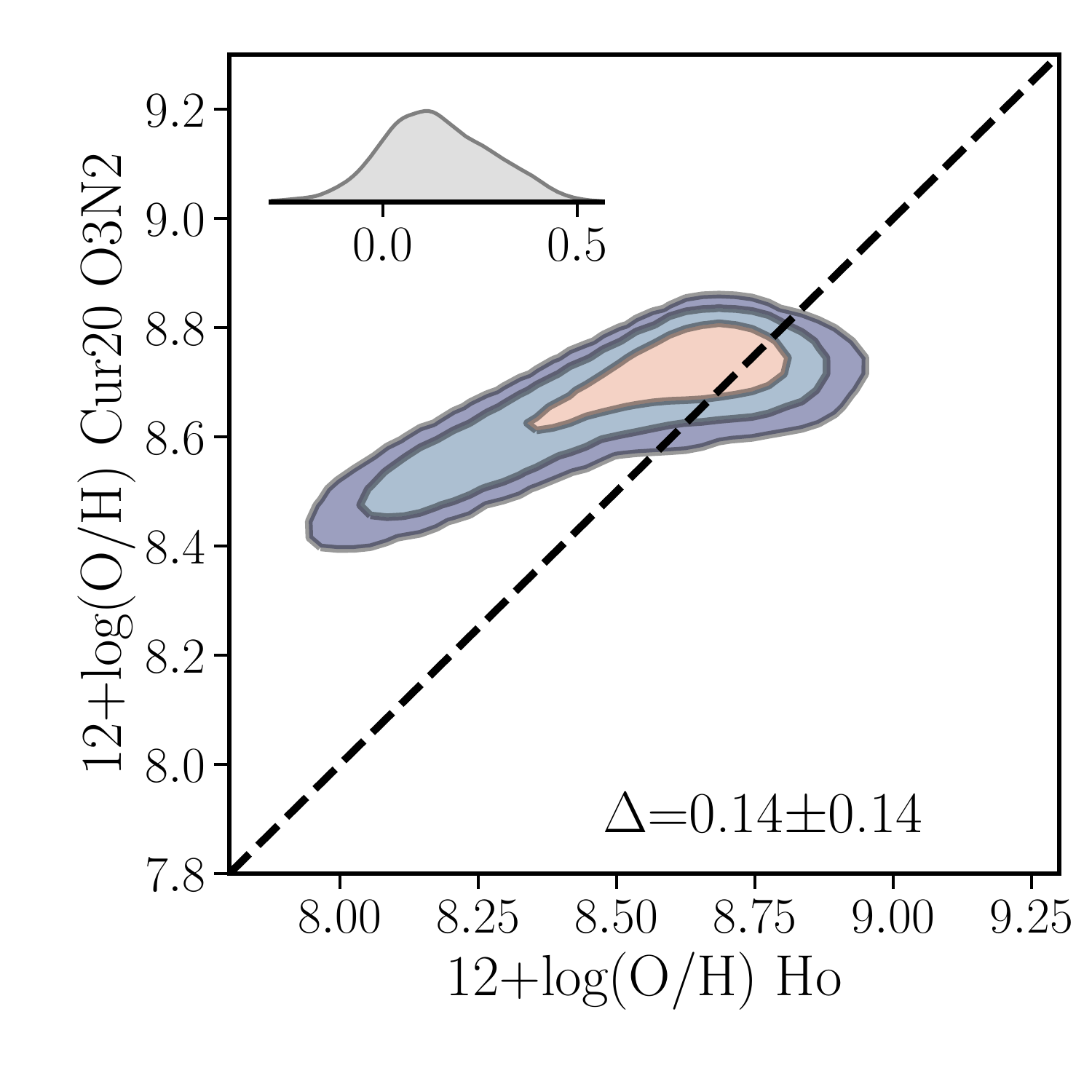}
\includegraphics[width=4.4cm,clip,trim=0 10 0 10]{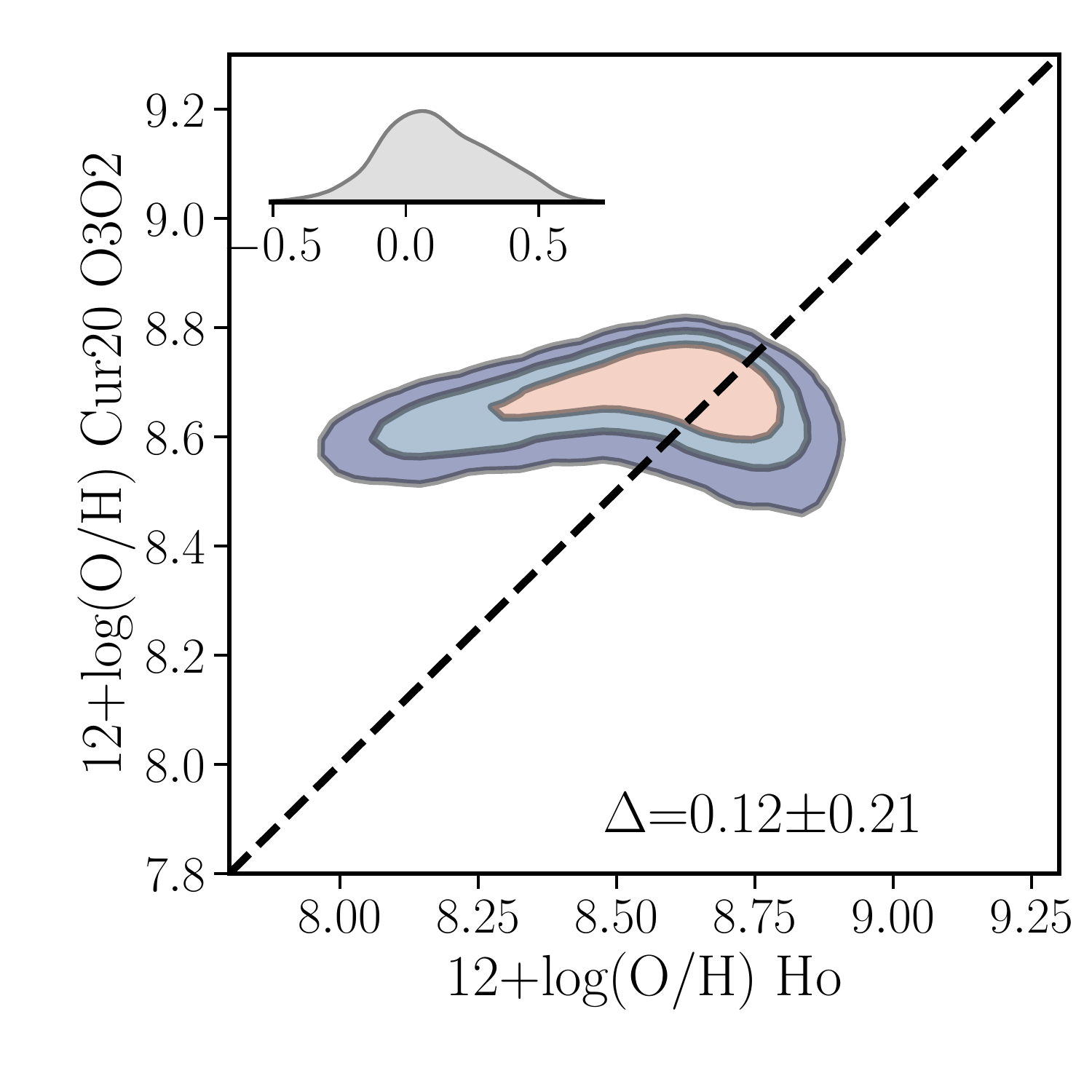}
\includegraphics[width=4.4cm,clip,trim=0 10 0 10]{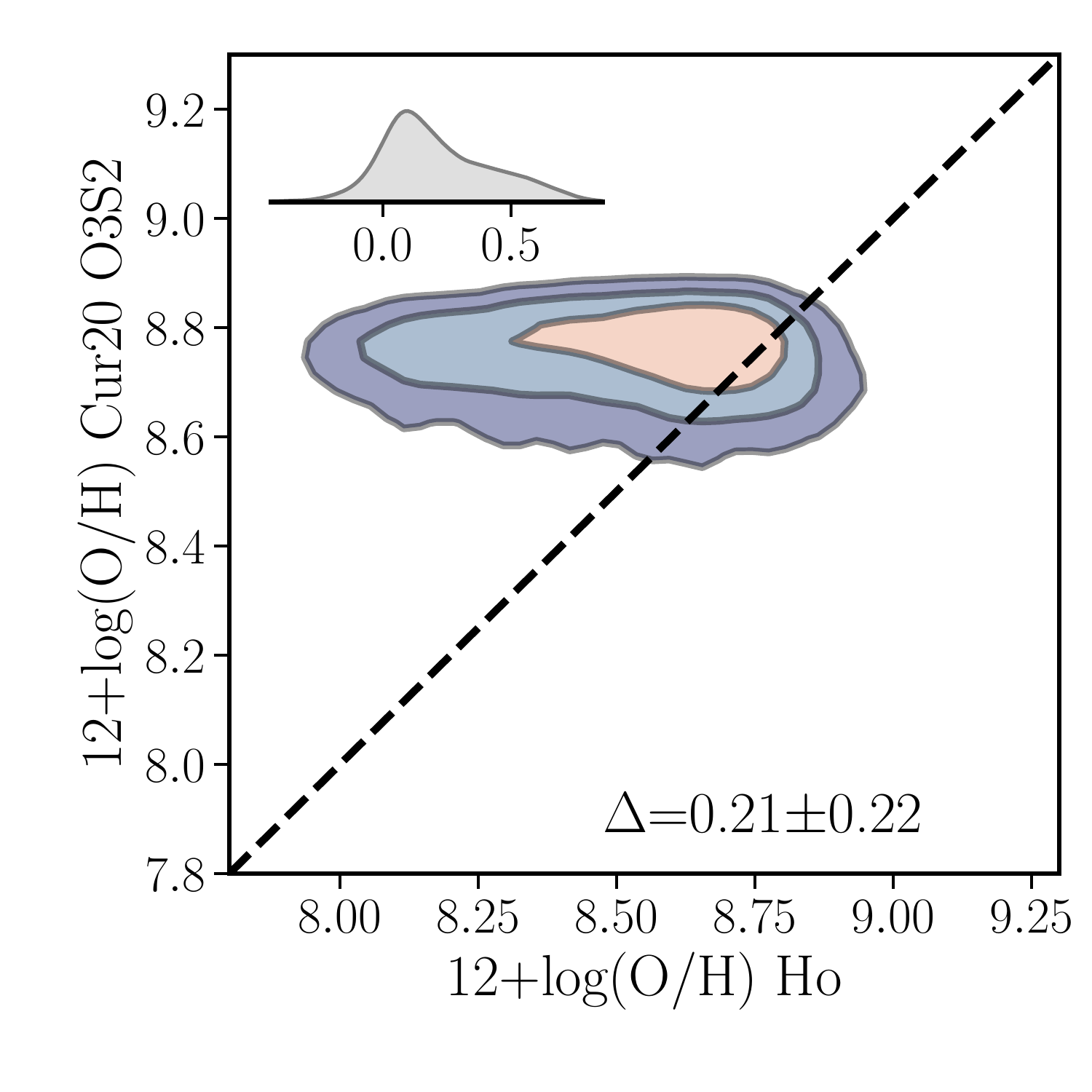}
\includegraphics[width=4.4cm,clip,trim=0 10 0 10]{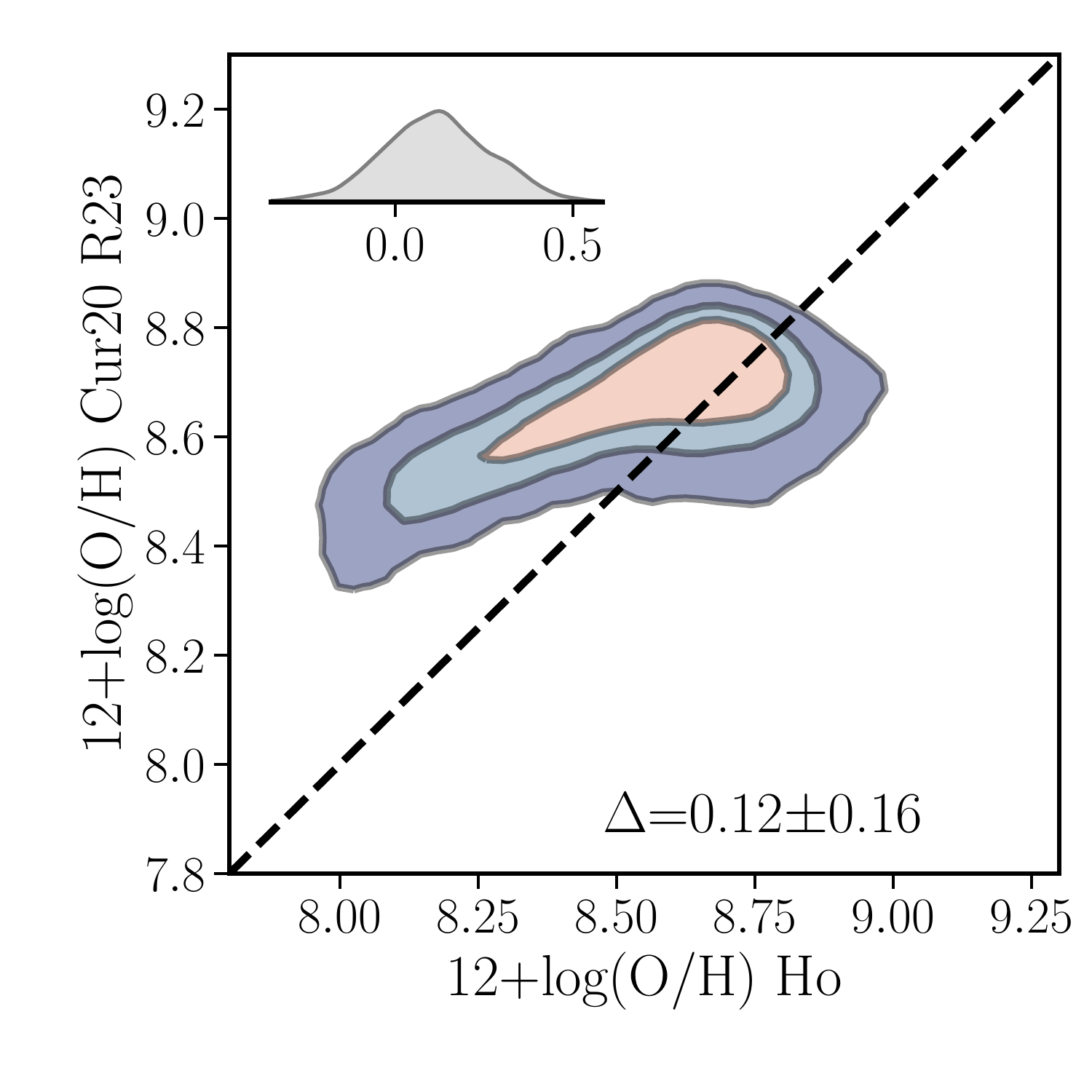}
\includegraphics[width=4.4cm,clip,trim=0 10 0 10]{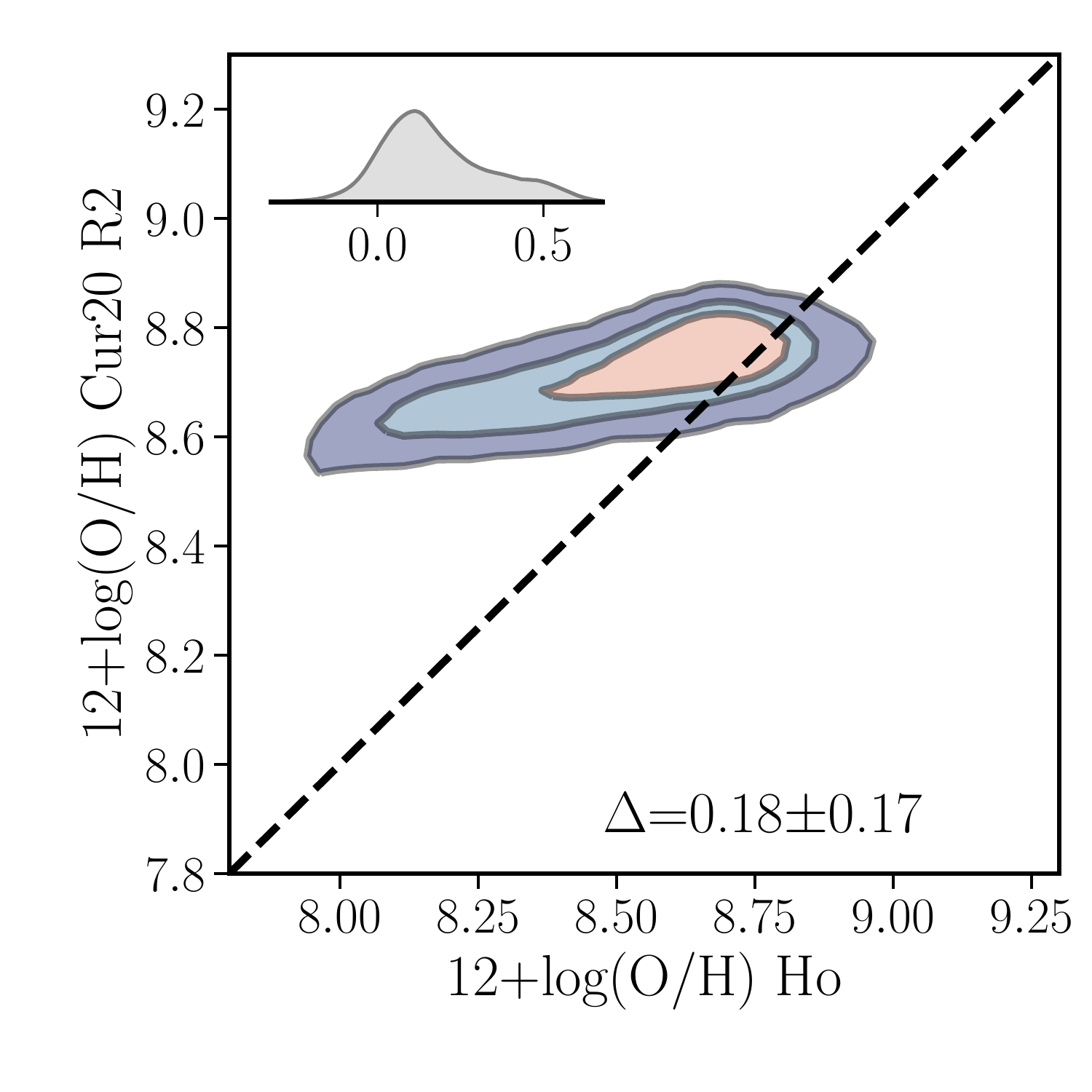}
\includegraphics[width=4.4cm,clip,trim=0 10 0 10]{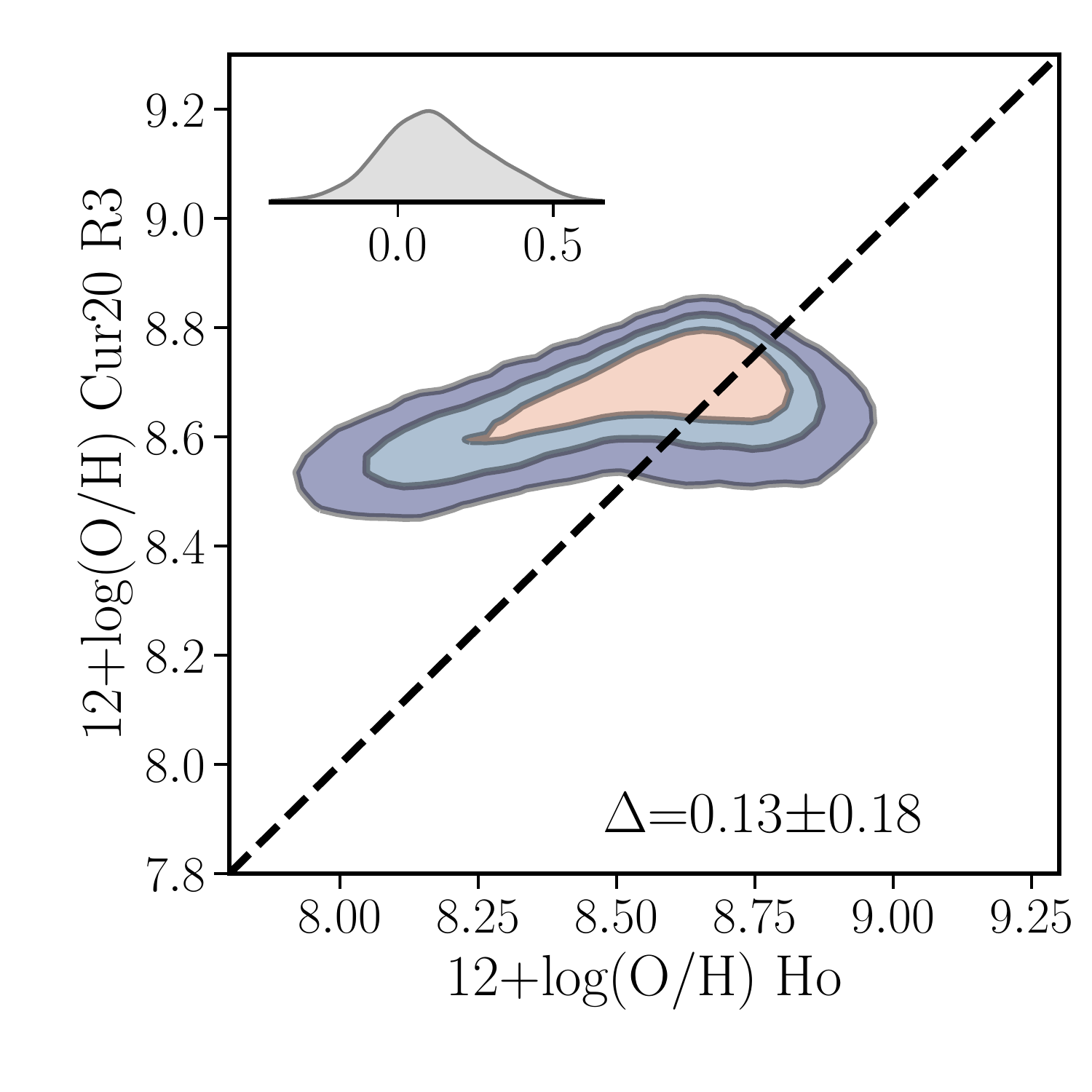}
\includegraphics[width=4.4cm,clip,trim=0 10 0 10]{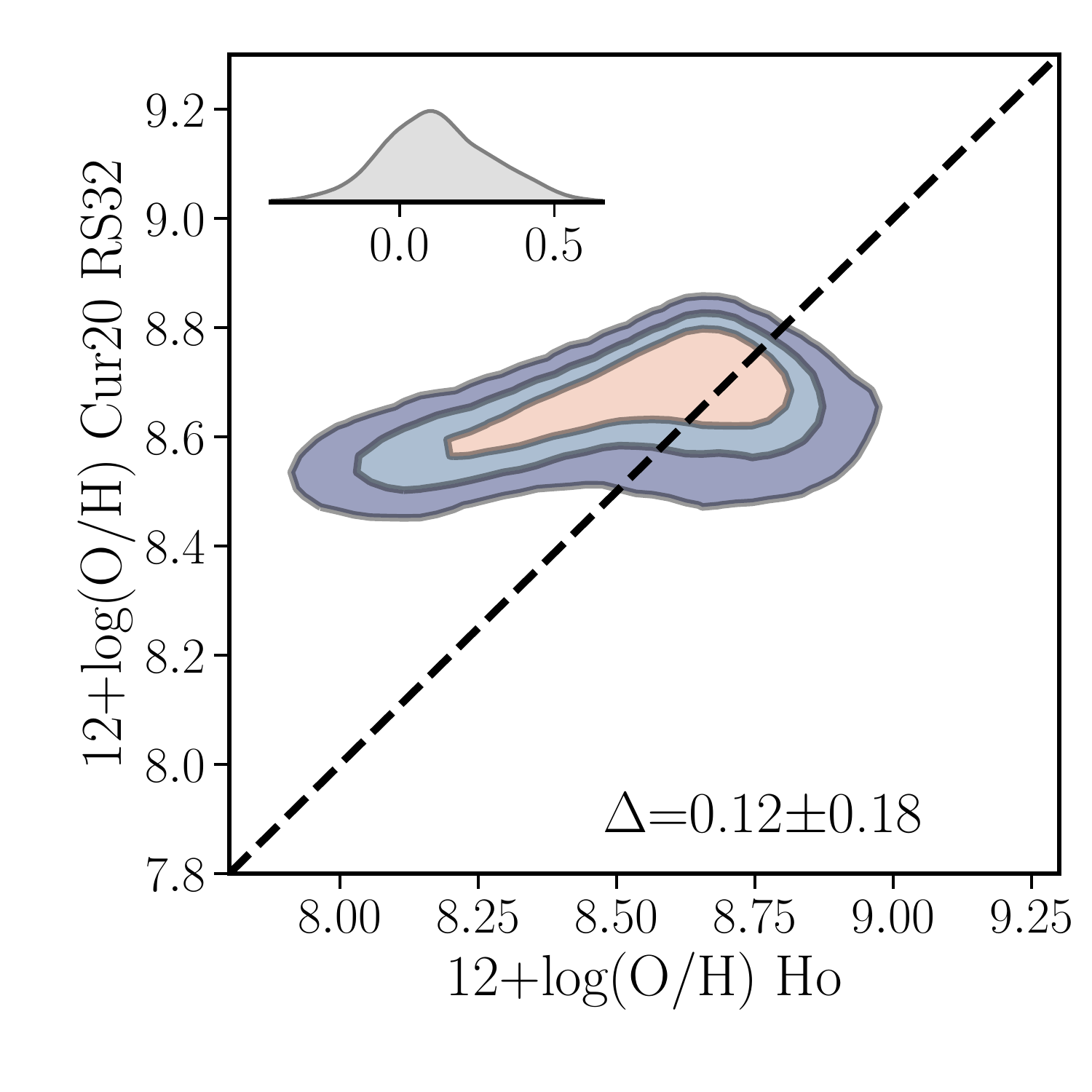}
\includegraphics[width=4.4cm,clip,trim=0 10 0 10]{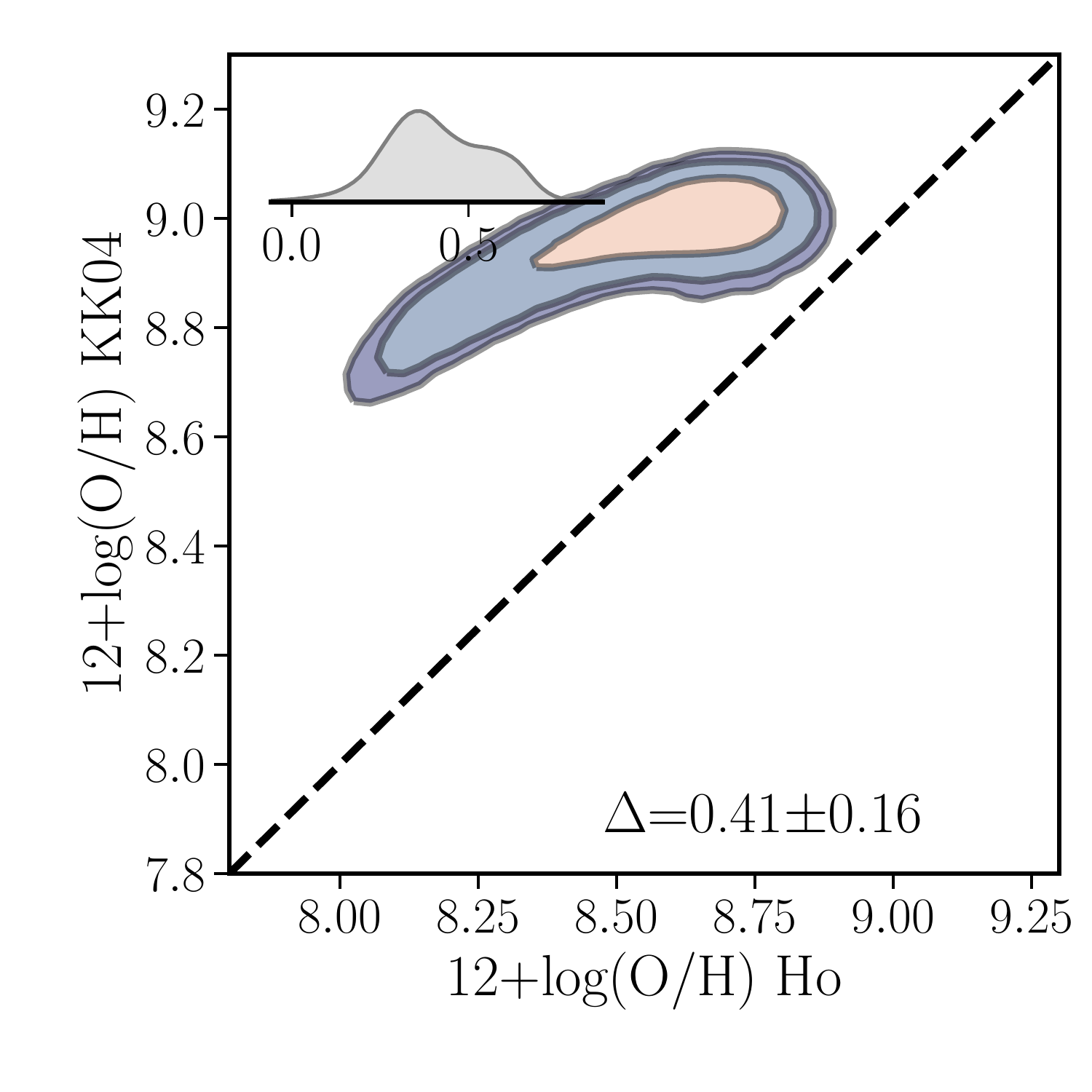}
\includegraphics[width=4.4cm,clip,trim=0 10 0 10]{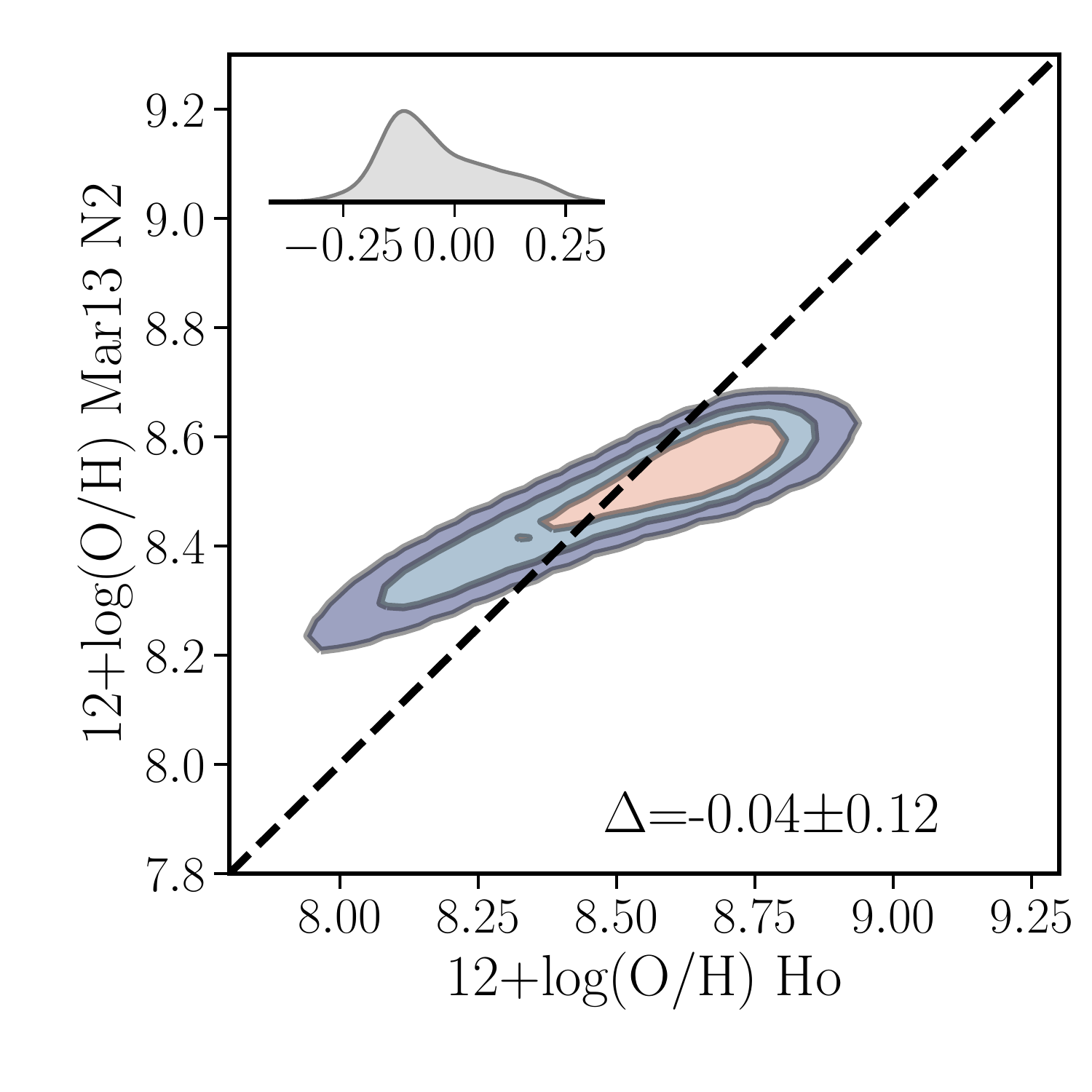}
\includegraphics[width=4.4cm,clip,trim=0 10 0 10]{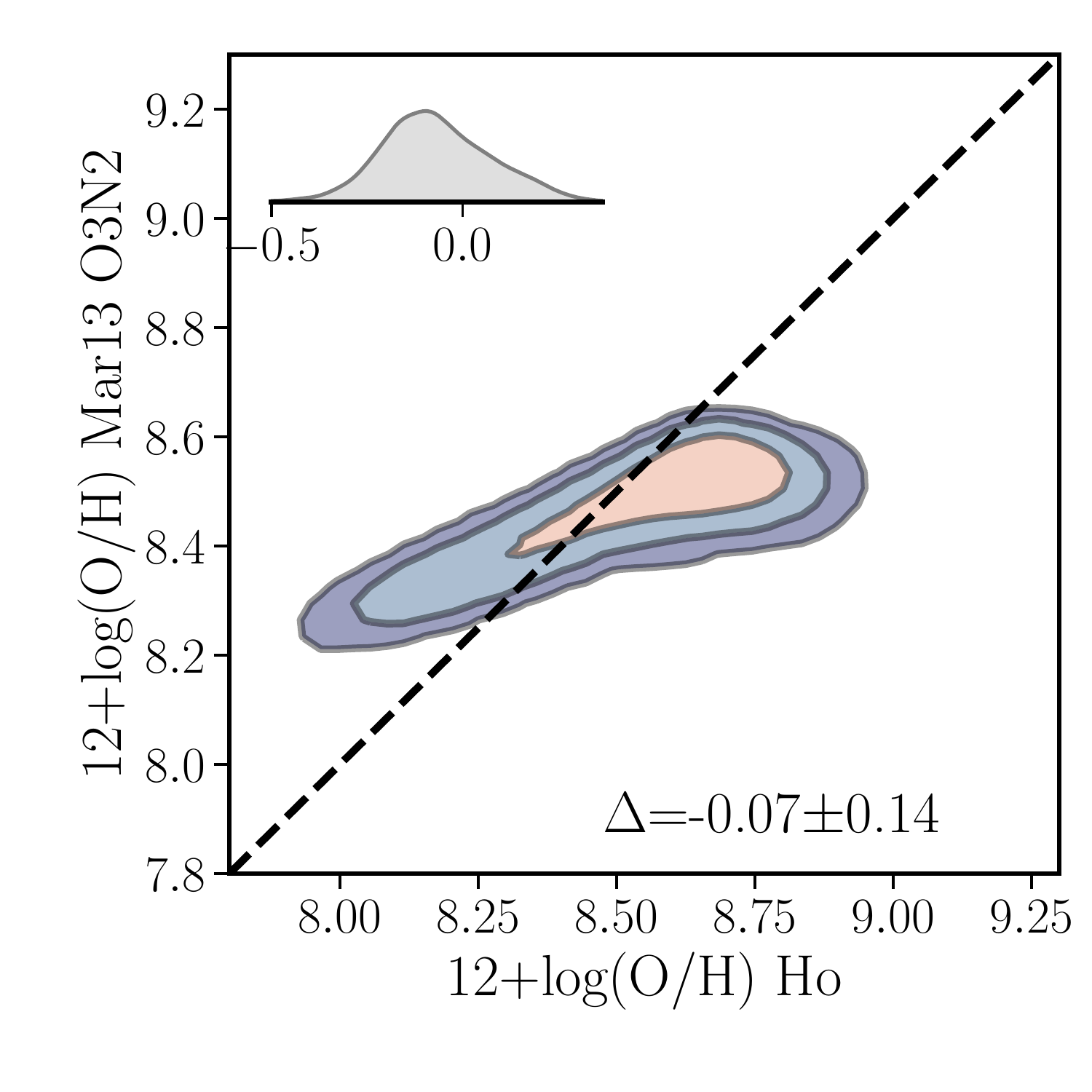}
\includegraphics[width=4.4cm,clip,trim=0 10 0 10]{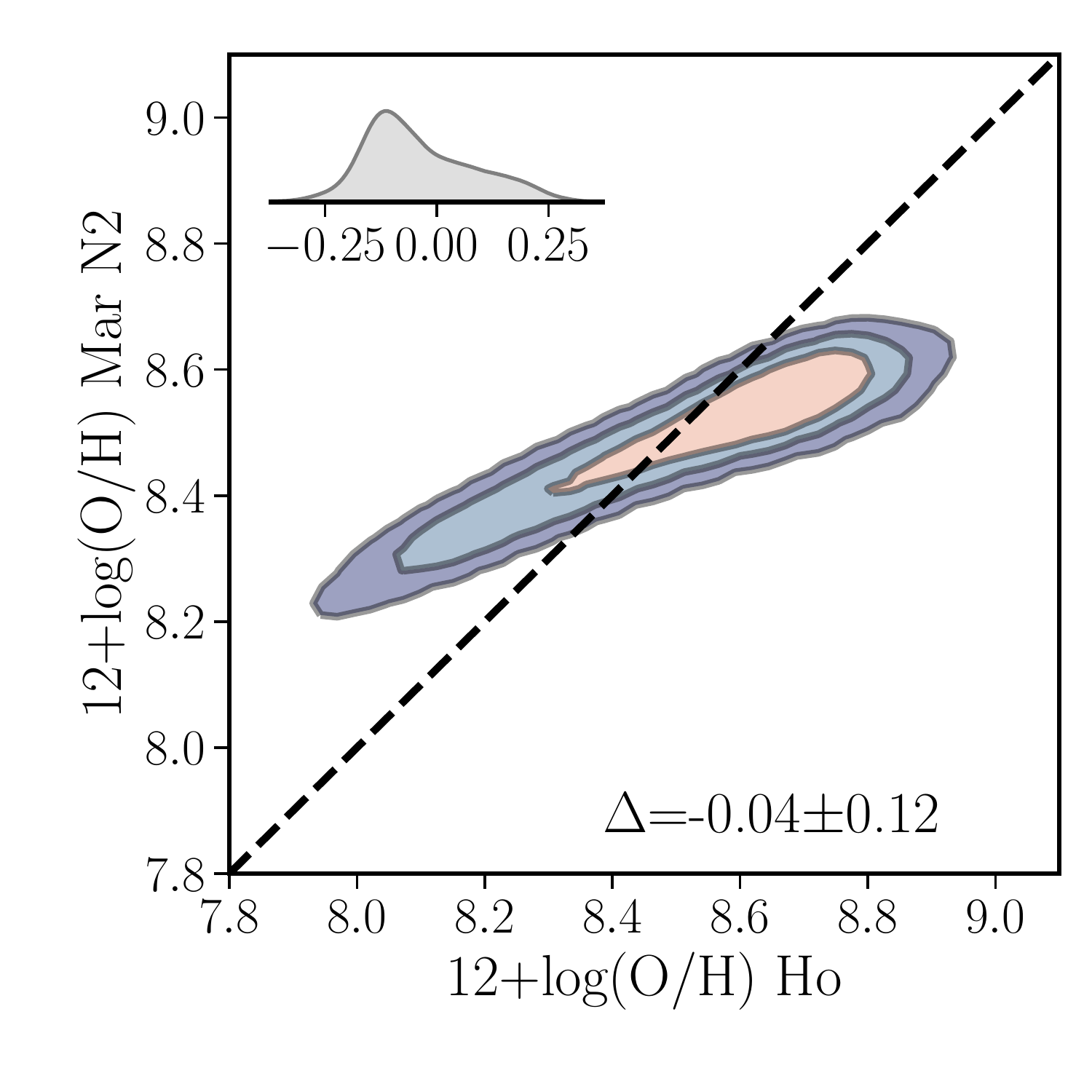}
\includegraphics[width=4.4cm,clip,trim=0 10 0 10]{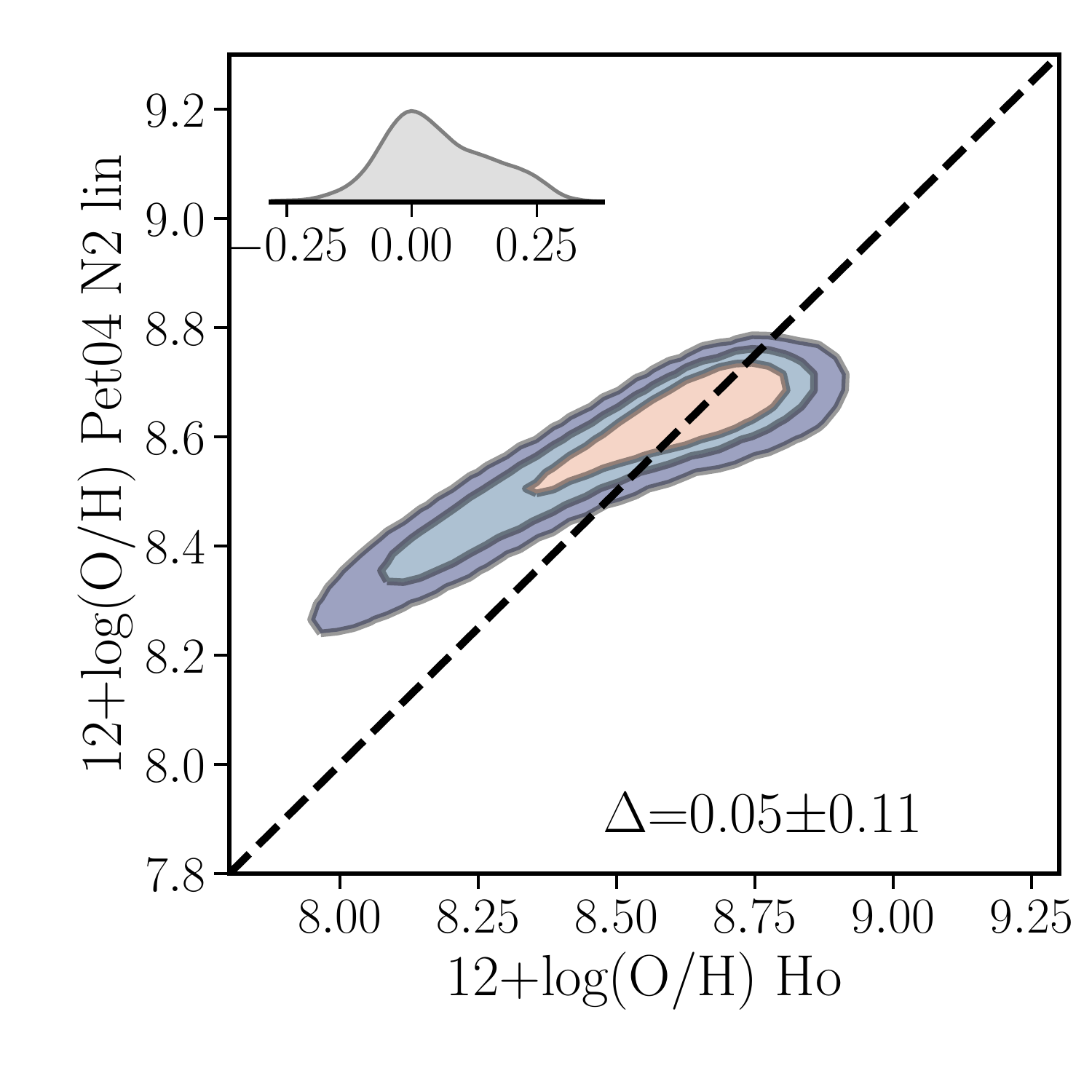}
\includegraphics[width=4.4cm,clip,trim=0 10 0 10]{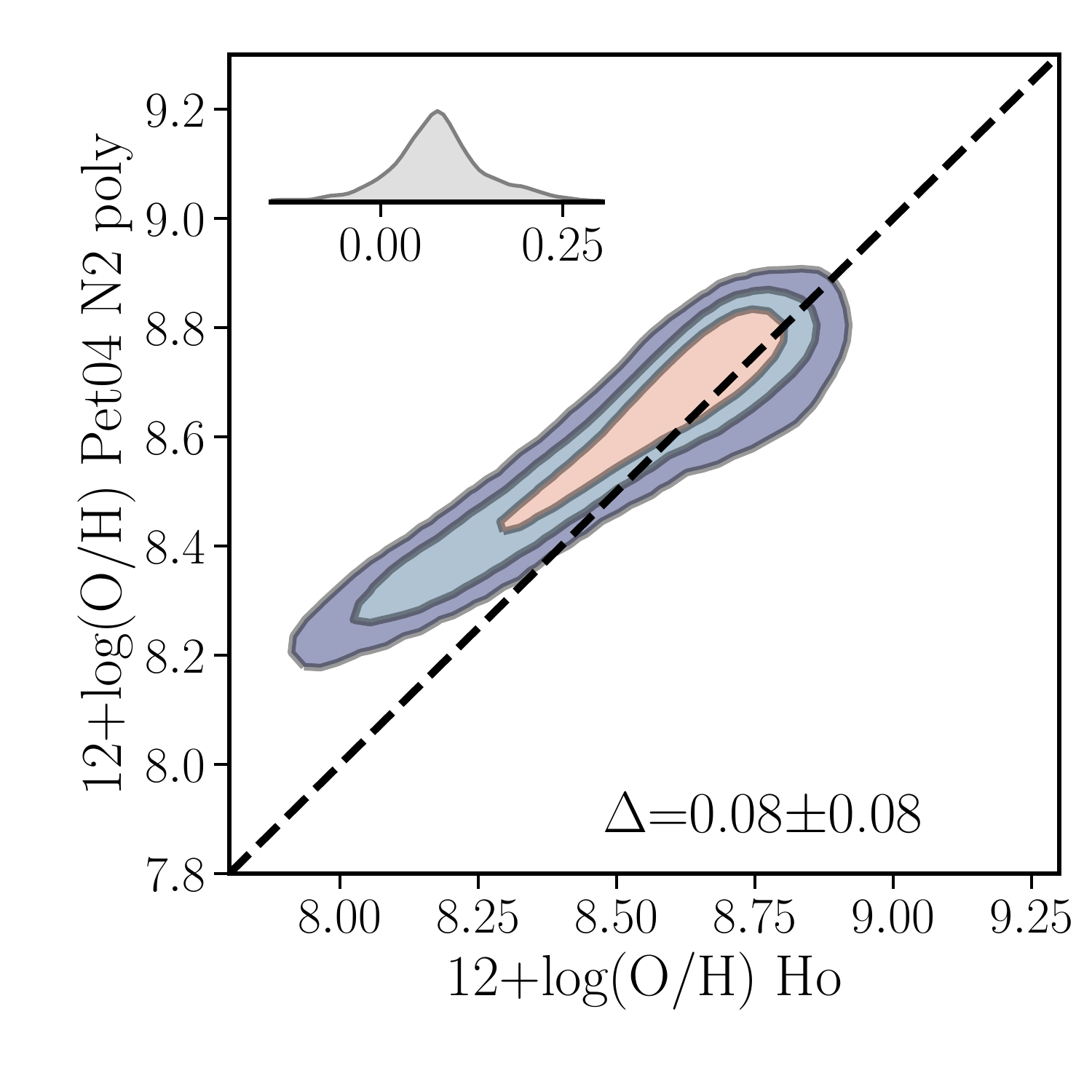}
\includegraphics[width=4.4cm,clip,trim=0 10 0 10]{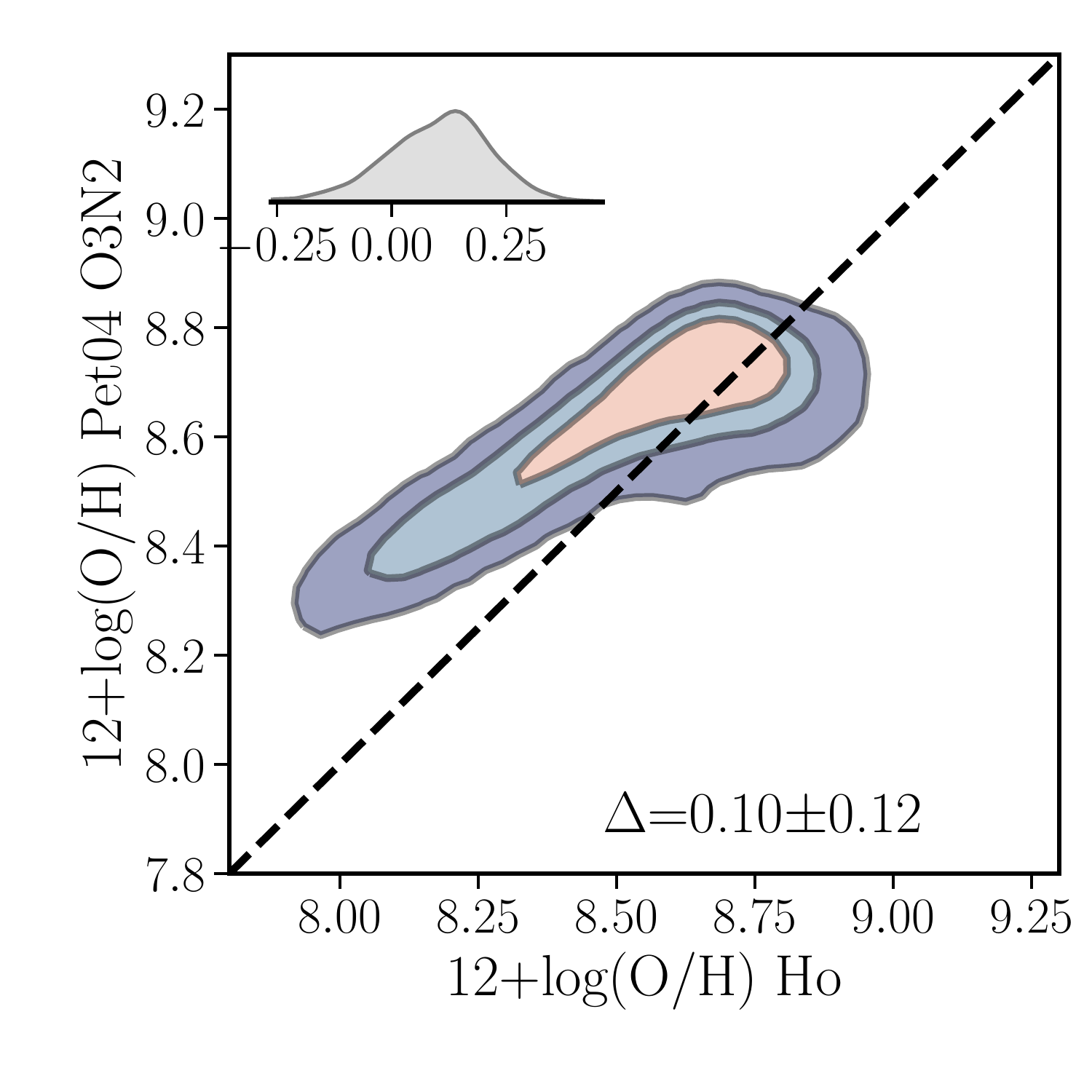}
\includegraphics[width=4.4cm,clip,trim=0 10 0 10]{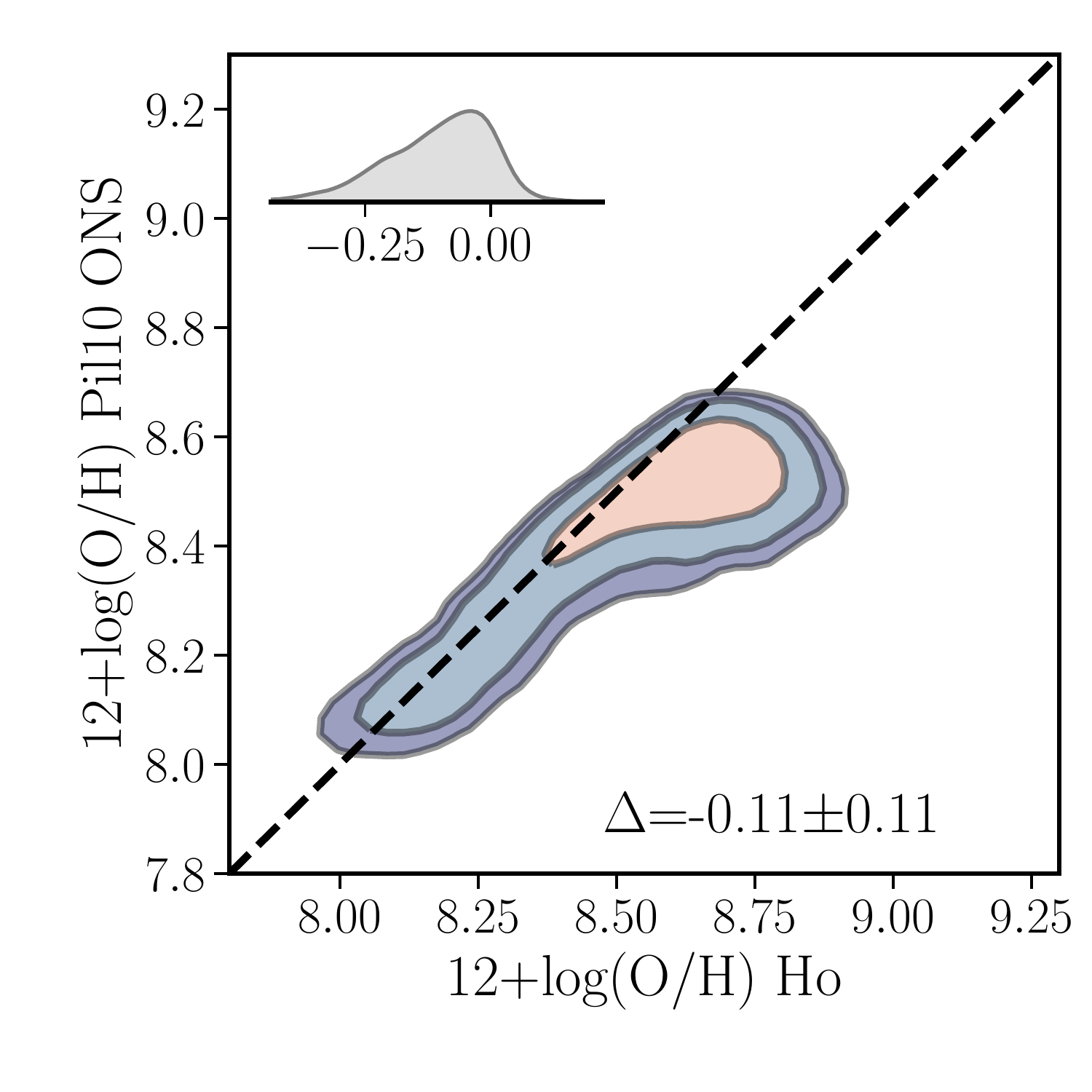}
\includegraphics[width=4.4cm,clip,trim=0 10 0 10]{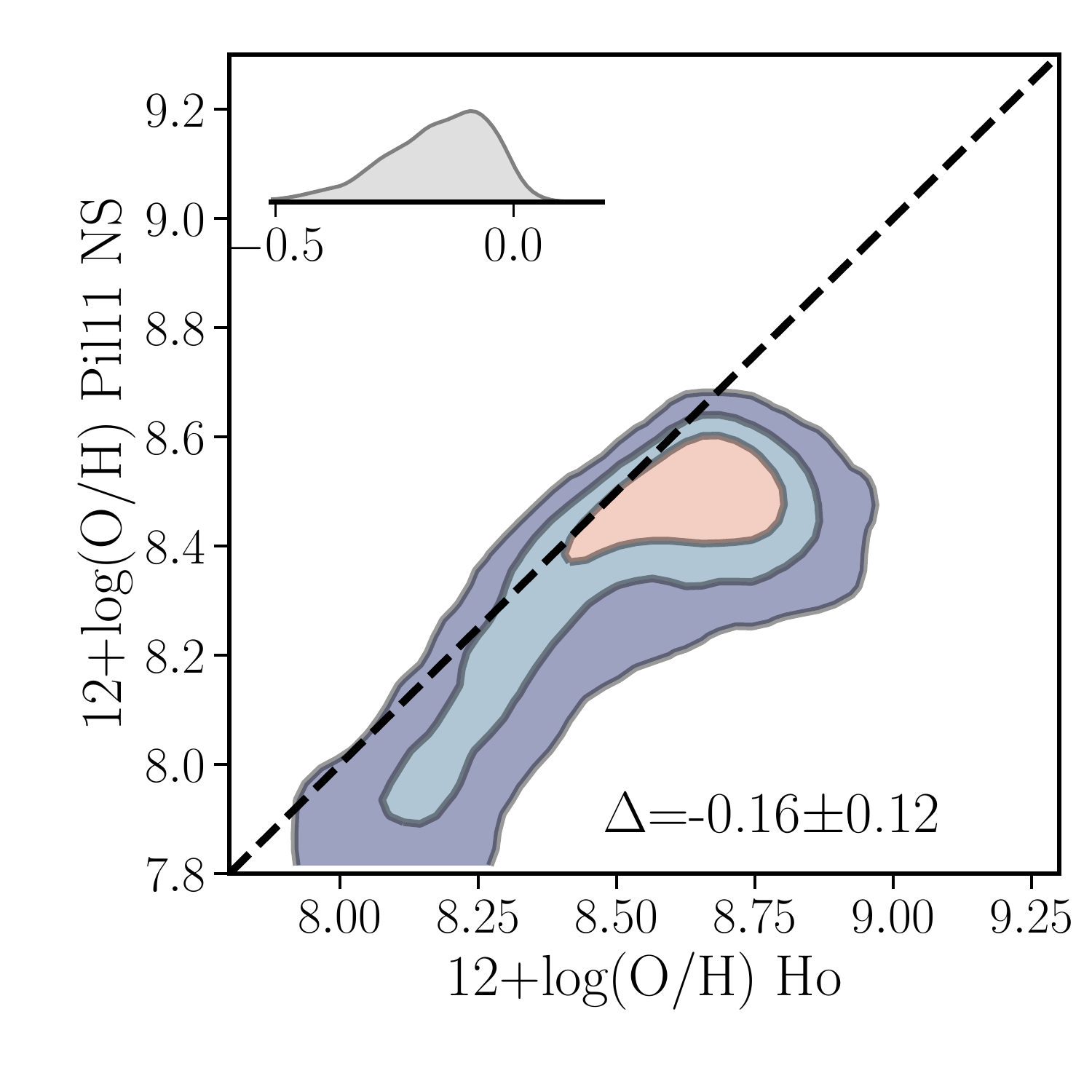}
\includegraphics[width=4.4cm,clip,trim=0 10 0 10]{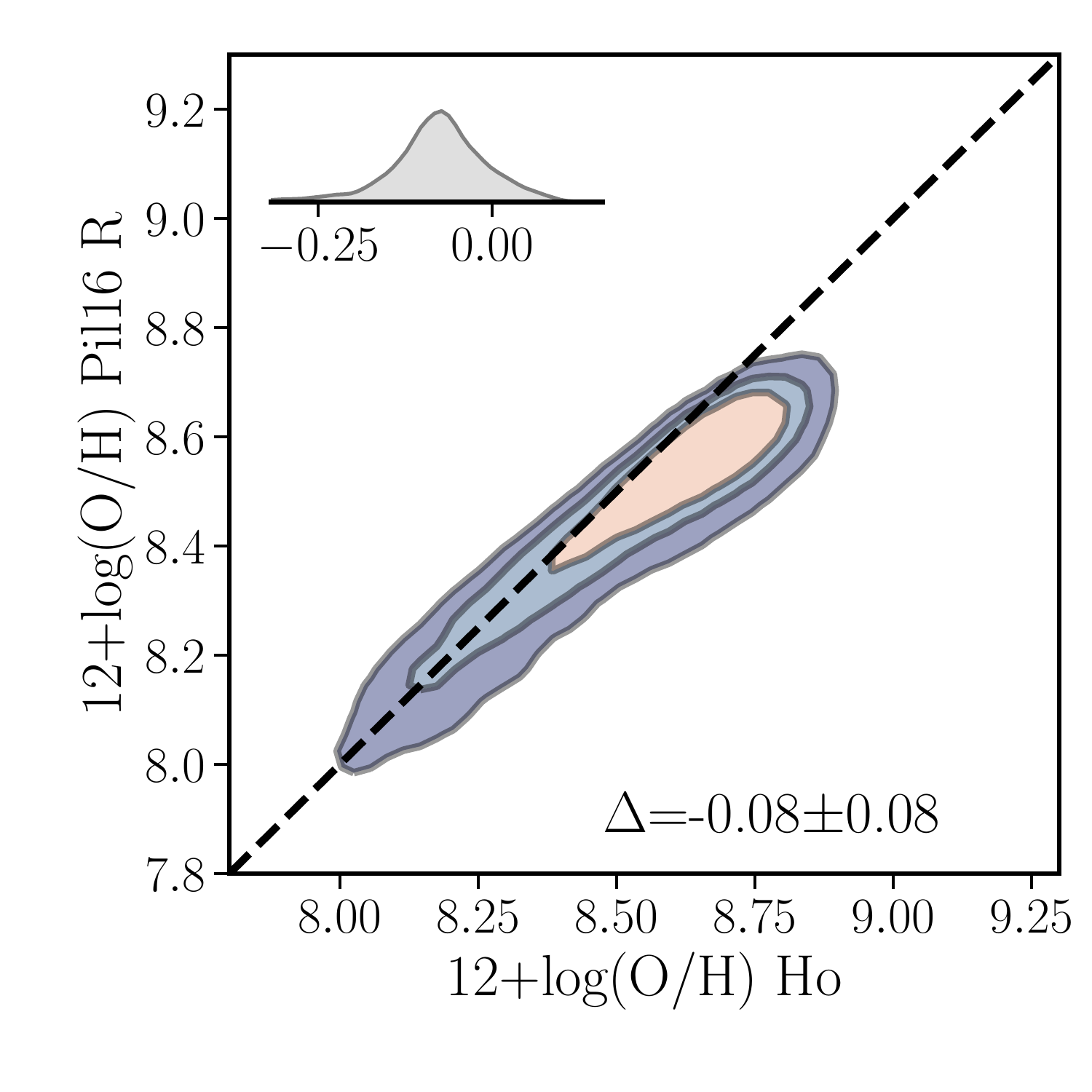}
\includegraphics[width=4.4cm,clip,trim=0 10 0 10]{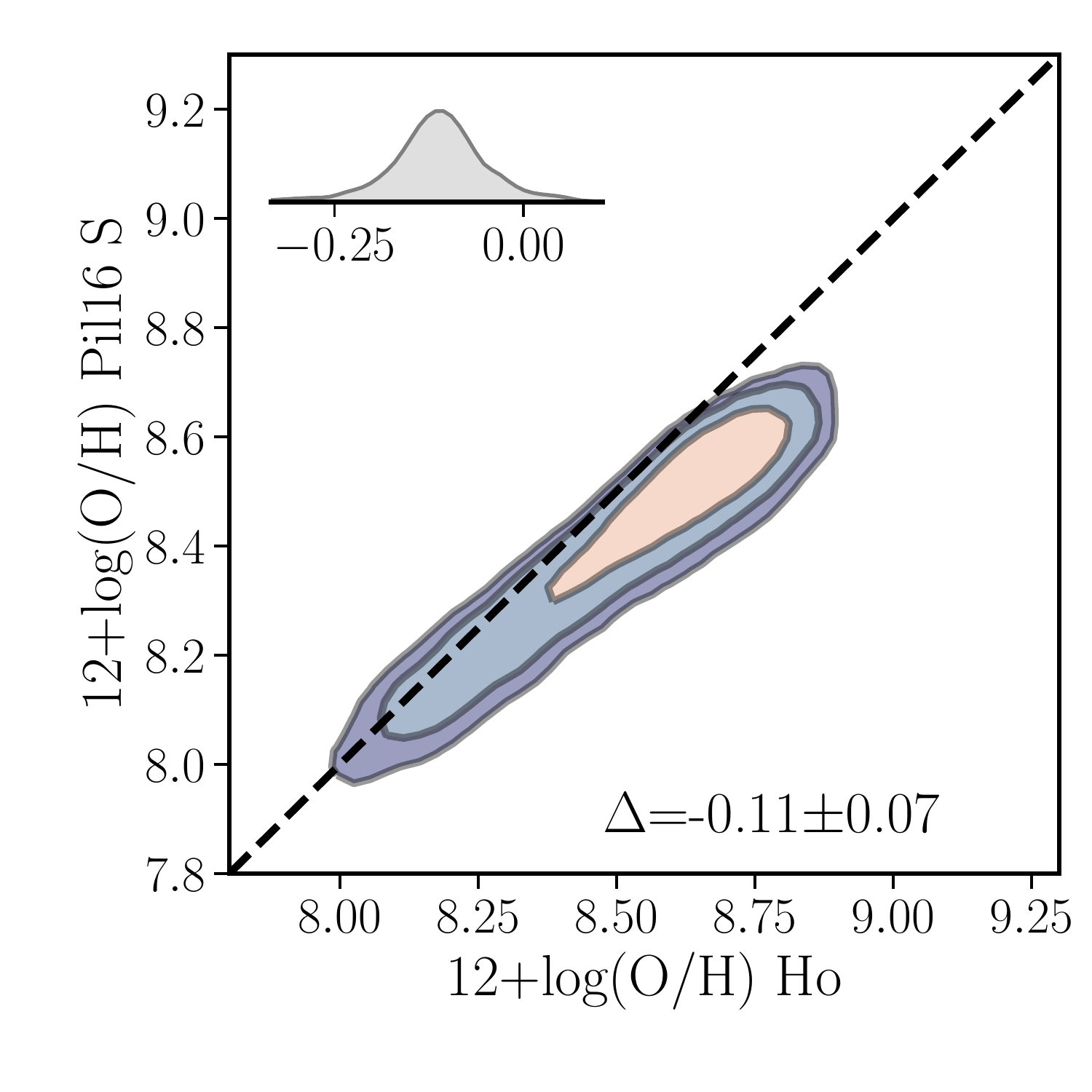}
 \endminipage
 \caption{Comparison between the oxygen abundances, 12+log(O/H), derived at the effective radius using different calibrators (one for each panel) listed in the final catalog as a function of the values derived using the \citet{ho19} one (adopted as fiducial one in 
 an arbitrary way). We adopt the same format as the one described in Fig. \ref{fig:comp_ssp_DR15} for the different panels in the figure.}
 \label{fig:comp_OH}
\end{figure*}

Figure \ref{fig:comp_OH} shows the comparison between the values derived using the complete list of different oxygen abundance calibrators included in the final catalog, as a function of those derived using the \citet{ho19} one, once excluded those already discussed in Sec. \ref{sec:cat_elines} (Fig. \ref{fig:comp_fewOH}).

%%%%%%%%%%%%%%%%%%%%%%%%%%%%%%%%%%%%%%%%%%%%%%%%%%%%%%%%%%%%%%%%%%%%%%%5
%%%%%%%%%%%%%%%%%%%%%%%%%%%%%%%%%%%%%%%%%%%%%%%%%%%%%%%%%%%%%%%%%%%%%%%5

%---------------------- MATCH_CUBES --------------------------------------------------- %
\begin{table}
\begin{center}
\caption{Repeated observations.}
\begin{tabular}{rrcrrc}
\hline\hline
\multicolumn{2}{c}{plate-ifudsgn} & Gal.& \multicolumn{2}{c}{plate-ifudsgn} & Gal.\\ 
\hline
10506-12702 & 10504-9102 & y  & 8953-9102 &  8479-6101 & n \\
11006-6103  &  8480-9102 & n  & 8953-9102 &  8480-6102  & n\\
8479-9101   & 11006-6103 & n  & 8983-3703 &  7495-12704 & y \\
8587-3703   &  8479-3702 & n  & 9029-12702& 11941-1901 & n\\
8587-3703   &  8480-3703 & n  & 9051-3701 &  8479-3702 & n\\
8587-3704   &  8479-3701 & n  & 9051-3701 &  8480-3703 & n \\
8587-3704   &  8480-3702 & n  & 9051-3702 &  8479-3701 & n \\
8587-6102   &  8479-6104 & n  & 9051-3702 &  8480-3702  & n\\
8587-6102   &  8480-6101 & n  & 9051-3704 &  8479-6101 & n \\
8587-6103   &  8479-6101 & n  & 9051-3704 &  8480-6102 & n \\
8587-6103   &  8480-6102 & n  & 9051-6104 &  8479-6102  & n\\
8587-6104   &  8479-6102 & n  & 9051-6104 &  8480-6103 & n \\
8587-6104   &  8480-6103 & n  & 9051-9102 &  8479-6104 & n \\
8952-12703  &  7495-12705& y  & 9051-9102 &  8480-6101 & n \\
8953-3701   &  8479-3701 & n  & 9194-12702&  9193-12702 & y\\
8953-3701   &  8480-3702 & n  & 9673-12703& 10142-12701 & n \\
8953-3703   &  8479-3702 & n  & 9674-12703& 10143-12701 & n \\
8953-3703   &  8480-3703 & n  & 9674-9102 & 10143-12705  & n\\
8953-6102   &  8479-6104 & n  & 9872-1901 &  7443-1901 & y\\
8953-6102   &  8480-6101 & n  & 9872-3702 &  7443-3701 & y\\
8953-6104   &  8479-6102 & n  & 9872-3703 &  7443-1902 & y\\
8953-6104   &  8480-6103 & n  & 9877-12704&  8479-12701& n\\
9877-12704  &  8480-12705 & n &          &             &  \\
\hline
\hline
\end{tabular}\label{tab:match}
\end{center}
\end{table}
%---------------------- MATCH_CUBES --------------------------------------------------- %

\section{Repeated Observations}
\label{app:rep}

Table \ref{tab:match} lists the 45 cubes that corresponds to repeated observations of the same object. Of them, only 7 corresponds to galaxies analyzed by \pyp. In one case (manga-9194-12702/manga-9194-12702) the offset between the two pointings is large enough to significantly affect the properties derived from the SDSS imaging data (e.g., $\Delta i\sim$0.1 mag). For the remaining six cases we can perform a direct comparison between the data and the parameters derived by our analysis. Despite the fact that the number statistics is very low, this comparison is very relevant since it is the best one available to estimate the real errors, not only on the physical quantities derived by \pyp, but in the absolute and relative spectrophotometric calibration of the data too.

For the comparison between the repeated observations we selected three parameters to explore the differences in the photometry: (i) the V-band magnitudes and (ii) B-R colors extracted from the MaNGA datacubes, and (iii) the stellar mass derived from this photometric values. In addition we select the same parameters that we adopted to compare between the DR15 and DR17 results (Sec. \ref{sec:comp_DR15}, Fig. \ref{fig:comp_ssp_DR15} and \ref{fig:comp_gas_DR15}), and with the DAP results (Sec. \ref{sec:comp_DAP}, Fig. \ref{fig:comp_ind_DAP}). Three additional parameters have been included for their relevance: (i) the oxygen abundance \citep[using the][calibrator]{ho19}, (ii) the nitrogen-to-oxygen relative abundance \citep[using the previous oxygen abundance and a calibrator for the nitrogen proposed by][]{pilyugin16} , and (iii) the ionization parameter \citep[using the calibrator by][]{mori16}, the three of them measured at the effective radius. Table \ref{tab:diff} lists the mean value and the standard deviation of the difference between the values reported in each repeated observation for the selected subset of parameters.

The agreement between the photometric parameters is very good, with mean offsets and standard deviations around them of the order of a few percent. This is the case of both the V-band magnitude and the B-R color. If these values were representative of the full dataset (which we cannot assure due to the very low number statistics), they would indicate an extremely accurate and precise absolute and relative (blue-to-red) spectrophotometric calibration. These results agree with the expectations and earlier reports on the quality of the MaNGA spectrophotometric calibration \citep[e.g.][]{renbin16, renbin16b}. These photometric errors are propagated through all the derived quantities irrespectively of the adopted analysis. For instance, in the case of the stellar mass derived using purely photometric information we found an offset of just $\sim$0.01 dex, but a scatter of $\sim$0.03 dex.

As expected the differences in the stellar properties derived from the decomposition of the stellar population using \pyp\ are larger than those found for the purely photometric quantities. The standard deviation of the difference for the stellar mass is of the order of a 10\%, which is the result of the propagation of the photometric error ($\sim$3\% in M$_{\star,phot}$) and the uncertainties in the derivation of the mass-to-light ratio ($\sim$7\%). Thus, the uncertainty introduced by the methodology is three times larger than the one introduced by just the photometric errors. Again, this result is in agreement with the expectations based on the comparison between DR15 and DR17 results discussed in Sec. \ref{sec:comp_DR15}. As a consequence the errors in the determination of the SFR based on the stellar population analysis are larger than the ones reported for M$_\star$, since this estimation relies on the same analysis but requires the estimation of the fraction of youngest stellar populations (what it is more imprecise). The differences in the estimation of the main properties of the stellar population (\ageLW, \metLW and A$_{\rm V,\star}$), are of the order of a few percent, ranging between a $\sim$3\% and a $\sim$10\%. These values agree with the expectations based on simulations for our code \citep{pipe3d,pypipe3d}. Finally, the differences in the kinematics parameters for the stellar populations are of the same order of the ones of the ones found when comparing the analysis for DR15 and DR17 datasets. Therefore, we consider that these errors are most probably associated to the ability to extract the corresponding kinematics information for this kind of data (i.e., signal-to-nose, spectral resolution) using the current methodology.

%---------------------- DIFF_MATCH --------------------------------------------------- %
\begin{table}
\begin{center}
\caption{Comparison between repeated observations.}
\begin{tabular}{llr}
\hline\hline
Parameter & Name & Difference \\
\hline
\multicolumn{3}{c}{Photometric parameters}\\
\hline
$V$-band mag     & {\tt V\_band\_mag}    & 0.035$\pm$0.037 mag\\
B-R color       & {\tt B-R}             & 0.030$\pm$0.015 mag\\
M$_{\star,phot}$& {\tt log\_Mass\_phot} & 0.011$\pm$0.026 dex\\
\hline
\multicolumn{3}{c}{Stellar population parameters}\\
\hline
M$_\star$&{\tt log\_Mass}        & 0.032$\pm$0.105 dex\\
$\Upsilon_\star$&{\tt ML\_avg}          & 0.052$\pm$0.067 dex\\
SFR$_\star$&{\tt log\_SFR\_ssp}    & 0.159$\pm$0.138 dex\\
\ageLW & {\tt Age\_LW\_Re\_fit} & -0.054$\pm$0.061 dex\\
\metLW & {\tt ZH\_LW\_Re\_fit}  & 0.044$\pm$0.026 dex\\
A$_{\rm V,\star}$ &{\tt Av\_ssp\_Re}      & 0.098$\pm$0.068 dex\\
$\sigma_{\star,cen}$&{\tt vel\_disp\_ssp\_cen} & -17.5$\pm$22.9 km\ s$^{-1}$\\
$v_{\star,2Re}$&{\tt vel\_ssp\_2} & -27.9$\pm$59.9 km\ s$^{-1}$\\
$\lambda_{Re}$ & {\tt Lambda\_Re}  & 0.134$\pm$0.166\\
\hline
\multicolumn{3}{c}{Emission line parameters}\\
\hline
H$\alpha$ flux &{\tt F\_Ha\_cen}  & -0.050$\pm$0.067 dex\\
EW(H$\alpha$)  &{\tt EW\_Ha\_cen} & -0.065$\pm$0.128 dex\\
H$\alpha$/H$\beta$ & {\tt Ha\_Hb\_cen} & 0.006$\pm$0.032 dex\\
$[$OIII$]$/H$\beta$ & {\tt log\_OIII\_Hb\_cen} & -0.036$\pm$0.083 dex\\
$[$NII$]$/H$\alpha$ & {\tt log\_NII\_Ha\_cen} & 0.020$\pm$0.053 dex\\
$[$SII$]$/H$\alpha$ & {\tt log\_SII\_Ha\_cen} & -0.016$\pm$0.040 dex\\
A$_{\rm V,gas}$ &{\tt Av\_gas\_Re} & -0.169$\pm$0.237 mag\\
SFR$_{H\alpha}$ & {\tt log\_SFR\_Ha} & -0.041$\pm$0.097 dex\\
M$_{\rm gas}$ &{\tt log\_Mass\_gas} & -0.003$\pm$0.008 dex\\
12+log(O/H) &{\tt OH\_Ho\_Re\_fit} & 0.018$\pm$0.024 dex\\
log(N/O)    & {\tt NO\_Pil16\_Ho\_R\_Re\_fit} & -0.004$\pm$0.022 dex\\
log(U)      & {\tt U\_Mor16\_O23\_fs\_Re\_fit} & -0.092$\pm$0.135 dex\\
\hline
\multicolumn{3}{c}{Stellar Indices}\\
\hline
Fe$_{5270}$ & {\tt Fe5270\_Re\_fit}&-0.035$\pm$0.356 \AA\\
Fe$_{5335}$ & {\tt Fe5335\_Re\_fit}&0.352$\pm$0.510 \AA\\
Mg$b$ & {\tt Mgb\_Re\_fit} & 0.115$\pm$0.350 \AA\ \\
H$_\delta$ &{\tt Hd\_Re\_fit} & -0.038$\pm$0.362 \AA\ \\
H$_\beta$& {\tt Hb\_Re\_fit} & 0.048$\pm$0.247 \AA\ \\
D4000 & {\tt D4000\_Re\_fit1} & -0.018$\pm$0.020\\
%{\tt nsa\_mstar} & -0.024$\pm$0.054\\
%{\tt g-r} NSA& -0.015$\pm$0.034\\
\hline
\hline
\end{tabular}\label{tab:diff}
\end{center}
\end{table}
%---------------------- MATCH_CUBES --------------------------------------------------- %

Once more, the emission line properties present differences that in general are larger than expected from the pure propagation of the photometric errors. In general extensive/integrated quantities, like the H$\alpha$ flux or the integrated SFR$_{H\alpha}$ present larger differences ($\sim$7-10\%) than intensive/relative ones, like the different line ratio ($\sim$3-8\%). Significant deviations form this picture are the results for the EW(H$\alpha$) and the A$_{\rm V,gas}$, which present large average differences and scatters ($\sim$10-13\%). Finally, the differences in the estimation of the oxygen and nitrogen-to-oxygen abundances are much lower than for the rest of the explored parameters, most probably due to the fact that they are derived based on a radial gradient analysis (Sec. \ref{sec:int}) and both parameters use to present well-defined radial gradients. On the other hand the ionization parameter presents the largest differences among the different explored parameters for the ionized gas ($\sim$10-14\%).

For the stellar indices we report differences of the order of those found when comparing the results between our analysis and the DAP (Sec. \ref{sec:comp_DAP}, Fig. \ref{fig:comp_ind_DAP}), although they are slightly lower. For instance for the H$\beta$ and Fe$_{5270}$ indices we report an scatter of $\sim$0.25-0.36\AA\ with a very small average difference ($\sim$0.05\AA), while the comparison with the DAP results in a scatter of $\sim$0.43-0.55\AA\ and a clear offset, at least for H$\beta$ ($\sim$0.21\AA). However, for other indices, like Fe$_{5335}$ we find a larger offset in the current comparison ($\sim$0.35\AA\ vs $\sim$0.13\AA) and a similar scatter ($\sim$0.5-0.6\AA). Due to the limited number of objects with repeated observations it is difficult to asset if the reported differences are really larger than the ones found when comparing with the DAP.

Despite of the low number statistics, we consider that this comparison is one of the best gauges of the uncertainties of the derived parameters. In general the results agree with the expectations and with previous comparisons, and they highlight the limitations and real uncertainties of the adopted methodology and the current dataset. We recommend that these uncertainties are considered in any further discussion based on the current analysis.

\section{Properties included in the catalog}
\label{app:cat}

Table \ref{tab:cat} lists the integrated, characteristics and aperture extracted properties included in the delivered the catalog of properties derived by the analysis described along this article. For each property it is included the column of the fits-table extension in which it is stored ($\#$ col.), the adopted name for the parameter, its units, and a short description of the delivered quantity.

\begin{table*}
\begin{center}
\caption{Integrated and characteristic parameters delivered for each analyzed datacube.}
\begin{tabular}{llll}
\hline\hline
\# col. & parameter name & units & Description  \\ 
\hline
0 & name & -- & manga-plate-ifudsgn unique name \\ 
1 & plate & -- & Plate ID of the MaNGA cube \\ 
2 & ifudsgn & -- & IFU bundle ID of the MaNGA cube \\ 
3 & plateifu & -- & Code formed by the plate and ifu names \\ 
4 & mangaid & -- & MaNGA ID \\ 
5 & objra & degree & RA of the object \\ 
6 & objdec & degree & DEC of the object \\ 
7 & log\_SFR\_H$\alpha$& log(M$_\odot$ yr$^{-1}$) & Integrated star-formation rate derived from the integrate H$\alpha$ luminosity \\ 
8 & FoV & -- & Ratio between the diagonal radius of the cube and Re \\
9 & Re\_kpc & kpc & Effective radius in kpc \\ 
10 & e\_log\_Mass & log(M$_\odot$) & Error of the integrated stellar mass { in logratihm-scale} \\
11 & e\_log\_SFR\_H$\alpha$& log(M$_\odot$ yr$^{-1}$) & Error of the Integrated star-formation rate derived from the H$\alpha$ luminosity \\
12 & log\_Mass & log(M$_\odot$) & Integrated stellar mass in units of the solar mass in logarithm scale \\ 
13 & log\_SFR\_ssp & log(M$_\odot$ yr$^{-1}$) & Integrated SFR derived from the SSP analysis t$<$32Myr \\ 
14 & log\_NII\_Ha\_cen & -- & Logarithm of the [NII]6583/H$\alpha$ line ratio in the central aperture \\ 
15 & e\_log\_NII\_Ha\_cen & -- & Error in the logarithm of the [NII]6583/H$\alpha$ line ratio \\ 
16 & log\_OIII\_Hb\_cen & -- & Logarithm of the [OIII]5007/H$\beta$ line ratio in the central aperture \\ 
17 & e\_log\_OIII\_Hb\_cen & -- & Error in the logarithm of the [OIII]5007/H$\beta$ line ratio  \\ 
18 & log\_SII\_Ha\_cen & -- & Logarithm of the [SII]6717+6731/H$\alpha$ line ratio in the central aperture \\
19 & e\_log\_SII\_Ha\_cen & -- & Error in the logarithm of the [SII]6717/H$\alpha$ line ratio  \\ 
20 & log\_OII\_Hb\_cen & -- & Logarithm of the [OII]3727/H$\beta$ line ratio in the central aperture\\
21 & e\_log\_OII\_Hb\_cen & -- & Error in the logarithm of the [OII]3727/H$\beta$ line ratio \\ 
22 & EW\_Ha\_cen & \AA\ & EW of H$\alpha$ in the central aperture \\ 
23 & e\_EW\_Ha\_cen & \AA\ & Error of the EW of H$\alpha$ in the central aperture \\ 
24 & ZH\_LW\_Re\_fit & dex & { LW metallicity of the stellar population at Re, normalized to the solar value, in logarithm scales} \\ 
25 & e\_ZH\_LW\_Re\_fit & dex & Error in the luminosity weighted metallicity of the stellar population \\ 
26 & alpha\_ZH\_LW\_Re\_fit & dex/Re & Slope of the gradient of the LW metallicity of the stellar population \\ 
27 & e\_alpha\_ZH\_LW\_Re\_fit & dex/Re & Error of the slope of the gradient of the LW log-metallicity \\ 
28 & ZH\_MW\_Re\_fit & dex & { MW metallicity of the stellar population at Re, normalized to the solar value, in logarithm scales} \\ 
29 & e\_ZH\_MW\_Re\_fit & dex & Error in the Mass weighted metallicity of the stellar population \\
30 & alpha\_ZH\_MW\_Re\_fit & dex/Re & Slope of the gradient of the MW log-metallicity of the stellar population \\ 
31 & e\_alpha\_ZH\_MW\_Re\_fit & dex/Re & Error of the slope of the gradient of the MW log-metallicity \\ 
32 & Age\_LW\_Re\_fit & log(yr) & Luminosity weighted age of the stellar population in logarithm scale \\ 
33 & e\_Age\_LW\_Re\_fit & log(yr) & Error in the luminosity weighted age of the stellar population \\ 
34 & alpha\_Age\_LW\_Re\_fit & log(yr)/Re & Slope of the gradient of the LW log-age of the stellar population \\ 
35 & e\_alpha\_Age\_LW\_Re\_fit & log(yr)/Re & Error of the slope of the gradient of the LW log-age of the stellar population\\ 
36 & Age\_MW\_Re\_fit & log(yr) & Mass weighted age of the stellar population in logarithm \\ 
37 & e\_Age\_MW\_Re\_fit & log(yr) & Error in the Mass weighted age of the stellar population \\ 
38 & alpha\_Age\_MW\_Re\_fit & { log(yr)}/Re & Slope of the gradient of the MW log-age of the stellar population \\ 
39 & e\_alpha\_Age\_MW\_Re\_fit & { log(yr)}/Re & Error of the slope of the gradient of the MW log-age of the stellar population \\ 
40 & Re\_arc & arcsec & Adopted effective radius in arcsec \\ 
41 & DL & { Mpc} & Adopted luminosity distance \\ 
42 & DA & { Mpc} & Adopted angular-diameter distance \\ 
43 & PA & degrees & Adopted position angle in degrees \\ 
44 & ellip & -- & Adopted ellipticity \\ 
45 & log\_Mass\_gas & log(M$_\odot$) & Integrated gas mass in units of the solar mass in logarithm scale \\ 
46 & vel\_sigma\_Re & -- & Velocity/dispersion ratio for the stellar populations within 1.5 Re \\ 
47 & e\_vel\_sigma\_Re & -- & Error in the velocity/dispersion ratio for the stellar population \\ 
48 & log\_SFR\_SF & log(M$_\odot$ yr$^{-1}$) & Integrated SFR using only the spaxels compatible with SF \\ 
49 & log\_SFR\_D\_C & log(M$_\odot$ yr$^{-1}$) & Integrated SFR diffuse corrected \\ 
\hline 
\end{tabular}\label{tab:cat}
\end{center}
\end{table*}

\addtocounter{table}{-1}

\begin{table*}
\begin{center}
\caption{Integrated and characteristic parameters {\it (continue})}
\begin{tabular}{llll}
\hline\hline
\# col. & parameter name & units & Description  \\ 
\hline
50 & OH\_O3N2\_cen & dex & Oxygen abundance using the calibrator O3N2 at the central aperture \\ 51 & e\_OH\_O3N2\_cen & dex & Error in Oxygen abundance using the calibrator O3N2 at the central aperture \\ 
52 & OH\_N2\_cen & dex & Oxygen abundance using the calibrator N2 at the central aperture \\ 53 & e\_OH\_N2\_cen & dex & Error in Oxygen abundance using the calibrator N2 at the central aperture \\ 
54 & OH\_ONS\_cen & dex & Oxygen abundance using the calibrator ONS at the central aperture \\ 55 & e\_OH\_ONS\_cen & dex & Error in Oxygen abundance using the calibrator ONS at the central aperture \\ 
56 & OH\_R23\_cen & dex & Oxygen abundance using the calibrator R23 at the central aperture \\ 57 & e\_OH\_R23\_cen & dex & Error in Oxygen abundance using the calibrator R23 at the central aperture \\ 
58 & OH\_pyqz\_cen & dex & Oxygen abundance using the calibrator pyqz at the central aperture \\ 59 & e\_OH\_pyqz\_cen & dex & Error in Oxygen abundance using the calibrator pyqz at the central aperture \\ 
60 & OH\_t2\_cen & dex & Oxygen abundance using the calibrator t2 at the central aperture \\ 61 & e\_OH\_t2\_cen & dex & Error in Oxygen abundance using the calibrator t2 at the central aperture \\ 
62 & OH\_M08\_cen & dex & Oxygen abundance using the calibrator M08 at the central aperture \\ 63 & e\_OH\_M08\_cen & dex & Error in Oxygen abundance using the calibrator M08 at the central aperture \\ 
64 & OH\_T04\_cen & dex & Oxygen abundance using the calibrator T04 at the central aperture \\ 65 & e\_OH\_T04\_cen & dex & Error in Oxygen abundance using the calibrator T04 at the central aperture \\ 
66 & OH\_dop\_cen & dex & Oxygen abundance using the calibrator dop at the central aperture \\ 67 & e\_OH\_dop\_cen & dex & Error in Oxygen abundance using the calibrator dop at the central aperture \\ 
68 & OH\_O3N2\_EPM09\_cen & dex & Oxygen abundance using the calibrator O3N2\_EPM09 at the central aperture \\ 69 & e\_OH\_O3N2\_EPM09\_cen & dex & Error in Oxygen abundance using the calibrator O3N2\_EPM09 at the central aperture \\ 
70 & log\_OI\_Ha\_cen & dex & Logarithm of the [OI]6301/H$\alpha$ within a central aperture  \\ 71 & e\_log\_OI\_Ha\_cen & dex & Error in the logarithm of the [OI]6301/H$\alpha$ within a central aperture  \\ 
72 & Ha\_Hb\_cen & -- & Ratio between the flux of Ha$\alpha$ and H$\beta$ within a central aperture \\ 73 & e\_Ha\_Hb\_cen & -- & Error of Ha\_Hb\_cen \\ 
74 & log\_NII\_Ha\_Re & dex & Logarithm of the [NII]6583/H$\alpha$ line ratio at 1Re \\ 75 & e\_log\_NII\_Ha\_Re & dex & Error in the logarithm of the [NII]6583/H$\alpha$ ratio at 1Re \\ 
76 & log\_OIII\_Hb\_Re & dex & Logarithm of the [OIII]5007/H$\beta$ line ratio at 1Re \\ 77 & e\_log\_OIII\_Hb\_Re & dex & Error in the logarithm of the [OIII]5007/H$\beta$ ratio at 1Re \\ 
78 & log\_SII\_Ha\_Re & dex & Logarithm of the [SII]6717+6731/H$\alpha$ ratio at 1Re \\ 79 & e\_log\_SII\_Ha\_Re & dex & Error in the logarithm of the [SII]6717/H$\alpha$ ratio at 1Re \\ 
80 & log\_OII\_Hb\_Re & dex & Logarithm of the [OII]3727/H$\beta$ ratio at 1Re \\ 81 & e\_log\_OII\_Hb\_Re & dex & Error in the logarithm of the [OII]3727/H$\beta$ at 1Re \\ 
82 & log\_OI\_Ha\_Re & dex & Logarithm of the [OI]6301/H$\alpha$ ratio at 1Re \\ 83 & e\_log\_OI\_Ha\_Re & dex & Error in the logarithm of the [OI]6301/H$\alpha$ ratio at 1Re \\ 
84 & EW\_Ha\_Re & \AA\ & EW of H$\alpha$ at 1Re \\ 85 & e\_EW\_Ha\_Re & \AA\ & Error of the EW of H$\alpha$ at 1Re \\ 
86 & Ha\_Hb\_Re & -- & Ratio between the flux of H$\alpha$ and H$\beta$ within at 1Re \\ 87 & e\_Ha\_Hb\_Re & -- & Error of Ha\_Hb\_Re \\ 
88 & log\_NII\_Ha\_ALL & dex & Logarithm of the [NII]6583/H$\alpha$ line ratio within the entire FoV \\ 89 & e\_log\_NII\_Ha\_ALL & dex & Error in the logarithm of the [NII]6583/$\alpha$ ratio within the entire FoV \\ 
90 & log\_OIII\_Hb\_ALL & dex & Logarithm of the [OIII]5007/H$\beta$ line ratio within the FoV \\ 91 & e\_log\_OIII\_Hb\_ALL & dex & Error in the logarithm of the [OIII]5007/H$\beta$ ratio within the FoV \\ 
92 & log\_SII\_Ha\_ALL & dex & Logarithm of the [SII]6717+6731/H$\alpha$ ratio within the entire FoV \\ 93 & e\_log\_SII\_Ha\_ALL & dex & Error in the logarithm of the [SII]6717/H$\alpha$ ratio within the FoV \\ 
94 & log\_OII\_Hb\_ALL & dex & Logarithm of the [OII]3727/H$\beta$ ratio within the entire FoV \\ 95 & e\_log\_OII\_Hb\_ALL & dex & Error in the logarithm of the [OII]3727/H$\beta$ ratio within the FoV \\ 
96 & log\_OI\_Ha\_ALL & dex & Logarithm of the [OI]6301/H$\alpha$ ratio within the entire FoV \\ 97 & e\_log\_OI\_Ha\_ALL & dex & Error in the logarithm of the [OI]6301/H$\beta$ ratio within the FoV \\ 
98 & EW\_Ha\_ALL & \AA\ & EW of H$\alpha$ within the entire FoV \\ 99 & e\_EW\_Ha\_ALL & \AA\ & Error of the EW of H$\alpha$ within the FoV \\ 
\hline 
\end{tabular}%\label{tab:cat}
\end{center}
\end{table*}

\addtocounter{table}{-1}

\begin{table*}
\begin{center}
\caption{Integrated and characteristic parameters {\it (continue})}
\begin{tabular}{llll}
\hline\hline
\# col. & parameter name & units & Description  \\ 
\hline
100 & Ha\_Hb\_ALL & -- & Ratio between the flux of H$\alpha$ and H$\beta$ within the entire FoV \\ 101 & Sigma\_Mass\_cen & log(M$_\odot$ pc$^{-2}$) & Stellar Mass surface density in the central aperture \\ 
102 & e\_Sigma\_Mass\_cen & log(M$_\odot$ pc$^{-2}$) & Error in Stellar Mass density in the central aperture \\ 103 & Sigma\_Mass\_Re & log(M$_\odot$ pc$^{-2}$) & Stellar Mass surface density at 1Re \\ 
104 & e\_Sigma\_Mass\_Re & log(M$_\odot$ pc$^{-2}$) & Error in Stellar Mass density at 1Re \\ 105 & Sigma\_Mass\_ALL & log(M$_\odot$ pc$^{-2}$) & Average Stellar Mass surface density within the entire FoV \\ 
106 & e\_Sigma\_Mass\_ALL & log(M$_\odot$ pc$^{-2}$) & Error in the average Stellar Mass surface density within the FoV \\ 107 & T30 & Gyr & Lookback time at which the galaxy has formed 30\% of its stellar mass \\ 
108 & ZH\_T30 & dex & Stellar metallicity, normalized by the solar value, in logarithm-scale at T30   time \\ 109 & ZH\_Re\_T30 & dex & Stellar metallicity, normalized by the solar value, in logarithm-scale at Re at T30   time \\ 
110 & a\_ZH\_T30 & dex/Re & Slope of the ZH gradient at T30   time \\ 111 & T40 & Gyr & Lookback time at which the galaxy has formed 40\% of its stellar mass \\ 
112 & ZH\_T40 & dex & Stellar metallicity, normalized by the solar value, in logarithm-scale at T40   time \\ 113 & ZH\_Re\_T40 & dex & Stellar metallicity, normalized by the solar value, in logarithm-scale at Re at T40   time \\ 
114 & a\_ZH\_T40 & dex/Re & Slope of the ZH gradient at T40   time \\ 115 & T50 & Gyr & Lookback time at which the galaxy has formed 50\% of its stellar mass \\ 
116 & ZH\_T50 & dex & Stellar metallicity, normalized by the solar value, in logarithm-scale at T50   time \\ 117 & ZH\_Re\_T50 & dex & Stellar metallicity, normalized by the solar value, in logarithm-scale at Re at T50   time \\ 
118 & a\_ZH\_T50 & dex/Re & Slope of the ZH gradient at T50   time \\ 119 & T60 & Gyr & Lookback time at which the galaxy has formed 60\% of its stellar mass \\ 
120 & ZH\_T60 & dex & Stellar metallicity, normalized by the solar value, in logarithm-scale at T60   time \\ 121 & ZH\_Re\_T60 & dex & Stellar metallicity, normalized by the solar value, in logarithm-scale at Re at T60   time \\ 
122 & a\_ZH\_T60 & dex/Re & Slope of the ZH gradient at T60   time \\ 123 & T70 & Gyr & Lookback time at which the galaxy has formed 70\% of its stellar mass \\ 
124 & ZH\_T70 & dex & Stellar metallicity, normalized by the solar value, in logarithm-scale at T70   time \\ 125 & ZH\_Re\_T70 & dex & Stellar metallicity, normalized by the solar value, in logarithm-scale at Re at T70   time \\ 
126 & a\_ZH\_T70 & dex/Re & Slope of the ZH gradient at T70   time \\ 127 & T80 & Gyr & Lookback time at which the galaxy has formed 80\% of its stellar mass \\ 
128 & ZH\_T80 & dex & Stellar metallicity, normalized by the solar value, in logarithm-scale at T80   time \\ 129 & ZH\_Re\_T80 & dex & Stellar metallicity, normalized by the solar value, in logarithm-scale at Re at T80   time \\ 
130 & a\_ZH\_T80 & dex/Re & Slope of the ZH gradient at T80   time \\ 131 & T90 & Gyr & Lookback time at which the galaxy has formed 90\% of its stellar mass \\ 
132 & ZH\_T90 & dex & Stellar metallicity, normalized by the solar value, in logarithm-scale at T90   time \\ 133 & ZH\_Re\_T90 & dex & Stellar metallicity, normalized by the solar value, in logarithm-scale at Re at T90   time \\ 
134 & a\_ZH\_T90 & dex/Re & Slope of the ZH gradient at T90   time \\ 135 & T95 & Gyr & Lookback time at which the galaxy has formed 95\% of its stellar mass \\ 
136 & ZH\_T95 & dex & Stellar metallicity, normalized by the solar value, in logarithm-scale at T95   time \\ 137 & ZH\_Re\_T95 & dex & Stellar metallicity, normalized by the solar value, in logarithm-scale at Re at T95   time \\ 
138 & a\_ZH\_T95 & dex/Re & Slope of the ZH gradient at T95   time \\ 139 & T99 & Gyr & Lookback time at which the galaxy has formed 99\% of its stellar mass \\ 
140 & ZH\_T99 & dex & Stellar metallicity, normalized by the solar value, in logarithm-scale at T99   time \\ 141 & ZH\_Re\_T99 & dex & Stellar metallicity, normalized by the solar value, in logarithm-scale at Re at T99   time \\ 
142 & a\_ZH\_T99 & dex/Re & Slope of the ZH gradient at T99   time \\ 143 & log\_Mass\_gas\_Av\_gas\_OH & log(M$_\odot$) & { Integrated gas mass using the calibrator log\_Mass\_gas\_Av\_gas\_OH, in logarithm-scale}\\ 
144 & log\_Mass\_gas\_Av\_ssp\_OH & log(M$_\odot$) & { Integrated gas mass using the calibrator log\_Mass\_gas\_Av\_ssp\_OH, in logarithm-scale} \\ 145 & vel\_ssp\_2 & km s$^{-1}$ & Stellar velocity at 2 Re \\ 
146 & e\_vel\_ssp\_2 & km/s & Error in vel\_ssp\_2 \\ 147 & vel\_Ha\_2 & km s$^{-1}$ & H$\alpha$ velocity at 2 Re \\ 
148 & e\_vel\_Ha\_2 & km/s & Error in vel\_Ha\_2 \\ 149 & vel\_ssp\_1 & km s$^{-1}$ & Stellar velocity at 1 Re \\ 
\hline 
\end{tabular}%\label{tab:cat}
\end{center}
\end{table*}

\addtocounter{table}{-1}

\begin{table*}
\begin{center}
\caption{Integrated and characteristic parameters{\it (continue})}
\begin{tabular}{llll}
\hline\hline
\# col. & parameter name& units & Description  \\ 
\hline
150 & e\_vel\_ssp\_1 & km/s & Error of vel\_ssp\_1 \\ 151 & vel\_Ha\_1 & km/s & H$\alpha$ velocity at 1 Re \\ 
152 & e\_vel\_Ha\_1 & km/s & Error of e\_vel\_Ha\_1 \\ 153 & log\_SFR\_ssp\_100Myr & log(M$_\odot$ yr$^{-1}$) & Integrated SFR derived from the SSP analysis for t$<$100Myr \\ 
154 & log\_SFR\_ssp\_10Myr & log(M$_\odot$ yr$^{-1}$) & Integrated SFR derived from the SSP analysis for t$<$10Myr \\ 155 & vel\_disp\_Ha\_cen & km/s & H$\alpha$ Velocity dispersion at the central aperture \\ 
156 & vel\_disp\_ssp\_cen & km/s & Stellar velocity dispersion at the central aperture \\ 157 & vel\_disp\_Ha\_1Re & km/s & H$\alpha$ velocity dispersion at 1 Re \\ 
158 & vel\_disp\_ssp\_1Re & km/s & Stellar velocity at 1 Re \\ 159 & log\_Mass\_in\_Re & log(M$_\odot$) & Integrated stellar mass within one optical Re \\ 
160 & ML\_int & M$_\odot$/L$_\odot$ & V-band mass-to-light ratio from integrated quantities \\ 161 & ML\_avg & M$_\odot$/L$_\odot$ & V-band mass-to-light ratio averaged across the FoV \\ 
162 & F\_Ha\_cen & 10$^{-16}$ \flux & Flux intensity of H$\alpha$ in the central aperture \\ 163 & e\_F\_Ha\_cen & 10$^{-16}$ \flux & Error of F\_Ha\_cen \\ 
164 & R50\_kpc\_V & kpc & { Radius at which it is integrated half of the light in the V-band within the FoV} \\ 165 & e\_R50\_kpc\_V & kpc & Error of R50\_kpc\_V \\ 
166 & R50\_kpc\_Mass & kpc & { Radius at which is integrated half of the stellar mass within the FoV } \\ 167 & e\_R50\_kpc\_Mass & kpc & Error of R50\_kpc\_Mass \\ 
168 & log\_Mass\_corr\_in\_R50\_V & log(M$_\odot$) & Integrated stellar mass within { R50\_kpc\_V} \\ 169 & e\_log\_Mass\_corr\_in\_R50\_V & log(M$_\odot$) & Error of log\_Mass\_corr\_in\_R50\_V \\ 
170 & log\_Mass\_gas\_Av\_gas\_log\_log & log(M$_\odot$) & Molecular gas mass derived from the dust extinction { using the log-log calibration} \\ 171 & Av\_gas\_Re & mag & Dust extinction in the V-band derived from the H$\alpha$/H$\beta$ line ratio { at 1 Re} \\ 
172 & e\_Av\_gas\_Re & mag & Error of Av\_gas\_Re\\ 173 & Av\_ssp\_Re & mag & Dust extinction in the V-band derived from the analysis of the stellar population \\ 
174 & e\_Av\_ssp\_Re & mag & Error of the Av\_ssp\_Re \\ 175 & Lambda\_Re & -- & Specific angular momentum ($\lambda$-parameter) for the stellar populations { within 1 Re} \\ 
176 & e\_Lambda\_Re & -- & { Error of the Lambda\_Re parameter} \\ 177 & nsa\_redshift & -- & -- \\ 
178 & nsa\_mstar &  log(M$_\odot$) & Stellar Mass derived by the NSA collaboration \\ 179 & nsa\_inclination & deg & Inclination derived by the NSA collaboration \\ 
  180 & flux\_[OII]3726.03\_Re\_fit & 10$^{-16}$ \flux  & Flux intensity of [OII]3726.03 at 1Re \\ 181 & e\_flux\_[OII]3726.03\_Re\_fit & 10$^{-16}$ \flux & Error in the flux intensity of [OII]3726.03 at 1Re \\ 
182 & flux\_[OII]3726.03\_alpha\_fit & 10$^{-16}$ \flux & Flux intensity of [OII]3726.03 at 1Re \\ 183 & e\_flux\_[OII]3726.03\_alpha\_fit & 10$^{-16}$ \flux & Error in the flux intensity of [OII]3726.03 at 1Re \\ 
184 & flux\_[OII]3728.82\_Re\_fit & 10$^{-16}$ \flux & Flux intensity of [OII]3728.82 at 1Re \\ 185 & e\_flux\_[OII]3728.82\_Re\_fit & 10$^{-16}$ \flux & Error in the flux intensity of [OII]3728.82 at 1Re \\ 
186 & flux\_[OII]3728.82\_alpha\_fit & 10$^{-16}$ \flux & Flux intensity of [OII]3728.82 at 1Re \\ 187 & e\_flux\_[OII]3728.82\_alpha\_fit & 10$^{-16}$ \flux & Error in the flux intensity of [OII]3728.82 at 1Re \\ 
188 & flux\_HI3734.37\_Re\_fit & 10$^{-16}$ \flux & Flux intensity of HI3734.37 at 1Re \\ 189 & e\_flux\_HI3734.37\_Re\_fit & 10$^{-16}$ \flux & Error in the flux intensity of HI3734.37 at 1Re \\ 
190 & flux\_HI3734.37\_alpha\_fit & 10$^{-16}$ \flux & Flux intensity of HI3734.37 at 1Re \\ 191 & e\_flux\_HI3734.37\_alpha\_fit & 10$^{-16}$ \flux & Error in the flux intensity of HI3734.37 at 1Re \\ 
192 & flux\_HI3797.9\_Re\_fit & 10$^{-16}$ \flux & Flux intensity of HI3797.9 at 1Re \\ 193 & e\_flux\_HI3797.9\_Re\_fit & 10$^{-16}$ \flux & Error in the flux intensity of HI3797.9 at 1Re \\ 
194 & flux\_HI3797.9\_alpha\_fit & 10$^{-16}$ \flux & Flux intensity of HI3797.9 at 1Re \\ 195 & e\_flux\_HI3797.9\_alpha\_fit & 10$^{-16}$ \flux & Error in the flux intensity of HI3797.9 at 1Re \\ 
196 & flux\_HeI3888.65\_Re\_fit & 10$^{-16}$ \flux & Flux intensity of HeI3888.65 at 1Re \\ 197 & e\_flux\_HeI3888.65\_Re\_fit & 10$^{-16}$ \flux & Error in the flux intensity of HeI3888.65 at 1Re \\ 
198 & flux\_HeI3888.65\_alpha\_fit & 10$^{-16}$ \flux & Flux intensity of HeI3888.65 at 1Re \\ \hline 
\end{tabular}%\label{tab:cat}
\end{center}
\end{table*}

\addtocounter{table}{-1}

\begin{table*}
\begin{center}
\caption{Integrated and characteristic parameters{\it (continue})}
\begin{tabular}{llll}
\hline\hline
\# col. & parameter name& units & Description  \\ 
\hline
199 & e\_flux\_HeI3888.65\_alpha\_fit & 10$^{-16}$ \flux & Error in the flux intensity of HeI3888.65 at 1Re \\ 
200 & flux\_HI3889.05\_Re\_fit & 10$^{-16}$ \flux & Flux intensity of HI3889.05 at 1Re \\ 201 & e\_flux\_HI3889.05\_Re\_fit & 10$^{-16}$ \flux & Error in the flux intensity of HI3889.05 at 1Re \\ 
202 & flux\_HI3889.05\_alpha\_fit & 10$^{-16}$ \flux & Flux intensity of HI3889.05 at 1Re \\ 203 & e\_flux\_HI3889.05\_alpha\_fit & 10$^{-16}$ \flux & Error in the flux intensity of HI3889.05 at 1Re \\ 
204 & flux\_HeI3964.73\_Re\_fit & 10$^{-16}$ \flux & Flux intensity of HeI3964.73 at 1Re \\ 205 & e\_flux\_HeI3964.73\_Re\_fit & 10$^{-16}$ \flux & Error in the flux intensity of HeI3964.73 at 1Re \\ 
206 & flux\_HeI3964.73\_alpha\_fit & 10$^{-16}$ \flux & Flux intensity of HeI3964.73 at 1Re \\ 207 & e\_flux\_HeI3964.73\_alpha\_fit & 10$^{-16}$ \flux & Error in the flux intensity of HeI3964.73 at 1Re \\ 
208 & flux\_[NeIII]3967.46\_Re\_fit & 10$^{-16}$ \flux & Flux intensity of [NeIII]3967.46 at 1Re \\ 209 & e\_flux\_[NeIII]3967.46\_Re\_fit & 10$^{-16}$ \flux & Error in the flux intensity of [NeIII]3967.46 at 1Re \\ 
210 & flux\_[NeIII]3967.46\_alpha\_fit & 10$^{-16}$ \flux & Flux intensity of [NeIII]3967.46 at 1Re \\ 211 & e\_flux\_[NeIII]3967.46\_alpha\_fit & 10$^{-16}$ \flux & Error in the flux intensity of [NeIII]3967.46 at 1Re \\ 
212 & flux\_CaII3968.47\_Re\_fit & 10$^{-16}$ \flux & Flux intensity of CaII3968.47 at 1Re \\ 213 & e\_flux\_CaII3968.47\_Re\_fit & 10$^{-16}$ \flux & Error in the flux intensity of CaII3968.47 at 1Re \\ 
214 & flux\_CaII3968.47\_alpha\_fit & 10$^{-16}$ \flux & Flux intensity of CaII3968.47 at 1Re \\ 215 & e\_flux\_CaII3968.47\_alpha\_fit & 10$^{-16}$ \flux & Error in the flux intensity of CaII3968.47 at 1Re \\ 
216 & flux\_Hepsilon3970.07\_Re\_fit & 10$^{-16}$ \flux & Flux intensity of H$\epsilon$3970.07 at 1Re \\ 217 & e\_flux\_Hepsilon3970.07\_Re\_fit & 10$^{-16}$ \flux & Error in the flux intensity of H$\epsilon$3970.07 at 1Re \\ 
218 & flux\_Hepsilon3970.07\_alpha\_fit & 10$^{-16}$ \flux & Flux intensity of H$\epsilon$3970.07 at 1Re \\ 219 & e\_flux\_Hepsilon3970.07\_alpha\_fit & 10$^{-16}$ \flux & Error in the flux intensity of H$\epsilon$3970.07 at 1Re \\ 
220 & flux\_Hdelta4101.77\_Re\_fit & 10$^{-16}$ \flux & Flux intensity of H$\delta$4101.77 at 1Re \\ 221 & e\_flux\_Hdelta4101.77\_Re\_fit & 10$^{-16}$ \flux & Error in the flux intensity of H$\delta$4101.77 at 1Re \\ 
222 & flux\_Hdelta4101.77\_alpha\_fit & 10$^{-16}$ \flux & Flux intensity of H$\delta$4101.77 at 1Re \\ 223 & e\_flux\_Hdelta4101.77\_alpha\_fit & 10$^{-16}$ \flux & Error in the flux intensity of H$\delta$4101.77 at 1Re \\ 
224 & flux\_Hgamma4340.49\_Re\_fit & 10$^{-16}$ \flux & Flux intensity of H$\gamma$4340.49 at 1Re \\ 225 & e\_flux\_Hgamma4340.49\_Re\_fit & 10$^{-16}$ \flux & Error in the flux intensity of H$\gamma$4340.49 at 1Re \\ 
226 & flux\_Hgamma4340.49\_alpha\_fit & 10$^{-16}$ \flux & Flux intensity of H$\gamma$4340.49 at 1Re \\ 227 & e\_flux\_Hgamma4340.49\_alpha\_fit & 10$^{-16}$ \flux & Error in the flux intensity of H$\gamma$4340.49 at 1Re \\ 
228 & flux\_Hbeta4861.36\_Re\_fit & 10$^{-16}$ \flux & Flux intensity of H$\beta$4861.36 at 1Re \\ 229 & e\_flux\_Hbeta4861.36\_Re\_fit & 10$^{-16}$ \flux & Error in the flux intensity of H$\beta$4861.36 at 1Re \\ 
230 & flux\_Hbeta4861.36\_alpha\_fit & 10$^{-16}$ \flux & Flux intensity of H$\beta$4861.36 at 1Re \\ 231 & e\_flux\_Hbeta4861.36\_alpha\_fit & 10$^{-16}$ \flux & Error in the flux intensity of H$\beta$4861.36 at 1Re \\ 
232 & flux\_[OIII]4958.91\_Re\_fit & 10$^{-16}$ \flux & Flux intensity of [OIII]4958.91 at 1Re \\ 233 & e\_flux\_[OIII]4958.91\_Re\_fit & 10$^{-16}$ \flux & Error in the flux intensity of [OIII]4958.91 at 1Re \\ 
234 & flux\_[OIII]4958.91\_alpha\_fit & 10$^{-16}$ \flux & Flux intensity of [OIII]4958.91 at 1Re \\ 235 & e\_flux\_[OIII]4958.91\_alpha\_fit & 10$^{-16}$ \flux & Error in the flux intensity of [OIII]4958.91 at 1Re \\ 
236 & flux\_[OIII]5006.84\_Re\_fit & 10$^{-16}$ \flux & Flux intensity of [OIII]5006.84 at 1Re \\ 237 & e\_flux\_[OIII]5006.84\_Re\_fit & 10$^{-16}$ \flux & Error in the flux intensity of [OIII]5006.84 at 1Re \\ 
238 & flux\_[OIII]5006.84\_alpha\_fit & 10$^{-16}$ \flux & Flux intensity of [OIII]5006.84 at 1Re \\ 239 & e\_flux\_[OIII]5006.84\_alpha\_fit & 10$^{-16}$ \flux & Error in the flux intensity of [OIII]5006.84 at 1Re \\ 
240 & flux\_HeI5015.68\_Re\_fit & 10$^{-16}$ \flux & Flux intensity of HeI5015.68 at 1Re \\ 241 & e\_flux\_HeI5015.68\_Re\_fit & 10$^{-16}$ \flux & Error in the flux intensity of HeI5015.68 at 1Re \\ 
242 & flux\_HeI5015.68\_alpha\_fit & 10$^{-16}$ \flux & Flux intensity of HeI5015.68 at 1Re \\ 243 & e\_flux\_HeI5015.68\_alpha\_fit & 10$^{-16}$ \flux & Error in the flux intensity of HeI5015.68 at 1Re \\ 
244 & flux\_[NI]5197.9\_Re\_fit & 10$^{-16}$ \flux & Flux intensity of [NI]5197.9 at 1Re \\ 245 & e\_flux\_[NI]5197.9\_Re\_fit & 10$^{-16}$ \flux & Error in the flux intensity of [NI]5197.9 at 1Re \\ 
246 & flux\_[NI]5197.9\_alpha\_fit & 10$^{-16}$ \flux & Flux intensity of [NI]5197.9 at 1Re \\ 247 & e\_flux\_[NI]5197.9\_alpha\_fit & 10$^{-16}$ \flux & Error in the flux intensity of [NI]5197.9 at 1Re \\ 
248 & flux\_[NI]5200.26\_Re\_fit & 10$^{-16}$ \flux & Flux intensity of [NI]5200.26 at 1Re \\ 249 & e\_flux\_[NI]5200.26\_Re\_fit & 10$^{-16}$ \flux & Error in the flux intensity of [NI]5200.26 at 1Re \\ 
\hline 
\end{tabular}%\label{tab:cat}
\end{center}
\end{table*}

\addtocounter{table}{-1}

\begin{table*}
\begin{center}
\caption{Integrated and characteristic parameters{\it (continue})}
\begin{tabular}{llll}
\hline\hline
\# col. & parameter name& units & Description  \\ 
\hline
250 & flux\_[NI]5200.26\_alpha\_fit & 10$^{-16}$ \flux & Flux intensity of [NI]5200.26 at 1Re \\ 251 & e\_flux\_[NI]5200.26\_alpha\_fit & 10$^{-16}$ \flux & Error in the flux intensity of [NI]5200.26 at 1Re \\ 
252 & flux\_HeI5876.0\_Re\_fit & 10$^{-16}$ \flux & Flux intensity of HeI5876.0 at 1Re \\ 253 & e\_flux\_HeI5876.0\_Re\_fit & 10$^{-16}$ \flux & Error in the flux intensity of HeI5876.0 at 1Re \\ 
254 & flux\_HeI5876.0\_alpha\_fit & 10$^{-16}$ \flux & Flux intensity of HeI5876.0 at 1Re \\ 255 & e\_flux\_HeI5876.0\_alpha\_fit & 10$^{-16}$ \flux & Error in the flux intensity of HeI5876.0 at 1Re \\ 
256 & flux\_NaI5889.95\_Re\_fit & 10$^{-16}$ \flux & Flux intensity of NaI5889.95 at 1Re \\ 257 & e\_flux\_NaI5889.95\_Re\_fit & 10$^{-16}$ \flux & Error in the flux intensity of NaI5889.95 at 1Re \\ 
258 & flux\_NaI5889.95\_alpha\_fit & 10$^{-16}$ \flux & Flux intensity of NaI5889.95 at 1Re \\ 259 & e\_flux\_NaI5889.95\_alpha\_fit & 10$^{-16}$ \flux & Error in the flux intensity of NaI5889.95 at 1Re \\ 
260 & flux\_NaI5895.92\_Re\_fit & 10$^{-16}$ \flux & Flux intensity of NaI5895.92 at 1Re \\ 261 & e\_flux\_NaI5895.92\_Re\_fit & 10$^{-16}$ \flux & Error in the flux intensity of NaI5895.92 at 1Re \\ 
262 & flux\_NaI5895.92\_alpha\_fit & 10$^{-16}$ \flux & Flux intensity of NaI5895.92 at 1Re \\ 263 & e\_flux\_NaI5895.92\_alpha\_fit & 10$^{-16}$ \flux & Error in the flux intensity of NaI5895.92 at 1Re \\ 
264 & flux\_[OI]6300.3\_Re\_fit & 10$^{-16}$ \flux & Flux intensity of [OI]6300.3 at 1Re \\ 265 & e\_flux\_[OI]6300.3\_Re\_fit & 10$^{-16}$ \flux & Error in the flux intensity of [OI]6300.3 at 1Re \\ 
266 & flux\_[OI]6300.3\_alpha\_fit & 10$^{-16}$ \flux & Flux intensity of [OI]6300.3 at 1Re \\ 267 & e\_flux\_[OI]6300.3\_alpha\_fit & 10$^{-16}$ \flux & Error in the flux intensity of [OI]6300.3 at 1Re \\ 
268 & flux\_[NII]6548.05\_Re\_fit & 10$^{-16}$ \flux & Flux intensity of [NII]6548.05 at 1Re \\ 269 & e\_flux\_[NII]6548.05\_Re\_fit & 10$^{-16}$ \flux & Error in the flux intensity of [NII]6548.05 at 1Re \\ 
270 & flux\_[NII]6548.05\_alpha\_fit & 10$^{-16}$ \flux & Flux intensity of [NII]6548.05 at 1Re \\ 271 & e\_flux\_[NII]6548.05\_alpha\_fit & 10$^{-16}$ \flux & Error in the flux intensity of [NII]6548.05 at 1Re \\ 
272 & flux\_Halpha6562.85\_Re\_fit & 10$^{-16}$ \flux & Flux intensity of H$\alpha$6562.85 at 1Re \\ 273 & e\_flux\_Halpha6562.85\_Re\_fit & 10$^{-16}$ \flux & Error in the flux intensity of H$\alpha$6562.85 at 1Re \\ 
274 & flux\_Halpha6562.85\_alpha\_fit & 10$^{-16}$ \flux & Flux intensity of H$\alpha$6562.85 at 1Re \\ 275 & e\_flux\_Halpha6562.85\_alpha\_fit & 10$^{-16}$ \flux & Error in the flux intensity of H$\alpha$6562.85 at 1Re \\ 
276 & flux\_[NII]6583.45\_Re\_fit & 10$^{-16}$ \flux & Flux intensity of [NII]6583.45 at 1Re \\ 277 & e\_flux\_[NII]6583.45\_Re\_fit & 10$^{-16}$ \flux & Error in the flux intensity of [NII]6583.45 at 1Re \\ 
278 & flux\_[NII]6583.45\_alpha\_fit & 10$^{-16}$ \flux & Flux intensity of [NII]6583.45 at 1Re \\ 279 & e\_flux\_[NII]6583.45\_alpha\_fit & 10$^{-16}$ \flux & Error in the flux intensity of [NII]6583.45 at 1Re \\ 
280 & flux\_[SII]6716.44\_Re\_fit & 10$^{-16}$ \flux & Flux intensity of [SII]6716.44 at 1Re \\ 281 & e\_flux\_[SII]6716.44\_Re\_fit & 10$^{-16}$ \flux & Error in the flux intensity of [SII]6716.44 at 1Re \\ 
282 & flux\_[SII]6716.44\_alpha\_fit & 10$^{-16}$ \flux & Flux intensity of [SII]6716.44 at 1Re \\ 283 & e\_flux\_[SII]6716.44\_alpha\_fit & 10$^{-16}$ \flux & Error in the flux intensity of [SII]6716.44 at 1Re \\ 
284 & flux\_[SII]6730.82\_Re\_fit & 10$^{-16}$ \flux & Flux intensity of [SII]6730.82 at 1Re \\ 285 & e\_flux\_[SII]6730.82\_Re\_fit & 10$^{-16}$ \flux & Error in the flux intensity of [SII]6730.82 at 1Re \\ 
286 & flux\_[SII]6730.82\_alpha\_fit & 10$^{-16}$ \flux & Flux intensity of [SII]6730.82 at 1Re \\ 287 & e\_flux\_[SII]6730.82\_alpha\_fit & 10$^{-16}$ \flux & Error in the flux intensity of [SII]6730.82 at 1Re \\ 
288 & flux\_[ArIII]7135.8\_Re\_fit & 10$^{-16}$ \flux & Flux intensity of [ArIII]7135.8 at 1Re \\ 289 & e\_flux\_[ArIII]7135.8\_Re\_fit & 10$^{-16}$ \flux & Error in the flux intensity of [ArIII]7135.8 at 1Re \\ 
290 & flux\_[ArIII]7135.8\_alpha\_fit & 10$^{-16}$ \flux & Flux intensity of [ArIII]7135.8 at 1Re \\ 291 & e\_flux\_[ArIII]7135.8\_alpha\_fit & 10$^{-16}$ \flux & Error in the flux intensity of [ArIII]7135.8 at 1Re \\ 
292 & flux\_HI9014.91\_Re\_fit & 10$^{-16}$ \flux & Flux intensity of HI9014.91 at 1Re \\ 293 & e\_flux\_HI9014.91\_Re\_fit & 10$^{-16}$ \flux & Error in the flux intensity of HI9014.91 at 1Re \\ 
294 & flux\_HI9014.91\_alpha\_fit & 10$^{-16}$ \flux & Flux intensity of HI9014.91 at 1Re \\ 295 & e\_flux\_HI9014.91\_alpha\_fit & 10$^{-16}$ \flux & Error in the flux intensity of HI9014.91 at 1Re \\ 
296 & flux\_[SIII]9069.0\_Re\_fit & 10$^{-16}$ \flux & Flux intensity of [SIII]9069.0 at 1Re \\ 297 & e\_flux\_[SIII]9069.0\_Re\_fit & 10$^{-16}$ \flux & Error in the flux intensity of [SIII]9069.0 at 1Re \\ 
298 & flux\_[SIII]9069.0\_alpha\_fit & 10$^{-16}$ \flux & Flux intensity of [SIII]9069.0 at 1Re \\ 299 & e\_flux\_[SIII]9069.0\_alpha\_fit & 10$^{-16}$ \flux & Error in the flux intensity of [SIII]9069.0 at 1Re \\ 
\hline 
\end{tabular}%\label{tab:cat}
\end{center}
\end{table*}

\addtocounter{table}{-1}

\begin{table*}
\begin{center}
\caption{Integrated and characteristic parameters{\it (continue})}
\begin{tabular}{llll}
\hline\hline
\# col. & parameter name& units & Description  \\ 
\hline
300 & flux\_[FeII]9470.93\_Re\_fit & 10$^{-16}$ \flux & Flux intensity of [FeII]9470.93 at 1Re \\ 301 & e\_flux\_[FeII]9470.93\_Re\_fit & 10$^{-16}$ \flux & Error in the flux intensity of [FeII]9470.93 at 1Re \\ 
302 & flux\_[FeII]9470.93\_alpha\_fit & 10$^{-16}$ \flux & Flux intensity of [FeII]9470.93 at 1Re \\ 303 & e\_flux\_[FeII]9470.93\_alpha\_fit & 10$^{-16}$ \flux & Error in the flux intensity of [FeII]9470.93 at 1Re \\ 
304 & flux\_[SIII]9531.1\_Re\_fit & 10$^{-16}$ \flux & Flux intensity of [SIII]9531.1 at 1Re \\ 305 & e\_flux\_[SIII]9531.1\_Re\_fit & 10$^{-16}$ \flux & Error in the flux intensity of [SIII]9531.1 at 1Re \\ 
306 & flux\_[SIII]9531.1\_alpha\_fit & 10$^{-16}$ \flux & Flux intensity of [SIII]9531.1 at 1Re \\ 307 & e\_flux\_[SIII]9531.1\_alpha\_fit & 10$^{-16}$ \flux & Error in the flux intensity of [SIII]9531.1 at 1Re \\ 
308 & OH\_Mar13\_N2\_Re\_fit \footnote{are derived based on a linear fit to the abundance gradient} & dex & Oxygen abundance using the calibrator Mar13\_N2 at 1Re \\ 309 & e\_OH\_Mar13\_N2\_Re\_fit & dex & Error in Oxygen abundance using the calibrator Mar13\_N2 at 1Re \\ 
310 & OH\_Mar13\_N2\_alpha\_fit & dex/Re & Slope of the O/H gradient using the calibrator Mar13\_N2 \\ 311 & e\_OH\_Mar13\_N2\_alpha\_fit & dex/Re & Error in the slope of the O/H gradient using the calibrator Mar13\_N2 \\ 
312 & OH\_Mar13\_O3N2\_Re\_fit & dex & Oxygen abundance using the calibrator Mar13\_O3N2 at 1Re \\ 313 & e\_OH\_Mar13\_O3N2\_Re\_fit & dex & Error in Oxygen abundance using the calibrator Mar13\_O3N2 at 1Re \\ 
314 & OH\_Mar13\_O3N2\_alpha\_fit & dex/Re & Slope of the O/H gradient using the calibrator Mar13\_O3N2 \\ 315 & e\_OH\_Mar13\_O3N2\_alpha\_fit & dex/Re & Error in the slope of the O/H gradient using the calibrator Mar13\_O3N2 \\ 
316 & OH\_T04\_Re\_fit & dex & Oxygen abundance using the calibrator T04 at 1Re \\ 317 & e\_OH\_T04\_Re\_fit & dex & Error in Oxygen abundance using the calibrator T04 at 1Re \\ 
318 & OH\_T04\_alpha\_fit & dex/Re & Slope of the O/H gradient using the calibrator T04 \\ 319 & e\_OH\_T04\_alpha\_fit & dex/Re & Error in the slope of the O/H gradient using the calibrator T04 \\ 
320 & OH\_Pet04\_N2\_lin\_Re\_fit & dex & Oxygen abundance using the calibrator Pet04\_N2\_lin at 1Re \\ 321 & e\_OH\_Pet04\_N2\_lin\_Re\_fit & dex & Error in Oxygen abundance using the calibrator Pet04\_N2\_lin at 1Re \\ 
322 & OH\_Pet04\_N2\_lin\_alpha\_fit & dex/Re & Slope of the O/H gradient using the calibrator Pet04\_N2\_lin \\ 323 & e\_OH\_Pet04\_N2\_lin\_alpha\_fit & dex/Re & Error in the slope of the O/H gradient using the calibrator Pet04\_N2\_lin \\ 
324 & OH\_Pet04\_N2\_poly\_Re\_fit & dex & Oxygen abundance using the calibrator Pet04\_N2\_poly at 1Re \\ 325 & e\_OH\_Pet04\_N2\_poly\_Re\_fit & dex & Error in Oxygen abundance using the calibrator Pet04\_N2\_poly at 1Re \\ 
326 & OH\_Pet04\_N2\_poly\_alpha\_fit & dex/Re & Slope of the O/H gradient using the calibrator Pet04\_N2\_poly \\ 327 & e\_OH\_Pet04\_N2\_poly\_alpha\_fit & dex/Re & Error in the slope of the O/H gradient using the calibrator Pet04\_N2\_poly \\ 
328 & OH\_Pet04\_O3N2\_Re\_fit & dex & Oxygen abundance using the calibrator Pet04\_O3N2 at 1Re \\ 329 & e\_OH\_Pet04\_O3N2\_Re\_fit & dex & Error in Oxygen abundance using the calibrator Pet04\_O3N2 at 1Re \\ 
330 & OH\_Pet04\_O3N2\_alpha\_fit & dex/Re & Slope of the O/H gradient using the calibrator Pet04\_O3N2 \\ 331 & e\_OH\_Pet04\_O3N2\_alpha\_fit & dex/Re & Error in the slope of the O/H gradient using the calibrator Pet04\_O3N2 \\ 
332 & OH\_Kew02\_N2O2\_Re\_fit & dex & Oxygen abundance using the calibrator Kew02\_N2O2 at 1Re \\ 333 & e\_OH\_Kew02\_N2O2\_Re\_fit & dex & Error in Oxygen abundance using the calibrator Kew02\_N2O2 at 1Re \\ 
334 & OH\_Kew02\_N2O2\_alpha\_fit & dex/Re & Slope of the O/H gradient using the calibrator Kew02\_N2O2 \\ 335 & e\_OH\_Kew02\_N2O2\_alpha\_fit & dex/Re & Error in the slope of the O/H gradient using the calibrator Kew02\_N2O2 \\ 
336 & OH\_Pil10\_ONS\_Re\_fit & dex & Oxygen abundance using the calibrator Pil10\_ONS at 1Re \\ 337 & e\_OH\_Pil10\_ONS\_Re\_fit & dex & Error in Oxygen abundance using the calibrator Pil10\_ONS at 1Re \\ 
338 & OH\_Pil10\_ONS\_alpha\_fit & dex/Re & Slope of the O/H gradient using the calibrator Pil10\_ONS \\ 339 & e\_OH\_Pil10\_ONS\_alpha\_fit & dex/Re & Error in the slope of the O/H gradient using the calibrator Pil10\_ONS \\ 
340 & OH\_Pil10\_ON\_Re\_fit & dex & Oxygen abundance using the calibrator Pil10\_ON at 1Re \\ 341 & e\_OH\_Pil10\_ON\_Re\_fit & dex & Error in Oxygen abundance using the calibrator Pil10\_ON at 1Re \\ 
342 & OH\_Pil10\_ON\_alpha\_fit & dex/Re & Slope of the O/H gradient using the calibrator Pil10\_ON \\ 343 & e\_OH\_Pil10\_ON\_alpha\_fit & dex/Re & Error in the slope of the O/H gradient using the calibrator Pil10\_ON \\ 
344 & OH\_Pil11\_NS\_Re\_fit & dex & Oxygen abundance using the calibrator Pil11\_NS at 1Re \\ 345 & e\_OH\_Pil11\_NS\_Re\_fit & dex & Error in Oxygen abundance using the calibrator Pil11\_NS at 1Re \\ 
346 & OH\_Pil11\_NS\_alpha\_fit & dex/Re & Slope of the O/H gradient using the calibrator Pil11\_NS \\ 347 & e\_OH\_Pil11\_NS\_alpha\_fit & dex/Re & Error in the slope of the O/H gradient using the calibrator Pil11\_NS \\ 
348 & OH\_Cur20\_RS32\_Re\_fit & dex & Oxygen abundance using the calibrator Cur20\_RS32 at 1Re \\ 349 & e\_OH\_Cur20\_RS32\_Re\_fit & dex & Error in Oxygen abundance using the calibrator Cur20\_RS32 at 1Re \\ 
\hline 
\end{tabular}%\label{tab:cat}
\end{center}
\end{table*}

\addtocounter{table}{-1}

\begin{table*}
\begin{center}
\caption{Integrated and characteristic parameters{\it (continue})}
\begin{tabular}{llll}
\hline\hline
\# col. & parameter name& units & Description  \\ 
\hline
350 & OH\_Cur20\_RS32\_alpha\_fit & dex/Re & Slope of the O/H gradient using the calibrator Cur20\_RS32 \\ 351 & e\_OH\_Cur20\_RS32\_alpha\_fit & dex/Re & Error in the slope of the O/H gradient using the calibrator Cur20\_RS32 \\ 
352 & OH\_Cur20\_R3\_Re\_fit & dex & Oxygen abundance using the calibrator Cur20\_R3 at 1Re \\ 353 & e\_OH\_Cur20\_R3\_Re\_fit & dex & Error in Oxygen abundance using the calibrator Cur20\_R3 at 1Re \\ 
354 & OH\_Cur20\_R3\_alpha\_fit & dex/Re & Slope of the O/H gradient using the calibrator Cur20\_R3 \\ 355 & e\_OH\_Cur20\_R3\_alpha\_fit & dex/Re & Error in the slope of the O/H gradient using the calibrator Cur20\_R3 \\ 
356 & OH\_Cur20\_O3O2\_Re\_fit & dex & Oxygen abundance using the calibrator Cur20\_O3O2 at 1Re \\ 357 & e\_OH\_Cur20\_O3O2\_Re\_fit & dex & Error in Oxygen abundance using the calibrator Cur20\_O3O2 at 1Re \\ 
358 & OH\_Cur20\_O3O2\_alpha\_fit & dex/Re & Slope of the O/H gradient using the calibrator Cur20\_O3O2 \\ 359 & e\_OH\_Cur20\_O3O2\_alpha\_fit & dex/Re & Error in the slope of the O/H gradient using the calibrator Cur20\_O3O2 \\ 
360 & OH\_Cur20\_S2\_Re\_fit & dex & Oxygen abundance using the calibrator Cur20\_S2 at 1Re \\ 361 & e\_OH\_Cur20\_S2\_Re\_fit & dex & Error in Oxygen abundance using the calibrator Cur20\_S2 at 1Re \\ 
362 & OH\_Cur20\_S2\_alpha\_fit & dex/Re & Slope of the O/H gradient using the calibrator Cur20\_S2 \\ 363 & e\_OH\_Cur20\_S2\_alpha\_fit & dex/Re & Error in the slope of the O/H gradient using the calibrator Cur20\_S2 \\ 
364 & OH\_Cur20\_R2\_Re\_fit & dex & Oxygen abundance using the calibrator Cur20\_R2 at 1Re \\ 365 & e\_OH\_Cur20\_R2\_Re\_fit & dex & Error in Oxygen abundance using the calibrator Cur20\_R2 at 1Re \\ 
366 & OH\_Cur20\_R2\_alpha\_fit & dex/Re & Slope of the O/H gradient using the calibrator Cur20\_R2 \\ 367 & e\_OH\_Cur20\_R2\_alpha\_fit & dex/Re & Error in the slope of the O/H gradient using the calibrator Cur20\_R2 \\ 
368 & OH\_Cur20\_N2\_Re\_fit & dex & Oxygen abundance using the calibrator Cur20\_N2 at 1Re \\ 369 & e\_OH\_Cur20\_N2\_Re\_fit & dex & Error in Oxygen abundance using the calibrator Cur20\_N2 at 1Re \\ 
370 & OH\_Cur20\_N2\_alpha\_fit & dex/Re & Slope of the O/H gradient using the calibrator Cur20\_N2 \\ 371 & e\_OH\_Cur20\_N2\_alpha\_fit & dex/Re & Error in the slope of the O/H gradient using the calibrator Cur20\_N2 \\ 
372 & OH\_Cur20\_R23\_Re\_fit & dex & Oxygen abundance using the calibrator Cur20\_R23 at 1Re \\ 373 & e\_OH\_Cur20\_R23\_Re\_fit & dex & Error in Oxygen abundance using the calibrator Cur20\_R23 at 1Re \\ 374 & OH\_Cur20\_R23\_alpha\_fit & dex/Re & Slope of the O/H gradient using the calibrator Cur20\_R23 \\ 375 & e\_OH\_Cur20\_R23\_alpha\_fit & dex/Re & Error in the slope of the O/H gradient using the calibrator Cur20\_R23 \\ 
376 & OH\_Cur20\_O3N2\_Re\_fit & dex & Oxygen abundance using the calibrator Cur20\_O3N2 at 1Re \\ 377 & e\_OH\_Cur20\_O3N2\_Re\_fit & dex & Error in Oxygen abundance using the calibrator Cur20\_O3N2 at 1Re \\ 
378 & OH\_Cur20\_O3N2\_alpha\_fit & dex/Re & Slope of the O/H gradient using the calibrator Cur20\_O3N2 \\ 379 & e\_OH\_Cur20\_O3N2\_alpha\_fit & dex/Re & Error in the slope of the O/H gradient using the calibrator Cur20\_O3N2 \\ 
380 & OH\_Cur20\_O3S2\_Re\_fit & dex & Oxygen abundance using the calibrator Cur20\_O3S2 at 1Re \\ 381 & e\_OH\_Cur20\_O3S2\_Re\_fit & dex & Error in Oxygen abundance using the calibrator Cur20\_O3S2 at 1Re \\ 
382 & OH\_Cur20\_O3S2\_alpha\_fit & dex/Re & Slope of the O/H gradient using the calibrator Cur20\_O3S2 \\ 383 & e\_OH\_Cur20\_O3S2\_alpha\_fit & dex/Re & Error in the slope of the O/H gradient using the calibrator Cur20\_O3S2 \\ 
384 & OH\_KK04\_Re\_fit & dex & Oxygen abundance using the calibrator KK04 at 1Re \\ 385 & e\_OH\_KK04\_Re\_fit & dex & Error in Oxygen abundance using the calibrator KK04 at 1Re \\ 
386 & OH\_KK04\_alpha\_fit & dex/Re & Slope of the O/H gradient using the calibrator KK04 \\ 387 & e\_OH\_KK04\_alpha\_fit & dex/Re & Error in the slope of the O/H gradient using the calibrator KK04 \\ 
388 & OH\_Pil16\_R\_Re\_fit & dex & Oxygen abundance using the calibrator Pil16\_R at 1Re \\ 389 & e\_OH\_Pil16\_R\_Re\_fit & dex & Error in Oxygen abundance using the calibrator Pil16\_R at 1Re \\ 
390 & OH\_Pil16\_R\_alpha\_fit & dex/Re & Slope of the O/H gradient using the calibrator Pil16\_R \\ 391 & e\_OH\_Pil16\_R\_alpha\_fit & dex/Re & Error in the slope of the O/H gradient using the calibrator Pil16\_R \\ 
392 & OH\_Pil16\_S\_Re\_fit & dex & Oxygen abundance using the calibrator Pil16\_S at 1Re \\ 393 & e\_OH\_Pil16\_S\_Re\_fit & dex & Error in Oxygen abundance using the calibrator Pil16\_S at 1Re \\ 
394 & OH\_Pil16\_S\_alpha\_fit & dex/Re & Slope of the O/H gradient using the calibrator Pil16\_S \\ 395 & e\_OH\_Pil16\_S\_alpha\_fit & dex/Re & Error in the slope of the O/H gradient using the calibrator Pil16\_S \\ 
396 & OH\_Ho\_Re\_fit & dex & Oxygen abundance using the calibrator Ho at 1Re \\ 397 & e\_OH\_Ho\_Re\_fit & dex & Error in Oxygen abundance using the calibrator Ho at 1Re \\ 
398 & OH\_Ho\_alpha\_fit & dex/Re & Slope of the O/H gradient using the calibrator Ho \\ 399 & e\_OH\_Ho\_alpha\_fit & dex/Re & Error in the slope of the O/H gradient using the calibrator Ho \\ 
\hline 
\end{tabular}%\label{tab:cat}
\end{center}
\end{table*}

\addtocounter{table}{-1}

\begin{table*}
\begin{center}
\caption{Integrated and characteristic parameters{\it (continue})}
\begin{tabular}{llll}
\hline\hline
\# col. & parameter name& units & Description  \\ 
\hline
400 & U\_Dors\_O32\_Re\_fit & dex &  log(U), ionization parameter using the calibrator Dors\_O32 at 1Re \\ 401 & e\_U\_Dors\_O32\_Re\_fit & dex & Error in log(U) using the calibrator Dors\_O32 at 1Re \\ 
402 & U\_Dors\_O32\_alpha\_fit & dex/Re & Slope of the log(U) gradient using the calibrator Dors\_O32 \\ 403 & e\_U\_Dors\_O32\_alpha\_fit & dex/Re & Error in the slope of the log(U) gradient using the calibrator Dors\_O32 \\ 
404 & U\_Dors\_S\_Re\_fit & dex & log(U) ionization parameter using the calibrator Dors\_S at 1Re \\ 405 & e\_U\_Dors\_S\_Re\_fit & dex & Error in log(U) using the calibrator Dors\_S at 1Re \\ 
406 & U\_Dors\_S\_alpha\_fit & dex/Re & Slope of the log(U) gradient using the calibrator Dors\_S \\ 407 & e\_U\_Dors\_S\_alpha\_fit & dex/Re & Error in the slope of the log(U) gradient using the calibrator Dors\_S \\ 
408 & U\_Mor16\_O23\_fs\_Re\_fit & dex &  log(U) ionization parameter using the calibrator Mor16\_O23\_fs at 1Re \\ 409 & e\_U\_Mor16\_O23\_fs\_Re\_fit & dex & Error in log(U) using the calibrator Mor16\_O23\_fs at 1Re \\ 
410 & U\_Mor16\_O23\_fs\_alpha\_fit & dex/Re & Slope of the log(U) gradient using the calibrator Mor16\_O23\_fs \\ 411 & e\_U\_Mor16\_O23\_fs\_alpha\_fit & dex/Re & Error in the slope of the log(U) gradient using the calibrator Mor16\_O23\_fs \\ 
412 & U\_Mor16\_O23\_ts\_Re\_fit & dex &  log(U) ionization parameter using the calibrator Mor16\_O23\_ts at 1Re \\ 413 & e\_U\_Mor16\_O23\_ts\_Re\_fit & dex & Error in log(U) using the calibrator Mor16\_O23\_ts at 1Re \\ 
414 & U\_Mor16\_O23\_ts\_alpha\_fit & dex/Re & Slope of the log(U) gradient using the calibrator Mor16\_O23\_ts \\ 415 & e\_U\_Mor16\_O23\_ts\_alpha\_fit & dex/Re & Error in the slope of the log(U) gradient using the calibrator Mor16\_O23\_ts \\ 
416 & NH\_Pil16\_R\_Re\_fit & dex & Nitrogen abundance using the calibrator Pil16\_R at 1Re \\ 417 & e\_NH\_Pil16\_R\_Re\_fit & dex & Error in Nitrogen abundance using the calibrator Pil16\_R at 1Re \\ 
418 & NH\_Pil16\_R\_alpha\_fit & dex/Re & Slope of the N/H gradient using the calibrator Pil16\_R \\ 419 & e\_NH\_Pil16\_R\_alpha\_fit & dex/Re & Error in the slope of the N/H gradient using the calibrator Pil16\_R \\ 
420 & NO\_Pil16\_R\_Re\_fit & dex & N/O abundance using the calibrator Pil16\_R at 1Re \\ 421 & e\_NO\_Pil16\_R\_Re\_fit & dex & Error in N/O abundance using the calibrator Pil16\_R at 1Re \\ 
422 & NO\_Pil16\_R\_alpha\_fit & dex/Re & Slope of the N/O gradient using the calibrator Pil16\_R \\ 423 & e\_NO\_Pil16\_R\_alpha\_fit & dex/Re & Error in the slope of the N/O gradient using the calibrator Pil16\_R \\ 
424 & NO\_Pil16\_Ho\_R\_Re\_fit & dex & N/O abundance using the calibrator Pil16\_Ho\_R at 1Re \\ 425 & e\_NO\_Pil16\_Ho\_R\_Re\_fit & dex & Error in N/O abundance using the calibrator Pil16\_Ho\_R at 1Re \\ 
426 & NO\_Pil16\_Ho\_R\_alpha\_fit & dex/Re & Slope of the N/O gradient using the calibrator Pil16\_Ho\_R \\ 427 & e\_NO\_Pil16\_Ho\_R\_alpha\_fit & dex/Re & Error in the slope of the N/O gradient using the calibrator Pil16\_Ho\_R \\ 
428 & NO\_Pil16\_N2\_R2\_Re\_fit & dex & N/O abundance using the calibrator Pil16\_N2\_R2 at 1Re \\ 429 & e\_NO\_Pil16\_N2\_R2\_Re\_fit & dex & Error in N/O abundance using the calibrator Pil16\_N2\_R2 at 1Re \\ 
430 & NO\_Pil16\_N2\_R2\_alpha\_fit & dex/Re & Slope of the N/O gradient using the calibrator Pil16\_N2\_R2 \\ 431 & e\_NO\_Pil16\_N2\_R2\_alpha\_fit & dex/Re & Error in the slope of the N/O gradient using the calibrator Pil16\_N2\_R2 \\ 
432 & Ne\_Oster\_S\_Re\_fit & dex & electron density using the Oster\_S estimator at 1Re \\ 433 & e\_Ne\_Oster\_S\_Re\_fit & dex & Error in n\_e using the Oster\_S estimator at 1Re \\ 
434 & Ne\_Oster\_S\_alpha\_fit & dex/Re & Slope of the n\_e gradient using the Oster\_S estimator \\ 435 & e\_Ne\_Oster\_S\_alpha\_fit & dex/Re & Error in the slope of n\_e gradient using the Oster\_S estimator \\ 
436 & Hd\_Re\_fit & \AA\ & Value of the Hd stellar index at 1Re \\ 437 & e\_Hd\_Re\_fit & \AA\ & Error of the Hd stellar index at 1Re \\ 
438 & Hd\_alpha\_fit & \AA/Re & Slope of the gradient of the Hd index \\ 439 & e\_Hd\_alpha\_fit & \AA/Re & Error of the slope of the gradient of the Hd index \\ 
440 & Hb\_Re\_fit & \AA\ & Value of the H$\beta$ stellar index at 1Re \\ 441 & e\_Hb\_Re\_fit & \AA\ & Error of the H$\beta$ stellar index at 1Re \\ 
442 & Hb\_alpha\_fit & \AA/Re & Slope of the gradient of the H$\beta$ index \\ 443 & e\_Hb\_alpha\_fit & \AA/Re & Error of the slope of the gradient of the H$\beta$index \\ 
444 & Mgb\_Re\_fit & \AA\ & Value of the Mgb stellar index at 1Re \\ 445 & e\_Mgb\_Re\_fit & \AA\ & Error of the Mgb stellar index at 1Re \\ 
446 & Mgb\_alpha\_fit & \AA/Re & Slope of the gradient of the Mgb index \\ 447 & e\_Mgb\_alpha\_fit & \AA/Re & Error of the slope of the gradient of the Mgb index \\ 
448 & Fe5270\_Re\_fit & \AA\ & Value of the Fe5270 stellar index at 1Re \\ 449 & e\_Fe5270\_Re\_fit & \AA\ & Error of the Fe5270 stellar index at 1Re \\ 
\hline 
\end{tabular}%\label{tab:cat}
\end{center}
\end{table*}

\addtocounter{table}{-1}

\begin{table*}
\begin{center}
\caption{Integrated and characteristic parameters{\it (continue})}
\begin{tabular}{llll}
\hline\hline
\# col. & parameter name& units & Description  \\ 
\hline
450 & Fe5270\_alpha\_fit & \AA/Re & Slope of the gradient of the Fe5270 index \\ 451 & e\_Fe5270\_alpha\_fit & \AA/Re & Error of the slope of the gradient of the Fe5270 index \\ 
452 & Fe5335\_Re\_fit & \AA\ & Value of the Fe5335 stellar index at 1Re \\ 453 & e\_Fe5335\_Re\_fit & \AA\ & Error of the Fe5335 stellar index at 1Re \\ 
454 & Fe5335\_alpha\_fit & \AA/Re & Slope of the gradient of the Fe5335 index \\ 455 & e\_Fe5335\_alpha\_fit & \AA/Re & Error of the slope of the gradient of the Fe5335 index \\ 
456 & D4000\_Re\_fit1 & -- & Value of the D4000 stellar index at 1Re \\ 457 & e\_D4000\_Re\_fit & -- & Error of the D4000 stellar index at 1Re \\ 
458 & D4000\_alpha\_fit & -- & Slope of the gradient of the D4000 index \\ 459 & e\_D4000\_alpha\_fit & -- & Error of the slope of the gradient of the D4000 index \\ 
460 & Hdmod\_Re\_fit & \AA\ & Value of the Hdmod stellar index at 1Re \\ 461 & e\_Hdmod\_Re\_fit & \AA\ & Error of the Hdmod stellar index at 1Re \\ 
462 & Hdmod\_alpha\_fit & \AA/Re & Slope of the gradient of the Hdmod index \\ 463 & e\_Hdmod\_alpha\_fit & \AA/Re & Error of the slope of the gradient of the Hdmod index \\ 
464 & Hg\_Re\_fit & \AA\ & Value of the Hg stellar index at 1Re \\ 465 & e\_Hg\_Re\_fit & \AA\ & Error of the Hg stellar index at 1Re \\ 
466 & Hg\_alpha\_fit & \AA/Re & Slope of the gradient of the Hg index \\ 467 & e\_Hg\_alpha\_fit & \AA/Re & Error of the slope of the gradient of the Hg index \\ 
468 & u\_band\_mag & mag & u-band { magnitude derived from the original cube} \\ 469 & u\_band\_mag\_error & mag & error in  u-band { magnitude derived from the original cube} \\ 
470 & u\_band\_abs\_mag & mag & u-band abs. { magnitude derived from the original cube} \\ 471 & u\_band\_abs\_mag\_error & mag & error in u-band { magnitude derived from the original cube} \\ 
472 & g\_band\_mag & mag & g-band { magnitude derived from the original cube} \\ 473 & g\_band\_mag\_error & mag & error in g-band { magnitude derived from the original cube} \\ 
474 & g\_band\_abs\_mag & mag & g-band abs. { magnitude derived from the original cube} \\ 475 & g\_band\_abs\_mag\_error & mag & error in g-band abs. { magnitude derived from the original cube} \\ 
476 & r\_band\_mag & mag & r-band { magnitude derived from the original cube} \\ 477 & r\_band\_mag\_error & mag & error in r-band { magnitude derived from the original cube} \\ 
478 & r\_band\_abs\_mag & mag & r-band abs. { magnitude derived from the original cube} \\ 479 & r\_band\_abs\_mag\_error & mag & error in r-band { magnitude derived from the original cube} \\ 
480 & i\_band\_mag & mag & i-band { magnitude derived from the original cube} \\ 481 & i\_band\_mag\_error & mag & error in i-band { magnitude derived from the original cube} \\ 
482 & i\_band\_abs\_mag & mag & i-band abs. { magnitude derived from the original cube} \\ 483 & i\_band\_abs\_mag\_error & mag & error in i-band { magnitude derived from the original cube} \\ 
484 & B\_band\_mag & mag & B-band { magnitude derived from the original cube} \\ 485 & B\_band\_mag\_error & mag & error in B-band { magnitude derived from the original cube} \\ 
486 & B\_band\_abs\_mag & mag & B-band abs. { magnitude derived from the original cube} \\ 487 & B\_band\_abs\_mag\_error & mag & error in B-band { magnitude derived from the original cube} \\ 
488 & V\_band\_mag & mag & V-band { magnitude derived from the original cube} \\ 489 & V\_band\_mag\_error & mag & error in V-band { magnitude derived from the original cube} \\ 
490 & V\_band\_abs\_mag & mag & V-band abs. { magnitude derived from the original cube} \\ 491 & V\_band\_abs\_mag\_error & mag & error in V-band { magnitude derived from the original cube} \\ 
492 & RJ\_band\_mag & mag & R-band { magnitude derived from the original cube} \\ 493 & RJ\_band\_mag\_error & mag & error R-band { magnitude derived from the original cube} \\ 
494 & RJ\_band\_abs\_mag & mag & R-band abs. { magnitude derived from the original cube} \\ 495 & RJ\_band\_abs\_mag\_error & mag & error in R-band abs. { magnitude derived from the original cube} \\ 
496 & R50 & arcsec & { Radius at which it is integrated half of the light within the FoV in the g-band} \\ 497 & error\_R50 & arcsec & Error in R50 \\ 
498 & R90 & arcsec & { Radius at which it is integrated 90\%\ of the light within the FoV in the g-band} \\ 499 & error\_R90 & arcsec & error in R90 \\ 
\hline 
\end{tabular}%\label{tab:cat}
\end{center}
\end{table*}

\addtocounter{table}{-1}

\begin{table*}
\begin{center}
\caption{Integrated and characteristic parameters{\it (continue})}
\begin{tabular}{llll}
\hline\hline
\# col. & parameter name& units & Description  \\ 
\hline
500 & C & -- & R90/R50 concentration index \\ 501 & e\_C & -- & error in the concentration index \\ 
502 & B-V & mag & B-V color \\ 503 & error\_B-V & mag & error in the B-V color \\ 
504 & B-R & mag & B-R color \\ 505 & error\_B-R & mag & error in the B-R color \\ 
506 & log\_Mass\_phot & log(M$_\odot$) & stellar masses derived from { the photometric data within the FoV} \\ 507 & e\_log\_Mass\_phot & dex & error in the stellar masses derived from { the photometric data} \\ 
508 & V-band\_SB\_at\_Re & mag/arcsec$^2$ & V-band surface brightness at 1Re \\ 509 & error\_V-band\_SB\_at\_Re & mag/arcsec$^2$ & Error in the V-band { surface brightness} at 1Re \\ 
510 & V-band\_SB\_at\_R\_50 & mag/arcsec$^2$ & V-band surface brightness at R50 \\ 511 & error\_V-band\_SB\_at\_R\_50 & mag/arcsec$^2$ & Error in V-band surface brightness at R50 \\ 
512 & nsa\_sersic\_n\_morph & -- & NSA sersic index \\ 513 & u-g & mag & u-g NSA color \\ 
514 & g-r & mag & g-r NSA color \\ 515 & r-i & mag & r-i NSA color \\ 
516 & i-z & mag & i-z NSA color \\ 517 & P(CD) & -- & Probability of being a CD galaxy \\ 
518 & P(E) & -- & Probability of being a E galaxy \\ 519 & P(S0) & -- & Probability of being a S0 galaxy \\ 
520 & P(Sa) & -- & Probability of being a Sa galaxy \\ 521 & P(Sab) & -- & Probability of being a Sab galaxy \\ 
522 & P(Sb) & -- & Probability of being a Sb galaxy \\ 523 & P(Sbc) & -- & Probability of being a Sbc galaxy \\ 
524 & P(Sc) & -- & Probability of being a Sc galaxy \\ 525 & P(Scd) & -- & Probability of being a Scd galaxy \\ 
526 & P(Sd) & -- & Probability of being a Sd galaxy \\ 527 & P(Sdm) & -- & Probability of being a Sdm galaxy \\ 
528 & P(Sm) & -- & Probability of being a Sm galaxy \\ 529 & P(Irr) & -- & Probability of being a Irr galaxy \\ 
530 & best\_type\_n & -- & Best morphological type index based on the NN analysis \\ 531 & best\_type & -- & Morphological Type derived by the NN analysis \\ 
532 & nsa\_nsaid & -- & NSA ID \\ 533 & Vmax\_w & Mpc$^{-3}$ dex$^{-1}$ & Weight for the volume correction in volume \\ 
534 & Num\_w & -- & Weight of the volume correction in number \\ 535 & QCFLAG & -- & QC flat 0=good 2=bad $>$2 warning \\ 
\hline 
\end{tabular}%\label{tab:cat}
\end{center}
\end{table*}

%\include{SDSS17Pipe3D_v3_1_1.desc}

%%%%%%%%%%%%%%%%%%%%%%%%%%%%%%%%%%%%%%%%%%%%%%%%%%%%%%%%%%%%%%%%%%%
%%%%%%%%%%%%%%%%%  APPENDIX END %%%%%%%%%%%%%%%%%%%%%%%%%%%%%%%%%%%
%%%%%%%%%%%%%%%%%%%%%%%%%%%%%%%%%%%%%%%%%%%%%%%%%%%%%%%%%%%%%%%%%%%

\end{document}

%% file: AppendixSSP.tex
\section{The C\&B SSP models}\label{app:ssp}
 
A major revision of the BC03 stellar population synthesis models was introduced in \citet[][hereafter \cb models]{plat2019}.
Even though the \cb models have been used by several authors
\citep[e.g., ][]{senchyna2022,senchyna2021,werle2022,werle2020,orozco2022,gonz2021,mayya2020},
a detailed description of the ingredients entering these models is lacking in the literature.
For completeness and for the benefit of the reader we summarize in this appendix the characteristics of these \cb models.

The \cb models follow the PARSEC evolutionary tracks \citep{marigo2013,chen2015} for the 16 chemical compositions listed in Table~\ref{tab:parsec}.
The initial solar nebula had ${\rm Z}\,=\,0.014$ and the current sun has a surface abundance of ${\rm Z}\,=\,0.017$.
These state of the art evolutionary tracks follow the evolution of stars of mass from 0.1 to 600 M$_\odot$ using a fine grid of mass and time steps.
The tracks run from the main sequence to the Wolf-Rayet (WR) phase for massive stars and up to the thermally-pulsing asymptotic giant branch (TP-AGB) for stars
below 6 M$_\odot$. In the \cb models we follow the evolution of the post-asymptotic giant (pAGB) phase of intermediate and low mass stars according to \citet{m3b2016}.
A major source of uncertainty in building these models is the lack of spectra, either theoretical or empirical, for stars of all the metallicities listed in Table~\ref{tab:parsec} at
all evolutionary phases. For illustration purposes, the check marks in Table~\ref{tab:available} indicate the nominal value of [Z/Z$_\odot$] for which the different theoretical atlases
listed in the table have been computed. To build sensible population synthesis models it is then necessary to make compromises on what stellar spectra to use for each set of tracks.

Table~\ref{tab:assigned} indicates the value of [Z/Z$_\odot$] of the theoretical atlases or the range of [Z/Z$_\odot$] for the stellar libraries used to build each population synthesis model.
In Table~\ref{tab:spprop} we list the spectral properties of the different theoretical models used in the UV spectral range. Finally, Table~\ref{tab:xmilesi} indicates the stellar atlas(es) used in the population synthesis models vs. wavelength range.
From Table~\ref{tab:xmilesi} we see that in the visible range the standard \cb models use the MILES library from 3540.5 to 7350.2 \AA\ and the IndoUS library from 7350.2 to 9399.8 \AA.
A version of these models was computed using the Stelib library for comparison with the BC03 models.
The \cb models were computed for the \citet{chab03}, \citet{salpeter55} and \citet{Kroupa2001} IMFs for M$_{UP} =100$, 300 and 600 M$_\odot$.
Each simple stellar population (SSP) model provides 220 spectra computed at different time steps ranging from 0 to 14 Gyr. 

\subsection{The MaStar variation of the {\rm C\&B} models}

We use the first release of the MaNGA stellar library \citep[MaStar,][]{yan19} to build a test set of the \cb SSP models. This release of the MaStar library contains 8646 spectra of 3321 unique stars. The spectra cover the wavelength range from 3622 to 10354 \AA\ at a resolving power of $R \approx 1800$. Due to the lack of a uniform calibration of the astrophysical parameters for the stars in this library, we opted for a quick and simple manner to assign a MaStar spectrum to each star in the synthesis model.
For each MILES spectrum used in our spectral synthesis we searched for the closest matching spectrum (in log flux) in the MaStar library.
We then replaced the MILES and the IndoUS spectra (cf. lines 3 and 4 in Table~\ref{tab:xmilesi}) by the selected MaStar spectrum in the full range covered by the latter. 
We checked that there were no systematic differences between the colors and the line strength indices computed for both sets of models as a function of time.
%The MaStar SSP models are available for all the metallicities listed in
The MaStar SSP models were computed for all the metallicities listed in Table~\ref{tab:parsec}.
As its MILES counterpart, each MaStar SSP model contains 220 spectra at time steps from 0 to 14 Gyr. The sub-set of the MaStar C\&B templates adopted along this article in the {\tt pyPipe3D} format, are accesible through the web \footnote{\url{http://ifs.astroscu.unam.mx/pyPipe3D/templates/}}.

Recently \citet{mejia21} have determined the stellar atmospheric parameters for the latest release of MaStar, comprising over $22\,$k unique stars. 
At the moment we are working on the implementation of this fully calibrated MaStar library in the \cb models. 

% Table~\ref{parsec} lists the chemical composition of the different sets of PARSEC evolutionary tracks that have been used to build the CB19 spectral synthesis models. 
% Each set of tracks will be referred by its Z or [Z/Z$_\odot$] value.
\begin{table}
\begin{center}
\caption{PARSEC tracks$^t$ used in \cb models}
\begin{tabular}{ccccc}\hline\hline
X & Y & Z & Z/Z$_\odot$ & [Z/Z$_\odot$]\\ 
\hline
0.5840 & 0.3560 & 0.060  & 3.529 &  0.55 \\ % &  0.63
0.6390 & 0.3210 & 0.040  & 2.353 &  0.37 \\ % &  0.41
0.6680 & 0.3020 & 0.030  & 1.764 &  0.25 \\ % &  0.27
0.6960 & 0.2840 & 0.020  & 1.176 &  0.07 \\ % &  0.08
0.7040 & 0.2790 & 0.017  & 1.000 &  0.00 \\ % &  0.00
0.7130 & 0.2730 & 0.014  & 0.824 & -0.08 \\ % & -0.09
0.7230 & 0.2670 & 0.010  & 0.588 & -0.23 \\ % & -0.24
0.7290 & 0.2630 & 0.008  & 0.471 & -0.33 \\ % & -0.34
0.7350 & 0.2590 & 0.006  & 0.353 & -0.45 \\ % & -0.47
0.7400 & 0.2560 & 0.004  & 0.235 & -0.63 \\ % & -0.65
0.7460 & 0.2520 & 0.002  & 0.118 & -0.93 \\ % & -0.95
0.7490 & 0.2500 & 0.001  & 0.059 & -1.23 \\ % & -1.26
0.7505 & 0.2490 & 0.0005 & 0.029 & -1.53 \\ % & -1.56
0.7508 & 0.2490 & 0.0002 & 0.012 & -1.93 \\ % & -1.96
0.7509 & 0.2490 & 0.0001 & 0.006 & -2.23 \\ % & -2.26
0.7700 & 0.2300 & 0.0000 & 0.000 & -\inft\\   % & -5.00
\hline
\end{tabular}\label{tab:parsec}
\end{center}
\tablerefs{(t) \cite{chen2015,marigo2013}. We use tracks computed for $[\alpha/\mathrm{Fe}]\,=\,0$.}
\end{table}
%----------------------------------------------------------------

% Table~\ref{available} indicates with a checkmark the value of [Z/Z$_\odot$] for which several atlases of model spectra used in the CB17 models have been computed by different authors.
\begin{table*}
\begin{center}
\caption{[Z/Z$_\odot$] of model spectra available in different atlases}
\begin{tabular}{c|c|c|c|c|c|c|c|c|c}\hline\hline
[Z/Z$_\odot$] & Tlusty$^a$ & Tlusty$^b$ &Martins$^c$ & UVBlue$^d$ & Rauch$^e$ & WMBasic$^f$ & PoWR$^g$ & Aringer$^h$ & BaSeL$^i$ \\
& O & B & A & F,G,K & T$>$50kK & Hot MS & WR & C stars & 3.1 \\
\hline
 +0.50  &       &       &       & \chmk &       &       &       &       & \chmk \\
 +0.30  & \chmk & \chmk & \chmk & \chmk &       & \chmk &       &       &       \\
 +0.00  & \chmk & \chmk & \chmk & \chmk & \chmk & \chmk & \chmk & \chmk & \chmk \\
 -0.30  & \chmk & \chmk &       &       &       &       &       &       &       \\
 -0.40  &       &       &       &       &       & \chmk & \chmk &       &       \\
 -0.50  &       &       & \chmk & \chmk &       &       &       & \chmk & \chmk \\
 -0.70  & \chmk & \chmk &       &       &       & \chmk & \chmk &       &       \\
 -1.00  & \chmk & \chmk & \chmk & \chmk & \chmk &       &       & \chmk & \chmk \\
 -1.15  &       &       &       &       &       &       & \chmk &       &       \\
 -1.30  &       &       &       &       &       & \chmk &       &       &       \\
 -1.50  & \chmk &       &       & \chmk &       &       &       &       & \chmk \\
 -1.70  & \chmk &       &       &       &       &       &       &       &       \\
 -2.00  & \chmk &       &       & \chmk &       &       &       &       & \chmk \\
 -3.00  & \chmk &       &       &       &       &       &       &       &       \\
-\inft & \chmk & \chmk &       &       &       &       &       &       & \\
\hline
\end{tabular}\label{tab:available}
\end{center}
\tablerefs{
(a) \cite{lanz2003,lanzerr2003}.
(b) \cite{lanz2007}.
(c) \cite{martins2005}.
(d) \cite{rodmer2005}.
(e) \cite{rauch2003}.
(f) \cite{leitherer2010}.
(g) \cite{sander2012,hamann2006,hainich2014,hainich2015,todt2015,graefener2002,hamann2003}.
(h) \cite{aringer2009}.
(i) \cite{westera2002}. We use stellar models computed for $[\alpha/\mathrm{Fe}]\,=\,0$.}
\end{table*}
%----------------------------------------------------------------

% In Table~\ref{assigned} we indicate the specific subset of each stellar atlas listed in Table~\ref{available} used for each set of PARSEC tracks in Table~\ref{parsec}.
% Additionally, the last two columns of Table~\ref{assigned} indicate the range of [Z/Z$_\odot$] used to select spectra from the MILES and Stelib empirical spectral libraries
% for each set of PARSEC tracks.
\begin{table*}
\begin{center}
\caption{[Z/Z$_\odot$] of different stellar atlases assigned to each PARSEC track in \cb models}
\begin{tabular}{cc|c|c|c|c|c|c|c|c|c|c|c}\hline\hline
 & & Tlusty & Tlusty & Martins & UVBlue & Rauch$^m$ & WMBasic & PoWR & Aringer & BaSeL$^m$ & MILES$^j$/IndoUS$^k$ & Stelib$^l$ \\
Z & Z/Z$_\odot$ & O & B & A & F,G,K  & T$>$50kK & Hot MS & WR & C stars & 3.1 & Libraries & Library\\
\hline
0.060  &   0.55 & +0.30 & +0.30 & +0.30 & +0.50 & +0.00 & +0.30 & +0.00 & +0.00 & +0.50 & [+0.1,+0.6) & $>$ 0.06        \\
0.040  &   0.37 & +0.30 & +0.30 & +0.30 & +0.30 & +0.00 & +0.30 & +0.00 & +0.00 & intrp & [+0.1,+0.6) & $>$ 0.06        \\
0.030  &   0.25 & +0.30 & +0.30 & +0.30 & +0.30 & +0.00 & +0.30 & +0.00 & +0.00 & intrp & [+0.1,+0.6) & $>$ 0.06        \\
0.020  &   0.07 & +0.00 & +0.00 & +0.00 & +0.00 & +0.00 & +0.00 & +0.00 & +0.00 & intrp & [-0.1,+0.1) & [-0.06,+0.06] \\
0.017  &   0.00 & +0.00 & +0.00 & +0.00 & +0.00 & +0.00 & +0.00 & +0.00 & +0.00 & +0.00 & [-0.1,+0.1) & [-0.06,+0.06] \\
0.014  &  -0.08 & +0.00 & +0.00 & +0.00 & +0.00 & intrp & +0.00 & +0.00 & +0.00 & intrp & [-0.1,+0.1) & [-0.06,+0.06] \\
0.010  &  -0.23 & -0.30 & -0.30 & -0.50 & -0.50 & intrp & -0.40 & -0.40 & +0.00 & intrp & [-0.6,-0.1) & [-0.48,-0.06) \\
0.008  &  -0.33 & -0.30 & -0.30 & -0.50 & -0.50 & intrp & -0.40 & -0.40 & -0.50 & intrp & [-0.6,-0.1) & [-0.48,-0.06) \\
0.006  &  -0.45 & -0.70 & -0.70 & -0.50 & -0.50 & intrp & -0.40 & -0.40 & -0.50 & -0.50 & [-1.3,-0.6) & [-0.92,-0.48) \\
0.004  &  -0.63 & -0.70 & -0.70 & -0.50 & -0.50 & intrp & -0.70 & -0.70 & -0.50 & intrp & [-1.3,-0.6) & [-0.92,-0.48) \\
0.002  &  -0.93 & -1.00 & -1.00 & -1.00 & -1.00 & -1.00 & -0.70 & -0.70 & -1.00 & -1.00 & [-1.3,-0.6) & $<$ -0.92       \\
0.001  &  -1.23 & -1.50 & -1.00 & -1.00 & -1.00 & blbdy & -1.30 & -1.00 & -1.00 & intrp & [-1.3,-0.6) & $<$ -0.92       \\
0.0005 &  -1.53 & -1.70 & -1.00 & -1.00 & -1.50 & blbdy & -1.30 & -1.00 & -1.00 & -1.50 & $<$ -1.3      & $<$ -0.92       \\
0.0002 &  -1.93 & -2.00 & -1.00 & -1.00 & -2.00 & blbdy & -1.30 & -1.00 & -1.00 & intrp & $<$ -1.3      & $<$ -0.92       \\
0.0001 &  -2.23 & -2.00 & -1.00 & -1.00 & -2.00 & blbdy & -1.30 & -1.00 & -1.00 & -2.00 & $<$ -1.3      & $<$ -0.92       \\
0.0000 & -\inft &-\inft &-\inft & -1.00 & -2.00 & blbdy & -1.30 & -1.00 & -1.00 & -2.00 & $<$ -1.3      & $<$ -0.92 \\
\hline
\end{tabular}\label{tab:assigned}
\end{center}
\tablerefs{
(j) \cite{sanchez2006,falcon2011,prugniel2011}.
(k) \cite{valdes2004}.
(l) \cite{leborgne2003}.
(m) Entries marked $intrp$ indicate that the model is interpolated at the corresponding [Z/Z$_\odot$].
Entries marked $blbdy$ indicate that a blackbody curve is used to approximate the stellar emission.
We use stellar models computed for $[\alpha/\mathrm{Fe}]\,=\,0$.}
\end{table*}

% Table~\ref{spprop} lists the spectral characteristics of the  model spectra relevant in the UV range used in the CB19 models as published.
\begin{table*}
\begin{center}
\caption{Spectral properties of different stellar atlases relevant in the UV}
\begin{tabular}{l|c|c|c}\hline\hline
Stellar & Stellar & Wavelength &  \\
Library &Type & Range &  $R_{\textsc{STEP}} = \lambda / \Delta\lambda$ \\
\hline
TLUSTY         & O stars     &    45\AA - 300$\mu$m  &   26,000 -  38,000  \\
TLUSTY         & B stars     &    54\AA - 300$\mu$m  &  100,000 - 200,000  \\
Martins et al. & A stars     &     3000 - 7000\AA    &   10,000 - 23,000   \\
UVBlue         & F,G,K stars &    \ 850 - 4700\AA    &   50,000            \\
Rauch          & T$>$50kK      &\ \ \ \ 5 - 2000\AA    &   50 - 20,000       \\
WMBasic        & Hot MS      &    \ 900 - 1500\AA    &    2,040            \\
WMBasic        & Hot MS      &     1500 - 2998\AA    &    3,860            \\
PoWR           & WR stars    &    5\AA\ - 8$\mu$m    &   10,000 \\
\hline
\end{tabular}\label{tab:spprop}
\end{center}
\end{table*}

\begin{table*}
\begin{center}
\caption{Wavelength coverage vs. spectral atlas adopted in the \cb models}
\begin{tabular}{c|c|c|ccc|l}\hline\hline
Wavelength & Sampling & FWHM & & Points & &  \\
Range (\AA) & Step (\AA) & $\Delta\lambda$ (\AA) & N & Acum & \% & Stellar Libraries \\
\hline
\ \ \ \ \ 5.6 - 911                  &   0.9    &   2.0    & 1007 &  1007 &  6.0 & Tlusty, Martins et al., UVBlue, Rauch, WMBasic, PoWR \\
\ \ \ \ \ \ \ \ 911 - 3540.5         &   0.5    &   1.0    & 5259 &  6266 & 31.1 & Tlusty, Martins et al., UVBlue, Rauch, WMBasic, PoWR \\
\ \ 3540.5 - 7350.2                  &   0.9    &   2.5    & 4233 & 10499 & 25.0 & MILES                                                \\
\ \ \ 7350.2 - 9399.8                &   0.4    &1.2 - 1.0 & 5124 & 15623 & 30.3 & IndoUS                                           \\
\ \ \ \ \ \ \ \ \ 9410 - 36000$\mu$m & variable & variable & 1279 & 16902 &  7.6 & BaSeL 3.1, Aringer et al., IRTF libray$^n$, + Dusty models for TP-AGB stars$^o$\\
\hline
\end{tabular}\label{tab:xmilesi}
\end{center}
\tablerefs{
(n) \cite{rayner2009}.
(o) \cite{dusty2000,ragl2010}.}
\end{table*}